 \documentclass[twocolumn]{aastex62}

\shorttitle{Light Curve Analysis of Novae}
\shortauthors{Hachisu \& Kato}

\begin{document}

\title{A Light Curve Analysis of Recurrent and Very Fast Novae \\
in our Galaxy, Magellanic Clouds, and M31}

\author[0000-0002-0884-7404]{Izumi Hachisu}
\affil{Department of Earth Science and Astronomy, 
College of Arts and Sciences, The University of Tokyo,
3-8-1 Komaba, Meguro-ku, Tokyo 153-8902, Japan} 
\email{hachisu@ea.c.u-tokyo.ac.jp}

\author[0000-0002-8522-8033]{Mariko Kato}
\affil{Department of Astronomy, Keio University, 
Hiyoshi, Kouhoku-ku, Yokohama 223-8521, Japan} 

\begin{abstract}
We analyzed optical, UV, and X-ray light curves of 14 recurrent 
and very fast novae in our galaxy, Magellanic Clouds, and M31,
and obtained their distances and white dwarf (WD) masses.  
Among the 14 novae, we found that eight novae host very massive 
($\gtrsim 1.35~M_\sun$) WDs and are candidates of
Type Ia supernova (SN~Ia) progenitors. We confirmed that the same
timescaling law and time-stretching method as in galactic novae can
be applied to extra-galactic fast novae. 
We classify the four novae, V745~Sco, T~CrB, V838~Her, and V1534~Sco,
as the V745~Sco type (rapid-decline), the two novae, RS~Oph and V407~Cyg,
as the RS~Oph type (circumstellar matter(CSM)-shock), and the two novae,
U~Sco and CI~Aql, as the U~Sco type (normal-decline).
The $V$ light curves of these novae almost
overlap with each other in the same group, if we properly stretch 
in the time direction (timescaling law).
We apply our classification method to LMC, SMC, and M31 novae.
YY~Dor, LMCN~2009a, and SMCN~2016 belong to the normal-decline type, 
LMCN~2013 to the CSM-shock type, and LMCN~2012a and M31N~2008-12a 
to the rapid-decline type.  We obtained the distance of SMCN~2016 to be 
$d=20\pm2$~kpc, suggesting that SMCN~2016 is a member of our galaxy.
The rapid-decline type novae have very massive WDs of 
$M_{\rm WD}=1.37-1.385~M_\sun$ and are promising candidates of 
SN~Ia progenitors.  This type of novae are much fainter than 
the MMRD relations.
\end{abstract}

\keywords{novae, cataclysmic variables --- stars: individual 
(RS~Oph, U~Sco, V745~Sco, V1534~Sco) --- stars: winds}

\section{Introduction}
\label{introduction}
Type Ia supernovae (SNe~Ia) play a crucial role in astrophysics,
because they are used as standard candles to measure cosmic distances
\citep{perlm99, riess98}, and produce a major proportion of iron group elements
in the chemical evolution of galaxies \citep{koba98}. However, the progenitor
systems have not yet been identified observationally or theoretically 
\citep[see, e.g.,][for a review]{mao14}.
One of the promising evolutionary paths to SNe~Ia is the single-degenerate
(SD) scenario, in which a white dwarf (WD) accretes matter from 
a nondegenerate companion, grows in mass to near the Chandrasekhar
mass, and explodes as a SN~Ia triggered by central carbon burning 
\citep[e.g.,][]{nom82,hkn99}.
The WD mass approaches the Chandrasekhar mass just before a SN~Ia
explosion. One of the candidates for such SD systems is a recurrent nova,
because their optical light curves quickly decay and their supersoft
X-ray source (SSS) phases are very short compared with other classical novae
\citep[e.g.,][]{hac01kb, hac07kl, nes07, schw11, pagnotta14}.  
The nova theory suggests
that such very fast evolving novae harbor a very massive WD
close to the Chandrasekhar mass \citep{hac06kb, hac10k, kat14shn}.
In the present paper, we analyze optical, near-infrared (NIR), 
ultra-violet (UV), and supersoft X-ray light curves of recurrent novae
as well as very fast novae, and estimate their basic properties including 
the WD masses to study the possibility of SN~Ia progenitors. 

A classical nova is a thermonuclear runaway event in a mass-accreting
WD in a binary. A recurrent nova is a classical nova
with multiple recorded outbursts. Hydrogen ignites to trigger an outburst
after a critical amount of hydrogen-rich matter is accreted on the WD. 
The photospheric radius of the hydrogen-rich
envelope expands to red-giant (RG) size and the binary becomes bright 
in the optical range \citep[e.g.,][]{sta72, pri86, nar80}.
The hydrogen-rich envelope then emits strong winds 
\citep[e.g.,][]{kat94h, hac06kb}.  
After the maximum expansion of the pseudo-photosphere, 
it begins to shrink, and the nova optical emission declines.
Subsequently, UV emission dominates the spectrum
and finally, supersoft X-ray emission increases. 
The nova outburst ends when the hydrogen shell burning ends.
Various timescaling laws have been proposed to identify
a common pattern among the nova optical light curves 
\citep[see, e.g., introduction of][]{hac08kc}.

\citet{hac06kb} found that the optical and NIR
light curves of several novae follow a similar decline law.
Moreover, they found that the time-normalized light curves
are independent of the WD mass, chemical composition
of the ejecta, and wavelength. 
They called this property ``the universal decline law.''
The universal decline law was examined first in several well-observed novae
and later in many other novae ($\gtrsim 30$ novae) 
\citep{hac07k,hac09ka,hac10k,hac15k,hac16k,hac16kb,hac06b,hac07kl,
hac08kc,kat09hc,kat12h}.
\citet{hac06kb} defined a unique timescaling factor of $f_{\rm s}$ for
optical, UV, and supersoft X-ray light curves of a nova.
The shortest timescales (smallest $f_{\rm s}$) 
correspond to the WD masses in the range $1.37-1.38~M_\sun$. Therefore,
the shortest $f_{\rm s}$ systems are candidates for immediate
progenitors of SNe~Ia \citep[e.g.,][]{hac01kb, hac10k, hac16k, hac16kb}.


\begin{figure*}
\plotone{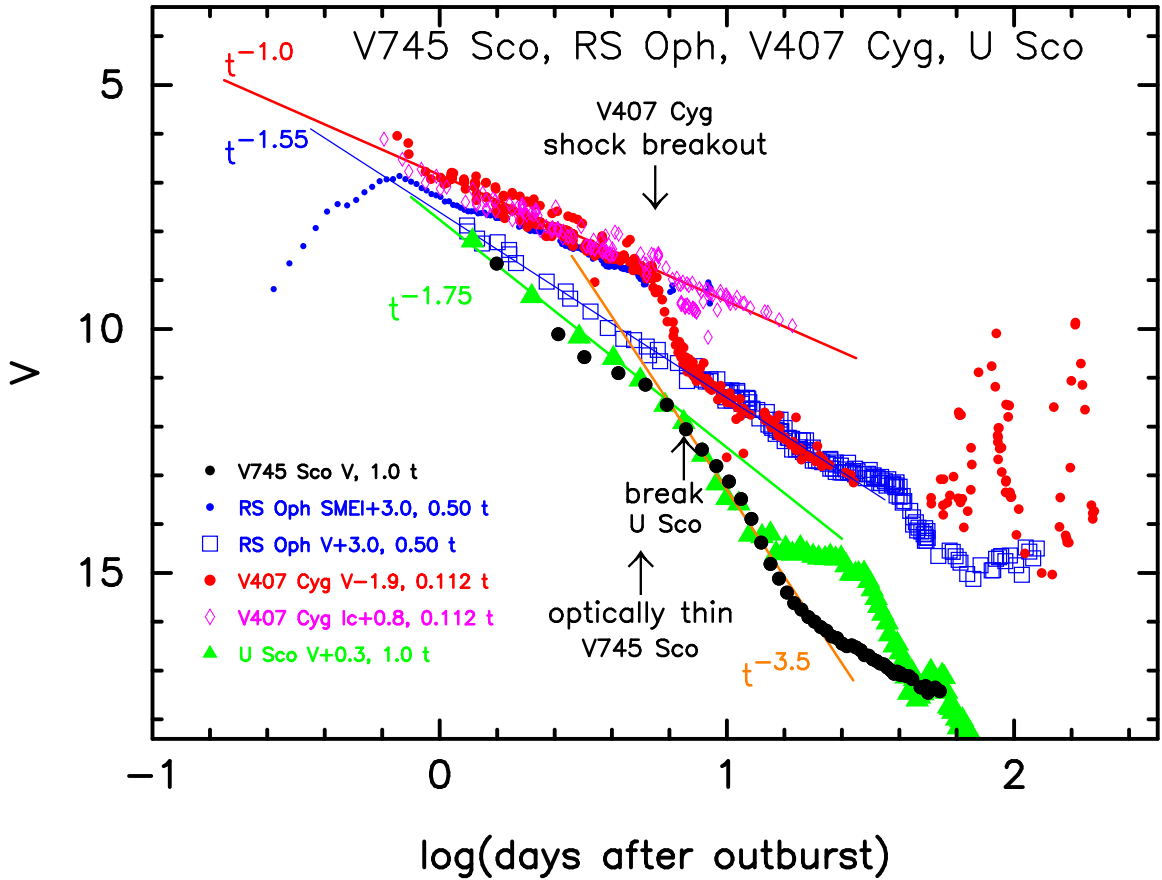}
\caption{
Three template $V$ light curves of the V745~Sco, RS~Oph, and U~Sco
type novae. We add the $V$ and $I_{\rm C}$ light curves of V407~Cyg and
the Solar Mass Ejection Imager (SMEI) magnitudes \citep{hou10}
of RS~Oph for comparison.
Each light curve is horizontally moved by $\Delta \log t = \log f_{\rm s}$
and vertically shifted by $\Delta V$ with respect to that of V745~Sco,
as indicated in the figure by, for example, ``RS~Oph V+3.0, 0.50 t,'' where
$\Delta V= 3.0$ and $f_{\rm s}=0.50$. The data are the same as those in
Figures \ref{v745_sco_v_bv_ub_xray_logscale},
\ref{v407_cyg_rs_oph_v_bv_ub_logscale}, and 
\ref{u_sco_v_bv_ub_xray_radio_logscale}.
The filled black circles represent the $V$ light curve of the V745~Sco
(2014) outburst. The filled red circles and open magenta diamonds 
denote the $V$ and $I_{\rm C}$ magnitudes of V407~Cyg, respectively.
The blue dots and open blue squares denote the SMEI and $V$ magnitudes 
of the RS~Oph (2006) outburst, respectively. 
The filled green triangles show the $V$ magnitudes of the 
U~Sco (2010) outburst. The solid red line represents the decline
trend of $F_\nu \propto t^{-1.0}$, solid blue line $F_\nu \propto t^{-1.55}$,
solid green lines $F_\nu \propto t^{-1.75}$, and solid orange line 
$F_\nu \propto t^{-3.5}$, where $t$ is the time from the outburst and
$F_\nu$ is the flux at the frequency $\nu$. See the text for more detail.
\label{v745_sco_u_sco_v407_cyg_rs_oph_v_template}}
\end{figure*}

However, there are exceptions of fast novae 
that do not follow the universal decline law. 
For example, V407~Cyg outbursted in 2010 \citep[e.g.,][]{mun11c}
and showed a decay trend of $F_\nu \propto t^{-1.0}$ 
in the early $V$ light curve, as shown in 
Figure \ref{v745_sco_u_sco_v407_cyg_rs_oph_v_template},
whereas the universal decline law of novae follows $F_\nu \propto t^{-1.75}$,
as in the early decay phase of U~Sco. Here, $F_\nu$ is the flux 
at the frequency $\nu$ and $t$ is the time from the outburst.
V745~Sco is a recurrent nova and does not show a substantial slope of
$F_\nu \propto t^{-1.75}$ in the early decay phase, but rather a steeper
decline of $F_\nu \propto t^{-3.5}$, as
shown in Figure \ref{v745_sco_u_sco_v407_cyg_rs_oph_v_template}.
We have to clarify the reasons for these exceptions 
and determine the WD mass based on our model light curve fitting.

Based on the universal decline law, \citet{hac10k} proposed a method
for obtaining the distance modulus in the $V$ band, $\mu_V\equiv (m-M)_V$,
of a nova. One nova light curve overlaps another with
a time-stretching factor of $f_{\rm s}$; the distance modulus
$(m-M)_V$ of the former nova is calculated from the distance modulus
of the latter. \citet{hac10k} called this procedure 
``the time-stretching method'' after the Type Ia supernova community
\citep{phi93}. 
This method has been confirmed for many classical novae in our galaxy
\citep{hac10k, hac15k, hac16k, hac16kb}.
However, we have not yet examined whether the time-stretching method is
applicable to very fast novae, including recurrent novae, 
both in our galaxy and extra-galaxies.

It has long been discussed that the brightness of a nova is related
with the nova speed class (the decay rate of the nova optical light curve).
Several empirical relations have been proposed so far 
\citep[e.g.,][]{sch57, del95, dow00}. Among them,
the maximum magnitude vs. rate of decline (MMRD)
relations of recurrent or very fast novae have sometimes been
questioned \citep[e.g.,][]{shara17}, based mainly on the result of \citet{kas11}. They claimed the discovery of a new 
rich class of fast and faint novae in M31. 
\citet{mun17} argued against the result of Kasliwal et al. and supported
the MMRD relations.
\citet{hac10k} theoretically examined the MMRD law on the basis of 
their universal decline law, and explained the reason for large scatter
of the nova MMRD distribution \citep[see also][]{hac15k, hac16k}.
They concluded that the main trend of the MMRD relation is governed
by the WD mass, i.e., the timescaling factor of $f_{\rm s}$, whereas
the peak brightness depends also on the initial envelope mass,
which is determined by the mass-accretion rate to the WD. This second
parameter causes large scatter around the main trend of the MMRD relations.
Thus, \citet{hac10k} reproduced the distribution of MMRD points 
summarized by \citet{dow00}. Hachisu \& Kato's (2010) study was 
based on the individual galactic classical novae. 

In the present work, we study recurrent and very fast novae and examine
whether they follow a similar timescaling law to classical novae.
Do they follow the universal decline law?
Do they follow the MMRD relations?
If not, what is the reason? Do they belong to a new class of novae,
as claimed by \citet{kas11}? We choose 14 novae and 
obtain their WD masses, distances, and absorptions.
We also discuss the possibility that they are progenitors of SNe~Ia.
Among 10 galactic recurrent novae \citep[e.g.,][]{schaefer10a},
we select five galactic recurrent novae, V745~Sco, T~CrB, RS~Oph, U~Sco,
and CI~Aql, mainly because of their rich observational data.
We exclude the recurrent nova T~Pyx from our analysis because
it is not a very fast nova. Instead, we include 
three galactic classical novae, V838~Her, V1534~Sco, and V407~Cyg,
because these fast novae have similar features to recurrent novae
and have plenty of observational data.
This paper is organized as follows.
First, we identify three types of light curve shapes for recurrent and
very fast novae, V745~Sco, RS~Oph, and U~Sco types
(see Figure \ref{v745_sco_u_sco_v407_cyg_rs_oph_v_template}).  
In Section \ref{v745_cyg_type}, we analyze the first group, i.e., the 
V745~Sco type, V745~Sco, T~CrB, V838~Her, and V1534~Sco, as
the rapid-decline type. In Section \ref{rs_oph_type}, we examine 
two novae, RS~Oph and V407~Cyg, as the circumstellar matter 
(CSM)-ejecta shock 
(RS~Oph) type. In Section \ref{u_sco_type}, we study two novae, 
U~Sco and CI~Aql, as the normal-decline (U~Sco) type.
In Section \ref{novae_lmc_smc_m31}, we analyze several recurrent and
very fast novae in Large Magellanic Cloud (LMC), Small Magellanic Cloud (SMC), and M31, i.e., YY~Dor, LMC~N~2009a, 
LMC~N~2012a, LMC~N~2013, SMC~N~2016, and M31N~2008-12a.
The discussion and conclusions follow in Sections
\ref{discussion} and \ref{conclusions}, respectively.


\begin{figure}
\plotone{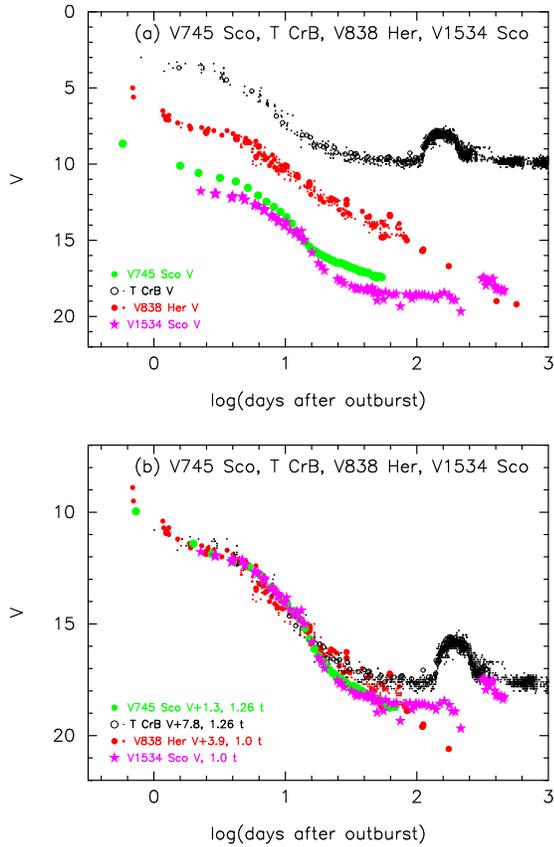}
\caption{
(a) $V$ and visual light curves of four novae, V745~Sco, T~CrB,
V838~Her, and V1534~Sco on a logarithmic timescale.
Large symbols show the $V$ magnitudes, and small dots denote the
visual magnitudes of each nova.
(b) The four $V$ light curves overlap each other if we properly 
squeeze/stretch the timescales and shift up/down the $V$ and visual
brightnesses.
\label{v1534_sco_v838_her_t_crb_v745_sco_v_template_no2}}
\end{figure}

\section{Timescaling Law of Rapid-decline Novae}
\label{v745_cyg_type}
We analyze the light curves of V745~Sco, T~CrB, V838~Her, and V1534~Sco,
in this order. These novae do not follow the universal decline law of 
$F_\nu\propto t^{-1.75}$, but rather decline much faster, as 
$F_\nu\propto t^{-3.5}$ in the early decline phase 
(see filled black circles in Figure
\ref{v745_sco_u_sco_v407_cyg_rs_oph_v_template}). 
Thus, we call them the rapid-decline (or V745~Sco) type novae.
V745~Sco, T~CrB, and V1534~Sco have a RG companion
whereas V838~Her has a main-sequence (MS) companion. V745~Sco and T~CrB are
recurrent novae. Despite these differences, they show similar
$V$ light curve shapes. 
Figure \ref{v1534_sco_v838_her_t_crb_v745_sco_v_template_no2}(a) 
shows the $V$ and visual light curves of these four novae on a logarithmic
timescale. At the first glance, these novae show different shapes.
However, V745~Sco and T~CrB have similar shapes in the initial 20 days, and
V838~Her and V1534~Sco decline similarly in the initial 15 days.
If we stretch the first two light curves in the time direction by a factor
of $1.26$, i.e., shift them rightward by $\Delta \log t= \log 1.26 = 0.1$, and
shift all four light curves in the vertical direction, 
these four nova light curves almost overlap with each other, as shown in
Figure \ref{v1534_sco_v838_her_t_crb_v745_sco_v_template_no2}(b). 
The vertical shifts $\Delta V(=1.3)$ and time-stretching factors 
$f_{\rm s}(=1.26)$ are shown in the figure, 
such as ``V745~Sco $V+1.3$, $1.26~t$'' for the V745~Sco light curve.
Note that the time-stretching factor of $f_{\rm s}$ corresponds to the 
horizontal shift by $\Delta \log t= \log f_{\rm s}$
in Figure \ref{v1534_sco_v838_her_t_crb_v745_sco_v_template_no2}(b). 


\begin{figure}
\includegraphics[height=12cm]{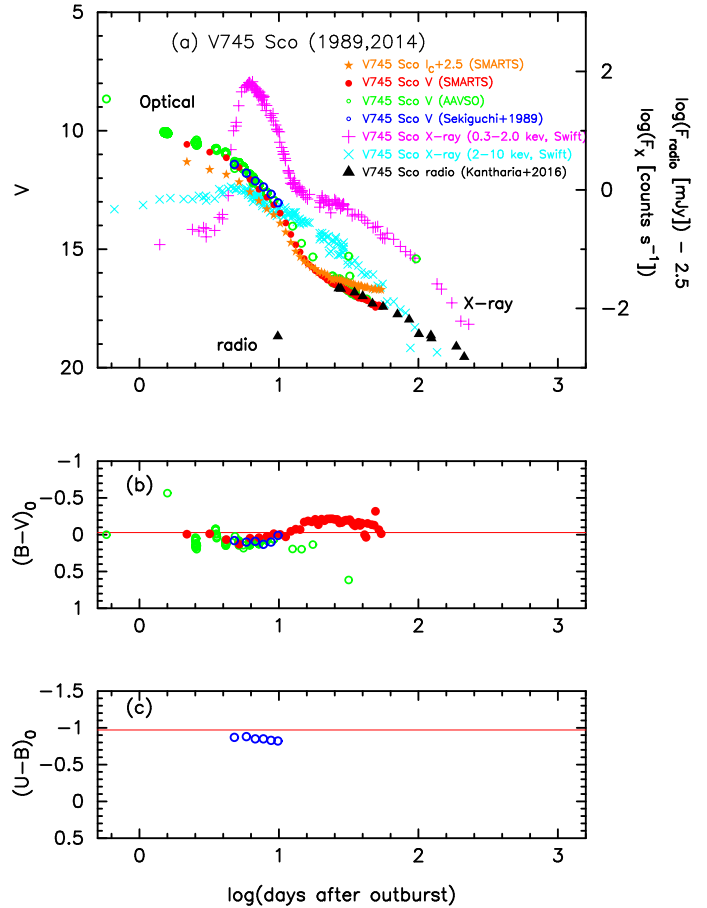}
\caption{
(a) The $V$ light curve of the V745~Sco 1989 \citep[open blue circles 
from][]{sek90} and 2014 (filled red circles from SMARTS and open green
circles from AAVSO) outbursts are plotted on a logarithmic timescale. 
We add the $I_C$ magnitudes (filled orange stars) taken from SMARTS.
We also include X-ray ($0.3-2.0$ keV with magenta pluses and $2.0-10$ keV
with cyan crosses) data taken from \citet{pag15} and radio (filled black 
triangles, GMRT at 610~MHz) data taken from \citet{kanth16}.
(b) The $(B-V)_0$ color curve of the V745~Sco 1989 and 2014 outbursts. 
The horizontal solid red line denotes the intrinsic color of optically
thick free-free emission, i.e., $(B-V)_0=-0.03$. 
(c) The $(U-B)_0$ color curve of the V745~Sco 1989 outburst. 
The horizontal solid red line denotes the intrinsic color of optically
thick free-free emission, i.e., $(B-V)_0=-0.97$. 
\label{v745_sco_v_bv_ub_xray_logscale}}
\end{figure}


\begin{figure*}
\plotone{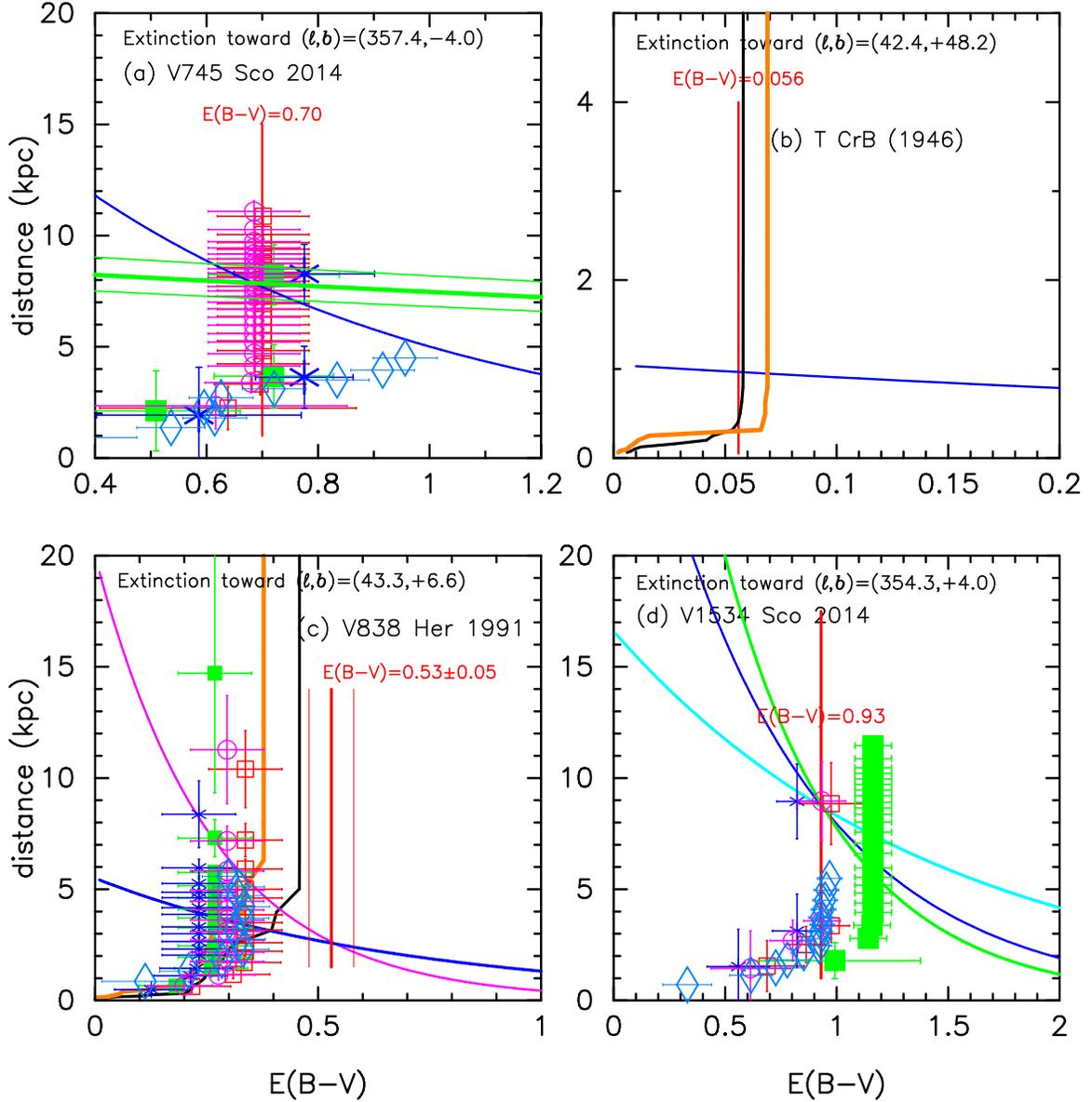}
\caption{
Distance-reddening relation toward (a) V745~Sco, (b) T~CrB, 
(c) V838~Her, and (d) V1534~Sco. The thick blue lines denote
(a) $(m-M)_V=16.6$, (b) $(m-M)_V=10.1$, (c) $(m-M)_V=13.7$,
and (d) $(m-M)_V=17.6$, together with Equation (\ref{v_distance_modulus}).
The vertical thick red lines denote 
the reddening of each nova, $E(B-V)$. Open red squares, filled green 
squares, blue asterisks, and open magenta circles with error bars
represent the relations of \citet{mar06} in the four directions
close to each nova.  The thick black and orange lines show 
the distance-reddening relations of \citet{gre15, gre18}, respectively. 
Open cyan-blue diamonds denote
the distance-reddening relation taken from \citet{ozd16}.
In panel (a), the thick green line flanked by thin green
lines indicates the distance-reddening relation of Equation 
(\ref{equation_distmod_IR}). In panel (c), the thick magenta line
denotes the distance-reddening relation given by the 1455~\AA\ 
light curve fitting together with Equation 
(\ref{uv1455_distance_modulus}). In panel (d), the green line denotes
$(m-M)_B= 18.55$ together with Equation 
(\ref{distance_modulus_relation_b}) and the cyan line indicates
$(m-M)_I= 16.1$ together with Equation (\ref{distance_modulus_relation_i}). 
\label{distance_reddening_v745_sco_t_crb_v838_her_v1534_sco_no2}}
\end{figure*}

\subsection{V745~Sco (1989, 2014)}
\label{v745_sco}
V745~Sco is a recurrent nova with three recorded outbursts in 1937, 
1989, and 2014 \citep[e.g.,][]{pag15}. 
Figure \ref{v745_sco_v_bv_ub_xray_logscale} shows the $V$ and $I_C$
light curves, X-ray ($0.3-2.0$ keV soft band and $2-10$ keV hard band)
light curves, radio (610~MHz) light curve, $(B-V)_0$, 
and $(U-B)_0$ color curves on a logarithmic timescale. 
This is similar to Figure 75 of \citet{hac16kb}, but we added 
the $I_C$ data from SMARTS, the radio data from \citet{kanth16}, and
the $V$ light curve from the American Association of Variable
Star Observers (AAVSO).
The 2014 outburst shows $m_{V,\rm max}=8.66$ (AAVSO), 
$t_2=2$~days, and $t_3=4$~days \citep[e.g.,][]{ori15}.


\begin{deluxetable*}{lllllrll}
\tabletypesize{\scriptsize}
\tablecaption{Extinctions, distance moduli, and distances of selected novae
\label{extinction_distance_various_novae}}
\tablewidth{0pt}
\tablehead{
\colhead{Object} & \colhead{Outburst} & \colhead{$E(B-V)$} 
& \colhead{$(m-M)_V$} & \colhead{$d$} & \colhead{$z$} 
& \colhead{$\log f_s$} & \colhead{Type} \\
  & Year &  &  &  (kpc) & (pc) &  &   
} 
\colnumbers
\startdata
CI~Aql & (2000)\tablenotemark{a} & 1.0 & 15.7 & 3.3 & $-47$ & $1.1$ & normal \\
T~CrB & (1946) & 0.056 & 10.1 & 0.96 & 715 & 0.0 & rapid \\
V407~Cyg & 2010 & 1.0 & 16.1 & 3.9 & $-33$ & 0.95 & CSM \\
YY~Dor & (2010) & 0.12 & 18.9 & 50.0 & --- & 0.6 & normal \\
V838~Her & 1991 & 0.53 & 13.7 & 2.6 & 310 & 0.1 & rapid \\
RS~Oph  & (2006) & 0.65 & 12.8 & 1.4 & 250 & 0.3 & CSM \\
U~Sco & (2010) & 0.26 & 16.3 & 12.6 & 4680 & 0.0 & normal \\
V745~Sco & (2014) & 0.70 & 16.6 & 7.8 & $-540$ & 0.0 & rapid \\
V1534~Sco & 2014 & 0.93 & 17.6 & 8.8 & 610 & 0.1 & rapid \\
LMCN~2009a & (2014) & 0.2 & 19.1 & 50.0 & --- & 0.8 & normal \\
LMCN~2012a & 2012 & 0.15 & 19.0 & 50.0 & --- & 0.1 & rapid \\
LMCN~2013 & 2013 & 0.12 & 18.9 & 50.0 & --- & 0.9 & CSM \\
SMCN~2016 & 2016 & 0.08 & 16.8 & 20.4 & $-13750$ & 0.6 & normal \\
M31N~2008-12a & (2015) & 0.30 & 24.8 & 780 & --- & 0.0 & rapid \\
\enddata
\tablenotetext{a}{The year in the parentheses denotes the analyzed 
outburst of each recurrent nova.} 
\end{deluxetable*}


\begin{deluxetable*}{llccccll}
\tabletypesize{\scriptsize}
\tablecaption{Decline rates of selected novae
\label{physical_properties_recurrent_novae}}
\tablewidth{0pt}
\tablehead{
\colhead{Object} & \colhead{$\log f_{\rm s}$} 
& \colhead{$t_2$} & \colhead{$t_3$} 
& \colhead{$M_{V,\rm max}$} & \colhead{$P_{\rm orb}$} 
& \colhead{Companion} & \colhead{Reference of $t_2$ and $t_3$} \\
\colhead{} & \colhead{} & \colhead{(day)} & \colhead{(day)} 
& \colhead{} & \colhead{(day)} & \colhead{} & \colhead{} 
} 
\colnumbers
\startdata
CI~Aql   & 1.1 & $25$  & $32$ & $-6.9$ & 0.618\tablenotemark{a} & subgiant & \citet{str10} \\
T~CrB   & 0.0 & $4$  & $6$ & $-7.6$ & 227.6\tablenotemark{b} & red-giant & \citet{str10} \\
V407~Cyg   & 0.95 & $5.9$  & $24$ & $-9.0$ & --- & Mira & \citet{mun11c}\\
YY~Dor  & 0.6 & $4.0$  & $10.9$  & $-8.2$ & --- & --- & \citet{wal12} \\
V838~Her & 0.1 & $1$ & $4$ & $-8.4$ & 0.2976\tablenotemark{c} & main-sequence & \citet{str10} \\
RS~Oph   & 0.3 & $7$ & $14$ & $-8.0$ & 453.6\tablenotemark{d} & red-giant & \citet{str10} \\
U~Sco    & 0.0 & $1.7$ & $3.6$ & $-8.7$ & 1.23\tablenotemark{e} & subgiant & \citet{schaefer11} \\
V745~Sco & 0.0 & $2$ & $4$ & $-7.9$ & --- & red-giant & \citet{ori15} \\
V1534~Sco & 0.1 & $5.6$ & $9.2$ & $-5.8$ & --- & red-giant & \citet{mun17} \\
LMC~N~2009a & 0.8 & $5.0$ & $10.4$ & $-8.5$ & 1.19\tablenotemark{f} & subgiant & \citet{bod16} \\
LMC~N~2012a & 0.1 & $1.1$ & $2.1$ & $-8.3$ & 0.802\tablenotemark{g} & subgiant & \citet{wal12}\\
LMC~N~2013 & 0.9 & $21$ & $47$ & $-7.4$ & --- & --- & present paper \\
SMC~N~2016 & 0.6 & $4.0$ & $7.8$ & $-8.3$ & --- & --- & \citet{aydi18} \\
M31N~2008-12a & 0.0 & $1.65$ & $2.37$ & $-6.25$ & --- & --- & \citet{dar16} \\ 
\enddata
\tablenotetext{a}{\citet{men95}.} 
\tablenotetext{b}{\citet{fek00}.} 
\tablenotetext{c}{\citet{ing92}, \citet{lei92}.} 
\tablenotetext{d}{\citet{bra09}.} 
\tablenotetext{e}{\citet{sch95r}.} 
\tablenotetext{f}{\citet{bod16}.} 
\tablenotetext{g}{\citet{schw15}, \citet{mro16}.} 
\end{deluxetable*}


\begin{deluxetable}{llllll}
\tabletypesize{\scriptsize}
\tablecaption{White dwarf masses of selected novae
\label{wd_mass_recurrent_novae}}
\tablewidth{0pt}
\tablehead{
\colhead{Object} & \colhead{$\log f_{\rm s}$} 
& \colhead{$M_{\rm WD}$} 
& \colhead{$M_{\rm WD}$} 
& \colhead{$M_{\rm WD}$} 
& \colhead{$M_{\rm WD}$} 
\\
\colhead{} & \colhead{} 
& \colhead{$f_{\rm s}$\tablenotemark{a}} 
& \colhead{UV~1455~\AA\tablenotemark{b}} 
& \colhead{$t_{\rm SSS-on}$\tablenotemark{c}} 
& \colhead{$t_{\rm SSS-off}$\tablenotemark{d}} 
\\
\colhead{} & \colhead{} & \colhead{($M_\sun$)} & \colhead{($M_\sun$)} 
& \colhead{($M_\sun$)} & \colhead{($M_\sun$)}  
} 
\colnumbers
\startdata
CI~Aql   & 1.1 & $1.18$  & --- & --- & --- \\
T~CrB   & 0.0 & $1.38$  & --- & --- & --- \\
V407~Cyg   & 0.95 & $1.22$  & --- & --- & --- \\
YY~Dor  & 0.6 & $1.29$  & ---  & --- & ---  \\
V838~Her & 0.1 & $1.37$ & $1.37$ & --- & ---  \\
RS~Oph   & 0.3 & $1.35$ & $1.35$ & $1.35$ & $1.35$  \\
U~Sco    & 0.0 & $1.37$ & --- & $1.37$ & $1.37$ \\
V745~Sco & 0.0 & $1.38$ & --- & $1.385$ & $1.385$ \\
V1534~Sco & 0.1 & $1.37$ & --- & --- & --- \\
LMC~N~2009a & 0.8 & $1.25$ & --- & $1.25$ & $1.25$ \\
LMC~N~2012a & 0.1 & $1.37$ & --- & --- & --- \\
LMC~N~2013 & 0.9 & $1.23$ & --- & --- & --- \\
SMC~N~2016 & 0.6 & $1.29$ & --- & --- & --- \\
M31N~2008-12a & 0.0 & $1.38$ & --- & $1.38$ & $1.38$ \\  
\enddata
\tablenotetext{a}{WD mass estimated from the $f_{\rm s}$ timescale.} 
\tablenotetext{b}{WD mass estimated from the UV~1455~\AA\  fit.} 
\tablenotetext{c}{WD mass estimated from the $t_{\rm SSS-on}$ fit.} 
\tablenotetext{d}{WD mass estimated from the $t_{\rm SSS-off}$ fit.} 
\end{deluxetable}

\subsubsection{Reddening and distance}
The galactic coordinates of V745~Sco are $(l,b)=(357\fdg3584, -3\fdg9991)$.
The NASA/IPAC galactic dust absorption 
map\footnote{http://irsa.ipac.caltech.edu/applications/DUST/} 
gives $E(B-V)=0.71\pm0.02$
toward V745~Sco, which is based on the galactic dust extinction of
\citet{schl11}. \citet{ban14} suggested that the extinction for V745~Sco
is $E(B-V)=0.70$ on the basis of the galactic dust extinction of \citet{schl11}
and \citet{mar06}. \citet{ori15} fitted the X-ray spectrum 10 days after
the discovery of the 2014 outburst with a model spectrum and obtained
the hydrogen column density of 
$N_{\rm H}=(6.9\pm0.9) \times 10^{21}$ cm$^{-2}$, which corresponds to
the extinction of $E(B-V)=N_{\rm H}/6.8 \times 10^{21}$ cm$^{-2}=1.0\pm0.1$
\citep{gue09} or $E(B-V)= N_{\rm H}/8.3 \times 10^{21}= 0.83\pm0.1$
\citep{lis14}. \citet{hac16kb} adopted $E(B-V)=0.70\pm0.1$,
mainly from the results of \citet{ban14} and the NASA/IPAC galactic
dust absorption map. We adopt $E(B-V)=0.70\pm0.1$, pursuant to \citet{hac16kb}.

The intrinsic colors are obtained from
\begin{equation}
(B-V)_0 = (B-V) - E(B-V),
\label{dereddening_eq_bv}
\end{equation}
\begin{equation}
(U-B)_0 = (U-B) - 0.64 E(B-V),
\label{dereddening_eq_ub}
\end{equation}
where the factor of $0.64$ is taken from \citet{rie85}.
The intrinsic color $(B-V)_0$ of AAVSO and \citet{sek90} are
systematically 0.1 and 0.3 mag redder than those of SMARTS, the data
of which are taken from \citet{pag15}.
Therefore, we shift them up by 0.1 and 0.3 mag, respectively, in Figure 
\ref{v745_sco_v_bv_ub_xray_logscale}(b).

There are sometimes substantial differences among colors of novae
observed at different observatories.  Such differences could originate
from, for example, slight differences in response functions
of each filter, different comparison stars, and so on.   
We do not know the exact origin of the difference in this nova, but
we suppose that the $(B-V)_0$ color is close to $-0.03$
in the early decline phase before the nebular phase.  This is because
nova spectra are dominated by optically thick free-free emission
(continuum) and its intrinsic color $(B-V)_0$ is $-0.03$ 
\citep[$F_\nu \propto \nu^{2/3}$, see][]{hac14k}. 
The color of SMARTS is close to this value in the very early phase.
We regard that the intrinsic colors of SMARTS are reasonable
but the colors of AAVSO and \citet{sek90} deviate from those
of SMARTS.  Thus, we correct the intrinsic $(B-V)_0$ colors of
AAVSO and Sekiguchi et al. by 0.1 and 0.3 mag, respectively.

The distance to V745~Sco was estimated by \citet{scha09} to be $d=7.8$~kpc
from the orbital period of $P_{\rm orb}=510\pm20$~days and the
corresponding Roche lobe size. However,
this period was not confirmed by \citet{mro14}, and thus, we do not use
the method based on the Roche lobe size.

\citet{mro14} detected semi-regular pulsations of
the RG companion (with periods of 136.5~days and 77.4~days).
\citet{hac16kb} estimated the distance from the period-luminosity
relation for pulsating red-giants, i.e.,
\begin{equation}
M_K= -3.51 \times (\log P ({\rm day})- 2.38)-7.25,
\label{equation_MkP}
\end{equation}
with an error of $\sim0.2$ mag \citep{whitelock08}.
\citet{hac16kb} obtained the absolute $K$ magnitude of $M_K=-6.39 \pm 0.2$
for the fundamental 136.5-day pulsation from the data of \citet{mro14}.
Adopting an average $K$ mag of $m_K=8.33$~mag \citep{hoa02}, \citet{hac16kb}
obtained $d=7.8\pm0.8$ kpc for $E(B-V)=0.70$ from
\begin{equation}
(m - M)_K= 0.353 \times E(B-V) + 5 \log ~(d/1~{\rm kpc}) +10,
\label{equation_distmod_IR}
\end{equation}
where they adopted the reddening law of $A_K=0.353 \times E(B-V)$
\citep{car89}. Their new distance is accidentally identical to 
Schaefer's value of 7.8~kpc. The vertical distance from
the galactic plane is approximately $z=-540$~pc, being
significantly above the scale height of galactic matter distribution
\citep[$z=\pm125$~pc, see, e.g.,][]{mar06}.
Thus, it is likely that V745~Sco belongs to the galactic bulge.

We plot the distance-reddening relation of Equation 
(\ref{equation_distmod_IR}) with a thick green line 
flanked by thin green lines in Figure
\ref{distance_reddening_v745_sco_t_crb_v838_her_v1534_sco_no2}(a).
In the same figure, we add the reddening of $E(B-V)=0.70$, indicated by the 
vertical thick red line. The thick green line and
red line cross at $d=7.8$~kpc. \citet{hac16kb} obtained the distance
modulus in the $V$ band, $(m-M)_V=16.6\pm 0.2$, which was calculated from
\begin{equation}
(m-M)_V= 5 \log (d/{\rm 10~pc}) + 3.1 E(B-V),
\label{v_distance_modulus}
\end{equation}
where the factor of $3.1$ is taken from \citet{rie85}.
We also plot this distance-reddening relation of Equation
(\ref{v_distance_modulus}) with the thick blue line in Figure
\ref{distance_reddening_v745_sco_t_crb_v838_her_v1534_sco_no2}(a).
Note that this relation of $(m-M)_V=16.6$ is not an independent
relation of $(m-M)_K= m_K - M_K =8.33-(-6.39)=14.72$
because it is just derived from Equation
(\ref{v_distance_modulus}) together with $E(B-V)=0.70$ and $d=7.8$~kpc.

After Hachisu \& Kato's (2016b) paper was published,
\citet{ozd16} obtained distance-reddening relations toward 46
novae based on the unique position of red clump giants
in the color-magnitude diagram. Their distance-$E(J-K)$ relation
toward V745~Sco
is plotted in their Figure C2.
They estimated the distance to V745~Sco to be $d=3.5\pm0.85$~kpc
for the assumed reddening of $E(B-V)=0.84\pm0.15$ (or
$E(J-K)=0.44\pm0.07$). This distance of $d=3.5$~kpc is
much shorter than our distance of $d=7.8$~kpc.
We converted their distance-$E(J-K)$ relation to a distance-$E(B-V)$ 
relation using the conversion of $E(J-K)=0.524 E(B-V)$ \citep{rie85},
and plotted it with the open cyan-blue diamonds in Figure
\ref{distance_reddening_v745_sco_t_crb_v838_her_v1534_sco_no2}(a). 

We also plot the distance-reddening relations given by 
\citet{mar06}, who published a three-dimensional (3D) dust extinction map
of our galaxy in the direction of $-100\fdg0 \le l \le 100\fdg0$
and $-10\fdg0 \le b \le +10\fdg0$ with grids of $\Delta l=0\fdg25$
and $\Delta b=0\fdg25$, where $(l,b)$ are the galactic coordinates.
Figure \ref{distance_reddening_v745_sco_t_crb_v838_her_v1534_sco_no2}(a) 
shows four relations in the directions close to V745~Sco, i.e.,
$(l, b)=(357\fdg25, -4\fdg00)$ (open red squares),
$(357\fdg50, -4\fdg00)$ (filled green squares),
$(357\fdg25, -3\fdg75)$ (blue asterisks),
and $(357\fdg50, -3\fdg75)$ (open magenta circles), with error bars.
The closest one is that of the open red squares.
The reddening saturates to $E(B-V)\sim0.7\pm0.1$ at the distance
of $d\sim3-4$~kpc. This is consistent with the result of
the NASA/IPAC galactic two-dimensional (2D) dust absorption map,
$E(B-V)=0.71\pm0.02$. The distance-reddening relation given by \citet{ozd16}
consistently follows the relation given by \citet{mar06} until $E(B-V)=0.7$,
but does not saturate at the distance of $d\sim3-4$~kpc. Their reddening
linearly increases to $E(B-V)=1.0$. 
This is the reason why \citet{ozd16} obtained a much smaller distance
for V745~Sco.

The angular resolution of the map of Marshall et al. is 0\fdg25=15\farcm\ 
Marshall et al. used only giants in their analysis, and thus the dust map 
they produced has little information for the nearest kiloparsec.
The effective resolution of the NASA/IPAC galactic 2D dust absorption
map depends on the few-arcminute resolution of the IRAS $100~\mu$m
image and the $\sim$1 degree COBE/DIRBE dust temperature map, which
is considerably larger than the molecular cloud structure observed 
in the interstellar medium.
\citet{ozd16} used red clump giants. The number density of red clump giants
is smaller than that of giants that Marshall et al. used.
Therefore, the angular resolution of \citet{ozd16} could be less than
that of \citet{mar06}. Although \citet{ozd16} showed that 
the distance-reddening relation toward WY~Sge is not significantly
different for the four resolutions of 0\fdg3, 0\fdg4, 0\fdg5, and 0\fdg8, 
we adopt $E(B-V)=0.70$ for V745~Sco in the present study, following
\citet{mar06} and the NASA/IPAC galactic 2D dust absorption map.
We will demonstrate that this reddening of $E(B-V)=0.70$ is
reasonable by comparing various properties of V745~Sco with those of
other novae. We summarize various properties of V745~Sco
in Tables \ref{extinction_distance_various_novae}, 
\ref{physical_properties_recurrent_novae}, and 
\ref{wd_mass_recurrent_novae}.


\begin{figure}
\includegraphics[height=8.5cm]{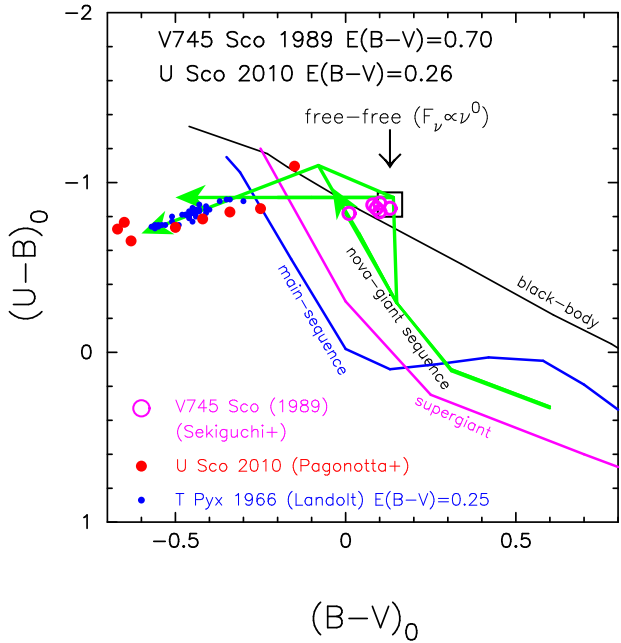}
\caption{
Color-color diagrams of V745~Sco (1989), U~Sco (2010), and T~Pyx (1966).
These color-color diagrams are similar to Figure 36 of \citet{hac16kb},
but we reanalyzed the data.  The open black square denotes the position
of optically thin free-free emission ($F_\nu\propto \nu^0$).
See the text for sources of observational data.
\label{color_color_diagram_t_pyx_v_745_sco_v407_cyg_u_sco_outburst}}
\end{figure}

\subsubsection{Color-color diagram}
\label{color_color_diagram_v745_sco}
Figure \ref{color_color_diagram_t_pyx_v_745_sco_v407_cyg_u_sco_outburst}
shows the color-color diagram of V745~Sco (1989) 
in outburst for the reddening of $E(B-V)=0.70$.
We add other two tracks of U~Sco (2010) and T~Pyx (1966) 
for comparison.  There are only six data (magenta open circles)
of V745~Sco but five of the six data are located at or near
the point of $(B-V)_0=+0.13$ and $(U-B)_0=-0.87$ denoted by
the open black square labeled ``free-free ($F_\nu\propto \nu^0$).''
This point corresponds to the position of optically thin free-free
emission \citep{hac14k}.  These positions of the data in the color-color
diagram are consistent with the fact that the ejecta had already been
optically thin when these data were obtained.  This is consistent with
the value of $E(B-V)=0.70$.


\begin{figure*}
\plotone{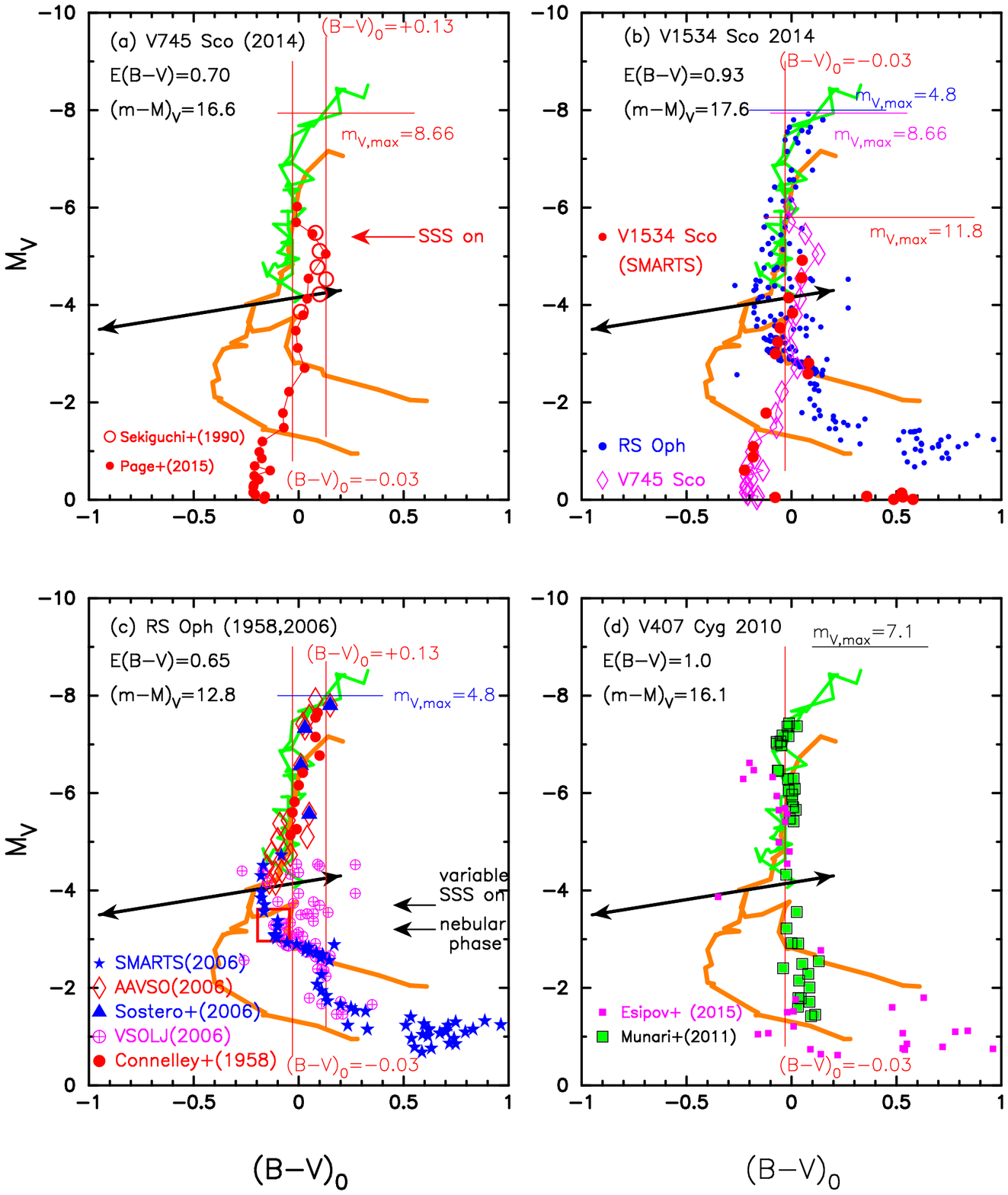}
\caption{
Color-magnitude diagrams of (a) V745~Sco, (b) V1534~Sco, (c) RS~Oph, 
and (d) V407~Cyg. The thick green lines show the template track
of V1668~Cyg and the thick orange lines represent that of LV~Vul
\citep[taken from][]{hac16kb}.  The two-headed black arrow indicates
the locations of inflection (for slow novae) or turning point of 
color-magnitude tracks \citep{hac16kb}.  
The vertical thin red lines indicate the color of optically thick,
$(B-V)_0=-0.03$, and optically thin, $(B-V)_0=+0.13$, free-free 
emission. See the text for more detail.
\label{hr_diagram_v745_sco_v1534_sco_rs_oph_v407_cyg_outburst}}
\end{figure*}

\subsubsection{Color-magnitude diagram}
Using $E(B-V)=0.70$ and $(m-M)_V=16.6$ ($d=7.8$~kpc), we plot the 
color-magnitude diagram, $(B-V)_0$-$M_V$, of V745~Sco in Figure
\ref{hr_diagram_v745_sco_v1534_sco_rs_oph_v407_cyg_outburst}(a).
The green lines denote the template track
of V1668~Cyg and the orange lines represent that of LV~Vul
\citep[taken from][]{hac16kb}.   The two-headed black arrow indicates
the locations of inflection (for slow novae) or turning point of 
color-magnitude tracks \citep{hac16kb}. 
In the very early phase, the track of V745~Sco follows the vertical
solid red line of $(B-V)_0=-0.03$, which is the intrinsic color of
optically thick free-free emission \citep[see, e.g.,][]{hac14k}.
This trend is similar to other classical novae such as V1668~Cyg.

The color increases to $(B-V)_0=+0.13$ four days after the discovery;
the discovery date is JD~2456695.194 \citep[e.g.,][]{pag15}. This is the 
intrinsic color of optically thin free-free emission \citep{hac14k}.
The color change from $(B-V)_0=-0.03$ to $(B-V)_0=+0.13$ indicates
that the ejecta became transparent (optically thin) after this date. 
The color then gradually becomes blue and reaches $(B-V)_0=-0.2$
in the very late phase. This is because strong emission
lines begin to contribute to the $B$ band flux \citep{hac14k}. 
The intrinsic color of V745~Sco is redder than $(B-V)_0 \gtrsim -0.2$
throughout the outburst. This is a common property among symbiotic nova 
systems with a RG companion, as shown in Figures 
\ref{hr_diagram_v745_sco_v1534_sco_rs_oph_v407_cyg_outburst}(b),
(c), and (d).


\begin{figure}
\includegraphics[height=7cm]{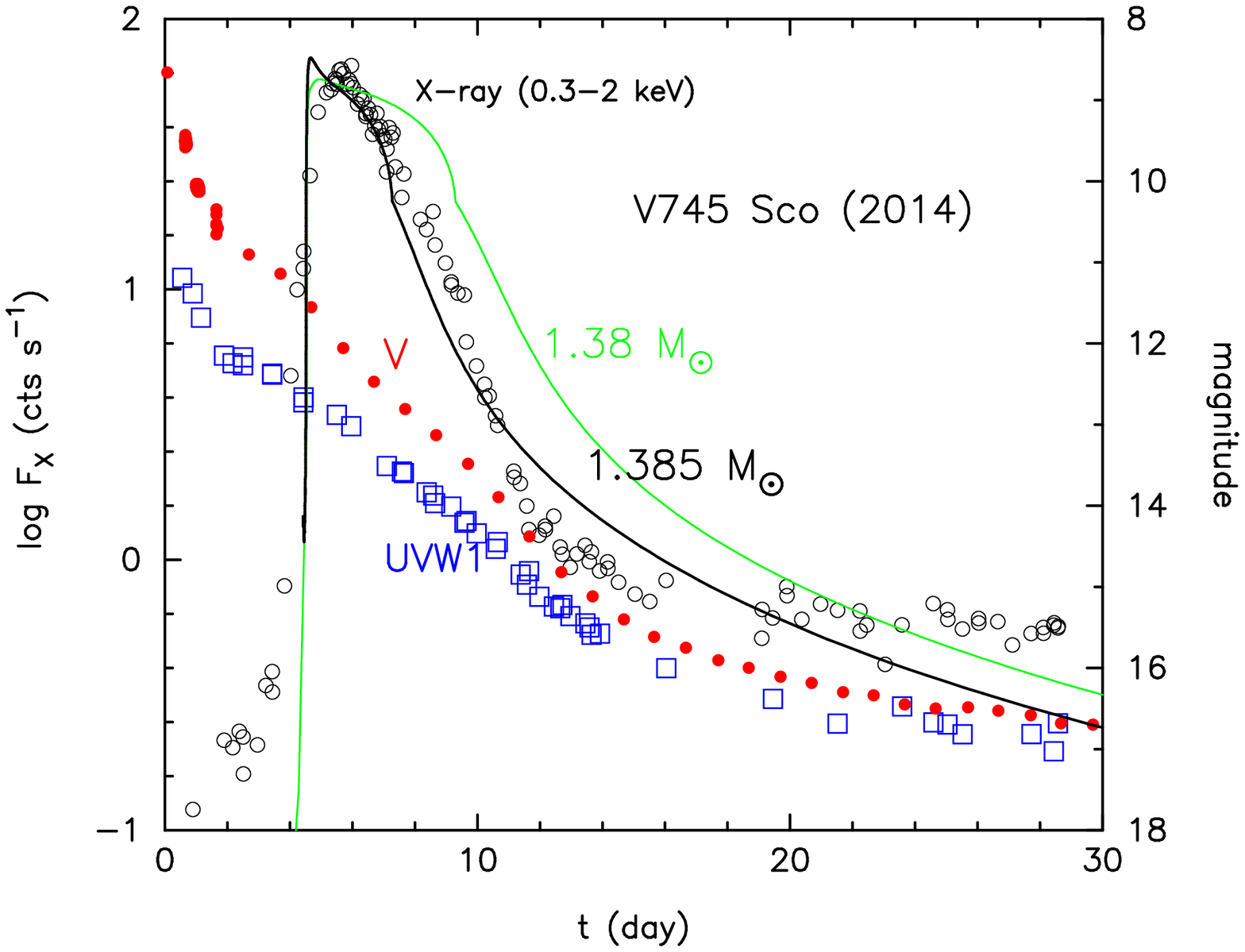}
\caption{
Model light curve fitting of the supersoft X-ray flux ($0.3-2.0$ keV denoted
by open black circles) for the V745~Sco 2014 outburst.
The start of day (horizontal axis) is the discovery date (JD~2456695.194). 
We also include the $V$ (filled red circles) and UVW1 (open blue squares)
magnitudes. All the observational data are taken from \citet{pag15}.
A $1.385~M_\sun$ WD model (solid black line) reasonably reproduces
the observed supersoft X-ray flux rather than a $1.38~M_\sun$ WD model
(solid green line). 
\label{v745_sco_xray_m1385}}
\end{figure}

\subsubsection{Model light curve of supersoft X-ray}
The SSS phase of V745~Sco started just four days 
after the discovery (Figure \ref{v745_sco_xray_m1385}).  
If the optical brightness reaches its maximum at the discovery
date, this turn-on time of SSS is the earliest record 
($t_{\rm SSS-on}=4$ days) among novae.  The second shortest record
($t_{\rm SSS-on}=6$ days) is the 1-yr recurrence
period nova, M31N~2008-12a \citep{hen15}. \citet{kat17sh} modeled
time-evolutions of such very-short-timescale novae and obtained a 
reasonable light curve fit with a $1.38~M_\sun$ WD for M31N~2008-12a.
The earlier appearance of SSS for V745~Sco suggests a more massive WD
than that of M31N~2008-12a.  We have modeled time-evolutions similar
to the calculation of \citet{kat17sh} and obtained a reasonable fit 
(solid black line) with the supersoft X-ray light curve (open black
circles), as shown in Figure \ref{v745_sco_xray_m1385}. 
We plot the supersoft X-ray flux ($0.3-2.0$ keV count rates of
the {\it Swift} XRT) as well as the $V$ magnitude and {\it Swift}
UVW1 magnitude light curves. All the observational data are 
taken from \citet{pag15}. Our obtained WD mass is $1.385~M_\sun$,
more massive than $1.38~M_\sun$ for M31N~2008-12a.
It is unlikely that these WDs were born as massive as they are
\citep[$1.38~M_\sun$ or $1.385~M_\sun$, see, e.g.,][]{doh15}.
We suppose that these WDs have grown in mass.

The X-ray flux of our model is calculated from blackbody spectra of
the photospheric temperature ($T_{\rm ph}$) and radius ($R_{\rm ph}$),
and the detailed flux itself is thus not so accurate, but the duration of
the SSS phase (rise and decay of the flux) is reasonably reproduced.
\citet{nom82} showed that, if a WD has a carbon-oxygen core
and its mass reaches $\sim1.38~M_\sun$ or more, 
the WD explodes as a SN~Ia. 
Therefore, V745~Sco is one of the most promising candidates
of SN~Ia progenitors.


\begin{figure}
\plotone{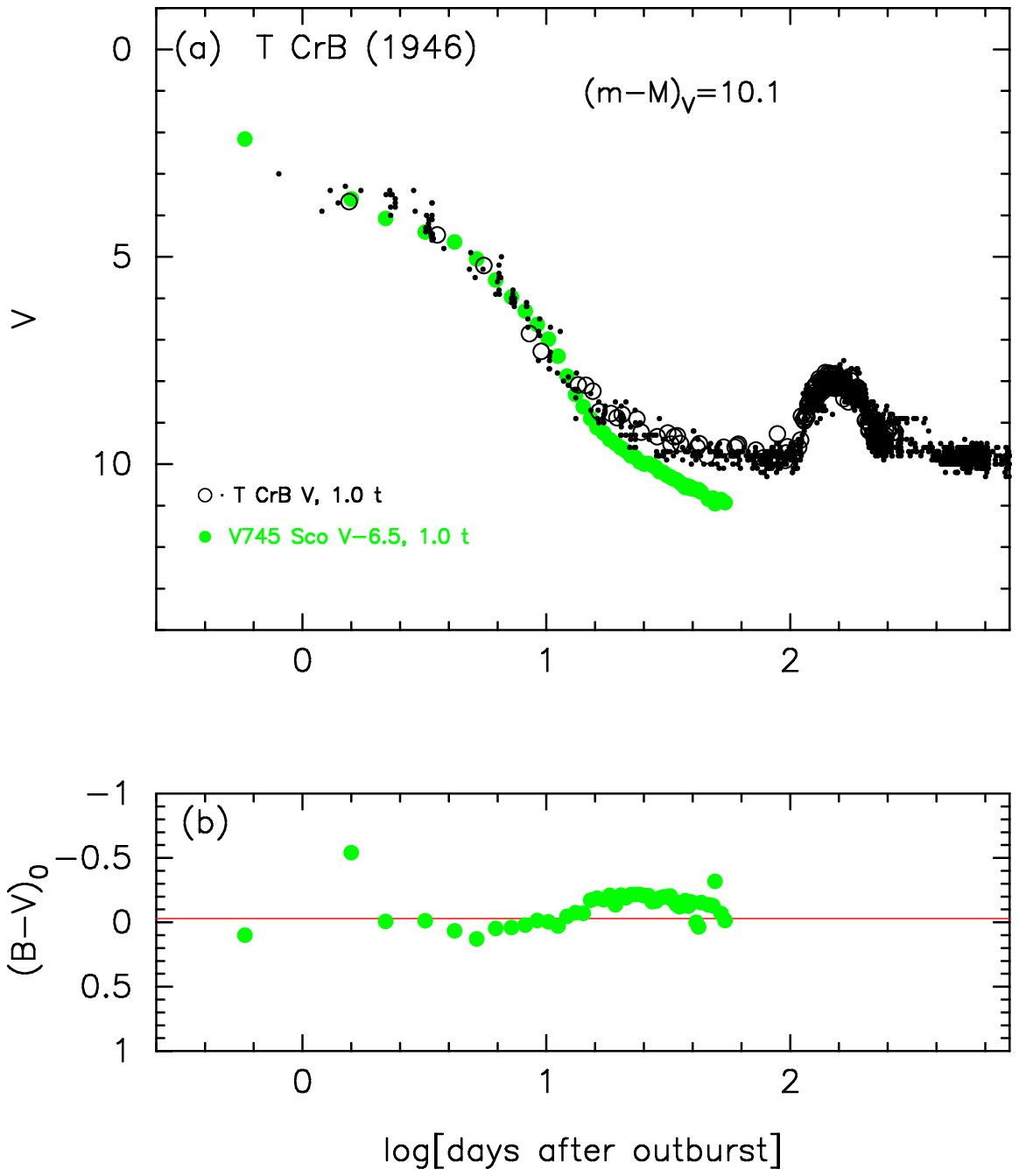}
\caption{
(a) The $V$ (open black circles) and visual (black dots) light curves 
of the T~CrB 1946 outburst are plotted on a logarithmic timescale
as well as the light/color curves of the V745~Sco 2014 outburst
(filled green circles). The timescales of these two novae are
the same, i.e., the timescaling factor of T~CrB is $f_{\rm s}=1.0$
with respect to that of V745~Sco.
The $V$ light curve of V745~Sco is shifted up by 6.5 mag.
The vertical shift and timescaling are written 
as ``V745 Sco V$-$6.5, 1.0 t'' in the figure.
(b) $(B-V)_0$ color curve of the V745~Sco 2014 outburst.  
The horizontal solid red line denotes the color of optically thick
free-free emission, i.e., $(B-V)_0=-0.03$. 
\label{t_crb_v745_sco_v_bv_ub_logscale}}
\end{figure}

\subsection{T~CrB (1946)}
\label{t_crb}
T~CrB is also a recurrent nova with two recorded outbursts in 1866 and 
1946 \citep[e.g.,][]{schaefer10a}. The orbital period was obtained
to be $P_{\rm orb}=227.6$~days \citep{fek00}.
Figure \ref{t_crb_v745_sco_v_bv_ub_logscale} shows the $V$ and visual
light curves of T~CrB on a logarithmic timescale.
The $V$ and visual light curves show a smooth decline and
a late secondary maximum.
The origin of the secondary maximum was discussed by \citet{hac01kb}
in terms of an irradiated tilting disk.
They also estimated the WD mass of T~CrB to be 
$M_{\rm WD}=1.37\pm0.01 ~M_\sun$ from the model light curve fitting.
The peak $V$ magnitude of T~CrB reaches $m_{V, \rm max}=2.5$ 
\citep{schaefer10a}. The decline rate is characterized 
by $t_2=4$ and $t_3=6$~days \citep[e.g.,][]{str10}.
Figure \ref{t_crb_v745_sco_v_bv_ub_logscale} also shows
the $V$ light curve of V745~Sco on the same timescale, but its brightness
is shifted up by 6.5 mag. These two light curves are very similar 
until $\sim10$ days after the outbursts. 

\subsubsection{Distance and reddening}
The distance was estimated by various authors 
\citep[e.g.,][]{bai81, har93, bel98, schaefer10a}, and summarized by
\citet{schaefer10a} to be $d=900\pm200$~pc.
In the present paper, we adopt $d=960\pm150$~pc after \citet{bel98}.
T~CrB shows an ellipsoidal variation in the quiescent light curve,
suggesting that the RG companion almost fills its Roche lobe.
Assuming that the RG companion just fills its Roche lobe,
\citet{bel98} obtained the absolute brightness of the RG companion
and estimated the distance.

For the reddening toward T~CrB, whose
galactic coordinates are $(l,b)=(42\fdg3738, +48\fdg1647)$,
the NASA/IPAC galactic 2D dust absorption
map gives $E(B-V)=0.056\pm0.003$ toward T~CrB.
If we adopt $E(B-V)=0.056$ toward T~CrB, then we calculate
the distance modulus in the $V$ band to be $(m-M)_V=10.1\pm0.3$
from Equation (\ref{v_distance_modulus}).
We plot the distance-reddening relations (black and orange lines)
of \citet{gre15, gre18}, respectively, in Figure
\ref{distance_reddening_v745_sco_t_crb_v838_her_v1534_sco_no2}(b).
\citet{gre15} published data for the galactic 3D
extinction map, which covers a wide range of the galactic
coordinates (over three quarters of the sky)
with grids of 3\farcm4 to 13\farcm7 and
a maximum distance resolution of 25\%.
Note that the values of $E(B-V)$ reported by Green et al. 
could have an error of $0.05-0.1$ mag compared 
with the 2D dust extinction maps.
The reddening saturates at the distance of $0.7$~kpc and the set of
$d=0.96$~kpc and $E(B-V)=0.056$ is consistent with the relation of 
\citet{gre15}.  Then, the vertical distance from the galactic plane 
is approximately $z=+715$~pc.  We summarize various properties of T~CrB
in Tables \ref{extinction_distance_various_novae}, 
\ref{physical_properties_recurrent_novae}, and \ref{wd_mass_recurrent_novae}.

\subsubsection{Absolute magnitudes and diversity in the late phase}
Figure \ref{t_crb_v745_sco_v_bv_ub_logscale}(a)
shows that the global decline
timescales of T~CrB and V745~Sco are very similar and their $V$ light
curves almost overlap with each other if we shift the $V$ light curve of 
V745~Sco vertically up by 6.5 mag. 
The $V$ light curves of the two novae deviate in the late phase
of the outbursts.
This is caused by different contributions of the RG companions and
accretion disks.  

The distance modulus in the $V$ band
is calculated to be $(m-M)_V=10.1\pm 0.3$ 
from Equation (\ref{v_distance_modulus}),
together with $d=960\pm150$~pc and $E(B-V)=0.056\pm0.003$.
For T~CrB and V745~Sco, we have the following relation
\begin{eqnarray}
(m&-&M)_{V, \rm T~CrB} = 10.1 \pm 0.3\cr 
&=& (m - M + \Delta V)_{V, \rm V745~Sco} \cr
&=& (16.6 \pm 0.2) + (-6.5\pm 0.1) = 10.1 \pm 0.3, 
\label{distance_modulus_t_crb_v745_sco}
\end{eqnarray}
where we adopt $(m-M)_{V, \rm V745~Sco}=16.6\pm 0.2$, as obtained in
Section \ref{v745_sco}, and the vertical shift of the V745~Sco $V$
light curve is $\Delta V=-6.5\pm 0.1$ because we change $\Delta V$ in 
steps of $0.1$ mag and searched for the best overlap.
This clearly shows that the two novae have the same absolute magnitudes
within the ambiguity of $\pm0.3$ mag when the timescales are the same.
The difference in the apparent magnitudes is due to the differences
in the distance and absorption.  

In general, there are two sources of $(m-M)_V$ ambiguities in our fitting
procedure: one is the
error in $(m-M)_V$ of the template nova and the other is the ambiguity of
vertical fit of $\Delta V$.  For the vertical fit, we change $\Delta V$
in steps of 0.1 mag and search for the best overlap by eye.
Its error is typically 0.1 mag (sometimes 0.2 mag) unless the $V$ data
are substantially scattered.  The $(m-M)_V$ ambiguity of the template
nova is dependent on each template (typically 0.2 or 0.3 mag).   
Thus, the errors of distance moduli $(m-M)_V$ are 0.2 or 0.3 mag
unless otherwise specified.


\begin{figure}
\includegraphics[height=8.5cm]{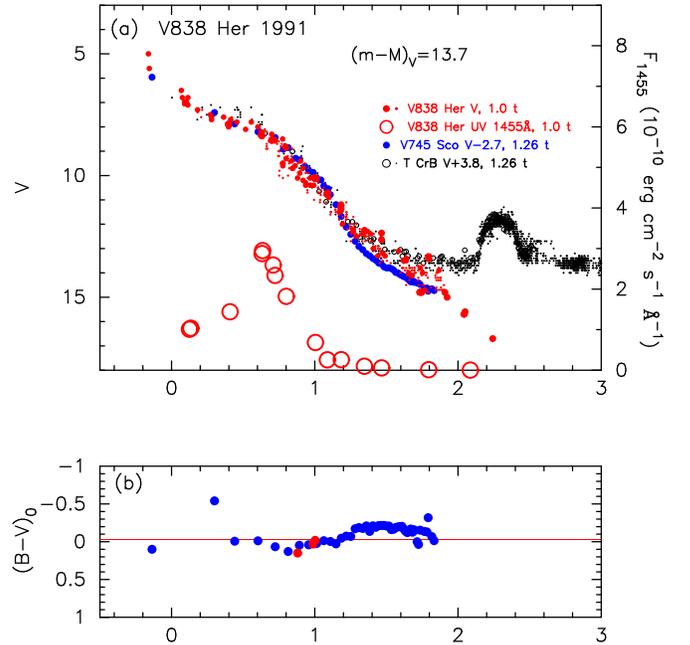}
\caption{
Same as Figure \ref{t_crb_v745_sco_v_bv_ub_logscale},
but we plot the visual (red dots) and $V$ (filled red circles),
UV~1455~\AA\ flux (large open red circles), and $(B-V)_0$ color curves 
(filled red circles) of V838~Her (red symbols) as well as 
the visual and $V$ light curves of T~CrB (black symbols), and 
the $V$ light curve and $(B-V)_0$ color curve of V745~Sco (blue symbols).
The timescales of T~CrB and V745~Sco are stretched by a factor
of $1.26$, i.e., the timescaling factor of V838~Her is $f_{\rm s}=1.26$
with respect to that of V745~Sco.
(a) 
The $V$ light curve of T~CrB is shifted down by 3.8 mag and
that of V745~Sco is shifted up by 2.7 mag. These shifts are indicated by
``T CrB V+3.8, 1.26 t'' and   ``V745 Sco V$-$2.7, 1.26 t'' in
the figure.
(b) $(B-V)_0$ color curves of V838~Her and V745~Sco. 
\label{v838_her_t_crb_v745_sco_v_bv_ub_logscale}}
\end{figure}


\begin{figure}
\includegraphics[height=12cm]{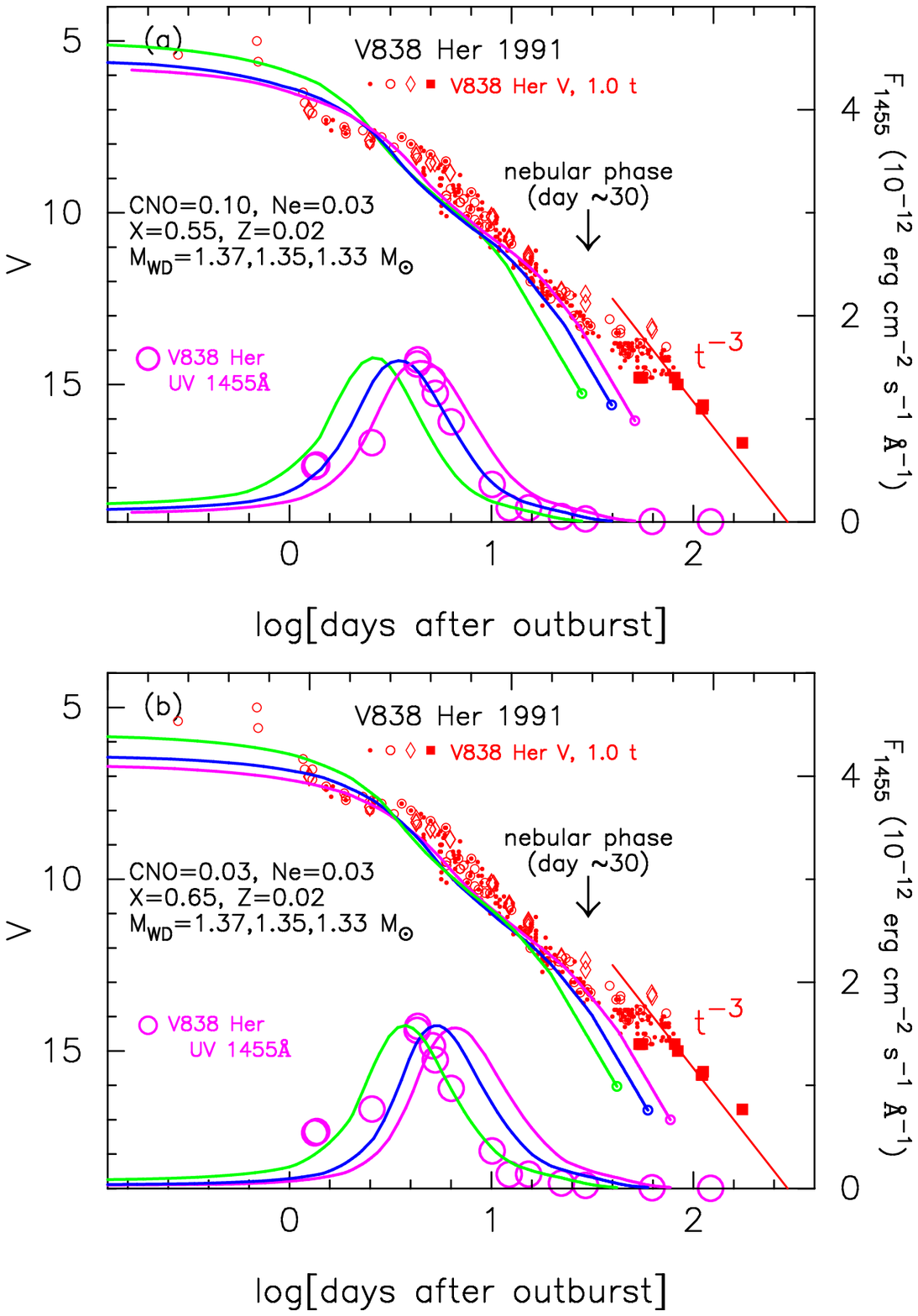}
\caption{
$V$ and UV~1455~\AA\ light curve fittings for V838~Her.
The sources of optical and UV~1455~\AA\ data are the same as those
in Figure 8 of \citet{kat09hc}.
(a) We plot three WD mass models of $1.37~M_\sun$ (green), 
$1.35~M_\sun$ (blue), and $1.33~M_\sun$ (magenta) for the
chemical composition of Ne nova 2 \citep{hac10k}.
The open circles at the right edge of the model light curves denote
the epoch when the optically thick winds stop.
We add a straight solid red line labeled ``$t^{-3}$'' in the figure
from the theoretical point of view.  
If the ejecta were expanding homologously, i.e., free expansion,
the flux would evolve as $F_\nu \propto t^{-3}$.
(b) We plot three WD mass models of $1.37~M_\sun$ (green), 
$1.35~M_\sun$ (blue), and $1.33~M_\sun$ (magenta), 
but for the different chemical composition of Ne nova 3 \citep{hac16k}.
\label{all_mass_v838_her_x55z02o10ne03_x65z02o03ne03_no2}}
\end{figure}


\begin{figure}
\includegraphics[height=12cm]{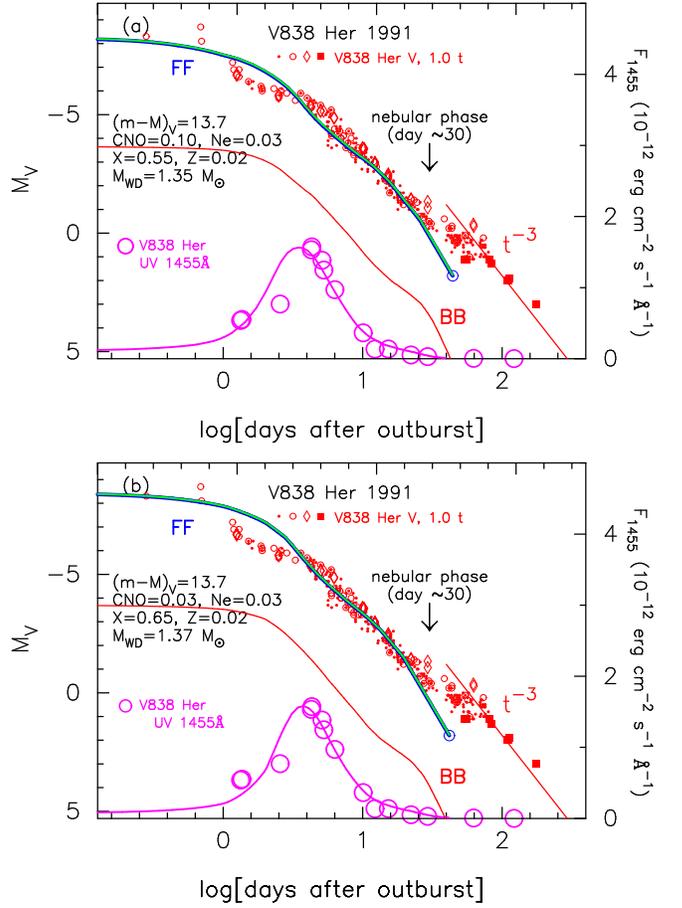}
\caption{
Same as Figure \ref{all_mass_v838_her_x55z02o10ne03_x65z02o03ne03_no2},
but we plot the best-fit models in the absolute $V$ magnitude:
(a) the $1.35~M_\sun$ WD 
with the chemical composition of Ne nova 2 and
(b) the $1.37~M_\sun$ WD with the chemical composition of Ne nova 3.
Three $V$ light curves are plotted for
the blackbody flux (solid red line labeled ``BB''), free-free flux 
(solid blue line labeled ``FF''), and the total flux (solid green)
of blackbody and free-free emission. The open blue circle at the 
right end of free-free flux denotes the epoch when the optically thick winds
stop.   
\label{all_mass_v838_her_x55z02o10ne03_x65z02o03ne03_fix_no2}}
\end{figure}

\subsection{V838~Her 1991}
\label{v838_her}
V838~Her is a very fast nova with $t_2=1$~day and $t_3=4$~days,
and its peak brightness reached $m_{V, \rm max}=5.3$ \citep{str10}. 
The orbital period was obtained to be
$P_{\rm orb}=0.2976$~days \citep{ing92, lei92}.
Figure \ref{v838_her_t_crb_v745_sco_v_bv_ub_logscale} shows the $V$ and
visual light curves of V838~Her on a logarithmic timescale.
This figure also shows the $V$ light and $B-V$ color curves of V745~Sco and
the $V$ and visual light curves of T~CrB to demonstrate the resemblance
among the three novae.  
The reddening toward V838~Her was estimated by several authors
\citep[e.g.,][]{mat93, har94, van96, kat09hc} to lie between 
$E(B-V)=0.3-0.7$.
Of these values, we adopt $E(B-V)=0.53\pm0.05$ after \citet{kat09hc}, which was
obtained from the 2175~\AA\ feature of the UV spectra of V838~Her.  
\citet{kat09hc} also derived
the distance of $d=2.7\pm0.5$~kpc from the UV~1455~\AA\ flux fitting and
the WD mass of $M_{\rm WD}=1.35\pm0.02 ~M_\sun$ from the model
light curve fitting.

\subsubsection{Model light curve fitting}
We briefly explain the $V$ light curve and UV~1455~\AA\  
light curve fitting of novae.  Nova spectra are dominated 
by the free-free emission the after optical maximum 
\citep[e.g.,][]{gal76, enn77}. This free-free emission comes
from optically thin plasma outside the photosphere.
\citet{hac06kb} calculated free-free emission model light curves of
novae and showed that theoretical light curves reproduce well observed
NIR/optical light curves of several classical novae from near the peak
to the nebular phase. These free-free emission model light curves are
calculated from the nova evolution models based on the optically thick
wind theory \citep{kat94h}. Their numerical models
provide the photospheric temperature ($T_{\rm ph}$), radius ($R_{\rm ph}$),
velocity ($v_{\rm ph}$), and wind mass-loss rate ($\dot M_{\rm wind}$)
of a nova hydrogen-rich envelope (mass of $M_{\rm env}$)
for a specified WD mass ($M_{\rm WD}$) and
chemical composition of the hydrogen-rich envelope.
The free-free emission model light curves are calculated 
by Equations (9) and (10) in \citet{hac06kb}.

     Envelope chemical composition of their model is given by the form
of $(X, Y, X_{\rm CNO}, X_{\rm Ne}, Z)$.  Here, $X$ is the hydrogen content,
$Y$ the helium content, $X_{\rm CNO}$ the abundance of carbon,
nitrogen, and oxygen, $X_{\rm Ne}$ the neon content,
and $Z= 0.02$ the heavy element (heavier than helium) content by weight,
in which carbon, nitrogen, oxygen, and neon are also included
with the solar composition ratios \citep{hac06kb}.
The chemical composition of the V838~Her ejecta was estimated by 
\citet{van96}, \citet{van97}, and \citet{schw07a}. 
The results are summarized in Table 2 of \citet{kat09hc}.
In this study, we calculated two cases, ``Ne nova 2'' ($X=0.55$, 
$Y=0.30$, $Z=0.02$, $X_{\rm CNO}=0.10$, and $X_{\rm Ne}=0.03$)
\citep{hac10k} and ``Ne nova 3''
($X=0.65$, $Y=0.27$, $Z=0.02$, $X_{\rm CNO}=0.03$, and $X_{\rm Ne}=0.03$)
\citep{hac16k},
both of which are close to the estimates mentioned above. 

In Figure \ref{all_mass_v838_her_x55z02o10ne03_x65z02o03ne03_no2}(a),
we plot three WD mass models: $1.37~M_\sun$ (green), $1.35~M_\sun$ (blue),
and $1.33~M_\sun$ (magenta) for Ne nova 2.
The open circles at the right edge of model light curves denote
the epoch when the optically thick winds stop. We tabulate these three 
free-free emission model light curves in Table 
\ref{light_curves_of_novae_ne2}. A more massive WD evolves faster.
The UV~1455~\AA\ band is an emission-line-free narrow band (20~\AA\ width
centered at 1455~\AA) invented by \citet{cas02} based on the {\it IUE}
spectra of novae. This band represents continuum flux at UV well
and is useful for model light curve fitting.
\citet{hac06kb, hac10k, hac15k, hac16k} calculated UV~1455~\AA\
model light curves for various WD masses
and chemical compositions of the hydrogen-rich envelope,
assuming blackbody emission at the photosphere.
In Figure \ref{all_mass_v838_her_x55z02o10ne03_x65z02o03ne03_no2}(a),
the best-fit model among the three WD mass models is the 
$1.35~M_\sun$ WD model (solid blue lines).

In Figure \ref{all_mass_v838_her_x55z02o10ne03_x65z02o03ne03_no2}(b),
we also plot three WD mass models: $1.37~M_\sun$ (green), 
$1.35~M_\sun$ (blue), and $1.33~M_\sun$ (magenta),
but for the different chemical composition of Ne nova 3.
We tabulate these three free-free emission model light curves
in Table \ref{light_curves_of_novae_ne3}.
For this chemical composition, the best-fit model 
among the three WD mass models is the $1.37~M_\sun$ WD model 
(solid green lines).

We add a straight solid red line labeled ``$t^{-3}$'' in these figures
from the theoretical point of view. After the optically thick winds stop
at the open circle (right edge of the free-free model $V$ light curve),
the ejecta mass $M_{\rm ej}$ is virtually constant with time
because of no mass supply from the WD.
If the ejecta were expanding homologously, i.e., free expansion,
the free-free emission flux would evolve as
$F_\nu \propto t^{-3}$  \citep[see, e.g.,][]{woo97, hac06kb}.
We plot this trend as a guide in the later decline phase.

After the optically thick winds stop, the photosphere quickly shrinks,
the photospheric temperature increases to emit high-energy photons,
and the nova enters the nebular phase. This nova entered the nebular
phase approximately 30 days after the outburst, i.e., 
at least by UT 1991 April 23 
\citep[e.g.,][]{wil94, van96}. In this case, the neon forbidden lines
had grown to become comparable in strength to H$\alpha$.  
In the nebular phase,
strong emission lines contribute to the $V$ magnitude.  As a result,
the model light curve deviates significantly 
from the observational $V$ light curve,
as shown in Figure \ref{all_mass_v838_her_x55z02o10ne03_x65z02o03ne03_no2},
because the model $V$ light curve does not include the effect of
emission lines, but rather represents only continuum (free-free) emission.

Fitting our model to the observed UV~1455~\AA\ flux,
we have the following relation: 
\begin{equation}
-2.5\log \left( F_{\lambda}^{\rm obs}/F_{\lambda}^{\rm mod}\right)
= R_\lambda E(B-V) + 5\log \left( {{d} \over {10~{\rm kpc}}} \right),
\label{uv1455_distance_modulus}
\end{equation}
where $F_{\lambda}^{\rm mod}$ is the model flux at the distance of
$d=10$~kpc, $F_{\lambda}^{\rm obs}$ is the observed flux,
the absorption is calculated from $A_\lambda=R_\lambda E(B-V)$, and
$R_\lambda=8.3$ for $\lambda=1455$~\AA\
\citep{sea79}.
For the $1.35~M_\sun$ WD model in Figure
\ref{all_mass_v838_her_x55z02o10ne03_x65z02o03ne03_no2}(a),
$F_{1455}^{\rm obs}=5.0$ and $F_{1455}^{\rm mod}=21.0$,
in units of $10^{-12}$~erg~cm$^{-2}$~s$^{-1}$~\AA$^{-1}$,
at the upper bound of Figure
\ref{all_mass_v838_her_x55z02o10ne03_x65z02o03ne03_no2}(a), whereas
for the $1.37~M_\sun$ WD model in Figure
\ref{all_mass_v838_her_x55z02o10ne03_x65z02o03ne03_no2}(b),
$F_{1455}^{\rm obs}=5.0$ and $F_{1455}^{\rm mod}=20.0$.
Substituting $E(B-V)=0.53$ into Equation 
(\ref{uv1455_distance_modulus}), both the cases give $d=2.6-2.7$~kpc. 
This value is consistent with the estimate by \citet{kat09hc}.
The distance calculated by our model fitting hardly depends 
on the assumed chemical composition.
The distance modulus in the $V$ band is calculated from Equation 
(\ref{v_distance_modulus}) to be $(m-M)_V=13.7\pm 0.3$ for V838~Her.
The vertical distance from the galactic plane is approximately 
$z=+310$~pc, because the galactic coordinates of V838~Her are 
$(l,b)=(43\fdg3155, +6\fdg6187)$. We summarize various properties of 
V838~Her in Tables \ref{extinction_distance_various_novae},
\ref{physical_properties_recurrent_novae}, and 
\ref{wd_mass_recurrent_novae}.

Assuming that $(m-M)_V=13.7$, we plot the absolute $V$ magnitude 
light curve of V838~Her in Figure
\ref{all_mass_v838_her_x55z02o10ne03_x65z02o03ne03_fix_no2}.
The total flux (solid green line) 
of the $V$ band consists of the two parts; one is
the flux of free-free emission (solid blue line labeled ``FF''), 
which comes from optically thin 
plasma outside the photosphere, and the other is the photospheric 
emission (solid red line labeled ``BB''),
which is approximated by blackbody emission. It is clear that
the flux of free-free emission is much larger than that of 
blackbody emission. Thus, the free-free emission dominates
the continuum spectra of very fast novae.  

If we average the hydrogen contents of the three results mentioned above,
i.e., \citet{van96}, \citet{van97}, and \citet{schw07a},
we obtain $X=(0.78 + 0.59 + 0.562)/3= 0.644$.  This value is close
to our case of Ne nova 3 ($X=0.65$, $Y=0.27$, $Z=0.02$, 
$X_{\rm CNO}=0.03$, and $X_{\rm Ne}=0.03$).  Therefore, we adopt
the $1.37~M_\sun$ WD for V838~Her.


\startlongtable
\begin{deluxetable}{llll}
\tabletypesize{\scriptsize}
\tablecaption{Free-free Light Curves of Ne Novae 2\tablenotemark{a}
\label{light_curves_of_novae_ne2}}
\tablewidth{0pt}
\tablehead{
\colhead{$m_{\rm ff}$} &
\colhead{1.33$M_\sun$} &
\colhead{1.35$M_\sun$} &
\colhead{1.37$M_\sun$} \\
\colhead{(mag)} &
\colhead{(day)} &
\colhead{(day)} &
\colhead{(day)} 
}
\colnumbers
\startdata
  4.500     & 0.0 & 0.0 & 0.0  \\
  4.750     & 0.4617     & 0.6000     & 0.4350     \\
  5.000     & 0.8397     &  1.019     & 0.7630     \\
  5.250     &  1.173     &  1.354     &  1.074     \\
  5.500     &  1.575     &  1.673     &  1.385     \\
  5.750     &  1.992     &  2.019     &  1.694     \\
  6.000     &  2.396     &  2.399     &  2.007     \\
  6.250     &  2.786     &  2.742     &  2.328     \\
  6.500     &  3.189     &  3.075     &  2.628     \\
  6.750     &  3.591     &  3.408     &  2.871     \\
  7.000     &  4.001     &  3.736     &  3.137     \\
  7.250     &  4.403     &  4.041     &  3.380     \\
  7.500     &  4.810     &  4.346     &  3.606     \\
  7.750     &  5.235     &  4.643     &  3.819     \\
  8.000     &  5.717     &  4.959     &  4.044     \\
  8.250     &  6.355     &  5.326     &  4.295     \\
  8.500     &  7.079     &  5.766     &  4.594     \\
  8.750     &  7.949     &  6.299     &  4.943     \\
  9.000     &  8.922     &  6.945     &  5.373     \\
  9.250     &  10.05     &  7.683     &  5.857     \\
  9.500     &  11.45     &  8.533     &  6.443     \\
  9.750     &  13.13     &  9.533     &  7.097     \\
  10.00     &  14.94     &  10.76     &  7.918     \\
  10.25     &  16.83     &  12.15     &  8.828     \\
  10.50     &  18.68     &  13.51     &  9.748     \\
  10.75     &  20.56     &  14.85     &  10.62     \\
  11.00     &  22.30     &  16.25     &  11.40     \\
  11.25     &  23.73     &  17.56     &  12.23     \\
  11.50     &  25.19     &  18.75     &  13.10     \\
  11.75     &  26.75     &  20.01     &  14.03     \\
  12.00     &  28.39     &  21.35     &  15.01     \\
  12.25     &  30.13     &  22.76     &  16.05     \\
  12.50     &  31.97     &  24.26     &  17.15     \\
  12.75     &  33.93     &  25.84     &  18.31     \\
  13.00     &  35.99     &  27.52     &  19.55     \\
  13.25     &  38.18     &  29.30     &  20.86     \\
  13.50     &  40.50     &  31.19     &  22.24     \\
  13.75     &  42.96     &  33.18     &  23.71     \\
  14.00     &  45.57     &  35.30     &  25.27     \\
  14.25     &  48.32     &  37.54     &  26.91     \\
  14.50     &  51.24     &  39.91     &  28.66     \\
  14.75     &  54.34     &  42.42     &  30.50     \\
  15.00     &  57.61     &  45.08     &  32.46     \\
\hline
X-ray\tablenotemark{b}
 & 23.4  & 14.3 & 7.80  \\
\hline
$\log f_{\rm s}$\tablenotemark{c}
 & $+0.10$  & 0.0 & $-0.12$  \\
\hline
$M_{\rm w}$\tablenotemark{d}
 & $2.2$ & $1.8$  & $1.4$  \\
\enddata
\tablenotetext{a}{Chemical composition of the envelope is assumed
to be that of Ne nova 2 in Table 2 of \citet{hac16k}.}
\tablenotetext{b}{Duration of supersoft X-ray phase in units of days.}
\tablenotetext{c}{Stretching factor with respect to V838~Her
UV~1455~\AA\ observation in Figure 
\ref{all_mass_v838_her_x55z02o10ne03_x65z02o03ne03_no2}.}
\tablenotetext{d}{Absolute magnitudes at the bottom point 
of free-free emission light curve (open circles) in Figures
\ref{all_mass_v838_her_x55z02o10ne03_x65z02o03ne03_no2}(a)
and \ref{all_mass_v838_her_x55z02o10ne03_x65z02o03ne03_fix_no2}(a)
by assuming $(m-M)_V = 13.7$  (V838~Her).  The absolute $V$ magnitude
is calculated from $M_V=m_{\rm ff} -15.0 + M_{\rm w}$.}
\end{deluxetable}


\startlongtable
\begin{deluxetable}{llll}
\tabletypesize{\scriptsize}
\tablecaption{Free-free Light Curves of Ne Novae 3\tablenotemark{a}
\label{light_curves_of_novae_ne3}}
\tablewidth{0pt}
\tablehead{
\colhead{$m_{\rm ff}$} &
\colhead{1.33$M_\sun$} &
\colhead{1.35$M_\sun$} &
\colhead{1.37$M_\sun$} \\
\colhead{(mag)} &
\colhead{(day)} &
\colhead{(day)} &
\colhead{(day)} 
}
\colnumbers
\startdata
  3.750     & 0.0 & 0.0 & 0.0 \\
  4.000     & 0.4317     & 0.4570     & 0.3855     \\
  4.250     & 0.9427     & 0.9240     & 0.9118     \\
  4.500     &  1.481     &  1.388     &  1.420     \\
  4.750     &  2.023     &  1.914     &  1.889     \\
  5.000     &  2.594     &  2.477     &  2.324     \\
  5.250     &  3.173     &  3.008     &  2.752     \\
  5.500     &  3.686     &  3.426     &  3.140     \\
  5.750     &  4.176     &  3.861     &  3.498     \\
  6.000     &  4.665     &  4.270     &  3.870     \\
  6.250     &  5.128     &  4.653     &  4.180     \\
  6.500     &  5.621     &  5.042     &  4.487     \\
  6.750     &  6.130     &  5.430     &  4.798     \\
  7.000     &  6.622     &  5.842     &  5.058     \\
  7.250     &  7.129     &  6.232     &  5.320     \\
  7.500     &  7.646     &  6.621     &  5.605     \\
  7.750     &  8.221     &  7.041     &  5.930     \\
  8.000     &  8.886     &  7.557     &  6.296     \\
  8.250     &  9.698     &  8.153     &  6.707     \\
  8.500     &  10.63     &  8.908     &  7.222     \\
  8.750     &  11.80     &  9.778     &  7.801     \\
  9.000     &  13.14     &  10.79     &  8.494     \\
  9.250     &  14.64     &  11.92     &  9.268     \\
  9.500     &  16.36     &  13.22     &  10.21     \\
  9.750     &  18.38     &  14.81     &  11.29     \\
  10.00     &  20.61     &  16.55     &  12.48     \\
  10.25     &  23.04     &  18.36     &  13.69     \\
  10.50     &  25.45     &  20.14     &  14.96     \\
  10.75     &  27.66     &  21.95     &  16.09     \\
  11.00     &  30.02     &  23.34     &  17.07     \\
  11.25     &  32.48     &  24.82     &  18.12     \\
  11.50     &  34.55     &  26.38     &  19.23     \\
  11.75     &  36.64     &  28.03     &  20.40     \\
  12.00     &  38.87     &  29.78     &  21.64     \\
  12.25     &  41.22     &  31.64     &  22.95     \\
  12.50     &  43.71     &  33.60     &  24.35     \\
  12.75     &  46.35     &  35.69     &  25.83     \\
  13.00     &  49.15     &  37.89     &  27.39     \\
  13.25     &  52.10     &  40.23     &  29.04     \\
  13.50     &  55.24     &  42.70     &  30.80     \\
  13.75     &  58.56     &  45.32     &  32.66     \\
  14.00     &  62.08     &  48.10     &  34.62     \\
  14.25     &  65.81     &  51.04     &  36.71     \\
  14.50     &  69.76     &  54.16     &  38.92     \\
  14.75     &  73.94     &  57.46     &  41.26     \\
  15.00     &  78.37     &  60.95     &  43.73     \\
\hline
X-ray\tablenotemark{b}
 & 41.6  & 24.1 & 12.4  \\
\hline
$\log f_{\rm s}$\tablenotemark{c}
 & 0.26  & 0.15 & 0.0  \\
\hline
$M_{\rm w}$\tablenotemark{d}
 & $2.5$ & $2.2$  & $1.8$  \\
\enddata
\tablenotetext{a}{Chemical composition of the envelope is assumed
to be that of Ne nova 3 in Table 2 of \citet{hac16k}.}
\tablenotetext{b}{Duration of supersoft X-ray phase in units of days.}
\tablenotetext{c}{Stretching factor with respect to V838~Her
 UV~1455~\AA\ observation in Figure 
\ref{all_mass_v838_her_x55z02o10ne03_x65z02o03ne03_no2}.}
\tablenotetext{d}{Absolute magnitudes at the bottom point 
of free-free emission light curve (open circles) in Figures
\ref{all_mass_v838_her_x55z02o10ne03_x65z02o03ne03_no2}(b)
and \ref{all_mass_v838_her_x55z02o10ne03_x65z02o03ne03_fix_no2}(b)
by assuming $(m-M)_V = 13.7$  (V838~Her).  The absolute $V$ magnitude
is calculated from $M_V=m_{\rm ff} -15.0 + M_{\rm w}$.}
\end{deluxetable}

\subsubsection{Timescaling law and time-stretching method}
We showed in the previous subsection
that the absolute magnitudes of the $V$ light curves
are the same for V745~Sco and T~CrB.
We plot the $V$ and visual magnitudes of V838~Her on a logarithmic
timescale in Figure \ref{v838_her_t_crb_v745_sco_v_bv_ub_logscale},
and plot the $V$ and visual magnitude light curves of V745~Sco
and T~CrB in the same figure but stretch their timescales
by a factor of $f_{\rm s}=1.26$.  We further shift the $V$ light
curve of V745~Sco up by $\Delta V= -2.7$ and that of T~CrB down
by $\Delta V= +3.8$.  We confirm that these three stretched $V$
light curves overlap.

\citet{hac10k} showed that, if the two nova light curves, called the template and the target,  
$(m[t])_{V,\rm target}$ and $(m[t])_{V,\rm template}$,
overlap each other after time-stretching by a factor of $f_{\rm s}$
in the horizontal direction and shifting vertically down by $\Delta V$, i.e.,
\begin{equation}
(m[t])_{V,\rm target} = \left((m[t \times f_{\rm s}])_V
+ \Delta V\right)_{\rm template},
\label{overlap_brigheness}
\end{equation}
their distance moduli in the $V$ band satisfy 
\begin{equation}
(m-M)_{V,\rm target} = \left( (m-M)_V
+ \Delta V\right)_{\rm template} - 2.5 \log f_{\rm s}.
\label{distance_modulus_formula}
\end{equation}
Here, $(m-M)_{V, \rm target}$ and $(m-M)_{V, \rm template}$ are
the distance moduli in the $V$ band
of the target and template novae, respectively.
For the set of V745~Sco, T~CrB, and V838~Her 
in Figure \ref{v838_her_t_crb_v745_sco_v_bv_ub_logscale}, 
Equations (\ref{overlap_brigheness}) and (\ref{distance_modulus_formula})
are satisfied, because we have the relation:   
\begin{eqnarray}
(m&-&M)_{V, \rm V838~Her} = 13.7\pm0.3 \cr
&=& (m - M + \Delta V)_{V, \rm V745~Sco} - 2.5 \log 1.26 \cr
&=& 16.6\pm0.2 -2.7\pm0.1 - 0.25 = 13.65 \pm0.2 \cr
&=& (m - M + \Delta V)_{V, \rm T~CrB} - 2.5 \log 1.26 \cr
&=& 10.1\pm0.3 + 3.8\pm0.1 - 0.25 = 13.65 \pm0.3,
\label{distance_modulus_v838_her_v745_sco}
\end{eqnarray}
where we adopt $(m-M)_{V, \rm V745~Sco}=16.6\pm0.2$ in Section \ref{v745_sco}
and $(m-M)_{V, \rm T~CrB}=10.1\pm0.3$ in Section \ref{t_crb}. 
This procedure was called ``the time-stretching method'' \citep{hac10k}.
See Appendix \ref{time-stretching_method} for more detail.

\subsubsection{Reddening and distance}
We check the distance and reddening toward V838~Her
based on distance-reddening relations.
Figure \ref{distance_reddening_v745_sco_t_crb_v838_her_v1534_sco_no2}(c)
shows various distance-reddening relations toward V838~Her.
Substituting $F_{1455}^{\rm obs}=5.0$ and $F_{1455}^{\rm mod}=20.0$
into Equation (\ref{uv1455_distance_modulus}), we plot the distance-reddening
relation (solid magenta line) for our UV~1455~\AA\ fit.
Even by substituting $F_{1455}^{\rm obs}=5.0$ and 
$F_{1455}^{\rm mod}=21.0$, we obtain a nearly overlapping magenta line.
The cross point between the vertical solid red line of $E(B-V)=0.53$ and
the solid magenta line gives a distance of 2.6~kpc. 
We also plot the distance-reddening relation calculated from Equation 
(\ref{v_distance_modulus}) with $(m-M)_V=13.7$ by the solid blue line.
The NASA/IPAC Galactic dust absorption
map gives $E(B-V)=0.368\pm0.003$ toward V838~Her.
The distance-reddening relations given by \citet{mar06}, \citet{gre15, gre18},
and \citet{ozd16} are roughly consistent with each other, i.e.,
$E(B-V)\sim0.3\pm0.1$ at the distance of $d=2.6$~kpc. 
If we adopt $E(B-V)=0.3$ toward V838~Her, we obtain
the distance of $d=6.5$~kpc from Equation (\ref{uv1455_distance_modulus}),
that is, the solid magenta line.
Then, the distance modulus in the $V$ band is calculated from Equation 
(\ref{v_distance_modulus}) to be $(m-M)_V=15.0$, which is inconsistent
with Equation (\ref{distance_modulus_v838_her_v745_sco}).

This large discrepancy can be understood as follows.
The 3D (2D) dust maps essentially give an averaged value
of a relatively broad region, and thus the pinpoint reddening could be
different from the value of the 3D (2D) dust maps, because
the resolutions of these dust maps are considerably larger
than molecular cloud structures observed in the interstellar medium,
as mentioned in Section \ref{v745_sco}.
The pinpoint estimate of the reddening toward V838~Her
was summarized by \citet{van96} to be $E(B-V)= 0.50\pm0.12$,
based on the Balmer decrement \citep{ing92,van96}, 
the equivalent width of \ion{Na}{1} interstellar lines \citep{lyn92},
the ratio of the UV flux above and below 2000~\AA\  \citep{sta92},
and the assumed intrinsic color at maximum light \citep{woo92}.
\citet{kat09hc} estimated the reddening to be $E(B-V)=0.53\pm0.05$ 
based on the 2175~\AA\ feature in the {\sl IUE} spectra of V838~Her.
When and only when we adopt $E(B-V)=0.53$, we obtain the distance
modulus in the $V$ band, $(m-M)_V=13.7$, which is consistent
with Equation (\ref{distance_modulus_v838_her_v745_sco}).


\begin{figure}
\includegraphics[height=8cm]{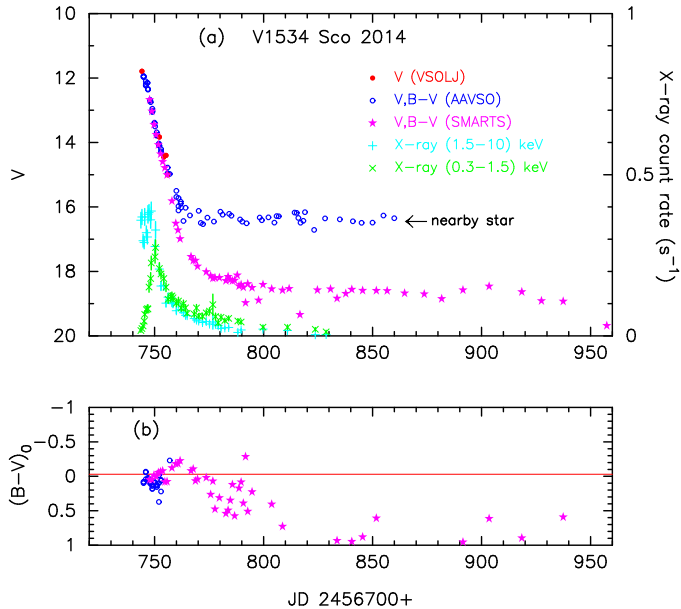}
\caption{
Optical light curve (a) and color curve (b) of V1534~Sco:
(a) A few $V$ data in the very early phase are taken from VSOLJ
(filled red circles).  
The $BV$ data are taken from AAVSO (open blue circles) and 
SMARTS (filled magenta stars). The AAVSO $V$ data saturate at
$V\sim 16.3$ owing to contamination of a nearby star.
The soft ($0.3-1.5$ keV, denoted by green crosses)
and hard ($1.5-10$ keV, denoted by cyan pluses) 
X-ray count rates are also plotted,
the data for which are taken from the {\it Swift} web site
\citep{eva09}. 
(b) The $(B-V)_0$ are dereddened with $E(B-V)=0.93$.  
\label{v1534_sco_v_bv_ub_color_curve}}
\end{figure}


\begin{figure}
\includegraphics[height=9.5cm]{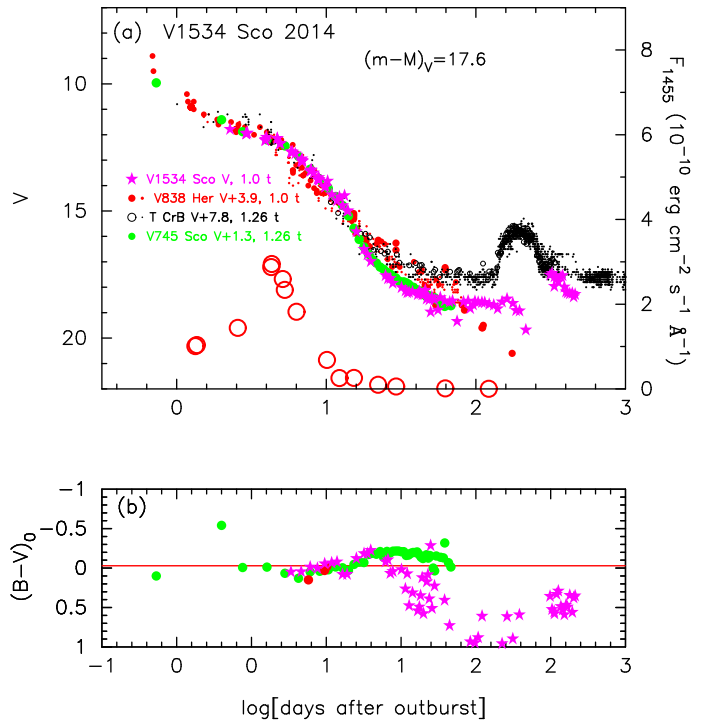}
\caption{
Same as Figure \ref{v838_her_t_crb_v745_sco_v_bv_ub_logscale},
but we plot the light/color curves of V1534~Sco as well as
V838~Her, T~CrB, and V745~Sco.
(a) The filled magenta stars denote the $V$ magnitudes of V1534~Sco,
the filled red circles (red dots) represent the $V$ (visual)
magnitudes of V838~Her, the black open circles (black dots) indicate the 
$V$ (visual) magnitude of T~CrB, and the filled green circles represent 
the $V$ magnitude of V745~Sco.  
(b) The $(B-V)_0$ colors of V1534~Sco are dereddened with $E(B-V)=0.93$.
\label{v1534_sco_v838_her_t_crb_v745_sco_v_bv_ub_logscale}}
\end{figure}


\begin{figure}
\includegraphics[height=13.0cm]{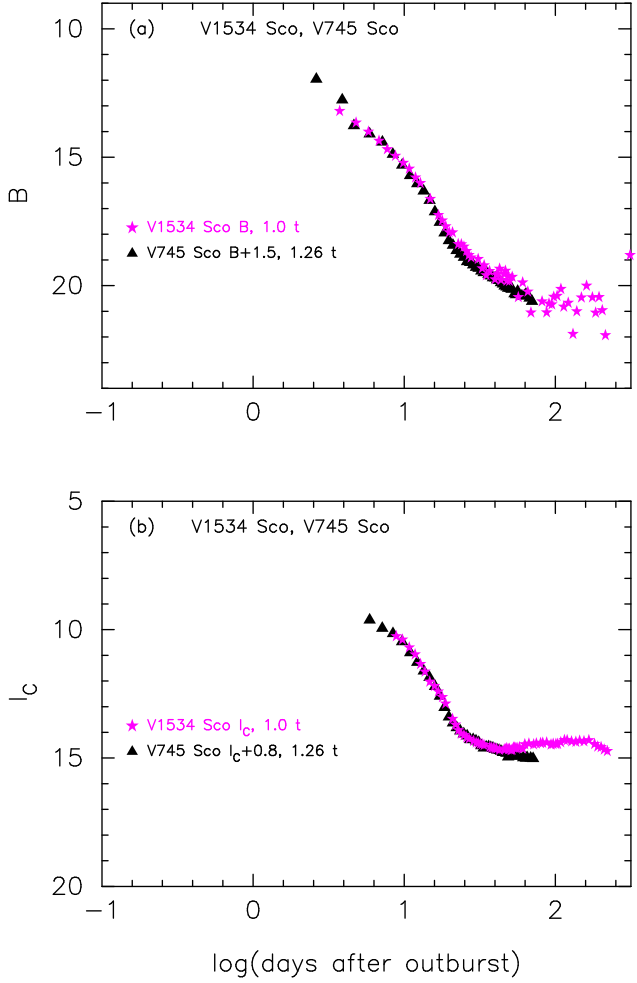}
\caption{
The (a) $B$ and (b) $I_C$ light curves of V1534~Sco and V745~Sco.
The filled magenta stars denote the $B$ and $I_C$ magnitudes of V1534~Sco
while the filled black triangles represent the $B$ and $I_C$ magnitudes
of V745~Sco.  The light curve of V745~Sco is stretched by $f_{\rm s}=1.26$
and shifted down by $\Delta B=1.5$ and $\Delta I_C=0.8$, as listed 
in the figure.
\label{v1534_sco_v745_sco_b_i_2fig}}
\end{figure}


\begin{figure}
\includegraphics[height=16.0cm]{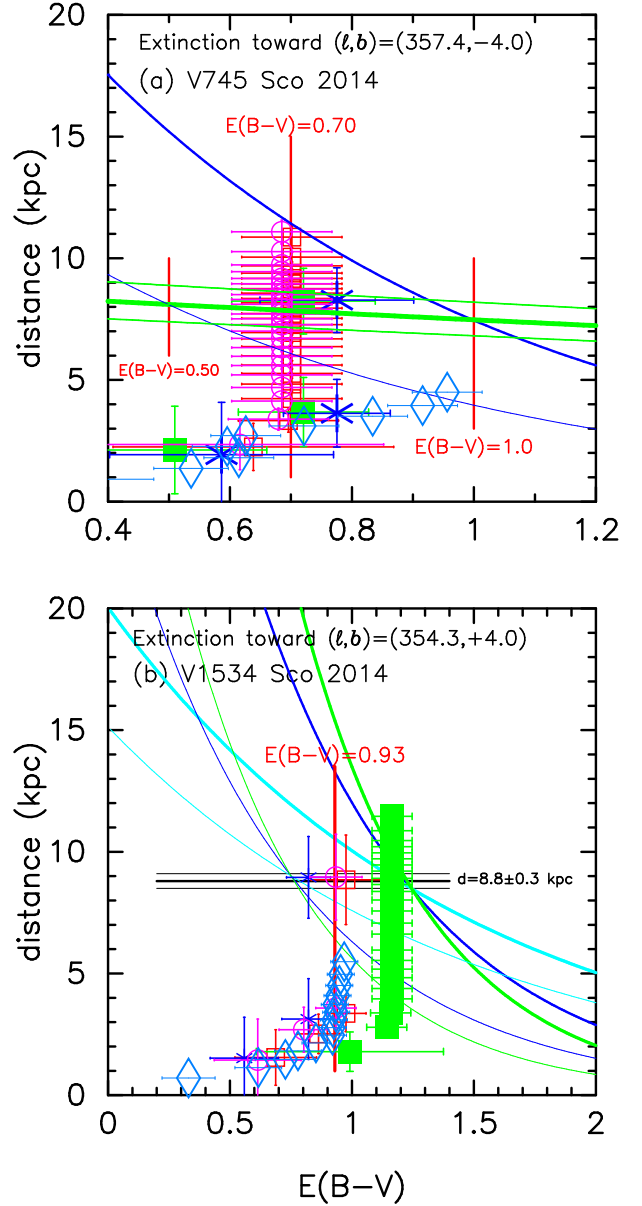}
\caption{
Distance-reddening relation toward (a) V745~Sco and (b) V1534~Sco.
Each symbol and lines are the same as those in Figure
\ref{distance_reddening_v745_sco_t_crb_v838_her_v1534_sco_no2}(a)
and (d), respectively.  
However, we plot various distance-reddening relations
assuming different reddenings of $E(B-V)=1.0$ (thick solid lines)
and $E(B-V)=0.5$ (thin solid lines) for V745~Sco for
comparison.  In panel (a), the thick/thin solid blue lines denote
$(m-M)_V=17.5$/16.1, which are just calculated from Equation
(\ref{v_distance_modulus}) together with $d=7.5$/8.1~kpc and 
$E(B-V)=1.0$/0.5, not a result of fitting.  
In panel (b), the thick/thin solid green, blue, and cyan lines denote
$(m-M)_B= 19.75$/17.85, $(m-M)_V= 18.5$/17.1, and $(m-M)_I= 16.5$/15.9,
respectively.  These three lines do not exactly 
but broadly cross at $d=8.5$/9.1~kpc and
$E(B-V)=1.24$/0.74, respectively.  See the text for detail.
\label{distance_reddening_v745_sco_v1534_sco_2fig_100}}
\end{figure}

\subsection{V1534~Sco 2014}
\label{v1534_sco}
V1534~Sco is a classical nova in a symbiotic system \citep{jos15}.
Figure \ref{v1534_sco_v_bv_ub_color_curve} shows (a) the $V$ and
X-ray fluxes and (b) $(B-V)_0$ color evolutions of V1534~Sco.
Here, $(B-V)_0$ are dereddened with $E(B-V)=0.93$ as explained below.
V1534~Sco reached $m_{V,\rm max}=11.8$
on JD~2456744.27 (UT 2014 March 27.77) from the data of the 
Variable Star Observers League of Japan (VSOLJ).
The AAVSO $V$ data becomes flat when the $V$ goes down to $V\sim16.3$.
This is an artifact due to the $V$ flux being contaminated by nearby stars
\citep[e.g.,][]{mun17}.  Therefore, we use only the data of SMARTS
\citep{wal12} in the following analysis.
The nova declined with $t_2=6\pm0.3$~day and was identified
as a He/N nova by \citet{jos15}. \citet{jos15} also obtained
the reddening of $E(B-V)=0.91$ from the empirical relations derived
by \citet{van87}, i.e., the intrinsic color of $(\bv)_0=0.23\pm0.06$
at maximum and $(\bv)_0=-0.02\pm0.04$ at time $t_2$.
From the NIR spectra of this nova, \citet{jos15}
concluded that the nova outbursted in a symbiotic system with an M5III 
($\pm$ two subclasses) RG companion and suggested that V1534~Sco
is a recurrent nova such as T~CrB, RS~Oph, and V745~Sco.
They estimated the distance to the nova as 8.1, 9.6, 13.0, 18.6, and 26.4~kpc
depending on the subclass, M3III, M4III, M5III, M6III, and M7III,
respectively.

\subsubsection{Timescaling law and time-stretching method}
Figure \ref{v1534_sco_v838_her_t_crb_v745_sco_v_bv_ub_logscale} shows
the $V$ light and $(B-V)_0$ color curves of V1534~Sco on a logarithmic
timescale. We add the light/color curves of the symbiotic recurrent novae
V745~Sco and T~CrB and the very fast nova V838~Her.
The $V$ light curves of four novae overlap each other, i.e.,
Equation (\ref{overlap_brigheness}) is satisfied.  
Here, we determine the horizontal shift of the V745~Sco $V$ light curve
with respect to V1534~Sco as $\Delta \log t = \log f_{\rm s}=0.10$. 
The positions of the other two novae are uniquely determined from Figure
\ref{v838_her_t_crb_v745_sco_v_bv_ub_logscale}.

We found no reliable distance or distance modulus to V1534~Sco
in the literature.
Therefore, we apply Equation (\ref{distance_modulus_formula}) to Figure
\ref{v1534_sco_v838_her_t_crb_v745_sco_v_bv_ub_logscale} and obtain 
the distance modulus in the $V$ band relative to the three novae as
\begin{eqnarray}
(m&-&M)_{V, \rm V1534~Sco} \cr 
&=& (m - M + \Delta V)_{V, \rm V745~Sco} - 2.5 \log 1.26 \cr
&=& 16.6\pm0.2 + 1.3\pm0.1 - 0.25 = 17.55\pm0.2 \cr
&=& (m - M + \Delta V)_{V, \rm T~CrB} - 2.5 \log 1.26 \cr
&=& 10.1\pm0.3 + 7.8\pm0.1 - 0.25 = 17.55\pm0.3 \cr
&=& (m - M + \Delta V)_{V, \rm V838~Her} - 2.5 \log 1.0 \cr
&=& 13.7\pm0.3 + 3.9\pm0.1 - 0.0 = 17.6\pm0.3,
\label{distance_modulus_v1534_sco_v745_sco}
\end{eqnarray}
where we adopt $(m-M)_{V, \rm V745~Sco}=16.6\pm0.2$ in Section \ref{v745_sco},
$(m-M)_{V, \rm T~CrB}=10.1\pm0.3$ in Section \ref{t_crb},
and $(m-M)_{V, \rm V838~Her}=13.7\pm0.3$ in Section \ref{v838_her}.
Thus, we obtain $(m-M)_V=17.6\pm0.3$ and $f_{\rm s}=1.26$ 
(against V745~Sco) for V1534~Sco.

Figure \ref{v1534_sco_v745_sco_b_i_2fig} shows the (a) $B$ and (b)
$I_C$ magnitudes of V1534~Sco and V745~Sco on a logarithmic timescale.
These two nova light curves overlap to each other.  Therefore, 
we again apply Equation (\ref{distance_modulus_formula}) to 
the $B$ magnitudes of V1534~Sco and V745~Sco in Figure
\ref{v1534_sco_v745_sco_b_i_2fig}(a), and obtain
\begin{eqnarray}
(m&-&M)_{B, \rm V1534~Sco} \cr 
&=& (m - M + \Delta B)_{B, \rm V745~Sco} - 2.5 \log 1.26 \cr
&=& 17.3\pm0.2 + 1.5\pm0.1 - 0.25 = 18.55\pm0.2,
\label{distance_modulus_v1534_sco_v745_sco_b}
\end{eqnarray}
where we adopt the absorption law of $A_B=4.1\times E(B-V)$ from
\citet{rie85} and use the distance-reddening relation of
\begin{equation}
(m-M)_B = 4.1\times E(B-V) + 5 \log (d/10~{\rm pc}),
\label{distance_modulus_relation_b}
\end{equation}
and $(m-M)_{B, \rm V745~Sco}=16.6+1.0\times 0.7= 17.3$.
Thus, we obtain $(m-M)_B=18.55\pm0.2$ for V1534~Sco.

We further apply Equation (\ref{distance_modulus_formula}) to 
the $I_C$ magnitudes of V1534~Sco and V745~Sco in Figure
\ref{v1534_sco_v745_sco_b_i_2fig}(b), and obtain
\begin{eqnarray}
(m&-&M)_{I, \rm V1534~Sco} \cr 
&=& (m - M + \Delta I_C)_{I, \rm V745~Sco} - 2.5 \log 1.26 \cr
&=& 15.55\pm0.2 + 0.8\pm0.1 - 0.25 = 16.1\pm0.2,
\label{distance_modulus_v1534_sco_v745_sco_i}
\end{eqnarray}
where we adopt the absorption law of $A_I=1.5\times E(B-V)$ from
\citet{rie85} and use the distance-reddening relation of
\begin{equation}
(m-M)_I = 1.5\times E(B-V) + 5 \log (d/10~{\rm pc}),
\label{distance_modulus_relation_i}
\end{equation}
and $(m-M)_{I, \rm V745~Sco}=16.6-1.6\times 0.7= 15.55$.
Thus, we obtain $(m-M)_I=16.1\pm0.2$ for V1534~Sco.

We plot the distance-reddening relations of Equations 
(\ref{distance_modulus_relation_b}), (\ref{v_distance_modulus}), 
and (\ref{distance_modulus_relation_i}) 
for $(m-M)_B=18.55$ (green line), $(m-M)_V=17.6$ (blue line),
and $(m-M)_I=16.1$ (cyan line) for V1534~Sco
in Figure \ref{distance_reddening_v745_sco_t_crb_v838_her_v1534_sco_no2}(d).
These three lines consistently cross at $d=8.8$~kpc and $E(B-V)=0.93$.
This demonstrates an independent consistency check of our distance
and reddening even if we assume the time-stretching method.

\subsubsection{Reddening and distance}
\label{reddening_distance_v1534sco}
We examine this result of $(m-M)_V=17.6$ and $E(B-V)=0.93$
from various points of view.
For the reddening toward V1534~Sco, whose
galactic coordinates are $(l,b)=(354\fdg3345, +3\fdg9915)$,
the NASA/IPAC Galactic dust absorption
map gives $E(B-V)=0.93\pm0.05$.
This value is close to the value of $E(B-V)=0.91$ given by \citet{jos15}
and consistent with our cross point of $d=8.8$~kpc and $E(B-V)=0.93$.
Therefore, we adopt $E(B-V)=0.93$ and further examine 
whether this value is reasonable or not.

Figure \ref{distance_reddening_v745_sco_t_crb_v838_her_v1534_sco_no2}(d)
shows various distance-reddening relations toward V1534~Sco.
The vertical solid red line denotes the reddening of $E(B-V)=0.93$. 
The solid green, blue, and cyan lines denote the distance moduli 
in the $B$, $V$, and $I_C$ bands, i.e., $(m-M)_B=18.55$,
$(m-M)_V=17.6$, and $(m-M)_I=16.1$.
These four lines cross at $E(B-V)=0.93$ and $d=8.8$~kpc. 
The relations of Marshall et al. (2006) are plotted
in four directions close to the direction of V1534~Sco:
$(l, b)=(354\fdg25, +3\fdg75)$ (open red squares),
$(354\fdg50, +3\fdg75)$ (filled green squares),
$(354\fdg25, +4\fdg00)$ (blue asterisks), and
$(354\fdg50, +4\fdg00)$ (open magenta circles).
The direction of V1534~Sco is midway between those of the blue asterisks
and open magenta circles.
The open cyan-blue diamonds show the relation of \citet{ozd16}, which
is roughly consistent with that of Marshall et al. until $d=6$~kpc. 
The cross point at $E(B-V)=0.93$ and $d=8.8$~kpc is consistent with
the relation of Marshall et al.
Then, the vertical distance from the galactic plane is approximately 
$z=+610$~pc.
Thus, it is likely that V1534~Sco belongs to the galactic bulge
\citep{mun17}.

\subsubsection{Color-magnitude diagram}
\label{color_magnitude_v1534sco}
Using $E(B-V)=0.93$ and $(m-M)_V=17.6$ ($d=8.8$~kpc), 
we plot the color-magnitude diagram of V1534~Sco in Figure
\ref{hr_diagram_v745_sco_v1534_sco_rs_oph_v407_cyg_outburst}(b).
The track of V1534~Sco (filled red circles) is located closely to
that of V745~Sco (open magenta diamonds).
These two tracks almost overlap apart from the difference
in the peak brightness:  V1534~Sco has $M_{V,\rm max}=11.8 -
17.6 = -5.8$, whereas V745~Sco reached $M_{V,\rm max}=8.66 -
16.6 = -7.94$.   
This overlap supports our derived values of $E(B-V)=0.93$ 
and $(m-M)_V=17.6$ ($d=8.8$~kpc). 
We conclude that the distance of $d=8.8\pm0.9$~kpc and the reddening of
$E(B-V)=0.93\pm0.05$ are reasonable.  
Thus, we confirm that Equations (\ref{overlap_brigheness}) and 
(\ref{distance_modulus_formula}) are satisfied for V1534~Sco.


\subsubsection{Consistency check with V745~Sco}
\label{consistency_check_v1534sco}
The three relations, i.e., $(m-M)_B=18.55$, $(m-M)_V=17.6$, 
and $(m-M)_I=16.1$ for V1534~Sco, consistently cross
at the point of $d=8.8$~kpc and $E(B-V)=0.93$ 
in the distance-reddening relation in Figure 
\ref{distance_reddening_v745_sco_t_crb_v838_her_v1534_sco_no2}(d).
These three distance moduli are calculated from Equations
(\ref{distance_modulus_v1534_sco_v745_sco_b}), 
(\ref{distance_modulus_v1534_sco_v745_sco}), 
and (\ref{distance_modulus_v1534_sco_v745_sco_i}) using
V745~Sco's $(m-M)_B=17.3$, $(m-M)_V=16.6$, and $(m-M)_I=15.55$.
However, these V745~Sco's values are calculated 
assuming the reddening of $E(B-V)=0.70$.
We check the dependency 
on the reddening of V745~Sco.

If we adopt a different value, for example, $E(B-V)=1.0$, we have
a different cross point as shown in Figure
\ref{distance_reddening_v745_sco_v1534_sco_2fig_100}(a).  Here 
we assume $(m-M)_K=8.33 - (-6.39)= 14.72$
(only this value is fixed from Equation
(\ref{equation_MkP}) and $m_K=8.33$).  Then, we calculate 
distance modulus in each band as
$(m-M)_B=(m-M)_K + 3.75 \times E(B-V) =18.5$, 
$(m-M)_V=(m-M)_K + 2.75 \times E(B-V)= 17.5$,
and $(m-M)_I=(m-M)_K + 1.15 \times E(B-V)= 15.9$ for V745~Sco.
Using these different values in Equations
(\ref{distance_modulus_v1534_sco_v745_sco_b}), 
(\ref{distance_modulus_v1534_sco_v745_sco}), 
and (\ref{distance_modulus_v1534_sco_v745_sco_i}), we obtain
$(m-M)_B=19.75$, $(m-M)_V=18.5$, and $(m-M)_I=16.5$ for V1534~Sco.
These new three lines do not exactly but broadly cross at $d=8.5$~kpc and
$E(B-V)=1.24$ as plotted (thick solid lines) in Figure
\ref{distance_reddening_v745_sco_v1534_sco_2fig_100}(b).
This reddening of $E(B-V)=1.24$ is much larger than the reddening
of $E(B-V)=0.91$ obtained by \citet{jos15} (or $E(B-V)=0.93$ of our
cross point in Figure
\ref{distance_reddening_v745_sco_t_crb_v838_her_v1534_sco_no2}(d)).

Similarly if we adopt a smaller reddening of $E(B-V)=0.50$ for V745~Sco,
we obtain $(m-M)_B=(m-M)_K + 3.75 \times E(B-V) =16.6$, 
$(m-M)_V=(m-M)_K + 2.75 \times E(B-V)= 16.1$,
and $(m-M)_I=(m-M)_K + 1.15 \times E(B-V)= 15.3$ for V745~Sco.
Then, we obtain $(m-M)_B=17.85$, $(m-M)_V=17.1$, 
and $(m-M)_I=15.9$ for V1534~Sco.
These three lines roughly cross at $d=9.1$~kpc and $E(B-V)=0.74$
as plotted (thin solid lines) in Figure
\ref{distance_reddening_v745_sco_v1534_sco_2fig_100}(b).
This value of $E(B-V)=0.74$ is much smaller than $E(B-V)\approx 0.9$.
Therefore, such a smaller value of $E(B-V)=0.50$ for V745~Sco is
not supported.

We have already discussed the distance and reddening 
in Sections \ref{reddening_distance_v1534sco} 
and \ref{color_magnitude_v1534sco},
and concluded that the reddening of $E(B-V)=0.93$ is
reasonable for V1534~Sco.
In other words, only the reddening of $E(B-V)\approx 0.7$ for V745~Sco
is consistent with the reddening of $E(B-V)\approx 0.9$ for V1534~Sco.
We should also note that the distance of V1534~Sco is well constrained
to $d=8.8\pm0.3$~kpc even for a wide range of $E(B-V)=0.5-1.0$ for V745~Sco. 
This analysis confirms that only the two sets of $d=7.8$~kpc and 
$E(B-V)=0.70$ for V745~Sco and $d=8.8$~kpc and $E(B-V)=0.93$ 
for V1534~Sco are consistent with each other in the distance-reddening
relations and color-magnitude diagram.


\begin{figure}
\includegraphics[height=8.5cm]{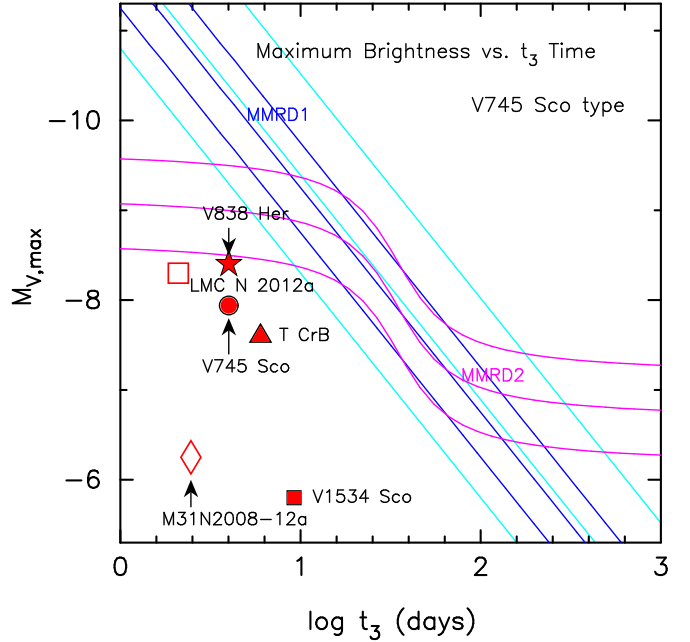}
\caption{
Maximum magnitude vs. rate of decline relations for the 
rapid-decline (V745~Sco) type novae. Each symbol is labeled with
a nova name. The filled symbols correspond to novae in our galaxy, whereas
the open symbols denote novae in extra-galaxies. 
Kaler-Schmidt's law \citep[labeled ``MMRD1,''][]{sch57}
is denoted by a blue line with two attendant lines corresponding to the cases of
$\pm 0.5$ mag brighter/fainter. Della Valle \& Livio's law 
\citep[labeled ``MMRD2,''][]{del95} 
is indicated by a magenta line flanked with $\pm 0.5$ mag
brighter/fainter lines.
The three cyan lines represent Equations
(\ref{theoretical_MMRD_relation_v1668_cyg_upper}),
(\ref{theoretical_MMRD_relation_v1668_cyg}), and 
(\ref{theoretical_MMRD_relation_v1668_cyg_lower}) in Appendix
\ref{mmrd}, from upper to lower. These cyan lines envelop the
galactic novae studied by \citet{dow00}, as shown in Figure
\ref{max_t3_scale_no4} of Appendix \ref{mmrd}.
We call this region the broad MMRD relation.
\label{max_t3_scale_v745_sco_type}}
\end{figure}

\subsection{Faint locations much below the MMRD relations}
\label{rapid_mmrd_relation}
It has been frequently discussed that very fast novae and 
recurrent novae sometimes deviate from the 
MMRD relations \citep[e.g.,][]{schaefer10a, hac10k, hac15k, hac16k}. 
Here, the two empirical MMRD relations are defined as
(Kaler-Schmidt's law, MMRD1) 
\begin{equation}
M_{V, {\rm max}} = -11.75 + 2.5 \log t_3,
\label{kaler-schmidt-law}
\end{equation}
by \citet{sch57}, and as (Della Valle \& Livio's law, MMRD2)
\begin{equation}
M_{V, {\rm max}} = -7.92 -0.81 \arctan \left(
{{1.32-\log t_2} \over {0.23}} \right),
\label{della-valle-livio-law}
\end{equation}
by \citet{del95}, where
$M_{V, {\rm max}}$ is the absolute $V$ magnitude at maximum,
($t_2$ or) $t_3$ is the period of days during which the nova decays 
by (two or) three magnitudes from the $V$ maximum.  We use 
$t_2\approx0.6 \times t_3$ \citep{hac06kb} to calculate $t_2$ from
$t_3$ in Equation (\ref{della-valle-livio-law}).
Kaler-Schmidt's law is denoted in Figure \ref{max_t3_scale_v745_sco_type}
by a blue solid line with two attendant blue solid lines,
corresponding to $\pm 0.5$ mag brighter/fainter cases.
Della Valle \& Livio's law is indicated by
a magenta solid line flanked with $\pm 0.5$ mag brighter/fainter cases.

As mentioned in Section \ref{introduction},
\citet{hac10k} theoretically examined the MMRD law on the basis of 
their universal decline law.
They showed that the main trend of the MMRD relation is governed
by the WD mass (timescaling factor of $f_{\rm s}$) and
the second parameter (the initial envelope mass, i.e., the ignition mass)
causes large scatter around the main trend of the MMRD relations.
\citet{hac10k} reproduced the distribution of MMRD points 
summarized by \citet{dow00}. We plot Hachisu \& Kato's results with 
the three cyan lines in Figure \ref{max_t3_scale_v745_sco_type},
which represent Equations
(\ref{theoretical_MMRD_relation_v1668_cyg_upper}),
(\ref{theoretical_MMRD_relation_v1668_cyg}), and 
(\ref{theoretical_MMRD_relation_v1668_cyg_lower}) in Appendix
\ref{mmrd}, from top to bottom.  These three cyan lines envelop
the MMRD points studied by \citet{dow00} as shown in Figure
\ref{max_t3_scale_no4} of Appendix \ref{mmrd}.  We call this region
the broad MMRD relation.

In Figure \ref{max_t3_scale_v745_sco_type}, we plot the 
MMRD points $(t_3, M_{V,\rm max})$ of the V745~Sco 
(rapid-decline) type novae in our galaxy, LMC, and M31, some of which (filled symbols)
are discussed in this section and the others (open symbols)
are examined in Section \ref{novae_lmc_smc_m31}. 
They are tabulated in Table \ref{physical_properties_recurrent_novae}.
All of them are far outside the broad MMRD relation (solid cyan lines).
It is clear that the MMRD relations cannot be applied to 
the rapid-decline type novae.


\begin{figure}
\includegraphics[height=5.5cm]{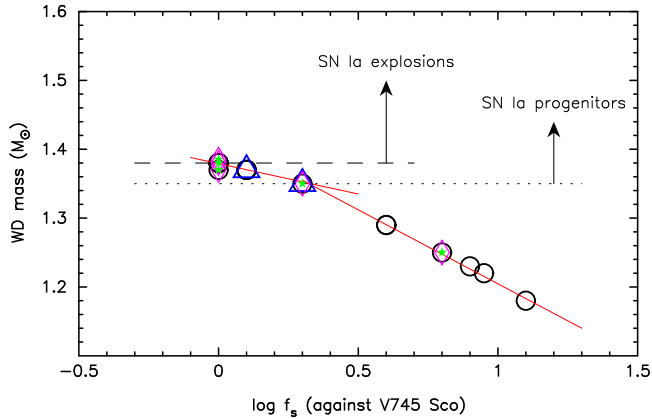}
\caption{
WD mass vs. timescaling factor $\log f_{\rm s}$.
The timescaling factor is measured against that of V745~Sco
($f_{\rm s}=1.0$ for V745~Sco).
The data are taken from Table \ref{wd_mass_recurrent_novae}.
See the text for more detail.
\label{timescale_wd_mass}}
\end{figure}

\subsection{WD mass vs. timescaling factor}
\label{wd_mass_vs_timescale}
Figure \ref{timescale_wd_mass} shows the WD mass against the timescaling
factor of $\log f_{\rm s}$, where the timescaling factor is measured 
based on that of V745~Sco ($f_{\rm s}=1.0$ for V745~Sco).
The WD mass of V745~Sco ($\log f_{\rm s}=0.0$)
is estimated to be $M_{\rm WD}=1.385~M_\sun$
from the X-ray model light curve fitting in Section \ref{v745_sco}.
The WD mass of M31N~2008-12a ($\log f_{\rm s}=0.0$)
was suggested to be $M_{\rm WD}=1.38~M_\sun$ from the X-ray and
optical model light curve fitting \citep[][see also Section
\ref{m31n2008_12a} below]{kat17sh}.
The WD mass of V838~Her ($\log f_{\rm s}=0.1$)
is obtained to be $M_{\rm WD}=1.37~M_\sun$ from the $V$ and UV~1455~\AA\  
model light curve fitting in Section \ref{v838_her}.
We plot these WD masses with open blue triangles (UV~1455~\AA),
open magenta diamonds ($t_{\rm SSS-on}$), and 
filled green stars ($t_{\rm SSS-off}$), as listed 
in Table \ref{wd_mass_recurrent_novae}. 

Assuming that the WD mass is linearly related to the timescaling
factor of $\log f_{\rm s}$ between $\log f_{\rm s}=0.0$ and 
$\log f_{\rm s}=0.3$ (and differently related between $\log f_{\rm s}=0.3$ 
and $\log f_{\rm s}=1.1$), we determine each WD mass of the novae
(large open black circles) as shown in Figure \ref{timescale_wd_mass}.     
We estimate the ambiguity of the WD mass determination
could be $\pm (0.01-0.02)~M_\sun$ from this linear relation.  
The main reason is the difference in the chemical composition of
hydrogen-rich envelope as shown in the case of V838~Her 
($1.37~M_\sun$ vs. $1.35~M_\sun$).

Among these 14 novae, we have already analyzed 4 novae in this section, i.e.,
T~CrB ($\log f_{\rm s}=0.0$, $1.38~M_\sun$, RG, $P_{\rm orb}=227.6$~days), 
V838~Her ($\log f_{\rm s}=0.1$, $1.37~M_\sun$, MS, $P_{\rm orb}=0.2976$~days), 
V745~Sco ($\log f_{\rm s}=0.0$, $1.385~M_\sun$, RG, $P_{\rm orb}=$unknown),
and V1534~Sco ($\log f_{\rm s}=0.1$, $1.37~M_\sun$, RG, 
$P_{\rm orb}=$unknown). 
It is unlikely that these WDs were born as massive 
as they are \citep[$\gtrsim 1.37~M_\sun$, see, e.g.,][]{doh15}.
We suppose that these WDs have grown in mass.
This strongly suggests further increases in 
the WD masses in these systems.

The WD masses of SN~Ia progenitors should be close to or exceed
the SN~Ia explosion mass of $M_{\rm Ia}=1.38~M_\sun$ \citep{nom82},
as mentioned in Section \ref{introduction}.
The typical mass-increasing rates of WDs 
are $\dot M_{\rm WD}\sim 1\times10^{-7}~M_\sun$~yr$^{-1}$ 
just below the stability line of hydrogen shell burning, that is,
in relatively short-recurrence-period novae 
\citep[see, e.g.,][]{kat17sh}.
The WD mass increases from $1.37~M_\sun$ to 
$M_{\rm Ia}=1.38~M_\sun$ and explodes as a SN~Ia. 
It takes approximately $t_{\rm Ia}\sim 0.01~M_\sun/
1\times10^{-7}~M_\sun$~yr$^{-1} = 1\times10^5$~yr, which is much shorter
than the evolution timescale of the donor (RG star or 
MS star).  Therefore, these WDs 
have grown in mass and will explode as a SN~Ia if the core
consists of carbon and oxygen.  

Note that V838 Her is identified as a neon nova because the
ejecta are enriched by neon.  The neon rich ejecta, however, do not
always mean that the underling WD has an oxygen-neon core.
A mass-increasing WD, like in some recurrent novae, develops
a helium layer underneath the hydrogen burning zone and experiences
periodic helium shell flashes \citep[e.g.,][]{wu17, kat17sh}.
Helium burning produces neon and other heavy elements that remain
after the helium shell flash and mixed into the freshly accreted
hydrogen-rich matter.  The next recurrent nova outburst could show
strong neon lines.  Thus, the neon nova identification should not be
directly connected to an oxygen-neon core when the WD mass is close to
the Chandrasekhar mass.  Considering this possibility and unlikely
born massive WD, we regard V838 Her is a candidate of SN Ia progenitors.

\subsection{Summary of the rapid-decline (V745~Sco) type novae}
\label{summary_v745_sco_novae}
We summarize the results on the four rapid-decline type novae. 
\begin{enumerate}
\item We analyzed four very fast novae including two recurrent novae,
V745~Sco, T~CrB, V838~Her, and V1534~Sco.  
We obtained the distances, distance moduli in the $V$ band,
and reddenings of the four novae
using various methods. The results are summarized in Table
\ref{extinction_distance_various_novae}. These novae are located
significantly above or below the scale height of galactic matter distribution
\citep[$z=\pm125$~pc, see, e.g.,][]{mar06}.
\item The $V$ light curves of the four novae almost overlap 
when we properly stretch their timescales by a factor of $f_{\rm s}$
and shift up or down their $V$ light curves by $\Delta V$ (see Figure
\ref{v1534_sco_v838_her_t_crb_v745_sco_v_template_no2}).
This means that these novae satisfy the timescaling law of 
Equation (\ref{overlap_brigheness}).
Utilizing the obtained distance moduli in the $V$ band and the 
time-stretching factor of $f_{\rm s}$, we confirm that 
these four novae satisfy the time-stretching method of
Equation (\ref{distance_modulus_formula}). 
The time-stretching method is applicable to the rapid-decline type
novae including recurrent novae.
\item All the four novae are substantially fainter than the MMRD
relations. In particular, V1534~Sco is located significantly below 
Kaler-Schmidt's law (MMRD1) and Della Valle \& Livio's law (MMRD2).
This means that the MMRD relations cannot be applied to the rapid-decline
type novae.
\item The WD mass of V745~Sco is estimated to be $M_{\rm WD}=1.385~M_\sun$
from our model light curve fitting with the supersoft X-ray light curve. 
This WD mass is more massive than $M_{\rm WD}=1.38~M_\sun$ of the 1-yr
recurrence period nova, M31N~2008-12a. This is consistent with the earlier
appearance of the SSS phase of V745~Sco ($t_{\rm SSS-on}\sim 4$~days)
than that of M31N~2008-12a ($t_{\rm SSS-on}\sim 6$~days).   
\item The WD mass of V838~Her is independently estimated 
to be $M_{\rm WD}=1.37~M_\sun$ from our model light curve fitting
with the UV~1455~\AA\ light curve. 
This is consistent with the timescaling factor $f_{\rm s}=1.26$ 
of V838~Her, which is slightly longer than $f_{\rm s}=1.0$ of V745~Sco
($M_{\rm WD}=1.385~M_\sun$), suggesting that its WD mass is
smaller than that of V745~Sco.  
\item The rapid-decline type novae have a timescaling factor of
$f_{\rm s}=1.0$ and $f_{\rm s}=1.26$. Therefore, their
WD masses are $M_{\rm WD}=1.38~M_\sun$ (or $1.385~M_\sun$)
and $M_{\rm WD}=1.37~M_\sun$, respectively.
It is unlikely that the WDs were born as massive as they are.
These WDs should have grown in mass after they were born.
This supports that the rapid-decline type novae are
immediate progenitors of SNe~Ia if their WDs have a carbon-oxygen core.
\end{enumerate}


\begin{figure}
\includegraphics[height=13cm]{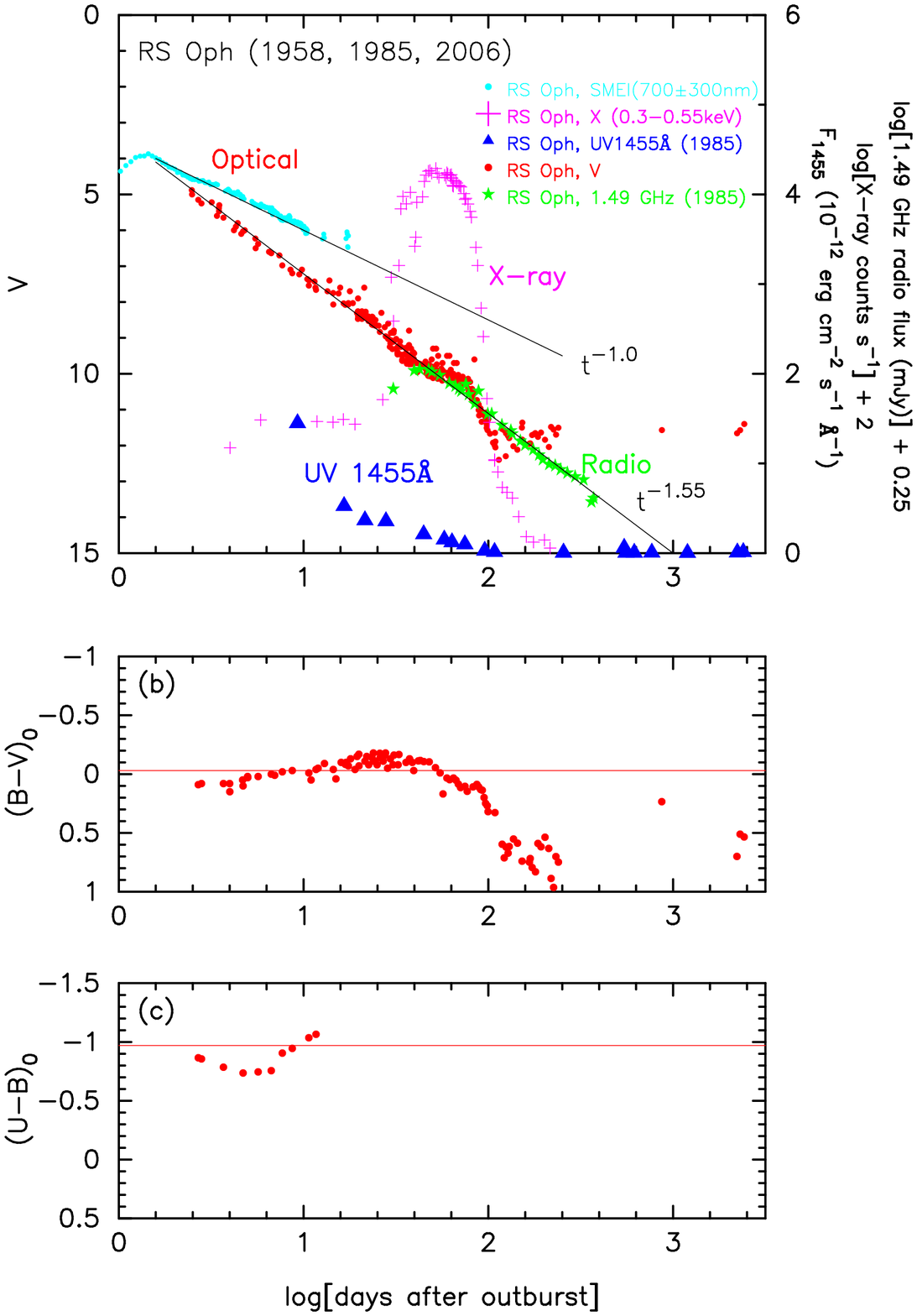}
\caption{
Same as Figure \ref{v745_sco_v_bv_ub_xray_logscale},
but for the RS~Oph 1958, 1985, and 2006 outbursts.
The colors are dereddened with $E(B-V)=0.65$.  
(a) The $V$ data (filled red circles) are taken from \citet{siv06},
AAVSO, and SMARTS for the 2006 outburst and from \citet{con58} 
for the 1958 outburst. The SMEI band data (filled cyan circles)
are from \citet{hou10} for the 2006 outburst.  
We also plot the supersoft X-ray \citep[$0.3-0.55$ keV,
magenta plus,][]{hac07kl} for the 2006 outburst, UV~1455~\AA\  band 
\citep[filled blue triangles,][]{cas02} and radio (1.49 GHz) band 
fluxes \citep[filled green stars,][]{hje86} for the 1985 outburst.
(b) The $B-V$ data (filled red circles) are from \citet{siv06},
AAVSO, and SMARTS for the 2006 outburst and from \citet{con58} 
for the 1958 outburst.
(c) The $U-B$ data (filled red circles) are taken from \citet{con58}
for the 1958 outburst.
\label{rs_oph_radio_v_ub_bv_uv_x_logscale_no4}}
\end{figure}

\section{Timescaling Law of CSM-shock Novae}
\label{rs_oph_type}
We analyze the light curves of RS~Oph and V407~Cyg and show that these 
two novae follow a timescaling law if we consider the interaction
between ejecta and CSM.
We call this group of novae the CSM-shock (RS~Oph) type novae.


\begin{figure*}
\plotone{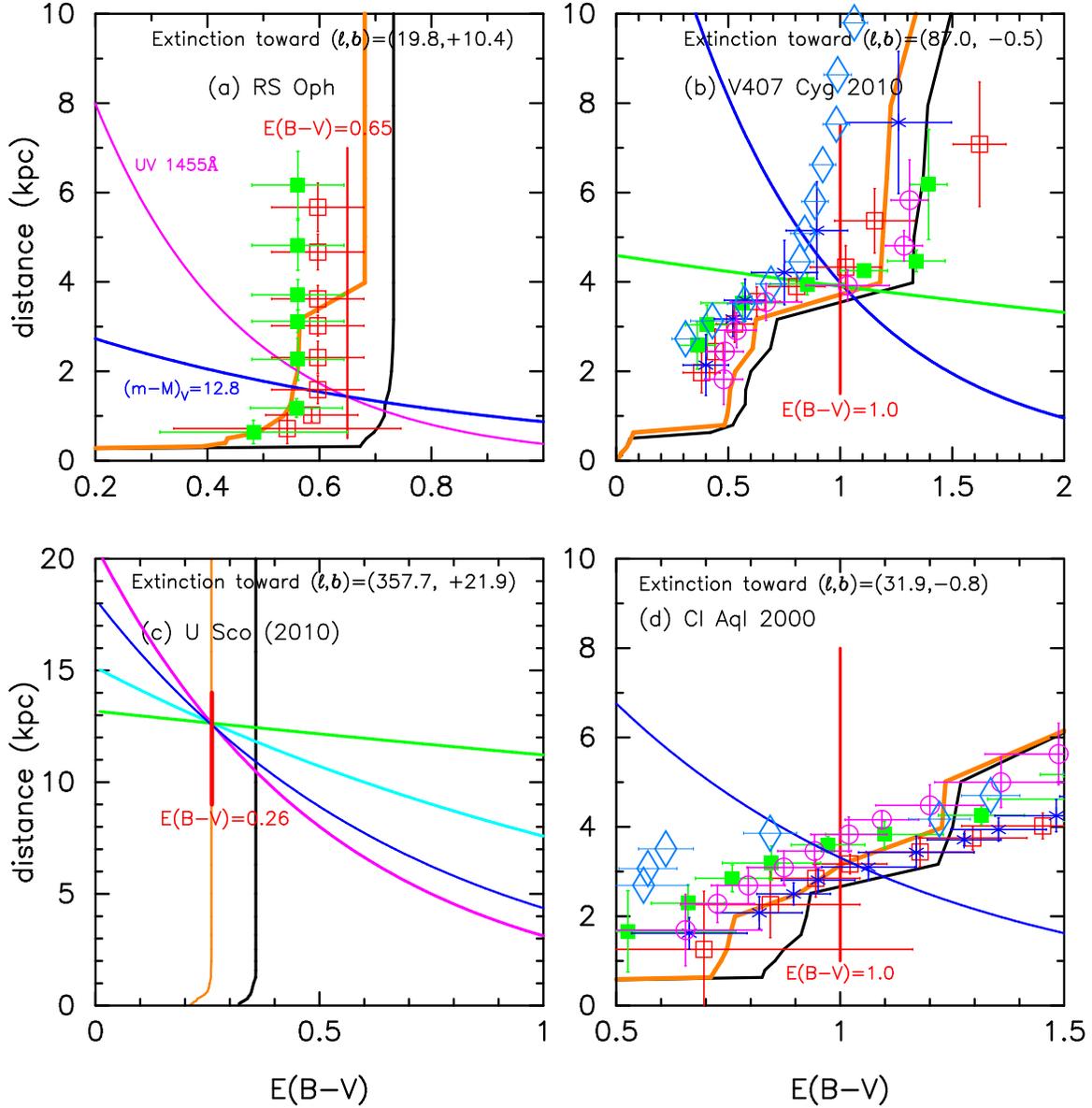}
\caption{
Same as Figure \ref{distance_reddening_v745_sco_t_crb_v838_her_v1534_sco_no2},
but for (a) RS~Oph, (b) V407~Cyg, (c) U~Sco, and (d) CI~Aql.
The thick solid blue lines denote
(a) $(m-M)_V=12.8$, (b) $(m-M)_V=16.1$, (c) $(m-M)_V=16.3$,
and (d) $(m-M)_V=15.7$.  In panel (c), we add three distance-reddening
relations of Equations (\ref{distance_modulus_relation_b}),
(\ref{distance_modulus_relation_i}), and 
(\ref{equation_distmod_IR}) for U~Sco, that is,
$(m-M)_B=16.6$ (magenta line), 
$(m-M)_I=15.9$ (cyan line), and $(m-M)_K=15.6$ (green line).
\label{distance_reddening_rs_oph_v407_cyg_u_sco_ci_aql_no2}}
\end{figure*}

\subsection{RS~Oph (2006)}
\label{rs_oph}
RS~Oph is a recurrent nova with six recorded outbursts
in 1898, 1933, 1958, 1967, 1985, and 2006 \citep[e.g.,][]{schaefer10a}.
The orbital period of 453.6 days was obtained by \citet{bra09}.
Figure \ref{rs_oph_radio_v_ub_bv_uv_x_logscale_no4} shows
(a) the $V$ magnitudes, SMEI magnitudes 
\citep{hou10}, radio (1.49 GHz) fluxes \citep{hje86}, UV~1455~\AA, 
and X-ray light curves, (b) $(B-V)_0$, and (c) $(U-B)_0$ color curves
of RS~Oph on a logarithmic timescale.  
The $V$ magnitude reaches $m_{V, \rm max}=4.8$
and declines with $t_2=6.8$ and $t_3=14$~days \citep{schaefer10a}.
The WD mass of RS~Oph was estimated to be $M_{\rm WD}=1.35\pm0.01~M_\sun$
by \citet{hac06b, hac07kl} from the model $V$ 
and supersoft X-ray light curve fittings.
Thus, we regard the WD mass of RS~Oph to be $1.35~M_\sun$.
\citet{hac01kb} argued that RS~Oph is a progenitor of SNe~Ia
because the WD mass is close to the SN~Ia explosion mass of 
$M_{\rm Ia}=1.38~M_\sun$ and now increases.
\citet{mik17} also reached a similar conclusion. 

The $V$ and radio light curves clearly show the trend of 
$F_\nu\propto t^{-1.55}$, as shown in
Figure \ref{rs_oph_radio_v_ub_bv_uv_x_logscale_no4}(a).
However, the SMEI magnitude light curve
has a different trend of $L_{\rm SMEI}\propto t^{-1.0}$,
where $L_{\rm SMEI}$ is the luminosity of the SMEI band.
This is because the SMEI magnitude is a wide-band (peak quantum
efficiency at 700 nm with a full width at half maximum (FWHM) of 300 nm) and include the flux
of very strong H$\alpha$ line, which mainly comes from the shock
interaction. RS~Oph has a RG companion and the companion star
emits cool slow winds \citep[$\sim40$~km~s$^{-1}$, see][]{iij09},
which form CSM
around the binary before the nova outburst.
The ejecta of the nova outburst have high velocity, up to 
$\sim4000$~km~s$^{-1}$, and collide with
the CSM, giving rise to strong shock \citep[e.g.,][]{sok06}. 
The shock interaction contributes to the H$\alpha$ line
and slows the decay of SMEI magnitude.
Such an interaction between ejecta and CSM was frequently observed in supernovae Type IIn, and
the relation $L_V \propto t^{-1.0}$ in the $V$-band luminosity
was calculated by \citet{mor13} for SN~2005ip.
We discuss this point in more detail in the next subsection on
V407~Cyg.

For the reddening and distance modulus toward RS~Oph, we adopt 
$E(B-V)=0.65\pm0.05$ and $(m-M)_V=12.8\pm0.2$ after \citet{hac16kb}.  
The distance is calculated to be $d=1.4\pm0.2$~kpc.
This reddening is roughly consistent with those obtained by \citet{sni87},
i.e., $E(B-V)=0.73\pm 0.06$ from the \ion{He}{2} line ratio of
1640\AA\  and 3203\AA, and $E(B-V)=0.73\pm 0.10$ from the 2715\AA\
interstellar dust absorption feature.
The NASA/IPAC galactic dust absorption map also gives 
$E(B-V)=0.64 \pm 0.03$ in the direction toward RS~Oph, whose
galactic coordinates are $(l, b)= (19\fdg7995, +10\fdg3721)$.

There are still debates on the distance to RS Oph
\citep[see, e.g.,][]{schaefer10a}.
\citet{hje86} estimated the distance to be 1.6 kpc from \ion{H}{1}
absorption-line measurements.  \citet{sni87} also obtained the distance of
1.6 kpc assuming the UV peak flux is equal to the Eddington luminosity.
\citet{har93} calculated a distance of 1290 pc from the
$K$-band luminosity.  \citet{hac01kb} obtained a smaller distance
of 0.6 kpc from the comparison of observed and theoretical UV fluxes
integrated for the wavelength region of 911-3250 \AA~.
They assumed blackbody radiation at the photosphere,
although the free-free flux is much larger than the blackbody flux
in this wavelength region.  \citet{hac06b} revised the distance to be 
$1.3-1.7$~kpc from the $y$ and $I_c$ band light curve fittings
in the late phase of the 2006 outburst.
\citet{obr06} estimated the distance of 1.6 kpc from VLBA mapping
observation with an expansion velocity indicated from emission line width.
\citet{mon06} estimated a shorter distance of $< 540$ pc assuming
that the IR interferometry size corresponds to the binary separation.
If we regard this IR emission region as a circumbinary disk, 
we get a much larger distance.  \citet{bar08} reviewed various
estimates and summarized that, for the 2006 outburst, the canonical
distance is $1.4^{+0.6}_{-0.2}$~kpc.  On the other hand,
\citet{scha09} proposed $d=4.2\pm0.9$~kpc assuming that the companion
fills its Roche lobe.  We do not think that this assumption is
supported by observation \citep[e.g.,][]{mur99}.
Therefore, our adopt value of $1.4\pm0.2$ kpc is roughly consistent
with many other estimates except for Schaefer's large value.

We plot the color-magnitude diagram of RS~Oph in Figure
\ref{hr_diagram_v745_sco_v1534_sco_rs_oph_v407_cyg_outburst}(c), 
the data of which are taken from \citet{con58} (filled red circles)
for the 1958 outburst, and AAVSO (open red diamonds), 
VSOLJ (encircled magenta pluses), SMARTS (blue stars), and
\citet{sos06ga, sos06gb} (filled blue triangles), 
\citet{sos06gc} (filled blue triangles) for the 2006 outburst.  

Figure \ref{distance_reddening_rs_oph_v407_cyg_u_sco_ci_aql_no2}(a)
shows various distance-reddening relations toward RS~Oph.
In the figure, we plot the vertical red line of $E(B-V)=0.65$,
the distance modulus in the $V$ band of $(m-M)_V=12.8$ (solid blue line), 
the UV~1455~\AA\ flux fitting (solid magenta line), 
the relations of Marshall et al. (2006):
$(l, b)= (19\fdg75, +10\fdg0)$ (open red squares)
and $(l, b)= (20\fdg0, +10\fdg0)$ (filled green squares),
and the relations of \citet{gre15, gre18} (solid black and orange lines,
respectively).
The three lines of $E(B-V)=0.65$, $(m-M)_V=12.8$, and UV~1455~\AA\    
flux fitting consistently cross at $E(B-V)=0.65$ and $d=1.4$~kpc. 
Then, the location of RS~Oph is approximately $z=+250$~pc
above the galactic plane.  These data are the same as those 
in Figure 15 of \citet{hac16kb}.  The relation of \citet{gre15}
(black line) gives a larger value of $E(B-V)=0.73$ for $d>2$~kpc.
However, the NASA/IPAC galactic dust absorption map gives 
$E(B-V)=0.64 \pm 0.03$ in the direction toward RS~Oph,
which is consistent with our value of $E(B-V)=0.65\pm0.05$.


\begin{figure}
\includegraphics[height=12.5cm]{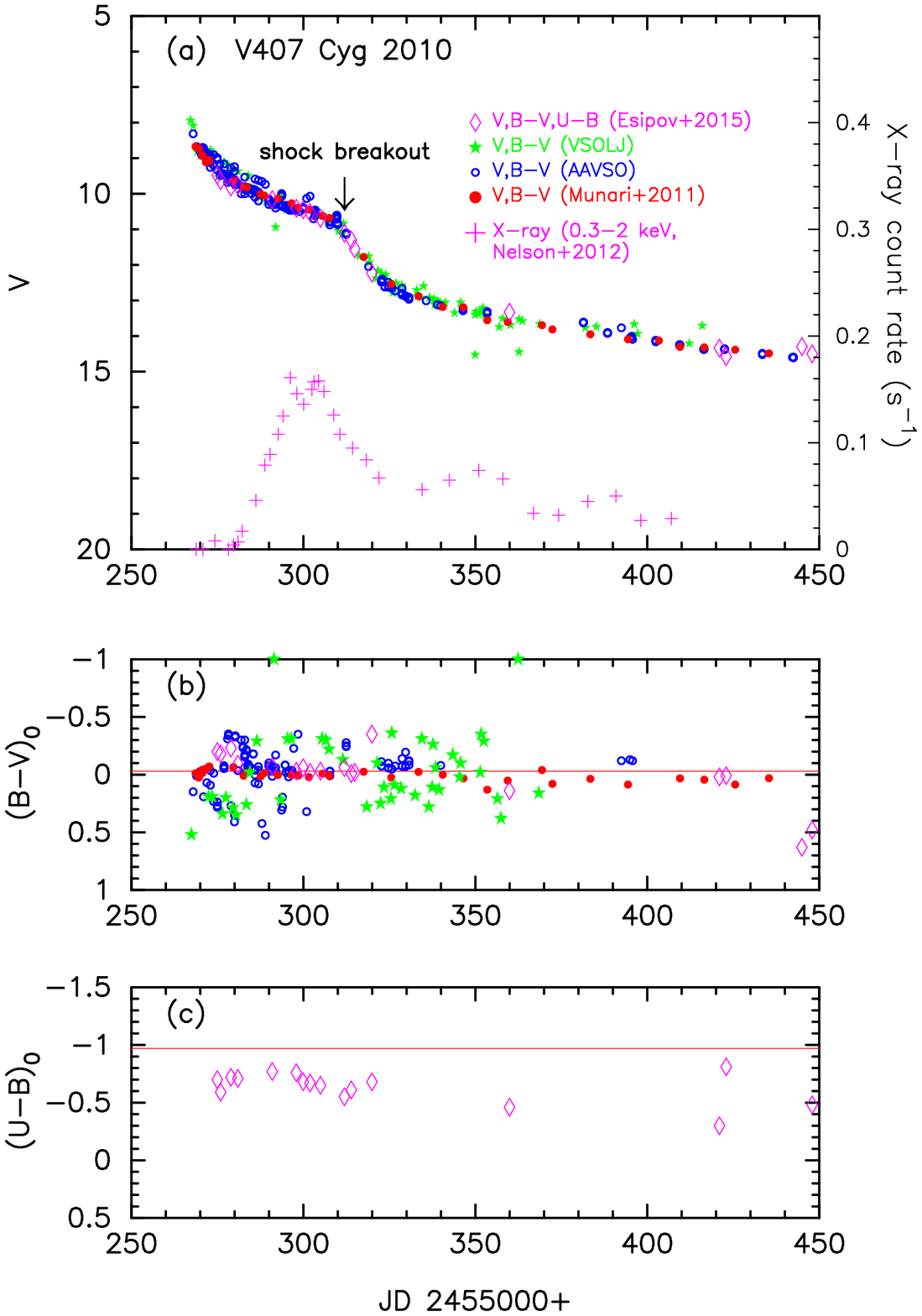}
\caption{
Same as Figure \ref{v1534_sco_v_bv_ub_color_curve}, but for V407~Cyg.
(a) The $UBV$ data (open magenta diamonds) are taken from \citet{esi15}.
The $BV$ data (filled green stars) are from VSOLJ.
The $BV$ data (open blue circles) are from AAVSO.  
The $BV$ data (filled red circles) are from \citet{mun11c}.
The X-rays ($0.3-2.0$ keV, denoted by magenta pluses) are from \citet{nel12a}.
The $V$ light curve shows a sharp decay around JD~2455310, which we
identify as the shock breakout.
(b) The $(B-V)_0$ are dereddened with $E(B-V)=1.0$.  
(c) The $(U-B)_0$ are dereddened with $E(B-V)=1.0$.  
The horizontal solid red line denotes the color of optically thick
free-free emission, i.e., $(U-B)_0=-0.97$. 
\label{v407_cyg_v_bv_ub_color_curve}}
\end{figure}


\begin{figure}
\includegraphics[height=12.5cm]{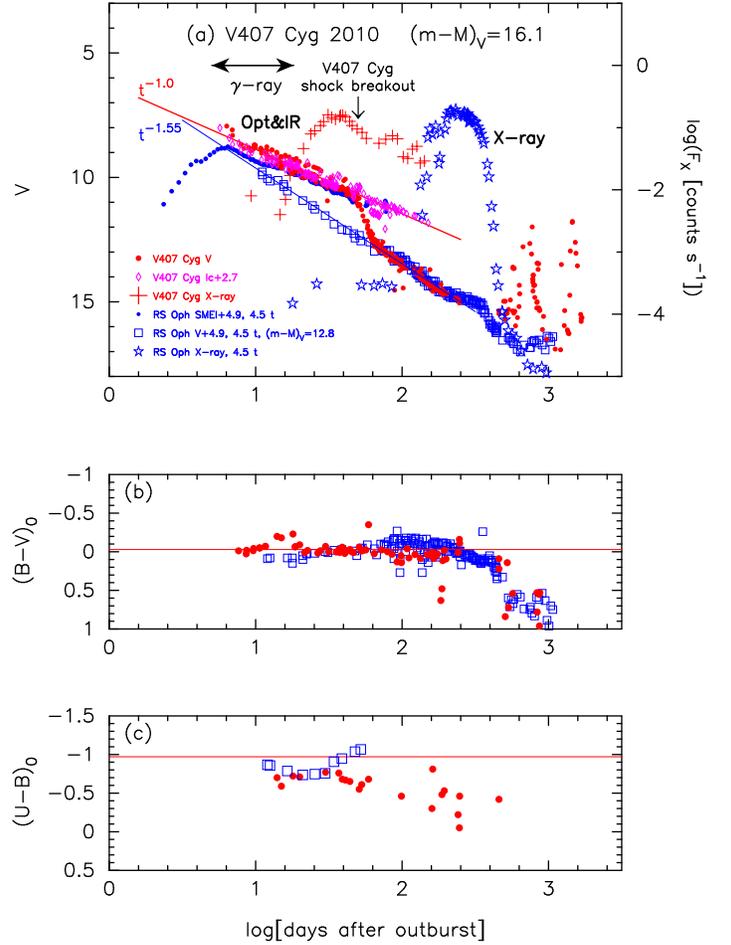}
\caption{
Same as Figure \ref{t_crb_v745_sco_v_bv_ub_logscale},
but we plot the light/color curves of V407~Cyg and RS~Oph.
(a) The filled red circles denote the $V$ magnitudes of V407~Cyg,
and the open blue squares represent the $V$ magnitudes of RS~Oph.
The $V$ data of V407~Cyg are taken from \citet{mun11c} and \citet{esi15}.
We add the $I_C$ magnitudes (open magenta diamonds taken from
AAVSO and VSOLJ) of V407~Cyg
and the SMEI magnitudes (filled blue circles) of RS~Oph.
We also add the X-ray fluxes of V407~Cyg (red pluses) and RS~Oph
(open blue stars). 
The (b) $(B-V)_0$ and (c) $(U-B)_0$ color curves.
The $(B-V)_0$ and $(U-B)_0$ are dereddened with $E(B-V)=1.0$ and
$E(B-V)=0.65$ for V407~Cyg and RS~Oph, respectively.
See the text for more detail. 
\label{v407_cyg_rs_oph_v_bv_ub_logscale}}
\end{figure}

\subsection{V407~Cyg 2010}
\label{v407_cyg}
V407~Cyg is a well-observed symbiotic nova \citep[e.g.,][]{mun90}, 
in which a WD accretes mass from a cool RG 
companion via the Roche-lobe overflow or stellar wind.
The RG companion of V407~Cyg is a Mira
with the pulsation period of 762.9 days in the $K$ band \citep{kol03}. 
SiO maser observations suggest that the Mira companion has already
reached a very late evolution stage of an AGB star \citep{deg11, cho15}.
The 2010 outburst of V407~Cyg was discovered on UT 2010 March 10.813
at $m_V=7.6$ \citep{nis10}.  
Figure \ref{v407_cyg_v_bv_ub_color_curve} shows (a) the $V$ and
X-ray light curves, (b) $(B-V)_0$, and (c) $(U-B)_0$ color curves of V407~Cyg.
Here, $(B-V)_0$ and $(U-B)_0$ are dereddened with $E(B-V)=1.0$,
as explained below. The nova reached $m_{V, \rm max}=7.1$
on UT 2010 March 10.8 (JD~2455266.3) and then declined with 
$t_2=5.9$ days and $t_3=24$ days in the $V$-band \citep{mun11c}.
The $V$ light curve sharply drops around JD~2455310. 

X-rays were observed with {\it Swift}, which are interpreted 
in terms of strong shock heating between the ejecta and circumstellar
cool wind \citep{nel12a}.  The strong shock possibly produced 
gamma-rays by accelerating high-energy particles.
V407~Cyg is the first gamma-ray-detected nova with 
{\it Fermi}/Large Area Telescope (LAT) \citep{abd10}. 
Since V407~Cyg, gamma-rays were observed
in six novae between 2010 and 2015, i.e., V1324~Sco, V959~Mon, 
V339~Del, V1369~Cen, V745~Sco, and V5668~Sgr \citep[e.g.,][]{ack14, morr17}.

\subsubsection{Reddening}
Many authors adopted the reddening of $E(B-V)=0.57$ and the distance of 
$d=2.7$~kpc after \citet{mun90}, who derived $E(B-V)=0.57$ by fitting 
the broad-band spectrum of V407~Cyg with an M6III model spectrum and 
$d=2.7$~kpc from the Mira period-luminosity relation.  
\citet{sho11} derived $E(B-V)= 0.45\pm0.05$ from the depth of the
diffuse interstellar absorption bands and proposed that the Mira looks
like an M8III rather than M6III type.  
\citet{iij15} argued that the diffuse interstellar bands cannot
give a reliable reddening because the different bands resulted in
very different values of the reddening; for example, 
$E(B-V)=0.15$ from the band at $\lambda=6613$~\AA\ but 
$E(B-V)=0.76$ from the band at $\lambda=6270$~\AA\ of his spectra.
\citet{iij15} obtained $E(B-V)=0.6$ from the color excess relation of 
$E(B-V)=(B-V) - (B-V)_0$ and the empirical relation of
the intrinsic color of
$(\bv)_0=-0.02\pm0.04$ at time $t_2$ \citep{van87}. He used the $B-V$
colors of VSOLJ/VSNET which are rather scattered and not reliable,
as clearly shown in Figure \ref{v407_cyg_v_bv_ub_color_curve}(b).
If we use the $B-V$ colors
of \citet{mun11c}, i.e., the filled red circles
in Figure \ref{v407_cyg_v_bv_ub_color_curve}(b), the empirical relation
of \citet{van87} gives a reddening value of $E(B-V)\sim 1.0$ 
at $t_2$-time.  In the present study, we adopt $E(B-V)=1.0$ which is examined in detail in Sections \ref{distance_reddening_v407_cyg},
\ref{cmd_v407_cyg}, and \ref{discussion_v407_cyg}.  

\subsubsection{Distance-reddening relation}
\label{distance_reddening_v407_cyg}
V407~Cyg is a symbiotic binary star system consisting of 
a mass-accreting hot WD and a cool Mira giant with
a pulsation period of 762.9 days in the $K$ band \citep{kol03}.
The period-luminosity relation of the LMC Miras has a bend 
at the pulsation period of $\sim400$ days \citep{ita11}. 
Beyond the bend, \citet{ita11} obtained the period-luminosity relation as
\begin{equation}
M_{K_{\rm s}} = (-6.850\pm0.901)\log P + 28.225\pm2.493 - \mu_{0,\rm LMC},
\label{mira_period_luminosity}
\end{equation}
where $P(>400$~days) is the pulsation period in days
and $\mu_{0,\rm LMC}$ is the distance modulus toward LMC. 
We adopt $\mu_{0,\rm LMC}=18.493\pm0.048$ \citep{pie13}.
Substituting $P=762.9$~days into Equation (\ref{mira_period_luminosity}),
we obtain the absolute $K_{\rm s}$ magnitude of $M_{K_{\rm s}}= -10.01$.
The average $K$ mag of V407~Cyg is $m_K=3.3$, and thus we have
\begin{equation}
(m-M)_K = 0.353 \times E(B-V) + 5 \log(d / 1{\rm ~kpc}) + 10 = 13.31,
\label{k_band_relation}
\end{equation}
where we adopt the reddening law of $A_K=0.353 \times E(B-V)$ 
\citep{car89}. We plot this distance-reddening relation
of $(m-M)_K=13.31$ using the thick green line in Figure  
\ref{distance_reddening_rs_oph_v407_cyg_u_sco_ci_aql_no2}(b).
Substituting $E(B-V)=1.0$ into Equation (\ref{k_band_relation}), 
we obtain the distance of $d=3.9$~kpc. Then, the location of V407~Cyg
is $z=-33$~pc below the galactic plane, because the galactic coordinates
of V407~Cyg are $(l,b)=(86\fdg9826, -0\fdg4820)$.
The distance modulus in the $V$ band is calculated to be $(m-M)_V=16.1$
from Equation (\ref{v_distance_modulus}).
In the same figure, we include the distance-reddening law of
$(m-M)_V=16.1$ (solid blue line) and the reddening of $E(B-V)=1.0$ 
(vertical solid red line).  

Figure \ref{distance_reddening_rs_oph_v407_cyg_u_sco_ci_aql_no2}(b)
also shows various 3D extinction maps toward V407~Cyg.
The relations of Marshall et al. (2006) are plotted
in four directions close to the direction of V407~Cyg:
$(l, b)=(86\fdg75,  -0\fdg25)$ (open red squares),
$(87\fdg00,  -0\fdg25)$ (filled green squares),
$(86\fdg75,  -0\fdg50)$ (blue asterisks), and
$(87\fdg00,  -0\fdg50)$ (open magenta circles).
The closest one is that of the open magenta circles.
We include the relations of \citet{gre15, gre18} (thick solid black and
orange lines, respectively)
and \citet{ozd16} (open cyan-blue diamonds). 
The 3D distance-reddening relations of Marshall et al. and Green et al.,
$(m-M)_V=16.1$, $(m-M)_K=13.31$, and $E(B-V)=1.0$ consistently
cross each other at the distance of $d=3.9$~kpc (and the reddening
of $E(B-V)=1.0$).
Thus, we finally confirm that the reddening of $E(B-V)=1.0$
and the distance modulus in the $V$ band of $(m-M)_V=16.1$ 
are reasonable.

\subsubsection{CSM-shock interaction}
Figure \ref{v407_cyg_rs_oph_v_bv_ub_logscale}(a) shows a comparison of
light curves of V407~Cyg and RS~Oph on a logarithmic timescale.
We stretch the timescale of RS~Oph by a factor of $f_{\rm s}=4.5$
and shift down both the $V$ and SMEI light curves of RS~Oph by 4.9 mag.
The $V$ light curve of RS~Oph (open blue squares) overlaps that of 
V407~Cyg (filled red circles) in the later phase, 
whereas the SMEI light curve of RS~Oph (blue dots) overlap 
the $V$ light curve of V407~Cyg (filled red circles).
The $V$ light curve of V407~Cyg decays as $F_\nu \propto t^{\alpha}$
($\alpha=-1.0$, denoted by the solid red line) in the early phase
(up to $t\sim45$~days).  

Cool winds from the Mira companion form CSM
around the binary \citep[e.g.,][]{moh12}.
The nova ejecta collide with the CSM
and form a strong shock \citep[e.g.,][]{orl12, pan15}.
\citet{sho11} and \citet{iij15} showed that the strong shock between
the ejecta and CSM contributes to the emission lines
and soft X-ray flux. Such an interaction between ejecta and CSM
was frequently observed in supernovae Type IIn. For example,
\citet{mor13} showed that $F_\nu \propto t^{\alpha}$ ($\alpha\approx-1.0$)
in the $V$-band for SN~2005ip. 
Thus, we interpret this early decay of $F_\nu \propto t^{-1}$ as the shock
interaction. The $I_C$ light curve of V407~Cyg also show a decline 
trend of $F_\nu \propto t^{-1.0}$, which is shifted down by 2.7 mag.
   
Considering the shock interaction, we propose the evolution
of V407~Cyg $V$ light curve as follows: in the early stage, just after
the optical maximum, the ejecta collide with the CSM and produce a strong
shock. This shock-heating contributes significantly to the $V$ brightness. 
Then, the shock broke out of the CSM approximately 45 days 
after the outburst (JD~2455310). Soon after the shock breakout,
the $V$ light curve decays as $F_\nu \propto t^{-1.55}$, as shown in
Figure \ref{v407_cyg_rs_oph_v_bv_ub_logscale}(a).
This $\alpha=-1.55$ is close to the universal decline law of 
$\alpha=-1.75$ \citep[see][]{hac06kb}.
 
RS~Oph decays as $F_\nu \propto t^{-1.55}$ in the $V$ band, as shown in
Figure \ref{rs_oph_radio_v_ub_bv_uv_x_logscale_no4}(a).
The CSM shock is much weaker in RS~Oph than in V407~Cyg, 
such that the $V$ light curve of RS~Oph is close to that of 
the universal decline law ($F_\nu \propto t^{-1.75}$)
showing no indication of strong shock interaction ($F_\nu \propto t^{-1.0}$).
The radio flux also follows the same decline trend of
$F_\nu \propto t^{-1.55}$ in the later phase.  
In other words, the shock interaction is strong enough to increase
the continuum flux to $F_\nu \propto t^{-1.0}$ in V407~Cyg, but
not enough in RS~Oph. In contrast, the SMEI light curve of RS~Oph
almost obeys $L_{\rm SMEI}\propto t^{-1.0}$ 
like the early decline trend of V407~Cyg, where $L_{\rm SMEI}$ 
is the SMEI band luminosity.
This is because the SMEI magnitude is a wide-band (peak quantum 
efficiency at 700 nm with an FWHM of 300 nm) and envelopes a 
very strong H$\alpha$ line, which mainly comes from the shock interaction.

It should be noted that the three rapid-decline novae, V745~Sco, T~CrB,
and V1534~Sco, also have a RG companion but do not show clear
evidence of shock-heating in their $V$ (or visual) light curves, because
no part shows $F_\nu\propto t^{-1}$.
Their $V$ light curves almost overlap to that of V838~Her, which has
a MS companion as mentioned in Section \ref{v838_her}.
The X-ray fluxes of V407~Cyg, RS~Oph, V745~Sco, and V1534~Sco 
were observed with {\it Swift}. Their origin could be shock-heating
in the very early phase, before the SSS phase started.  
\citet{mun18} showed no evidence of deceleration of ejecta in V1534~Sco.
These indicate that the shape of $V$ light curve changes from 
that of V407~Cyg to RS~Oph, and finally to V1534~Sco,
depending on the strength of shock interaction. 
V407~Cyg shows a strongest limit of shock interaction
while V1534~Sco corresponds to a weakest limit of shock.

\subsubsection{Timescaling law and time-stretching method}
\label{v407_cyg_timescaling_law}
As discussed in the previous subsection, the $V$
light curve of V407~Cyg essentially follows a similar decline law to
RS~Oph. We apply V407~Cyg and RS~Oph to
Equations (\ref{overlap_brigheness}) and (\ref{distance_modulus_formula})
and obtain the following relation 
\begin{eqnarray}
(m&-&M)_{V, \rm V407~Cyg} = 16.1 \cr
&=& (m - M + \Delta V)_{V, \rm RS~Oph} - 2.5 \log f_{\rm s} \cr
&=& 12.8 + \Delta V - 2.5 \log f_{\rm s},
\label{distance_modulus_v407_cyg_rs_oph_nofs}
\end{eqnarray}
where we adopt $(m-M)_{V, \rm RS~Oph}=12.8$ in Section \ref{rs_oph}.
We have the relation between $\Delta V$ and $f_{\rm s}$ as 
\begin{equation}
\Delta V - 2.5 \log f_{\rm s}=3.3,
\label{v407_cyg_rs_oph_dv_log_fs}
\end{equation}
where $\Delta V$ is the vertical shift and $\log f_{\rm s}$ is the
horizontal shift with respect to the original $V$ light curve of RS~Oph 
in Figure \ref{rs_oph_radio_v_ub_bv_uv_x_logscale_no4}(a).
If we choose an arbitrary $\Delta V$, 
these two $V$ light curves do not overlap. 
We search by eye for the best-fit value by changing 
$\Delta V$ in steps of $0.1$ mag and obtain
the set of $\Delta V$ and $\log f_{\rm s}$ for best overlap;
$\Delta V=4.9$ and $\log f_{\rm s}=0.65$ ($f_{\rm s}=4.5$),
as shown in Figure \ref{v407_cyg_rs_oph_v_bv_ub_logscale}.

\subsubsection{WD mass of V407~Cyg}
\label{wd_mass_v407_cyg}
Using the linear relation between $M_{\rm WD}$ and $\log f_{\rm s}$
in Figure \ref{timescale_wd_mass}, we obtain the WD mass of
$M_{\rm WD}=1.22~M_\sun$ for V407~Cyg (see also Table
\ref{wd_mass_recurrent_novae}).  Here, we use the linear
relation between $\log f_{\rm s}=0.3$ and $\log f_{\rm s}=1.1$ 
(right solid red line). Even if we assume the WD mass increases
at the rate of $\dot M_{\rm WD}=1\times10^{-7}~M_\sun$~yr$^{-1}$
as discussed in Section \ref{wd_mass_vs_timescale}, it takes
$t_{\rm Ia}=(1.38-1.22)~M_\sun/1\times10^{-7}M_\sun$~yr$^{-1}=
1.6\times10^{6}$~yr to explode as a SN~Ia.  
We do not expect that this high mass-accretion rate will continue
for such a long time,
because the Mira companion has already reached a very late
evolution stage of an AGB star as suggested by the SiO maser observations
\citep{deg11, cho15}.  
Therefore, we suppose that V407~Cyg is not a progenitor of SNe~Ia.

\subsubsection{Color-magnitude diagram}
\label{cmd_v407_cyg}
Using $E(B-V)=1.0$ and $(m-M)_V=16.1$, we plot the color-magnitude 
diagram of V407~Cyg in Figure
\ref{hr_diagram_v745_sco_v1534_sco_rs_oph_v407_cyg_outburst}(d).
The color evolves down along with the red solid line of
$(B-V)_0=-0.03$, which is the intrinsic color of optically thick winds
\citep[e.g.,][]{hac14k}. Note that the track of V407~Cyg is very
similar to and closely located to that of RS~Oph (Figure
\ref{hr_diagram_v745_sco_v1534_sco_rs_oph_v407_cyg_outburst}(c)).
Here, we adopt only the data of \citet{mun11c} and \citet{esi15}
because the other $B-V$ color data of the VSOLJ and AAVSO archives are 
rather scattered, as can be seen in Figure
\ref{v407_cyg_v_bv_ub_color_curve}(b).
This similarity again confirms that our adopted
values of $E(B-V)=1.0$ and $(m-M)_V=16.1$ ($d=3.9$~kpc) are reasonable. 

\subsubsection{Discussion on the distance}
\label{discussion_v407_cyg}
The distance to V407~Cyg was determined to be $d=2.7$~kpc by
\citet{mun90} or $d=1.9$~kpc by \citet{kol98}, based on the
absolute $K$ magnitudes of the Mira companion, which were
calculated from the period-luminosity relation of the Mira
variables \citep{gla82, fea89}.
\citet{mun90} assumed $E(B-V)=0.57$, $J=5.1$, $H=4.0$, $K=3.3$,
and the relations given by \citet{gla82},
whereas \citet{kol98} used $E(B-V)=0.40$, $L_{\rm cool, max}=4.7\times
10^3 L_\sun (d/{\rm kpc})^2$ from the spectral energy distribution (SED)
fitting with the cool RG and $L_{\rm cool, mid}=1.2\times
10^4 L_\sun$ from the period-luminosity relation of \citet{fea89}. 
However, a bend was recently found at $P\sim 400$~days
in the period-luminosity relation of Mira variables \citep[e.g.,][]{ita11}.
Above the bend, i.e., $P\gtrsim 400$~days, the intrinsic luminosity
of a Mira is much brighter than the old period-luminosity relation.
\citet{iij15} made his period-luminosity relation of Mira variables for
$P > 400$~days. Using the minimum $K$ magnitudes of V407~Cyg, he obtained
the distance of $5.3\pm1$~kpc, much larger than the old values of 2.7 and
1.9~kpc. Instead of Iijima's relation,
we adopted the period-luminosity relation obtained by \citet{ita11}
and estimated the distance modulus in the $K$-band, i.e., $(m-M)_K=13.31$
(thick solid green line in Figure
\ref{distance_reddening_rs_oph_v407_cyg_u_sco_ci_aql_no2}(b)).
This relation gives a reasonable cross point of $E(B-V)=1.0$ and
$d=3.9$~kpc with the distance modulus in the $V$-band, i.e., 
$(m-M)_V=16.1$ (solid blue line in Figure  
\ref{distance_reddening_rs_oph_v407_cyg_u_sco_ci_aql_no2}(b)). 
This cross point is consistent with the distance-reddening relation
given by \citet{mar06} (open magenta circles in Figure  
\ref{distance_reddening_rs_oph_v407_cyg_u_sco_ci_aql_no2}(b))
and by \citet{gre15, gre18} (solid black and orange lines). 
This consistency supports our new estimates of
$E(B-V)=1.0$ and $d=3.9$~kpc ($(m-M)_V=16.1$).


\begin{figure}
\includegraphics[height=8.5cm]{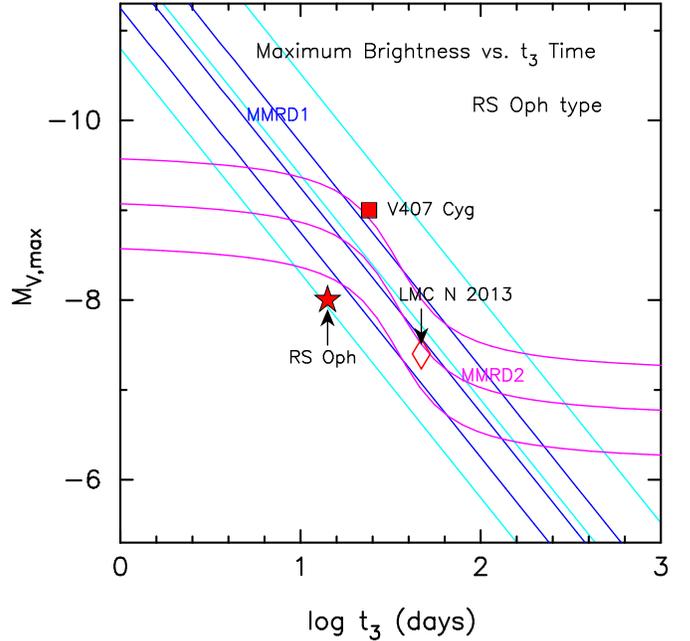}
\caption{
Same as Figure \ref{max_t3_scale_v745_sco_type},
but for the CSM-shock (RS~Oph) type novae.
The three novae, which are contaminated by shock-heating, are 
virtually consistent with the MMRD relations.
\label{max_t3_scale_rs_oph_type}}
\end{figure}

\subsection{Template light curves of CSM-shock type novae}
We have analyzed two novae, RS~Oph and V407~Cyg, both of which have
CSM that was shock-heated by the ejecta.  
The $V$ light curves of these novae were influenced by this shock-heating.
The decay of optical and NIR light curves obeys the $F_\nu \propto t^{-1}$
(or $L_{\rm SMEI}\propto t^{-1}$ for RS~Oph)
during the strong deceleration phase of the ejecta.  
After the shock breaks out of the CSM, the decay follows $F_\nu \propto t^{-1.55}$, 
which is close to the universal decline law of $F_\nu \propto t^{-1.75}$.
Thus, we propose the template $V$ light curves of the CSM-shock
type novae, as shown in Figure 
\ref{v745_sco_u_sco_v407_cyg_rs_oph_v_template},
which are represented by, and can be applied to,
both the V407~Cyg and RS~Oph $V$ light curves.

The three rapid-decline novae, V745~Sco, T~CrB, and V1534~Sco, 
have a RG companion, although they show no part of
shock-heating in the $V$ light curve.
We conclude that the effect of shock-heating
is strongest in V407~Cyg and becomes weaker in RS~Oph, V745~Sco,
and V1534~Sco, in that order. V407~Cyg behaves as
$F_\nu \propto t^{-1}$ in the early phase both of $V$ and
$I_C$ magnitudes; RS~Oph has no part of $F_\nu \propto t^{-1}$
in the $V$ magnitude but has a part of $F_\nu \propto t^{-1}$
in the SMEI magnitude, V745~Sco and V1534~Sco have no part of
$F_\nu \propto t^{-1}$ but emit X-ray (possibly shock-origin)
before the SSS phase.

\subsection{MMRD relation of CSM-shock type novae}
\label{mmrd_CSM}
We plot each MMRD point in Figure \ref{max_t3_scale_rs_oph_type}
for the CSM-shock type novae. In the figure, we include another
nova observed in LMC, which is analyzed in Section
\ref{novae_lmc_smc_m31} and categorized to the CSM-shock type.  
It seems that these novae broadly follow both the MMRD1 
(Kaler-Schmidt's law) and MMRD2 (Della Valle \& Livio's law),
although RS~Oph is located slightly below the MMRD relations.  
The light curves of V407~Cyg and LMC~N~2013 are contaminated 
by the flux from shock-heating and the slow decline rates 
make the $t_2$ and $t_3$ times longer.
If we take the SMEI light curve of RS~Oph, $F_\nu \propto t^{-1.0}$,
the $t_3$ time becomes longer and its MMRD point is calculated to be
$(t_3, M_{V, \rm max})=($60 days, $-8.0$), which is located on the upper
flanked line of the MMRD1 relation 
(blue line in Figure \ref{max_t3_scale_rs_oph_type}).  
Conversely, if there is no contribution from the CSM shock, V407~Cyg and 
LMC~N~2013 should be located at the rather left side of the MMRD1 relation
but still inside the cyan lines (broad MMRD region).
These left-lower side positions are consistent with the general trend of
normal-decline type novae which will be examined later 
in Section \ref{u_sco_type}.

\subsection{Summary of the CSM-shock (RS~Oph) type novae}
\label{summary_rs_oph_novae}
We summarize the results of the two CSM-shock type novae. 
\begin{enumerate}
\item We have estimated the distance, reddening, and distance modulus of
RS~Oph using various methods after \citet{hac16kb}. The results are
summarized in Table~\ref{extinction_distance_various_novae}.
\item We have also estimated the distance, reddening, and distance 
modulus of V407~Cyg using various methods, especially, from the new
Mira period-luminosity relation proposed by \citet{ita11}.
The results are summarized in Table \ref{extinction_distance_various_novae}.
\item The $V$ light curve of V407~Cyg decays as $F_\nu \propto t^{-1.0}$
in the early phase (up to $t\sim45$~days) and then sharply drops 
and obeys $F_\nu \propto t^{-1.55}$ like RS~Oph
after the shock breaks out of the CSM.
Thus, V407~Cyg follows a timescaling law similar to RS~Oph
except for the early shock interaction phase.
\item We confirm that these two novae satisfy the time-stretching method,
i.e., Equations (\ref{overlap_brigheness}) and 
(\ref{distance_modulus_formula}). 
\item These two novae broadly follow the MMRD relations, i.e.,
Kaler-Schmidt's law (MMRD1) and the law of Della Valle \& Livio (MMRD2).
\item 
The WD mass of RS~Oph is $M_{\rm WD}=1.35~M_\sun$ and 
very close to the critical mass of SN~Ia explosion,
$M_{\rm Ia}=1.38~M_\sun$ \citep{nom82}.  The numerical simulations 
show that the WD grows in mass \citep{hac01kb, kat17sh}.
This suggests that RS~Oph is a progenitor of SNe~Ia.  
\item The WD mass of V407~Cyg is estimated to be $M_{\rm WD}=1.22~M_\sun$
from the timescaling factor of $f_{\rm s}=8.9$ against V745~Sco
as listed in Table \ref{wd_mass_recurrent_novae}.
SiO maser observations suggest that the Mira companion has already
reached a very late evolution stage of an AGB star \citep{deg11, cho15}.
It is unlikely that the WD mass will increase to 
$M_{\rm Ia}=1.38~M_\sun$ during the remaining of life of the Mira companion.
Therefore, V407~Cyg is not a progenitor of SNe~Ia.
\end{enumerate}

\section{Timescaling Law of Normal-decline Novae}
\label{u_sco_type}
We analyze the light curves of the two recurrent novae, U~Sco and CI~Aql,
and show that they follow a timescaling law.
We call this group of novae the normal-decline (U~Sco) type novae.


\begin{figure}
\includegraphics[height=12cm]{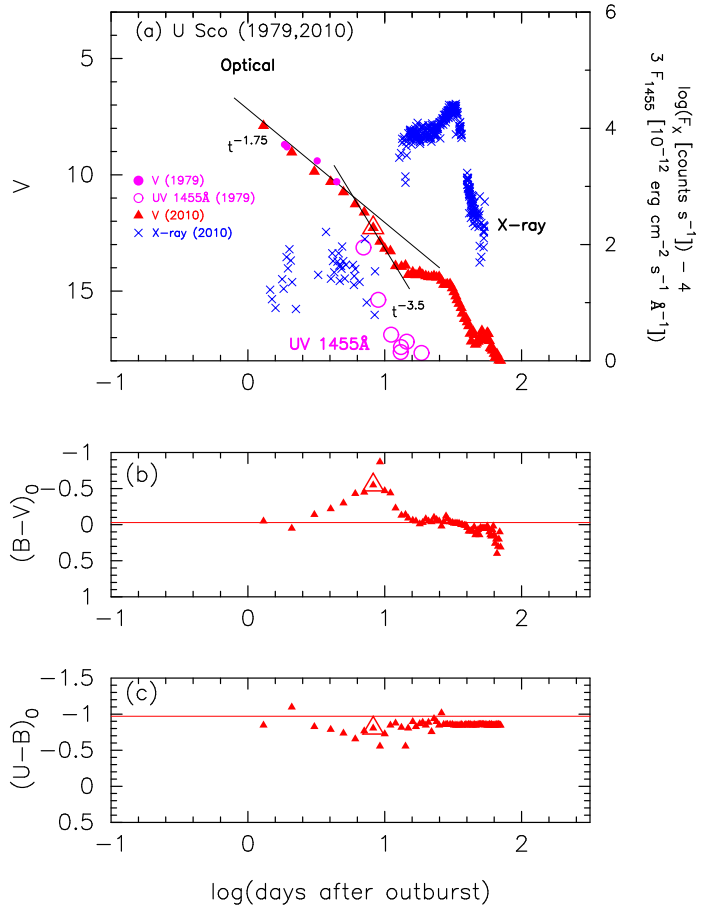}
\caption{
Light/color curves of the U~Sco 1979 and 2010 outbursts.
(a) The filled red triangles denote the $V$ magnitudes of the 
U~Sco (2010) outburst, and the filled and open magenta circles represent 
the $V$ and UV~1455~\AA\ fluxes of the U~Sco (1979) outburst, respectively.
The blue crosses are the X-ray ($0.3-10$ keV) count rates of 
the U~Sco (2010) outburst.  The (b) $(B-V)_0$ and (c) $(U-B)_0$ color
curves of the U~Sco (2010) outburst. The large red triangle indicates
the start of the nebular phase approximately eight days after the outburst.
See the text for detail.
\label{u_sco_v_bv_ub_xray_radio_logscale}}
\end{figure}


\begin{figure}
\includegraphics[height=12cm]{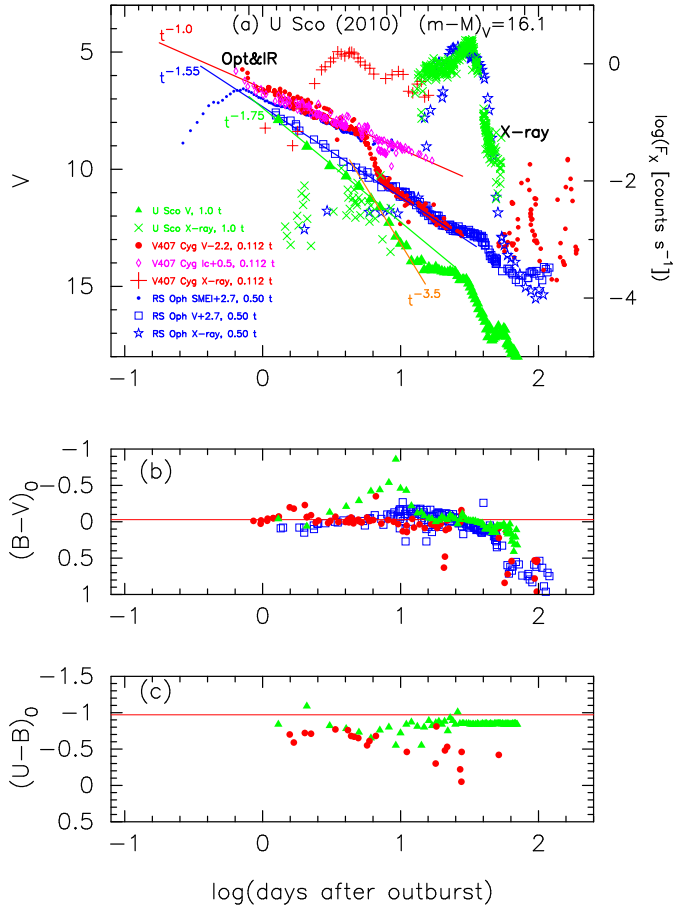}
\caption{
Same as Figure \ref{v407_cyg_rs_oph_v_bv_ub_logscale},
but for the light/color curves of U~Sco as well as RS~Oph and V407~Cyg.
(a) The filled green triangles denote the $V$ magnitudes of U~Sco,
the open blue squares represent the $V$ magnitudes of RS~Oph
and the filled red circles are the $V$ magnitudes of V407~Cyg.
We add the X-ray count rates of U~Sco (green crosses) as well as
RS~Oph (open blue stars) and V407~Cyg (red pluses).
The blue dots are the SMEI magnitudes of RS~Oph and open magenta
diamonds are the $I_C$ magnitude of V407~Cyg. 
The (b) $(B-V)_0$ and (c) $(U-B)_0$ color curves. 
\label{u_sco_v407_cyg_rs_oph_v_bv_ub_logscale}}
\end{figure}


\begin{figure}
\includegraphics[height=12cm]{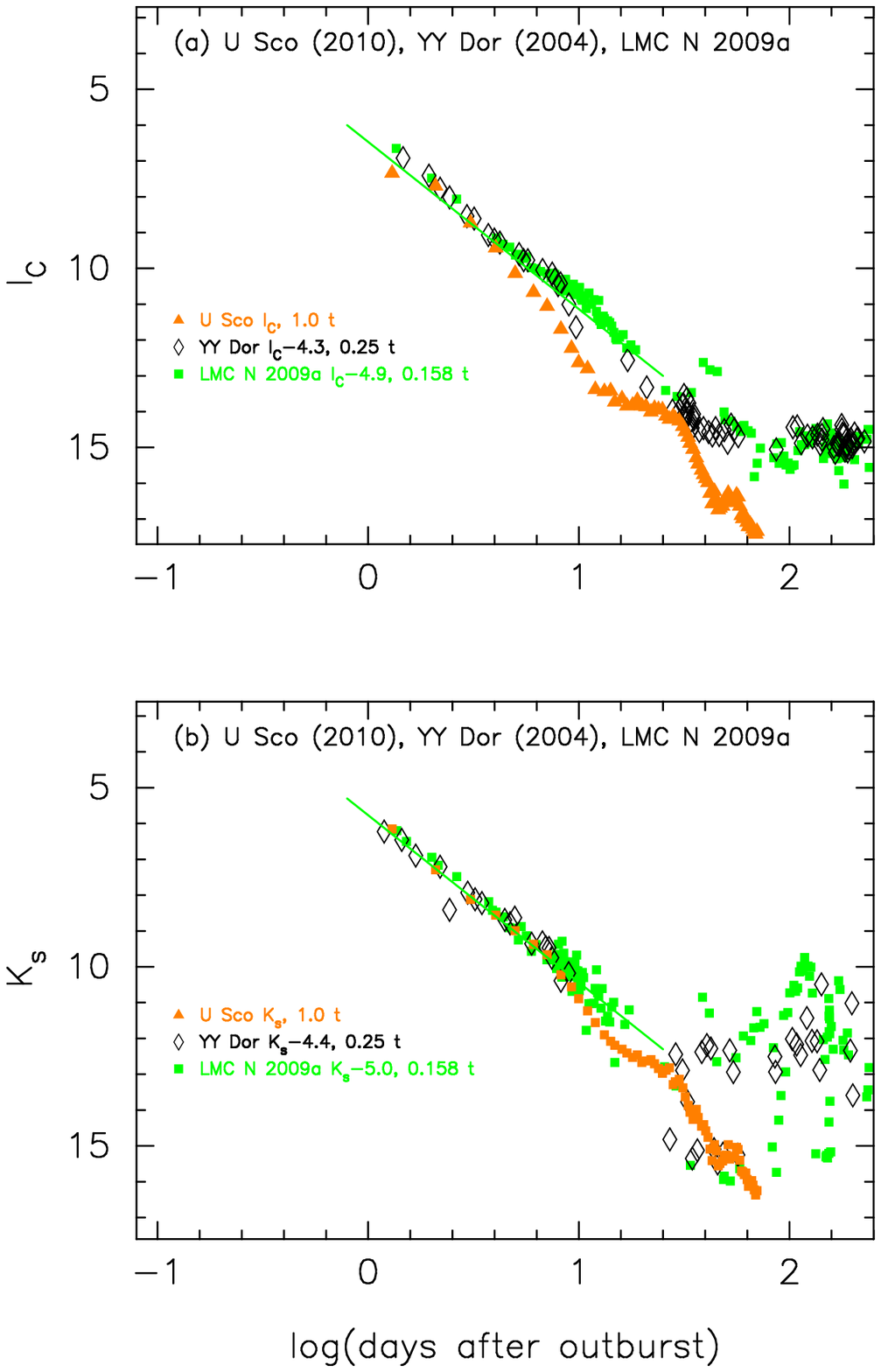}
\caption{
The (a) $I_C$ and (b) $K_s$ light curves of U~Sco (2010), 
YY~Dor (2004), and LMC~N~2009a.
Each light curve is horizontally stretched by $\log f_{\rm s}$
and vertically shifted by $\Delta I_C$ or $\Delta K_s$.
See the text for detail.
\label{u_sco_yy_dor_lmcn_2009a_i_k_logscale_2fig}}
\end{figure}


\begin{figure}
\includegraphics[height=12cm]{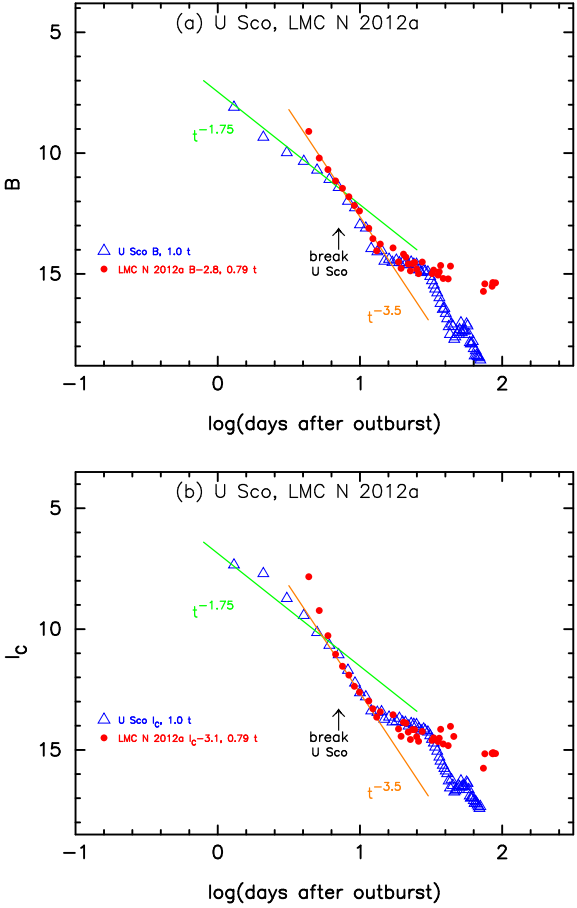}
\caption{
The (a) $B$ and (b) $I_C$ light curves of U~Sco (2010) and 
LMC~N~2012a.  The light curves of LMC~N~2012a
are horizontally stretched by $\log f_{\rm s}=0.1$
and vertically shifted by $\Delta B = -2.8$ and $\Delta I_C= -3.1$.
See the text for detail.
\label{u_sco_lmcn2012a_b_i_2fig}}
\end{figure}


\begin{figure*}
\plotone{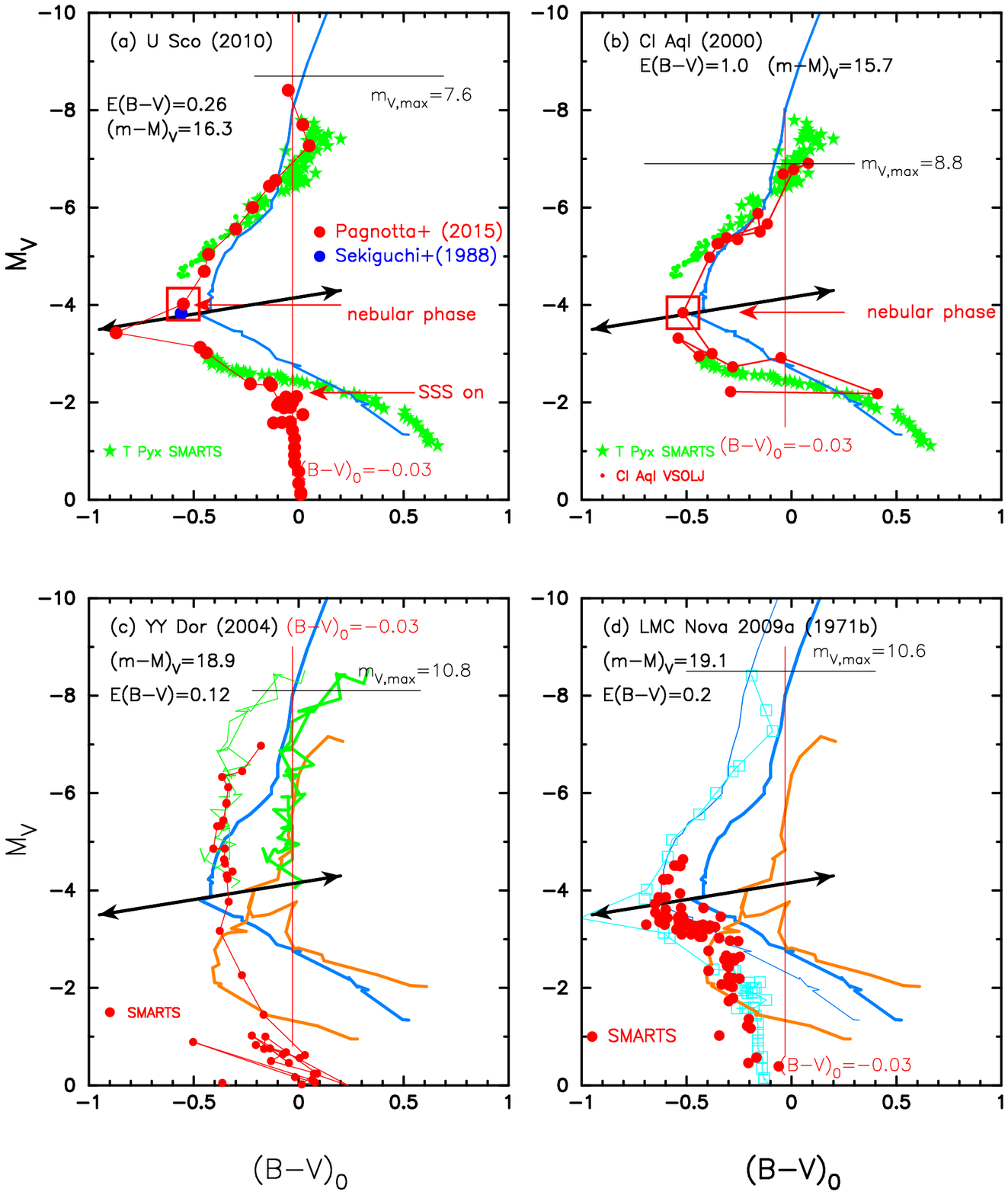}
\caption{
Same as Figure \ref{hr_diagram_v745_sco_v1534_sco_rs_oph_v407_cyg_outburst},
but for (a) U~Sco, (b) CI~Aql, (c) YY~Dor, and (d) LMC~N~2009a. 
The thick solid cyan-blue lines show the template track of V1500~Cyg.
The thick solid orange lines show the template track of LV~Vul.
The thick solid green lines show the template track of V1668~Cyg.
In panels (a) and (b), we add the data of the T~Pyx 2011 outburst
(filled green stars, taken from SMARTS), 
which are the same as those in Figures 16 and 19 of \citet{hac16kb}.
In panel (c), the thin solid green lines correspond to the track of
V1668~Cyg but blue-shifted by $\Delta (B-V)_0=-0.3$ mag. 
The track of YY~Dor (filled red circles connected with a solid red line)
almost overlaps with this blue-shifted V1668~Cyg track. In panel (d), 
the thin solid cyan-blue lines denote a $\Delta (B-V)_0=-0.2$ mag
blue-shifted track of V1500~Cyg. The open cyan squares
connected with thin solid cyan line represent a U~Sco
track blue-shifted by $\Delta (B-V)_0=-0.15$ mag.  
\label{hr_diagram_u_sco_ci_aql_yy_dor_lmcn2009a_outburst}}
\end{figure*}

\subsection{U~Sco (2010)}
\label{u_sco}
     U~Sco is a recurrent nova with ten recorded outbursts
in 1863, 1906, 1917, 1936, 1945, 1969, 1979, 1987, 1999, and 2010,
nearly every ten years \citep{schaefer10a}.
Figure \ref{u_sco_v_bv_ub_xray_radio_logscale} shows
(a) the $V$, UV~1455~\AA, and X-ray, (b) $(B-V)_0$, and 
(c) $(U-B)_0$ evolutions of the U~Sco (1979, 2010) outbursts
on a logarithmic timescale. Here, $(B-V)_0$ and $(U-B)_0$ of U~Sco 
are dereddened with $E(B-V)=0.26$, as explained below.
The $UBV$ data of the U~Sco (2010) outburst are taken from \citet{pagnotta15}.
The X-ray data of the U~Sco (2010) outburst are from 
the {\it Swift} web page \citep{eva09}.
The $V$ data of the U~Sco (1979) outburst are from IAU Circular No.3373(2)
and the UV~1455~\AA\ fluxes of the U~Sco (1979) outburst are 
calculated using the data taken from the INES archive 
data sever\footnote{http://sdc.laeff.inta.es/}.
The maximum brightness and decline rates of the 2010 outburst were
estimated to be $m_{V, \rm max}=7.6$, $t_2=1.7$ days, and $t_3=3.6$ days
by \citet{schaefer11b}.

U~Sco is one of the fastest novae. The WD mass should be very massive
and close to the Chandrasekhar mass. Its mass was estimated 
by \citet{tho01} to be
$M_{\rm WD}=1.55\pm 0.24 ~M_\sun$ from the primary orbital velocity of
$K_{\rm W}=93\pm10$~km~s$^{-1}$ and the secondary orbital velocity of
$K_{\rm R}=170\pm10$~km~s$^{-1}$. \citet{hkkm00} obtained the WD mass
of $M_{\rm WD}=1.37\pm 0.01 ~M_\sun$ for the U~Sco (1999) outburst based
on their model light curve fitting.  Using the $1.37~M_\sun$ WD model,
\citet{hac12k} reasonably reproduced the $V$ and supersoft X-ray
light curves for the U~Sco (2010) outburst. Thus, we regard the WD mass
of U~Sco to be $1.37~M_\sun$. \citet{hkkm00} argued that U~Sco is a
promising progenitor of SNe~Ia.  We summarize these results 
in Tables \ref{extinction_distance_various_novae},
\ref{physical_properties_recurrent_novae},
and \ref{wd_mass_recurrent_novae}.

\subsubsection{Break in the light curve slope}
The slope of nova decay changes from the universal decline 
of $F_\nu \propto t^{-1.75}$ to a more rapid decay of 
$F_\nu \propto t^{-3.5}$ by approximately eight days after
the outburst, at the large open red triangle, as shown in Figure 
\ref{u_sco_v_bv_ub_xray_radio_logscale}(a).
We call this change of slope the ``break,'' as shown in Figure
\ref{v745_sco_u_sco_v407_cyg_rs_oph_v_template}. 
Such a change of decay trend can be seen
in the model light curves (Figure
\ref{all_mass_v1668_cyg_x45z02c15o20_calib_universal_no2}
in Appendix \ref{model_light_curve}) when the wind mass-loss rate
sharply decreases and the photosphere quickly shrinks.
In the case of V1668~Cyg,
this change is coincident with the start
of the nebular phase, as shown in Figure
\ref{all_mass_v1668_cyg_x45z02c15o20_calib_universal_no2}.
Therefore, we consider this change as the transition to optically 
thin from optically thick of the ejecta.

Approximately 14 days after the outburst,
the $V$ light curve of U~Sco becomes flat, as shown in Figure 
\ref{u_sco_v_bv_ub_xray_radio_logscale}(a).
This plateau is due to the contribution from the irradiated accretion
disk \citep{hkkm00}. Approximately 30 days after the outburst, 
hydrogen shell-burning ended, as clearly shown by the drop in supersoft
X-ray flux in Figure \ref{u_sco_v_bv_ub_xray_radio_logscale}(a),
which corresponds to the end of the plateau phase
in the $V$ light curve.

\subsubsection{Timescaling law and time-stretching method}
\label{u_sco_timescaling_law}
Figure \ref{v745_sco_u_sco_v407_cyg_rs_oph_v_template} shows
that the $V$ light curves of V745~Sco and U~Sco almost overlap with
each other until 12 days after the outbursts.   We apply Equation 
(\ref{distance_modulus_formula}) to these two novae and obtain
\begin{eqnarray}
(m&-&M)_{V, \rm U~Sco} \cr
&=& (m - M + \Delta V)_{V, \rm V745~Sco} - 2.5 \log 1.0 \cr
&=& 16.6 - 0.3 + 0.0 = 16.3,
\label{distance_modulus_u_sco_v745_sco}
\end{eqnarray}
where we adopt $(m-M)_{V, \rm V745~Sco}=16.6$ in Section \ref{v745_sco}.
Thus, we have independently determined the distance moduli of 
U~Sco, RS~Oph, and V407~Cyg.
Figure \ref{u_sco_v407_cyg_rs_oph_v_bv_ub_logscale} shows 
the (a) $V$ and X-ray, (b) $(B-V)_0$, and (c) $(U-B)_0$ evolutions
of these three novae. We have already fixed the horizontal shifts
between RS~Oph and V407~Cyg in Figure \ref{v407_cyg_rs_oph_v_bv_ub_logscale}.
We further determine the horizontal shifts of these two novae with respect to
U~Sco by overlapping the ends of the SSS phase
between U~Sco (green crosses) and RS~Oph (open blue stars), as shown 
in Figure \ref{u_sco_v407_cyg_rs_oph_v_bv_ub_logscale}. 
Although these three $V$ light curves do not perfectly overlap with each other,
that is, do not perfectly satisfy Equation (\ref{overlap_brigheness}),
we apply Equation (\ref{distance_modulus_formula}) 
to U~Sco, RS~Oph, and V407~Cyg in Figure 
\ref{u_sco_v407_cyg_rs_oph_v_bv_ub_logscale}
and have the relation of
\begin{eqnarray}
(m&-&M)_{V, \rm U~Sco} = 16.3 \cr
&=& (m - M + \Delta V)_{V, \rm RS~Oph} - 2.5 \log 0.50 \cr
&=& 12.8 + 2.7 + 0.75 = 16.25 \cr
&=& (m - M + \Delta V)_{V, \rm V407~Cyg} - 2.5 \log 0.112 \cr
&=& 16.1 - 2.2 + 2.38 = 16.28,
\label{distance_modulus_u_sco_rs_oph}
\end{eqnarray}
where we adopt $(m-M)_{V, \rm RS~Oph}=12.8$ in Section \ref{rs_oph}
and $(m-M)_{V, \rm V407~Cyg}=16.1$ in Section \ref{v407_cyg}.

We check the distance modulus of U~Sco using the $I_C$ and
$K_s$ light curves of YY~Dor and LMC~N~2009a
in Figure \ref{u_sco_yy_dor_lmcn_2009a_i_k_logscale_2fig}
and the $B$ and $I_C$ light curves of LMC~N~2012a in Figure
\ref{u_sco_lmcn2012a_b_i_2fig}, 
which are studied later in Sections \ref{yy_dor}, \ref{lmcn_2009a},
and \ref{lmcn_2012a}, respectively.  
These three novae are members of LMC and
their distances are well constrained, i.e., $\mu_0\equiv 
(m-M)_0=18.5$ as mentioned later in Section \ref{novae_lmc_smc_m31}.

Both the $I_C$ and $K_s$ light curves in Figure
\ref{u_sco_yy_dor_lmcn_2009a_i_k_logscale_2fig} 
have a slope of $F_\nu\propto t^{-1.75}$ and overlap with the
U~Sco light curve for the horizontal shift
of $\log f_{\rm s}=0.6$ for YY~Dor as in Section \ref{yy_dor} and
$\log f_{\rm s}=0.8$ for LMC~N~2009a as in Section \ref{lmcn_2009a}.
Thus, we conclude that YY~Dor and LMC~N~2009a belong to the U~Sco
type (normal decline).  
Then, we apply Equation (\ref{distance_modulus_formula}) to these
three novae in Figure \ref{u_sco_yy_dor_lmcn_2009a_i_k_logscale_2fig}
and obtain
\begin{eqnarray}
(m&-&M)_{I, \rm U~Sco} \cr
&=& (m - M + \Delta I_C)_{I, \rm YY~Dor} - 2.5 \log 0.25 \cr
&=& 18.7 - 4.3 + 1.5 = 15.9 \cr
&=& (m - M + \Delta I_C)_{I, \rm LMC~N~2009a} - 2.5 \log 0.158 \cr
&=& 18.8 - 4.9 + 2.0 = 15.9,
\label{distance_modulus_i_u_sco_yy_dor_lmcn2009a}
\end{eqnarray}
where we adopt $(m-M)_{I, \rm YY~Dor}=18.5 + 1.5\times 0.12= 18.7$
from Section \ref{yy_dor}, 
$(m-M)_{I, \rm LMC~N~2009a}=18.5 + 1.5\times 0.2=18.8$
from Section \ref{lmcn_2009a}.
We also obtain for the $K_s$ band as
\begin{eqnarray}
(m&-&M)_{K, \rm U~Sco} \cr
&=& (m - M + \Delta K_s)_{K, \rm YY~Dor} - 2.5 \log 0.25 \cr
&=& 18.5 - 4.4 + 1.5 = 15.6 \cr
&=& (m - M + \Delta K_s)_{K, \rm LMC~N~2009a} - 2.5 \log 0.158 \cr
&=& 18.6 - 5.0 + 2.0 = 15.6,
\label{distance_modulus_k_u_sco_yy_dor_lmcn2009a}
\end{eqnarray}
where we adopt $(m-M)_{K, \rm YY~Dor}=18.5 + 0.35\times 0.12= 18.5$
from Section \ref{yy_dor}, 
$(m-M)_{K, \rm LMC~N~2009a}=18.5 + 0.35\times 0.2=18.6$
from Section \ref{lmcn_2009a}.

Figure \ref{u_sco_lmcn2012a_b_i_2fig}(a) shows the $B$ light curves
of U~Sco and LMC~N~2012a.  These two light curves overlap each other
in the $t^{-3.5}$ slope (and also in the plateau phase) 
for the horizontal shift $\log f_{\rm s}=0.1$ as obtained 
in Sections \ref{lmcn_2012a} and \ref{horizontal_shifts}.
Applying Equation (\ref{distance_modulus_formula}) to the two novae,
we obtain
\begin{eqnarray}
(m&-&M)_{B, \rm U~Sco} \cr
&=& (m - M + \Delta B)_{B, \rm LMC~N~2012a} - 2.5 \log 0.79 \cr
&=& 19.15 - 2.8 + 0.25 = 16.6, 
\label{distance_modulus_u_sco_lmcn2012a_b}
\end{eqnarray}
where we adopt 
$(m-M)_{B, \rm LMC~N~2012a}= (m-M)_{V, \rm LMC~N~2012a} + 1.0\times E(B-V)
= 19.0 + 0.15= 19.15$ in Section \ref{lmcn_2012a}, using
Equation (\ref{distance_modulus_relation_b}).

Similarly from Figures \ref{u_sco_lmcn2012a_b_i_2fig}(b),
we obtain
\begin{eqnarray}
(m&-&M)_{I, \rm U~Sco} \cr
&=& (m - M + \Delta I_C)_{I, \rm LMC~N~2012a} - 2.5 \log 0.79 \cr
&=& 18.76 - 3.1 + 0.25 = 15.91, 
\label{distance_modulus_v745_sco_u_sco_lmcn2012a_i}
\end{eqnarray}
where we adopt 
$(m-M)_{I, \rm LMC~N~2012a}= (m-M)_{V, \rm LMC~N~2012a} - 1.6\times E(B-V)
= 19.0 - 0.24= 18.76$ in Section \ref{lmcn_2012a}.
This value is consistent with that obtained in Equation
(\ref{distance_modulus_i_u_sco_yy_dor_lmcn2009a}).

Thus, we plot four distance-reddening relations of
Equation (\ref{distance_modulus_relation_b}),
(\ref{v_distance_modulus}),
(\ref{distance_modulus_relation_i}), and 
(\ref{equation_distmod_IR}) for U~Sco, that is,
$(m-M)_B=16.6$ (magenta line), $(m-M)_V=16.3$ (blue line), 
$(m-M)_I=15.9$ (cyan line), and $(m-M)_K=15.6$ (green line) in 
Figure \ref{distance_reddening_rs_oph_v407_cyg_u_sco_ci_aql_no2}(c).
The four distance-reddening relations cross each other
at $d=12.6$~kpc and $E(B-V)=0.26$.
Therefore, we adopt $(m-M)_V= 16.3$, $d=12.6$~kpc, and $E(B-V)=0.26$.

\subsubsection{Color-color diagram}
\label{color_color_diagram_u_sco}
We plot the color-color diagram of U~Sco in outburst in Figure
\ref{color_color_diagram_t_pyx_v_745_sco_v407_cyg_u_sco_outburst}
for the reddening of $E(B-V)=0.26$ as well as V745~Sco and T~Pyx.
\citet{hac16kb} discussed the reddening toward U~Sco based on the
color-color diagram in the outburst, the $UBV$ data of which are taken from
\citet{pagnotta15}, and concluded that the reddening of
$E(B-V)=0.35\pm0.05$ is consistent with the general track 
(green lines) of novae
in the intrinsic color-color diagram \citep{hac14k}. 
We reanalyzed the data in Figure
\ref{color_color_diagram_t_pyx_v_745_sco_v407_cyg_u_sco_outburst}
and concluded that the track of U~Sco is still consistent with
the general track (green lines) even for $E(B-V)=0.26$.
This confirms that the reddening toward U~Sco is $E(B-V)=0.26\pm0.05$.

We check the reddening value of $E(B-V)=0.26\pm0.05$ from the literature.
U~Sco is located at a high galactic latitude,
$(l,b)=(357\fdg6686,+21\fdg8686)$. 
Thus, the extinction toward U~Sco could be close to
the galactic 2D dust extinction. The NASA/IPAC galactic dust absorption
map gives $E(B-V)=0.32\pm0.04$ toward U~Sco. 
We plot the distance-reddening relations (black and orange lines)
given by \citet{gre15, gre18}, respectively.  
The cross point at $d=12.6$~kpc and $E(B-V)=0.26$ is consistent with
the orange line (revised version) of \citet{gre18}.

Unfortunately,
direct measurements of reddening show a large scatter between 0.1 and 0.35
\citep[e.g.,][for a summary]{schaefer10a}.
For example, \citet{bar81} obtained two different absorptions
toward U Sco in the 1979 outburst,
$E(B-V)\sim0.2$ from the line ratio of \ion{He}{2}
at an early phase of the outburst ($\sim12$ days after maximum)
and $E(B-V)\sim0.35$ from the Balmer line ratio at a late phase
of the outburst ($\sim 33-34$~days after maximum).  
The latter value
was obtained assuming the case B recombination.  It should be noted
that it could be incorrect if the case B condition is not satisfied
as mentioned by \citet{bar81}.
Thus, we may conclude that the value of $E(B-V)\sim0.26\pm0.05$ 
is broadly consistent with other various estimates.

\subsubsection{Color-magnitude diagram}
Figure \ref{hr_diagram_u_sco_ci_aql_yy_dor_lmcn2009a_outburst}(a)
shows the color-magnitude diagram (filled red circles) of U~Sco
for $E(B-V)=0.26$ and $(m-M)_V= 16.3$.
The filled red circles are taken from \citet{pagnotta15}
and the filled blue circle denotes the 1987 outburst taken from 
\citet{sek88}.  
We also plot the track (filled green stars) of the recurrent nova
T~Pyx (2011) in outburst, the data of which are taken from SMARTS
\citep{wal12}. The distance of T~Pyx is taken from \citet{sok13}
and the reddening of T~Pyx is taken from \citet{hac14k, hac16kb}, that is,
$d=4.8$~kpc, $E(B-V)=0.25$, and $(m-M)_V=14.2$.  
The track of U~Sco is closely located to that of T~Pyx. 
This similarity may support the set of $E(B-V)=0.26$
and $(m-M)_V=16.3$ for U~Sco.   

\citet{hac16kb} identified several types of tracks (shape and location)
in the color-magnitude diagram of nova outbursts.  They further found that
the V1500~Cyg and V1974~Cyg types of tracks show a large turn
from blueward to redward, as shown by the solid cyan-blue line in
Figure \ref{hr_diagram_u_sco_ci_aql_yy_dor_lmcn2009a_outburst}(a).
For many V1500~Cyg and V1974~Cyg types novae, they confirmed 
that the turning points are located on the two-headed black arrow 
in the color-magnitude diagram, which is given by Equation (5) of 
\citet{hac16kb}, and shown in Figure 
\ref{hr_diagram_u_sco_ci_aql_yy_dor_lmcn2009a_outburst}(a).
We identify the turning point (or cusp) of U~Sco as $(B-V)_0=-0.55$
and $M_V=-4.02$ (large open red square).
This point corresponds to the large open red triangle in Figure
\ref{u_sco_v_bv_ub_xray_radio_logscale}, i.e., the bend of
slope from $F_\nu \propto t^{-1.75}$ to $F_\nu \propto t^{-3.5}$ or
the start of the nebular phase.
The open red triangle is located 0.2 mag above the two-headed arrow,
but the data point (filled blue circle) of \citet{sek88} is almost
on the two-headed arrow.  This may support the trend that the turning
(or reflection) point is located on the two-headed arrow.

After the turning point,
U~Sco goes down further, crossing the two-headed arrow, and jumps
to the left (toward blue) up to $(B-V)_0\sim-1.0$.
We neglect this bluest point in our determination of the
turning point, because it could be affected by the first flare
in the 2010 outburst \citep[e.g.,][]{schaefer11, max12, anu13, pagnotta15}.
This sharp pulse is seen in the $(B-V)_0$ color evolution
in Figure \ref{u_sco_v_bv_ub_xray_radio_logscale}(b).  If we neglect
this epoch, we have a relatively smooth turning point at
$(B-V)_0=-0.55$ in Figure \ref{u_sco_v_bv_ub_xray_radio_logscale}(b).

Below $M_V > -3.0$, we suppose that the $B$ and $V$ magnitudes are affected
by a large irradiated disk around the WD \citep[e.g.,][]{hkkm00} and 
the $B-V$ color is contaminated by the disk radiation. This is the reason
why the color-magnitude track stays at $(B-V)_0\approx-0.0$
between $-2.0 \lesssim M_V \lesssim -0.0$.
In Figure \ref{hr_diagram_u_sco_ci_aql_yy_dor_lmcn2009a_outburst}(a),
we added the epoch when the SSS phase started
at $m_V=14.0$, approximately 14 days after the outburst \citep{schaefer11}.

\subsubsection{Distance and reddening}
Figure \ref{distance_reddening_rs_oph_v407_cyg_u_sco_ci_aql_no2}(c)
shows various distance-reddening relations.  The black and orange lines
show the distance-reddening relations calculated by \citet{gre15, gre18},
respectively.  The four (magenta, blue, cyan, and green) lines denote
the distance moduli in the four bands, $(m-M)_B=16.6$, $(m-M)_V=16.3$, 
$(m-M)_I=15.9$, and $(m-M)_K=15.6$, respectively.  These are calculated
from the distance moduli of LMC novae, YY~Dor, LMC~N~2009a, and LMC~N~2012a.
Therefore, we regard that these four are based on firm ground.
The four lines cross at $d=12.6$~kpc and $E(B-V)=0.26$ as already
mentioned in Section \ref{u_sco_timescaling_law}.
The reddening of $E(B-V)=0.26$ is just on the distance-reddening
relation (orange line) revised by \citet{gre18}.
From the consistency in the distance-reddening relation (cross point
and Green et al.'s reddening) and also in the color-color and
color-magnitude diagrams, we conclude that the set of 
$(m-M)_V=16.3\pm0.2$ and $E(B-V)=0.26\pm0.05$ ($d=12.6\pm2.0$~kpc) are
reasonable.

\subsubsection{Discussion on the distance}
The distance of $d=12.6\pm2.0$~kpc is larger than the distance
of $d=6-7$~kpc calculated by \citet{hkkm00, hkkmn00} from the nova
explosion/quiescent models.  Hachisu et al. calculated the nova brightness
from the WD photosphere and the photospheric surface of the accretion disk
with blackbody approximation. Their brightness did not
include a large contribution of free-free emission (and nebula emission)
from optically thin gaseous matter outside the WD photosphere.  
This is the reason why they obtained a much shorter distance.

\citet{schaefer10a} proposed a value
of $d=12\pm2$~kpc assuming totality at mid-eclipse and G5IV spectral type
of the companion star.  This is consistent with our estimate.
However, various spectral types of the companion star have
been reported. \citet{anu00} proposed K2IV, which was revised to 
K0$\sim$K1 by \citet{anu13}.  
\citet{mas12} suggested the spectral type not earlier 
than F3 and not later than G, but they could not confidently exclude 
an early K spectral type. We must be careful to use the spectral type
in determining the distance.


\begin{figure}
\includegraphics[height=11.3cm]{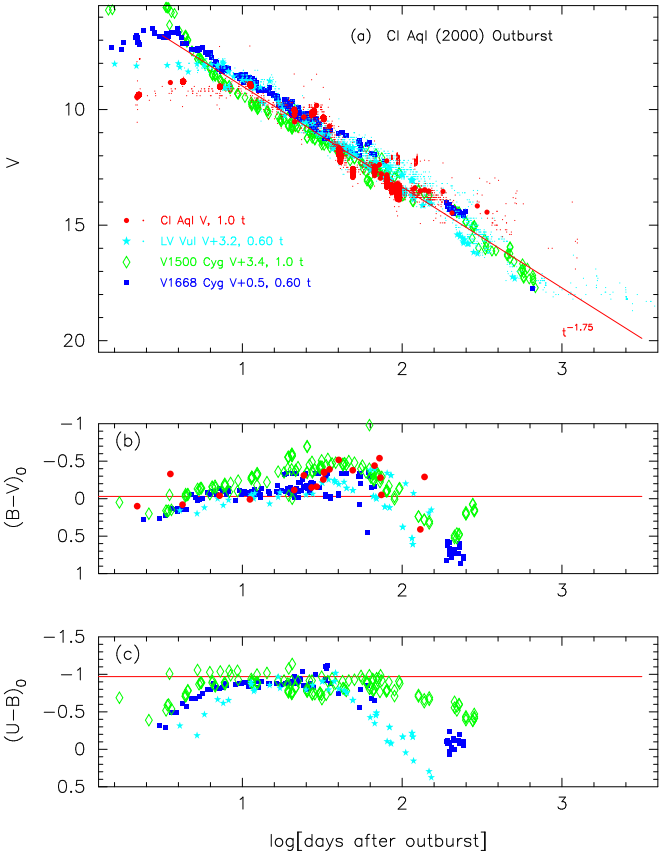}
\caption{
Light/color curves of the CI~Aql (2000) outburst
as well as LV~Vul, V1500~Cyg, and V1668~Cyg.
(a) The filled red circles and dots denote the $V$ and visual magnitudes
of CI~Aql, whereas the filled cyan stars, open green diamonds, and
filled blue squares represent the $V$ magnitudes of LV~Vul, V1500~Cyg,
and V1668~Cyg, respectively.  The data of LV~Vul, V1668~Cyg,
and V1500~Cyg are the same as those in Figures 4 and 1 of \citet{hac16kb},
and Figure 6 of \citet{hac14k}, respectively.  These novae broadly follow 
the universal decline law of $F_\nu\propto t^{-1.75}$
(solid red line) \citep{hac06kb}.
The (b) $(B-V)_0$ and (c) $(U-B)_0$ color curves.  
The colors of CI~Aql are dereddened with $E(B-V)=1.0$.
\label{ci_aql_lv_vul_V1500_cyg_v_bv_ub_logscale}}
\end{figure}


\begin{figure}
\includegraphics[height=13.5cm]{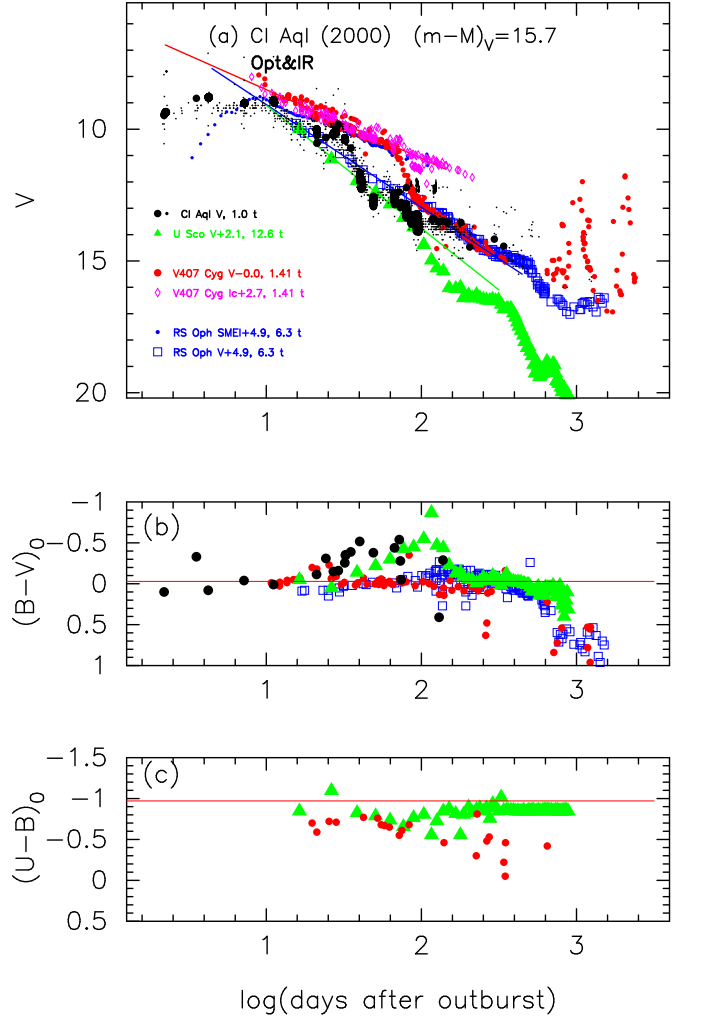}
\caption{
Same as Figure \ref{u_sco_v407_cyg_rs_oph_v_bv_ub_logscale},
but we plot the light/color curves of CI~Aql (2000) as well as 
U~Sco (2010), RS~Oph (2006), and V407~Cyg.
(a) The filled black circles and black dots denote the $V$ and visual
magnitudes of CI~Aql.
The (b) $(B-V)_0$ and (c) $(U-B)_0$ color curves.
The colors of CI~Aql are dereddened with $E(B-V)=1.0$. 
\label{ci_aql_u_sco_v407_cyg_rs_oph_v_bv_ub_logscale_no2}}
\end{figure}

\subsection{CI~Aql (2000)}
\label{ci_aql}
     CI~Aql is also a recurrent nova with three recorded outbursts
in 1917, 1941, and 2000 \citep[e.g.,][]{schaefer10a}.
The maximum brightness and decline rates of the 2000 outburst were
summarized as $m_{V, \rm max}=8.8$, $t_2=25$ days, and $t_3=32$ days
by \citet{str10}. The WD mass was estimated by \citet{hac03a} and
\citet{hac03ka} to be $M_{\rm WD}=1.2\pm0.05~M_\sun$ from the model
light curve fitting. The mass transfer rate and WD mass increasing
rate were estimated to be $\dot M_2\sim -1\times10^{-7} M_\sun$~yr$^{-1}$
and $\dot M_{\rm WD}=1.8\times 10^{-8}~M_\sun$~yr$^{-1}$, respectively, by \citet{hac03ka}.
It takes $t_{\rm Ia} > (1.38 - 1.2)~M_\sun/1.8
\times 10^{-8}~M_\sun$~yr$^{-1}  > 1\times10^7$~yr, during which time
the companion will lose $\sim 1.0~M_\sun$ and the mass ratio will be
reversed. Then the thermal timescale mass transfer cannot be maintained.
We may conclude that CI~Aql is not a progenitor of SNe~Ia.

\subsubsection{Universal decline law}
Figure \ref{ci_aql_lv_vul_V1500_cyg_v_bv_ub_logscale} shows
the light/color curves of CI~Aql on a logarithmic timescale.
This figure also shows the light/color curves of the classical novae,
LV~Vul, V1500~Cyg, and V1668~Cyg. Although the $V$ light curve
of CI~Aql (2000) shows a wavy structure, these four novae broadly follow
the universal decline law of $F_\nu\propto t^{-1.75}$ (solid red line).
The horizontal shifts of $\log f_{\rm s}$ are essentially determined
to overlap the $(B-V)_0$ color curves of these novae as shown in Figure
\ref{ci_aql_lv_vul_V1500_cyg_v_bv_ub_logscale}.  
Once the horizontal shift is fixed, the vertical shift of $V$ light curve
is uniquely determined.
When the $V$ light curves obey the universal decline law and overlap 
each other, i.e., satisfy Equation (\ref{overlap_brigheness}),
we can apply Equation (\ref{distance_modulus_formula}) to 
Figure \ref{ci_aql_lv_vul_V1500_cyg_v_bv_ub_logscale} and
obtain the relation:
\begin{eqnarray}
(m&-&M)_{V, \rm CI~Aql} \cr
&=& (m - M + \Delta V)_{V, \rm LV~Vul} - 2.5 \log 0.60 \cr
&=& 11.9 + 3.2 + 0.55 = 15.65 \cr
&=& (m - M + \Delta V)_{V, \rm V1668~Cyg} - 2.5 \log 0.60 \cr
&=& 14.6 + 0.5 + 0.55 = 15.65 \cr
&=& (m - M + \Delta V)_{V, \rm V1500~Cyg} - 2.5 \log 1.0 \cr
&=& 12.3 + 3.4 + 0.0 = 15.7,
\label{distance_modulus_ci_aql_lv_vul}
\end{eqnarray}
where we adopt $(m-M)_{V, \rm LV~Vul}=11.9$, 
$(m-M)_{V, \rm V1668~Cyg}=14.6$, and
$(m-M)_{V, \rm V1500~Cyg}=12.3$ in \citet{hac16kb}.
We searched for the best overlap of CI~Aql with V1668~Cyg and
LV~Vul but the vertical shifts of $\Delta V$ are not accurately
determined because of its wavy structure of the $V$ light curve. 
Thus, we obtain $(m-M)_V=15.7\pm 0.5$ for CI~Aql.

\subsubsection{Reddening and distance}
Figure \ref{distance_reddening_rs_oph_v407_cyg_u_sco_ci_aql_no2}(d) shows
various distance-reddening relations toward CI~Aql, whose 
galactic coordinates are $(l,b)=(31\fdg6876, -0\fdg8120)$.
The relations of \citet{mar06} are plotted
in four directions close to the direction of CI~Aql:
$(l, b)=(31\fdg50, -0\fdg75)$ (open red squares),
$(31\fdg75, -0\fdg75)$ (filled green squares),
$(31\fdg50, -1\fdg00)$ (blue asterisks), and
$(31\fdg75, -1\fdg00)$ (open magenta circles).
The closest one is that shown with filled green squares.
We added the relations of \citet{gre15, gre18} (black and orange lines)
and \citet{ozd16} (open cyan-blue diamonds). 
We adopt the reddening of $E(B-V)=1.0$ after \citet{hac16kb}.
The 3D distance-reddening relations of Marshall et al. and
Green et al. (orange line), $(m-M)_V=15.7$ 
(solid blue line), and $E(B-V)=1.0$ (vertical solid red line) 
consistently cross each other at the distance of $d=3.3$~kpc
and reddening of $E(B-V)=1.0$. Thus, we finally confirm 
that both the reddening of $E(B-V)=1.0$ and distance modulus
in the $V$ band of $(m-M)_V=15.7$ are reasonable.

\subsubsection{Color-magnitude diagram}
Figure \ref{hr_diagram_u_sco_ci_aql_yy_dor_lmcn2009a_outburst}(b)
shows the color-magnitude diagram of CI~Aql.
We identify the turning point (or cusp) to be $(B-V)_0=-0.52$
and $M_V=-3.84$, which is indicated by the large open red square.
This turning point is on the two-headed black arrow 
given by Equation (5) of \citet{hac16kb}. This property is common among
U~Sco and a number of classical novae, as discussed by \citet{hac16kb}.
The track of CI~Aql almost overlaps with that of T~Pyx. This again
supports the set of $E(B-V)=1.0$ and $(m-M)_V=15.7$ for CI~Aql.   


\subsubsection{Timescaling law and time-stretching method}
Figure \ref{ci_aql_u_sco_v407_cyg_rs_oph_v_bv_ub_logscale_no2} compares
CI~Aql with RS~Oph, V407~Cyg, and U~Sco.  
If we adopt $(m-M)_V=15.7$ for CI~Aql, the vertical shift of the $V$ light
curve and horizontal shift of $\log f_{\rm s}$ are uniquely determined.
We apply Equation
(\ref{distance_modulus_formula}) to Figure 
\ref{ci_aql_u_sco_v407_cyg_rs_oph_v_bv_ub_logscale_no2}(a)
to obtain the relation:
\begin{eqnarray}
(m&-&M)_{V, \rm CI~Aql} = 15.7 \cr
&=& (m - M + \Delta V)_{V, \rm RS~Oph} - 2.5 \log 6.3 \cr
&=& 12.8 + 4.9 - 2.0 = 15.7 \cr
&=& (m - M + \Delta V)_{V, \rm V407~Cyg} - 2.5 \log 1.41 \cr
&=& 16.1 - 0.0 - 0.38 = 15.72 \cr
&=& (m - M + \Delta V)_{V, \rm U~Sco} - 2.5 \log 12.6 \cr
&=& 16.3 + 2.1 - 2.75 = 15.65,
\label{distance_modulus_ci_aql_rs_oph}
\end{eqnarray}
where we adopt $(m-M)_{V, \rm RS~Oph}=12.8$ in Section \ref{rs_oph},
$(m-M)_{V, \rm V407~Cyg}=16.1$ in Section \ref{v407_cyg},
and $(m-M)_{V, \rm U~Sco}=16.3$ in Section \ref{u_sco}.
The $V$ light curve of CI~Aql does not accurately overlap with
the other three $V$ light curves, mainly because of the wavy shape.
However, the global trend follows the other three $V$ light curves.
Note that the $(B-V)_0$ color curve of CI~Aql roughly overlaps
with the trend of U~Sco.
In this sense, the time-stretching method of Equation
(\ref{distance_modulus_formula}) seems to be valid, even for
the wavy $V$ light curve of CI~Aql.


\begin{figure}
\includegraphics[height=8.5cm]{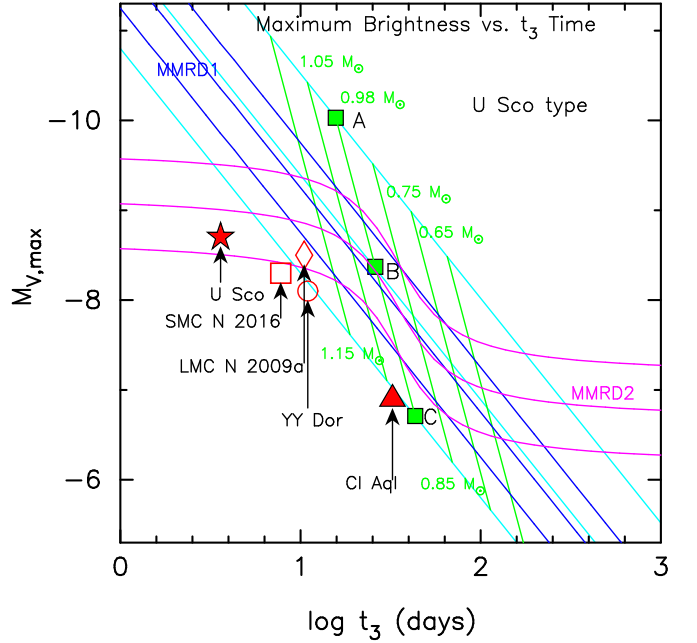}
\caption{
Same as Figure \ref{max_t3_scale_v745_sco_type},
but we plot U~Sco (filled red star), 
CI~Aql (filled red triangle), YY~Dor (open red circles), 
LMC~N~2009a (open red diamond), and
SMC~N~2016 (open red square).  
Here, we assume a distance of $20.4$~kpc for SMC~N~2016.
We also plot the MMRD lines (solid green lines) for
the same WD mass but with different initial envelope masses
(different ignition masses of the hydrogen-rich envelope).  
We plot 1.15, 1.05, 0.98, 0.85, 0.75, and $0.65~M_\sun$ WDs for
the chemical composition of CO nova 3.
The cyan line passing through point A (C) is a MMRD relation
for a much larger (smaller) initial envelope mass than that of point B. 
The slope of these cyan lines is the same as that of MMRD1.
See Appendix \ref{mmrd} for more detail.
\label{max_t3_scale_u_sco_type}}
\end{figure}

\subsection{MMRD relation of normal-decline type novae}
We plot each MMRD point of the two recurrent novae, U~Sco and CI~Aql,
in Figure \ref{max_t3_scale_u_sco_type}. We added the same type of
novae in LMC and SMC that are analyzed in Section \ref{novae_lmc_smc_m31}.
These normal-decline type novae are located slightly below or within
the lower bound of MMRD1 and MMRD2.
These locations can be understood in terms of 
small ignition masses, where the ignition mass is the hydrogen-rich
envelope mass at the start of a nova outburst. 

\citet{hac10k} theoretically examined the MMRD law for the galactic
classical novae on the basis of the universal decline law. 
They concluded that the main trend of the MMRD relation is governed
by the WD mass, i.e., the timescaling factor of $f_{\rm s}$ is the
main parameter. The second parameter is the ignition mass,
which determines the deviation from the main trend.
To show the dependences,
we plot the MMRD relations (solid green lines) 
in Figure \ref{max_t3_scale_u_sco_type} for
the same WD mass but with different initial envelope masses.
Here, we plot 1.15, 1.05, 0.98, 0.85, 0.75, and $0.65~M_\sun$ WDs for
the chemical composition of CO nova 3,
which are taken from 
Figure \ref{max_t3_scale_no4} of Appendix \ref{mmrd}. 

There are three points denoted by A, B, and C on the $0.98~M_\sun$
WD model in Figure \ref{max_t3_scale_u_sco_type}.  
The classical nova V1668~Cyg is located on point B. 
A nova on point A starts from a much larger envelope mass (smaller
mass-accretion rate) than that of V1668~Cyg whereas a nova on point C 
has a much smaller envelope mass at ignition (larger mass-accretion rate)
than that of V1668~Cyg even if they have 
the same WD mass as that of V1668~Cyg.
The three cyan lines passing through points, A, B, and C, 
represent the model MMRD relations derived in Appendix \ref{mmrd}.
These three cyan lines envelop the distribution of the classical novae
studied by \citet{dow00} as shown in Figure \ref{max_t3_scale_no4} of
Appendix \ref{mmrd}.
The upper bound cyan line represents the case of larger envelope 
mass at ignition and the lower bound cyan line indicates that of smaller
envelope mass at ignition for different WD masses.

In Figure \ref{max_t3_scale_u_sco_type}, the normal-decline type novae
are located almost on the lower bound of, or slightly below, the broad MMRD 
relation (cyan lines).  These locations are theoretically
understood in terms of small initial envelope masses,
but the location of U~Sco is slightly lower.
This may be related to the shorter recurrence period
of U~Sco ($\sim 10$ yr) than the other ($\sim 20$ yr or more).
The larger the mass-accretion rate is, the smaller the ignition mass is 
\citep[e.g.,][]{kat14shn}.   
Thus, the normal-decline type novae can be understood based on the
universal decline law ($F_\nu \propto t^{-1.75}$). 
We may conclude that the normal-decline type novae have
the same properties as those of classical novae that follow
the universal decline law and do not belong to the faint class 
claimed by \citet{kas11}.

\subsection{Summary of normal-decline (U~Sco) type novae}
\label{summary_u_sco_novae}
We have analyzed the two recurrent novae, U~Sco and CI~Aql.
The decay of optical and NIR light curves broadly follows the universal
decline of $F_\nu\propto t^{-1.75}$ \citep{hac06kb}.
If we properly stretch the timescales and shift up or down
the $V$ magnitudes, we showed that these novae broadly obey 
the timescaling law of Equation (\ref{overlap_brigheness})
and the time-stretching method of Equation (\ref{distance_modulus_formula})
is valid for these two novae.
Although the $V$ light curve of CI~Aql shows a wavy structure and does not 
exactly follow the universal decline law, we call these novae the normal
decline (or U~Sco) type novae.  

We summarize the results of the two normal-decline (U~Sco) type novae. 
\begin{enumerate}
\item We have estimated the distance, reddening, and distance modulus of
U~Sco from various literature and methods after \citet{hac16kb}.
The results are summarized in Table \ref{extinction_distance_various_novae}.
\item We have also estimated the distance, reddening, and distance 
modulus of CI~Aql from various literature and methods.
The results are summarized in Table \ref{extinction_distance_various_novae}.
\item The $V$ light curve of U~Sco decays as $F_\nu \propto t^{-1.75}$
in the early phase and then drops as $F_\nu \propto t^{-3.5}$. 
\item The $V$ light curves of CI~Aql shows a wavy structure but its
global trend roughly follows the universal trend of $F_\nu\propto t^{-1.75}$.
\item We confirm that these two novae broadly overlap, 
i.e., satisfy Equation (\ref{overlap_brigheness}) and the time-stretching
method, i.e., Equation (\ref{distance_modulus_formula}). 
\item These two novae are located at the lower bound of 
the law of Della Valle \& Livio (MMRD2).  CI~Aql is located at the lower side
of Kaler-Schmidt's law (MMRD1) but U~Sco is slightly fainter than
this trend.
\item The WD mass of U~Sco was estimated to be $M_{\rm WD}=1.37~M_\sun$
from the model light curve fitting \citep{hkkm00}.  This WD mass is so 
close to the explosion mass of SN~Ia, $M_{\rm Ia}=1.38~M_\sun$ \citep{nom82}.
The WD mass increases now \citep{hkkm00}.  
This suggests that U~Sco is a promising progenitor of SNe~Ia.
\item The WD mass of CI~Aql is estimated to be $M_{\rm WD}=1.2\pm0.05~M_\sun$
\citep{hac03ka} from the light curve fitting.  The WD mass increases now
\citep{hac03ka} but it takes $t_{\rm Ia} > (1.38 - 1.2)~M_\sun/1.8
\times 10^{-8}~M_\sun$~yr$^{-1}  > 1\times10^7$~yr, during which
the companion will lose $\sim 1.0~M_\sun$ and the mass ratio will be
reversed. The thermal timescale mass transfer cannot be maintained.
Therefore, CI~Aql is probably not a progenitor of SNe~Ia.
\end{enumerate}

\section{Novae in Magellanic Clouds and M31}
\label{novae_lmc_smc_m31}
This is the first time that we apply our time-stretching method to 
extra-galactic novae, i.e., LMC, SMC, and M31. Their distances
are generally well determined, and thus it is a good litmus test for 
our time-stretching method. We also observe if there are any
characteristic differences between these novae and galactic novae,
especially owing to the difference in the metallicity.
We analyze YY~Dor (2004), Nova~LMC~2009a, Nova LMC~2012a,
Nova~LMC~2013 in LMC, and Nova~SMC~2016 in SMC, and M31~N~2008-12a 
(2015) in M31. For LMC, we use the distance modulus of $\mu_0=18.5$
\citep{pie13} and the reddening of $E(B-V)=0.12$ \citep{imara07} and 
neglect the difference in the reddening inside LMC unless otherwise specified.


\begin{figure}
\includegraphics[height=13cm]{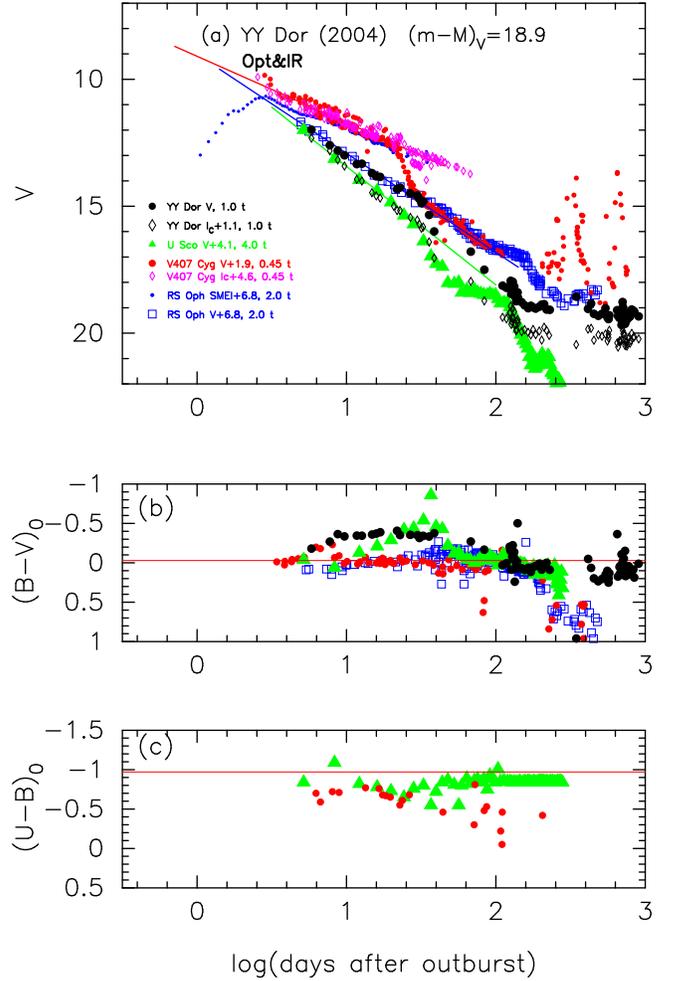}
\caption{
Same as Figure \ref{u_sco_v407_cyg_rs_oph_v_bv_ub_logscale},
but for the light/color curves of YY~Dor (2004) as well as U~Sco,
V407~Cyg, and RS~Oph. The data of YY~Dor are taken from SMARTS.
(a) The filled black circles denote the $V$ magnitudes of YY~Dor, whereas
the filled green triangles, filled red circles, and open blue squares
denote the $V$ magnitudes of U~Sco, V407~Cyg, and RS~Oph, respectively.
The $I_C$ magnitudes (open black diamonds) of YY~Dor are added.
The (b) $(B-V)_0$ and (c) $(U-B)_0$ color curves. The colors of 
YY~Dor are dereddened with $E(B-V)=0.12$. 
See the text for detail.
\label{yy_dor_v407_cyg_rs_oph_v_bv_ub_logscale_no2}}
\end{figure}

\subsection{YY~Dor (2004)}
\label{yy_dor}
YY~Dor is a recurrent nova in LMC with two recorded outbursts
in 1937 and 2004 \citep[e.g.,][]{bond04, shaf13, mas14}. 
The brightness reaches $m_{V, \rm max}=10.8$ \citep{lil04}
and the decline rates are estimated to be $t_2=4.0$ days and 
$t_3=10.9$ days \citep{wal12}. The distance modulus in the $V$ band
is $(m-M)_V= \mu_0 + A_V= 18.5 + 3.1\times 0.12 = 18.9$, where
$A_V= 3.1\times E(B-V)$ is the absorption in the $V$ band.

Figure \ref{yy_dor_v407_cyg_rs_oph_v_bv_ub_logscale_no2} shows
the (a) $V$ and $I_C$ light curves and 
(b) $(B-V)_0$ color curve of YY~Dor (2004).
The data of YY~Dor are taken from SMARTS \citep{wal12}.
We explain our procedure of overlapping the nova light curves. 
In Figure \ref{yy_dor_v407_cyg_rs_oph_v_bv_ub_logscale_no2}(a), 
the $V$ light curve of YY~Dor has a slope of $F_\nu\propto t^{-1.55}$
rather than $F_\nu\propto t^{-1.75}$.  Therefore, we overlap it to
the $V$ light curve of RS~Oph rather than that of U~Sco.
We adopt $(m-M)_{V, \rm YY~Dor}=18.9$ and apply Equation
(\ref{distance_modulus_formula}) to the set of YY~Dor and RS~Oph.
Then, we have
\begin{eqnarray}
(m&-&M)_{V, \rm YY~Dor} = 18.9 \cr
&=& (m - M + \Delta V)_{V, \rm RS~Oph} - 2.5 \log f_{\rm s} \cr
&=& 12.8 + \Delta V - 2.5 \log f_{\rm s},
\label{distance_modulus_yy_dor_u_sco}
\end{eqnarray}
where we adopt $(m-M)_{V, \rm RS~Oph}=12.8$ in Section \ref{rs_oph}.
In this case, we have the relation:
\begin{equation}
\Delta V - 2.5 \log f_{\rm s}=6.1.
\label{dv_log_fs}
\end{equation}
We change the value of $\Delta V$ in steps of 0.1 mag, obtain
$\log f_{\rm s}$ from Equation (\ref{dv_log_fs}), and plot
the $V$ light curve of YY~Dor for the set of $\Delta V$ (vertical shift)
and $\log f_{\rm s}$ (horizontal shift).
We search for the best-fit position of overlapping, as shown
in Figure \ref{yy_dor_v407_cyg_rs_oph_v_bv_ub_logscale_no2}(a). 
In this case, we finally obtain $\Delta V=6.8$ and $\log f_{\rm s}=0.3$
($f_{\rm s}=2.0$). After this procedure, the other two, V407~Cyg
and U~Sco, are fixed automatically because the overlapping positions of
U~Sco, V407~Cyg, and RS~Oph are already known in Figure
\ref{u_sco_v407_cyg_rs_oph_v_bv_ub_logscale}.
Thus, we obtain the vertical and horizontal shifts of 
$\Delta V$ and $\log f_{\rm s}$ for these novae.

We also plot the $I_C$ light curve (open black diamonds) of YY~Dor
in Figure \ref{yy_dor_v407_cyg_rs_oph_v_bv_ub_logscale_no2}(a)
in order to confirm that our horizontal fit is reasonable.
The $I_C$ light curve has a slope of $F_\nu\propto t^{-1.75}$
and its break of the slope coincides with the break of U~Sco
$V$ light curve for the horizontal shift of $\log f_{\rm s}=0.3$.
Thus, we conclude that YY~Dor belongs to the U~Sco
type (normal decline), although its $V$ light curve has a slope of
$F_\nu\propto t^{-1.55}$.

Now, we examine whether the time-stretching method is applicable to 
LMC novae. Using $(m-M)_V=18.9$ for YY~Dor, we apply Equation
(\ref{distance_modulus_formula}) to all the novae in Figure 
\ref{yy_dor_v407_cyg_rs_oph_v_bv_ub_logscale_no2}(a) 
and obtain the relation:
\begin{eqnarray}
(m&-&M)_{V, \rm YY~Dor} = 18.9 \cr
&=& (m - M + \Delta V)_{V, \rm RS~Oph} - 2.5 \log 2.0 \cr
&=& 12.8 + 6.8 - 0.75 = 18.85 \cr
&=& (m - M + \Delta V)_{V, \rm V407~Cyg} - 2.5 \log 0.45 \cr
&=& 16.1 + 1.9 + 0.88 = 18.88 \cr
&=& (m - M + \Delta V)_{V, \rm U~Sco} - 2.5 \log 4.0 \cr
&=& 16.3 + 4.1 - 1.5 = 18.9,
\label{distance_modulus_yy_dor_rs_oph_v407_cyg}
\end{eqnarray}
where we adopt $(m-M)_{V, \rm RS~Oph}=12.8$ in Section \ref{rs_oph},
$(m-M)_{V, \rm V407~Cyg}=16.1$ in Section \ref{v407_cyg}, and
$(m-M)_{V, \rm U~Sco}=16.3$ in Section \ref{u_sco}.
These values are close to $(m-M)_{V, \rm YY~Dor} = 18.9$.
Thus, we confirm that the set of YY~Dor, U~Sco, V407~Cyg, and RS~Oph
consistently overlap with each other in the early phase, i.e., satisfy
the timescaling law of Equation (\ref{overlap_brigheness}) and,
at the same time, satisfy
the time-stretching method of Equation (\ref{distance_modulus_formula}).

\begin{figure}
\includegraphics[height=13cm]{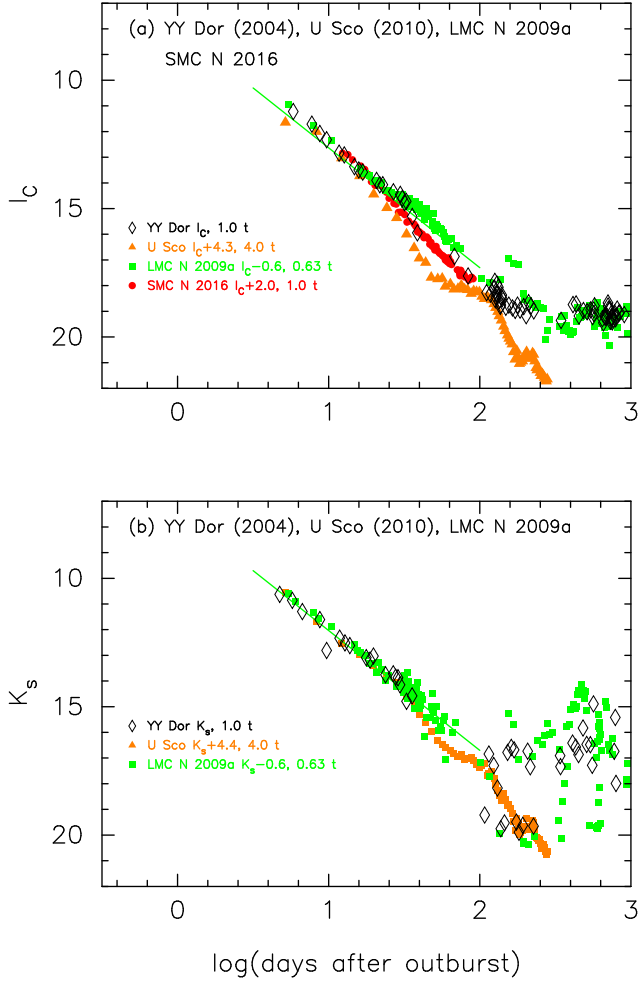}
\caption{
The (a) $I_C$ and (b) $K_s$ light curves of YY~Dor (2004), 
U~Sco (2010), and LMC~N~2009a.  We also add the $I_C$ light curve of
SMC~N~2016.
Each light curve is horizontally stretched by $\log f_{\rm s}$
and vertically shifted by $\Delta I_C$ (or $\Delta K_s$).
See the text for detail.
\label{yy_dor_lmcn_2009a_smcn2016_u_sco_i_k_logscale_2fig}}
\end{figure}

Figure \ref{yy_dor_lmcn_2009a_smcn2016_u_sco_i_k_logscale_2fig} shows 
the $I_C$ and $K_s$ light curves of YY~Dor, U~Sco, and LMC~N~2009a.
We also add the $I_C$ light curve of SMC~N~2016 in Figure
\ref{yy_dor_lmcn_2009a_smcn2016_u_sco_i_k_logscale_2fig}(a).
This figure is essentially the same as Figure
\ref{u_sco_yy_dor_lmcn_2009a_i_k_logscale_2fig}, but we plot
the other light curves against YY~Dor.
These three novae have a slope of $F_\nu\propto t^{-1.75}$
(green lines) in the $I_C$ and $K_s$ light curves and overlap with the same
scaling factors of $f_{\rm s}$ as that of the $V$ light curves.
The distance modulus in the $I_C$ band is calculated to be
$(m-M)_{I, \rm YY~Dor} = 18.5 + 1.5\times 0.12= 18.7$.
We apply Equation (\ref{distance_modulus_formula}) to all the novae
and obtain
\begin{eqnarray}
(m&-&M)_{I, \rm YY~Dor} = 18.7 \cr
&=& (m - M + \Delta I_C)_{I, \rm U~Sco} - 2.5 \log 4.0 \cr
&=& 15.9 + 4.3 - 1.5 = 18.7 \cr
&=& (m - M + \Delta I_C)_{I, \rm LMC~N~2009a} - 2.5 \log 0.63 \cr
&=& 18.8 - 0.6 + 0.5 = 18.7 \cr
&=& (m - M + \Delta I_C)_{I, \rm SMC~N~2016} - 2.5 \log 1.0 \cr
&=& 16.7 + 2.0 - 0.0 = 18.7,
\label{distance_modulus_i_yy_dor_u_sco_lmcn2009a}
\end{eqnarray}
where we adopt $(m-M)_{I, \rm U~Sco}=16.3 - 1.6\times 0.26= 15.9$
from Section \ref{u_sco}, 
$(m-M)_{I, \rm LMC~N~2009a}=18.5 + 1.5\times 0.2=18.8$
from Section \ref{lmcn_2009a}, and
$(m-M)_{I, \rm SMC~N~2016}=16.7$ from Section \ref{smcn_2016}
in advance.
Thus, we confirm that the set of YY~Dor, U~Sco, LMC~N~2009a, and SMC~N~2016
consistently overlap with each other in the early phase, i.e., satisfy
the timescaling law of Equation (\ref{overlap_brigheness}) and,
at the same time, satisfy
the time-stretching method of Equation (\ref{distance_modulus_formula}).

For the $K_s$ light curves, we similarly obtain
\begin{eqnarray}
(m&-&M)_{K, \rm YY~Dor} = 18.5 + 0.35\times 0.12=18.5 \cr
&=& (m - M + \Delta K_s)_{K, \rm U~Sco} - 2.5 \log 4.0 \cr
&=& 15.6 + 4.4 - 1.5 = 18.5 \cr
&=& (m - M + \Delta K_s)_{K, \rm LMC~N~2009a} - 2.5 \log 0.63 \cr
&=& 18.6 - 0.6 + 0.5 = 18.5,
\label{distance_modulus_k_yy_dor_u_sco_lmcn2009a}
\end{eqnarray}
where we adopt $(m-M)_{K, \rm U~Sco}=16.3 - 2.75\times 0.26= 15.6$
from Section \ref{u_sco}, 
$(m-M)_{K, \rm LMC~N~2009a}=18.5 + 0.35\times 0.2=18.6$
from Section \ref{lmcn_2009a}.
Thus, we confirm that the set of YY~Dor, U~Sco, and LMC~N~2009a
satisfy the timescaling law of Equation (\ref{overlap_brigheness}) and
the time-stretching method of Equation (\ref{distance_modulus_formula}).
Thus, we conclude that YY~Dor (and LMC~N~2009a) belong to the normal
decline (U~Sco) type.


We plot the MMRD point of YY~Dor in Figure \ref{max_t3_scale_u_sco_type}.
It lies in the lower side of the broad MMRD relations, close to the
other members of the normal-decline type novae. 

Figure \ref{hr_diagram_u_sco_ci_aql_yy_dor_lmcn2009a_outburst}(c)
shows the color-magnitude diagram of YY~Dor.  
We add a track (thin solid green lines) of V1668~Cyg shifted by 
$\Delta (B-V)=-0.3$ for comparison.  The color-magnitude track of YY~Dor
almost overlaps with the blue-shifted V1668~Cyg track. 
We suppose that the reason for this blue-shift is the lower metallicity
of LMC stars, [Fe/H]$=-0.55$ \citep[see, e.g.,][]{pia13}.

The WD mass of YY~Dor is estimated to be $M_{\rm WD}=1.29~M_\sun$ 
from the linear relation between $M_{\rm WD}$ and $\log f_{\rm s}$
in Figure \ref{timescale_wd_mass} (see Table \ref{wd_mass_recurrent_novae}
for the other WD masses).
Although the exact recurrence period and orbital period of the binary
are not known, we assume that the mass transfer rate and WD mass increasing
rate are similar to those of CI~Aql. Then,
the mass transfer rate and WD mass increasing
rate were estimated to be $\dot M_2\sim 1\times10^{-7} M_\sun$~yr$^{-1}$ and 
$\dot M_{\rm WD}=1.8\times 10^{-8}~M_\sun$~yr$^{-1}$ by \citet{hac03ka}.
It takes $t_{\rm Ia} > (1.38 - 1.29)~M_\sun/1.8
\times 10^{-8}~M_\sun$~yr$^{-1}  > 0.5\times10^7$~yr, during which
the companion will lose $\gtrsim 0.5~M_\sun$ and the mass ratio could be
reversed. The thermal timescale mass transfer cannot be maintained.
We may conclude that YY~Dor is not a progenitor of SNe~Ia.


\begin{figure}
\includegraphics[height=12cm]{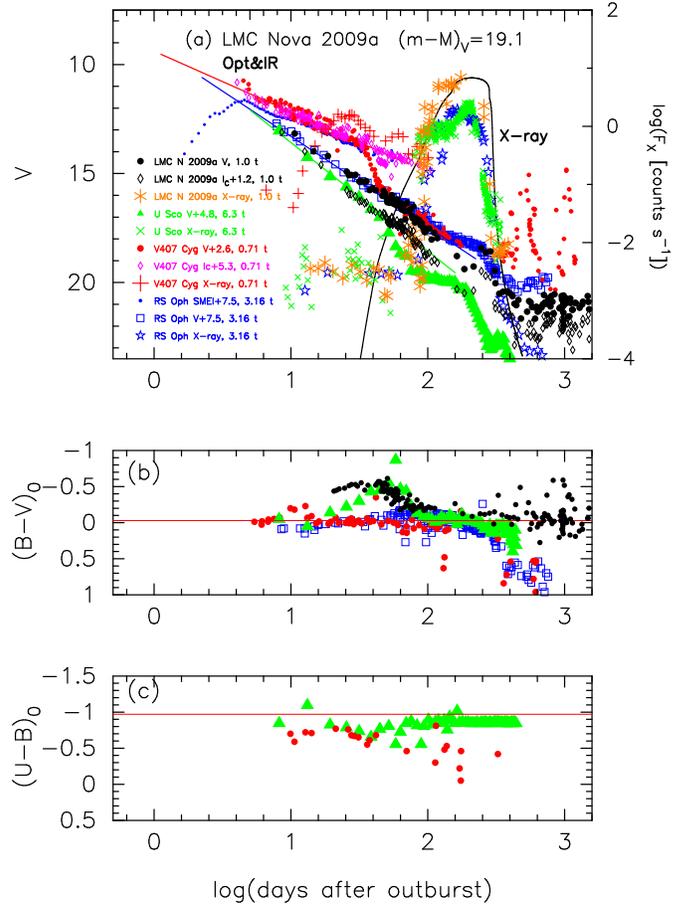}
\caption{
Same as Figure \ref{u_sco_v407_cyg_rs_oph_v_bv_ub_logscale},
but for LMC~N~2009a. We included U~Sco, V407~Cyg, and RS~Oph for comparison.
(a) The filled black circles denote the $V$ magnitudes and
orange asterisks represent the X-ray flux of LMC~N~2009a.
The $I_C$ magnitudes (open black diamonds) of LMC~N~2009a are added.
The solid black line is the model supersoft X-ray flux of a 
$1.25~M_\sun$ WD with the chemical composition of Ne nova 3.
The (b) $(B-V)_0$ and (c) $(U-B)_0$ color curves.  The colors of
LMC~N~2009a are dereddened with $E(B-V)=0.2$.
\label{lmcn2009a_u_sco_v407_cyg_rs_oph_v_bv_ub_logscale_no2}}
\end{figure}

\subsection{Nova~LMC~2009a}
\label{lmcn_2009a}
LMC~Nova~2009a is a recurrent nova with recorded outbursts in
1971 and 2009.  There was some confusion regarding the locations of the 1971
(1971b or 1971-08a) nova and 2009 (2009a or 2009-02a) nova.
As a result, the recurrence was once suspected 
\citep[e.g.,][]{shaf13, mro16}.   
\citet{bod16} remeasured the original plate of the 1971 outburst and
confirmed that the 1971 nova position is the same as the 2009 nova 
position within 0\farcs1.
The brightness reached $m_{\rm max}=10.6$ 
\citep[unfiltered Kodak Technical Pan films by][]{lil09a}.
The decline rates are estimated to be $t_2=5.0$ days and 
$t_3=10.4$ days in the $V$ band \citep{bod16}.  
We assume that the $V$ maximum also reached $m_{V, \rm max}=10.6$.
Figure \ref{lmcn2009a_u_sco_v407_cyg_rs_oph_v_bv_ub_logscale_no2} shows
the (a) $V$ light curve and (b) $(B-V)_0$ color curve of LMC~N~2009a.
The data of LMC~N~2009a are taken from SMARTS \citep{wal12}.
\citet{bod16} reported pan-chromatic observations of LMC~N~2009a,
the emergence of the SSS phase at $63-70$ days,
which was initially highly variable, and periodic modulations with
$P = 1.2$ days, which is most probably orbital in nature.
They identified the progenitor system,
i.e., the secondary is most likely a subgiant feeding a luminous 
accretion disk and suggested a WD mass of $1.1-1.3 ~M_\sun$
from the properties of the SSS phase.  
They also obtained the reddening $E(B-V)=A_V/3.1
= 0.6/3.1 = 0.2$ from the neutral hydrogen column density.  
Using this value, we calculate the intrinsic color of $(B-V)_0$
and plot them in Figure
\ref{lmcn2009a_u_sco_v407_cyg_rs_oph_v_bv_ub_logscale_no2}(b).
The color evolution of LMC~N~2009a is similar to that of U~Sco.  
The distance modulus in the $V$ band is calculated to be 
$(m-M)_V=\mu_0 + A_V=18.5 + 0.6 =19.1$.

Using $(m-M)_V=19.1$ for LMC~N~2009a, we confirm that the $V$ light
curves of LMC~N~2009a overlap to that of RS~Oph 
and satisfy Equation (\ref{overlap_brigheness}) as shown in Figure
\ref{lmcn2009a_u_sco_v407_cyg_rs_oph_v_bv_ub_logscale_no2}.
The slope of the $V$ light curve is close to that of RS~Oph rather
than U~Sco while the slope of $I_C$ light curve (open black diamonds)
is close to that of U~Sco.  Therefore, we conclude that LMC~N~2009a belongs
to the U~Sco (normal decline) type. 
In the figure, we adopt the horizontal shift $\log f_{\rm s}=0.50$ 
of RS~Oph against LMC~N~2009a by overlapping the end of SSS phase 
of RS~Oph (open blue stars) to that of LMC~N~2009a (orange asterisks).
Then, we apply Equation (\ref{distance_modulus_formula}) to Figure 
\ref{lmcn2009a_u_sco_v407_cyg_rs_oph_v_bv_ub_logscale_no2} 
and obtain the relation:
\begin{eqnarray}
(m&-&M)_{V, \rm LMCN~2009a} = 19.1 \cr
&=& (m - M + \Delta V)_{V, \rm RS~Oph} - 2.5 \log 3.16 \cr
&=& 12.8 + 7.5 - 1.25 = 19.05 \cr
&=& (m - M + \Delta V)_{V, \rm V407~Cyg} - 2.5 \log 0.71 \cr
&=& 16.1 + 2.6 + 0.38 = 19.08 \cr
&=& (m - M + \Delta V)_{V, \rm U~Sco} - 2.5 \log 6.3 \cr
&=& 16.3 + 4.8 - 2.0 = 19.1,
\label{distance_modulus_lmcn_2009a_rs_oph}
\end{eqnarray}
where we adopt $(m-M)_{V, \rm RS~Oph}=12.8$ in Section \ref{rs_oph},
$(m-M)_{V, \rm V407~Cyg}=16.1$ in Section \ref{v407_cyg}, and
$(m-M)_{V, \rm U~Sco}=16.3$ in Section \ref{u_sco}.
From the close agreement of $(m-M)_V=19.1$,
we confirm that the set of LMC~N~2009a, U~Sco, V407~Cyg, and RS~Oph follow
the timescaling law of Equation (\ref{overlap_brigheness}) 
and satisfy the time-stretching method of Equation
(\ref{distance_modulus_formula}).  

With the same timescaling factors, we have already showed that
the $I_C$ and $K_s$ light curves of LMC~N~2009a, YY~Dor, and U~Sco
overlap to each other and satisfy the time-stretching method of Equation
(\ref{distance_modulus_formula}) in Sections \ref{u_sco} and \ref{yy_dor}.
These $I_C$ and $K_s$ light curves have a slope of $F_\nu\propto t^{-1.75}$,
so we regard LMC~N~2009a as a normal decline (U~Sco) type.
We plot the MMRD point in Figure \ref{max_t3_scale_u_sco_type}.
It is located in the lower bound of the broad MMRD relation. 

Figure \ref{hr_diagram_u_sco_ci_aql_yy_dor_lmcn2009a_outburst}(d)
shows the color-magnitude diagram of LMC~N~2009a.  
The track of LMC~N~2009a is in good agreement with
the 0.2 mag blue-shifted V1500~Cyg and 0.15 mag blue-shifted 
U~Sco tracks.  The irradiated accretion disk contributes to 
the $B$, $V$, and $(B-V)$ below $M_V > -3$ and, as a result,
the $(B-V)_0$ color is almost constant. 
The LMC~N~2009a track overlaps to the 0.15 mag blue-shifted U~Sco track.
It should be noted that the turning point is located on the line of
Equation (5) of \citet{hac16kb}, i.e., the two-headed
black arrow, in Figure
\ref{hr_diagram_u_sco_ci_aql_yy_dor_lmcn2009a_outburst}(d).
We again suppose that the bluer
feature of the LMC~N~2009a track is due to the lower metallicity
of the LMC stars, as mentioned in Section \ref{yy_dor}.

The WD mass of LMC~N~2009a is estimated to be $M_{\rm WD}=1.25~M_\sun$ 
from the linear relation between $M_{\rm WD}$ and $\log f_{\rm s}$
in Figure \ref{timescale_wd_mass} (see Table \ref{wd_mass_recurrent_novae}
for the other WD masses).  We plot the X-ray flux (solid black line) of 
a $1.25~M_\sun$ WD with the chemical composition of Ne nova 3 in Figure 
\ref{lmcn2009a_u_sco_v407_cyg_rs_oph_v_bv_ub_logscale_no2}(a), which
broadly reproduces the observation.
If the orbital period is $1.2$~days, the companion is a subgiant, as in
U~Sco and CI~Aql. The exact recurrence period is not known, but it is 38 yr
or shorter. If we assume that the mass transfer rate and WD mass
increasing rate are similar to those of CI~Aql, it takes $t_{\rm Ia} > 
(1.38 - 1.25)~M_\sun/1.8 \times 10^{-8}~M_\sun$~yr$^{-1}  > 
0.72\times10^7$~yr, during which the companion will lose 
$\gtrsim 0.72~M_\sun$ and the mass ratio could be reversed. As a result,
the thermal timescale mass transfer cannot be maintained.  
We may conclude that LMC~N~2009a is not a progenitor of SNe~Ia.


\begin{figure}
\includegraphics[height=9cm]{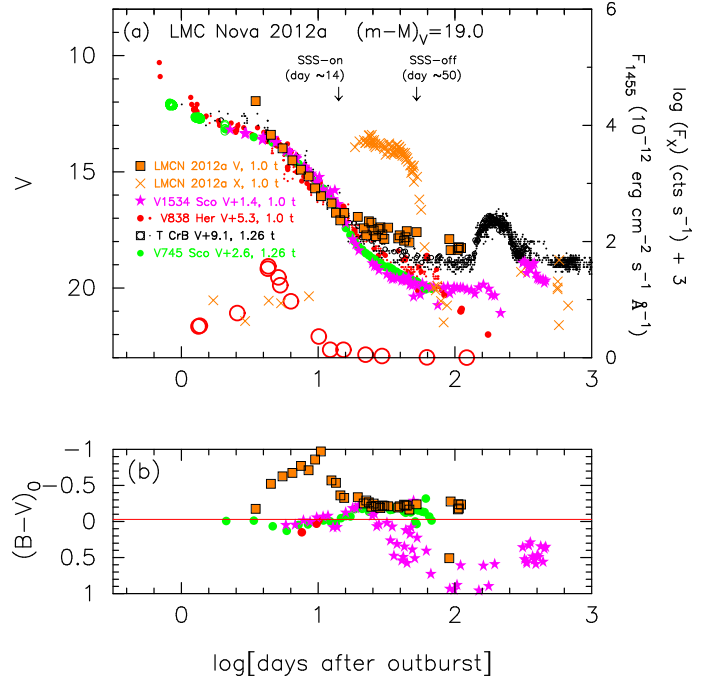}
\caption{
Same as Figure \ref{v1534_sco_v838_her_t_crb_v745_sco_v_bv_ub_logscale},
but for the light/color curves of LMC~N~2012a as well as V1534~Sco, V838~Her,
T~CrB, and V745~Sco.
(a) The filled orange squares with black outlines
denote the $V$ magnitudes of LMC~N~2012a.  The data are taken from SMARTS.
The magenta crosses represent the {\it Swift} XRT X-ray flux ($0.3-10$ keV).
The data are taken from the {\it Swift} web page \citep{eva09}. 
(b) The $(B-V)_0$ colors of LMC~N~2012a are dereddened with $E(B-V)=0.15$.
\label{lmcn2012a_v1534_sco_v838_her_t_crb_v745_sco_v_bv_ub_logscale}}
\end{figure}


\begin{figure}
\includegraphics[height=12.5cm]{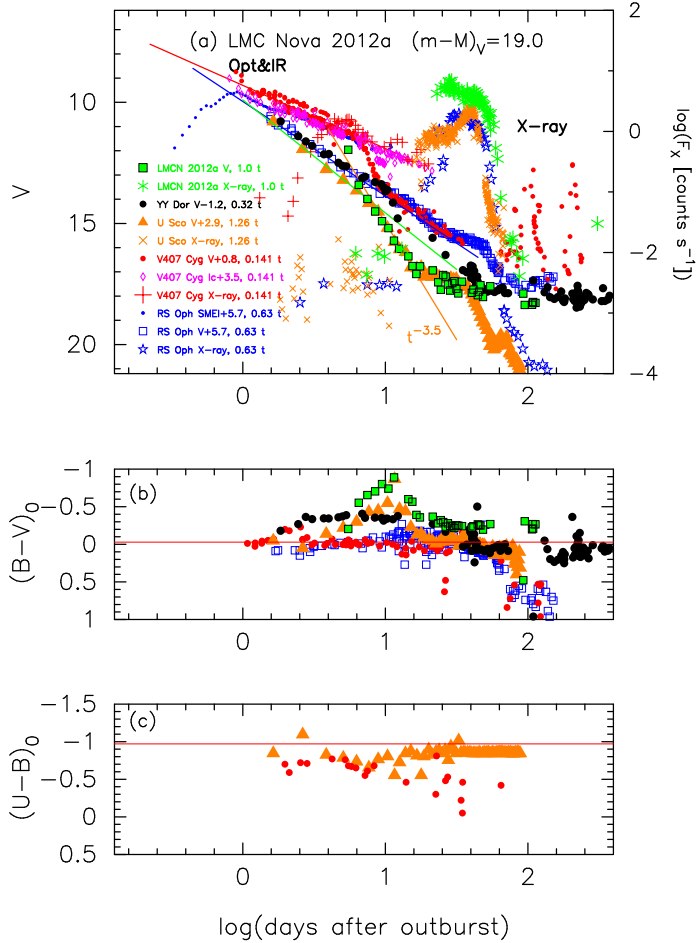}
\caption{
Same as Figure \ref{lmcn2009a_u_sco_v407_cyg_rs_oph_v_bv_ub_logscale_no2},
but for the light/color curves of LMC~N~2012a as well as YY~Dor,
U~Sco, V407~Cyg, and RS~Oph.
(a) The filled green squares with black outlines
denote the $V$ magnitudes of LMC~N~2012a.
The (b) $(B-V)_0$ and (c) $(U-B)_0$ color curves. 
The colors of LMC~N~2012a are dereddened with $E(B-V)=0.15$.
\label{lmcn2012a_yy_dor_v407_cyg_rs_oph_v_bv_ub_logscale}}
\end{figure}


\begin{figure*}
\plotone{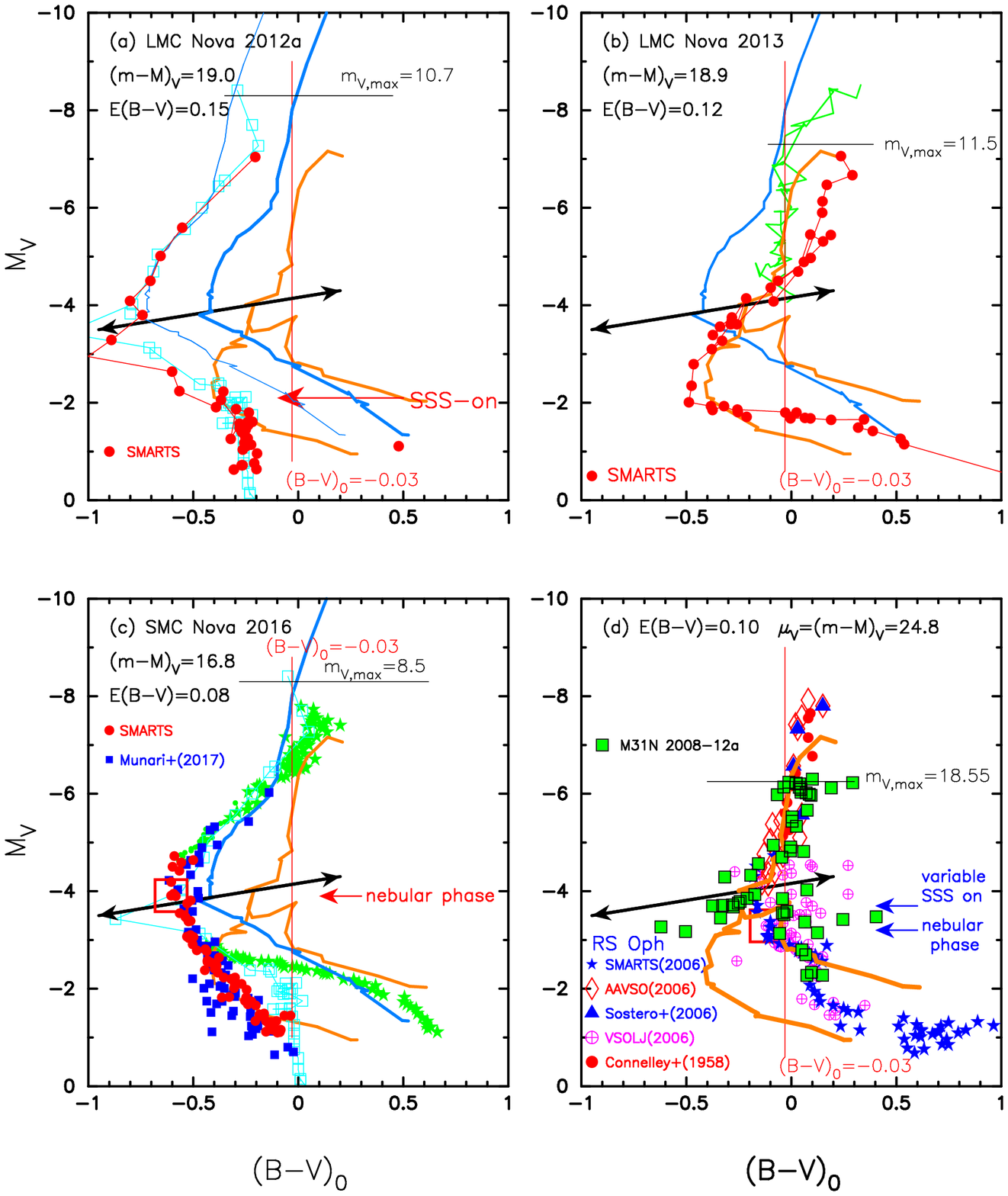}
\caption{
Same as Figure \ref{hr_diagram_v745_sco_v1534_sco_rs_oph_v407_cyg_outburst},
but for (a) LMC~N~2012a, (b) LMC~N~2013, (c) SMC~N~2016, and (d)
M31N~2008-12a. The thick solid cyan-blue lines show the template track
of V1500~Cyg, the thick solid orange lines show that of LV~Vul, and
the thick solid green lines show that of V1668~Cyg.
In panel (a), we include a $\Delta (B-V)_0=-0.3$ mag blue-shifted
track (thin solid cyan-blue line) of V1500~Cyg and 
$\Delta (B-V)_0=-0.25$ mag blue-shifted track (open cyan squares
connected with thin solid cyan line) of U~Sco (2010).  
In panel (c), we include the U~Sco track (open cyan squares
connected with thin solid cyan line) and T~Pyx track (filled green stars).
The large open red square denotes the start of the nebular phase, 23 days
after optical maximum.
In panel (d), we include the data of RS~Oph, which are the same as those
in Figure 16(a) of \citet{hac16kb}. The filled green squares with black outlines denote M31N~2008-12a. Other symbols are all for RS~Oph. 
\label{hr_diagram_lmcn2012a_lmcn2013_smcn2016_m31n2008_12a_outburst}}
\end{figure*}

\subsection{Nova~LMC~2012a}
\label{lmcn_2012a}
LMC~Nova~2012a reached maximum at 
$m_{\rm max}=10.7$ \citep[digital SLR camera by][]{sea12}.
The decline rates are estimated to be $t_2=1.1$ days and 
$t_3=2.1$ days in the $V$ band \citep{wal12}.  
We assumed that the $V$ maximum also reached $m_{V, \rm max}=10.7$.
Figure \ref{lmcn2012a_v1534_sco_v838_her_t_crb_v745_sco_v_bv_ub_logscale}
shows the (a) $V$ light curve and (b) $(B-V)_0$ color curve of LMC~N~2012a.
The data of LMC~N~2012a are taken from SMARTS \citep{wal12}.
\citet{schw15} reported the {\it Swift} and {\it Chandra} X-ray 
observations. The SSS phase started $13\pm5$ days 
and ended approximately 50 days after the discovery. They obtained the orbital
period of $19.24\pm 0.03$~hr (0.802 days) and the reddening of $E(B-V)=0.15$.
We adopt this reddening and calculated the distance modulus in the $V$ band
of $(m-M)_V=\mu_0 + A_V = 18.5 + 3.1\times 0.15 = 19.0$.

Using $(m-M)_V=19.0$ for LMC~N~2012a, we confirm that the $V$ light curve
of LMC~N~2012a overlaps well with the $V$ light curves of V1534~Sco, V838~Her,
T~CrB, and V745~Sco. These five novae satisfy 
Equation (\ref{overlap_brigheness}). Then, we apply Equation
(\ref{distance_modulus_formula}) to Figure 
\ref{lmcn2012a_v1534_sco_v838_her_t_crb_v745_sco_v_bv_ub_logscale}
and obtain the relation:
\begin{eqnarray}
(m&-&M)_{V, \rm LMCN~2012a} = 19.0 \cr
&=& (m - M + \Delta V)_{V, \rm V1534~Sco} - 2.5 \log 1.0 \cr
&=& 17.6 + 1.4 + 0.0 = 19.0 \cr
&=& (m - M + \Delta V)_{V, \rm V838~Her} - 2.5 \log 1.0 \cr
&=& 13.7 + 5.3 + 0.0 = 19.0 \cr
&=& (m - M + \Delta V)_{V, \rm T~CrB} - 2.5 \log 1.26 \cr
&=& 10.1 + 9.1 - 0.25 = 18.95 \cr
&=& (m - M + \Delta V)_{V, \rm V745~Sco} - 2.5 \log 1.26 \cr
&=& 16.6 + 2.6 - 0.25 = 18.95,
\label{distance_modulus_lmcn2012a_v745_sco}
\end{eqnarray}
where we adopt 
$(m-M)_{V, \rm V1534~Sco}=17.6$ in Section \ref{v1534_sco},
$(m-M)_{V, \rm V838~Her}=13.7$ in Section \ref{v838_her},
$(m-M)_{V, \rm T~CrB}=10.1$ in Section \ref{t_crb}, and
$(m-M)_{V, \rm V745~Sco}=16.6$ in Section \ref{v745_sco}.
We confirm that these novae satisfy 
Equation (\ref{distance_modulus_formula}). 
We regard LMC~N~2012a as a rapid-decline (V745~Sco) type nova.
We plot the MMRD point of
LMC~N~2012a in Figure \ref{max_t3_scale_v745_sco_type}.
The point is located closely to the other novae, and
significantly below MMRD1 but slightly below MMRD2.

Figure \ref{lmcn2012a_yy_dor_v407_cyg_rs_oph_v_bv_ub_logscale}
compares the light/color curves of LMC~N 2012a with different types of
novae, YY~Dor, U~Sco, V407~Cyg, and RS~Oph. 
In this figure, there is no exhibit of $F_\nu \propto t^{-1.75}$, 
but one can observe $F_\nu \propto t^{-3.5}$ 
in the $V$ light curve of LMC~N~2012a that overlaps with U~Sco.  
This is because the initial envelope mass
(ignition mass of hydrogen-rich envelope) is too small to reach
the universal decline trend of $F_\nu \propto t^{-1.75}$, which is  
explained in more detail in Section \ref{deviation_mmrd}.
From Figure \ref{lmcn2012a_yy_dor_v407_cyg_rs_oph_v_bv_ub_logscale},
we obtain the relation:
\begin{eqnarray}
(m&-&M)_{V, \rm LMCN~2012a} = 19.0 \cr
&=& (m - M + \Delta V)_{V, \rm YY~Dor} - 2.5 \log 0.32 \cr
&=& 18.9 - 1.2 + 1.25 = 18.95 \cr
&=& (m - M + \Delta V)_{V, \rm U~Sco} - 2.5 \log 1.26 \cr
&=& 16.3 + 2.9 - 0.25 = 18.95 \cr
&=& (m - M + \Delta V)_{V, \rm V407~Cyg} - 2.5 \log 0.141 \cr
&=& 16.1 + 0.8 + 1.13 = 19.03 \cr
&=& (m - M + \Delta V)_{V, \rm RS~Oph} - 2.5 \log 0.63 \cr
&=& 12.8 + 5.7 + 0.5 = 19.0,
\label{distance_modulus_lmcn_2012a_rs_oph}
\end{eqnarray}
where we adopt $(m-M)_{V, \rm YY~Dor}=18.9$ in Section \ref{yy_dor},
$(m-M)_{V, \rm U~Sco}=16.3$ in Section \ref{u_sco}, 
$(m-M)_{V, \rm V407~Cyg}=16.1$ in Section \ref{v407_cyg}, and 
$(m-M)_{V, \rm RS~Oph}=12.8$ in Section \ref{rs_oph}.
Thus, we suggest that Equation (\ref{distance_modulus_formula}) is
satisfied by the set of LMC~N~2012a, YY~Dor, U~Sco, V407~Cyg, and RS~Oph
even for the overlap in the part of $F_\nu \propto t^{-3.5}$. 
Comparing the timescaling factors between Figure
\ref{lmcn2012a_v1534_sco_v838_her_t_crb_v745_sco_v_bv_ub_logscale} and
Figure \ref{lmcn2012a_yy_dor_v407_cyg_rs_oph_v_bv_ub_logscale},
we bridge the timescales between the two groups, V745~Sco and U~Sco.
Then, we obtain the common timescaling factors $f_{\rm s}$
among the three groups, as discussed
in more detail in Section \ref{horizontal_shifts}. 

For the other bands, assuming the same timescaling factors,
we have already compared the $B$ and $I_C$
light curves of LMC~N~2012a with those of U~Sco in Figure
\ref{u_sco_lmcn2012a_b_i_2fig} of Section \ref{u_sco_timescaling_law}.
Using the time-stretching method, we have obtained the distance 
and reddening of U~Sco, because the distance and distance moduli
of LMC~N~2012a are well constrained.

Figure \ref{hr_diagram_lmcn2012a_lmcn2013_smcn2016_m31n2008_12a_outburst}(a)
shows the color-magnitude diagram of LMC~N~2012a.  
The track of LMC~N~2012a is in good agreement with
the 0.3 mag blue-shifted V1500~Cyg (thin solid cyan-blue line) 
and 0.25 mag blue-shifted U~Sco (open cyan squares connected with
cyan line) tracks.
The blue-shifted U~Sco track overlaps well with that of LMC~N~2012a,
including the nearly constant $(B-V)_0$ phase below $M_V > -2$.
The brightness and color of this SSS phase are dominated by the irradiated
accretion disk.
We regard that the lower metallicity environment in LMC
is the cause of the bluer position of the color-magnitude track.

The WD mass of LMC~N~2012a is estimated to be $M_{\rm WD}=1.37~M_\sun$ 
from the linear relation between $M_{\rm WD}$ and $\log f_{\rm s}$
in Figure \ref{timescale_wd_mass} (see Table \ref{wd_mass_recurrent_novae}
for the other WD masses).
It is unlikely that the WD was born as massive 
as $M_{\rm WD}\sim 1.37~M_\sun$ \citep{doh15}.
We suppose that the WD has grown in mass.
This strongly suggests further increase in 
the WD masses in these systems.
The orbital period is 0.802 days and the companion is a subgiant.
Although the recurrence period is not known, we assume that 
the LMC~N~2012a is a recurrent nova with a relatively short recurrence
period, because the WD had grown in mass and the companion is a 
lobe-filling subgiant.
If we further assume that the mass-increasing rate of WD 
is $\dot M_{\rm WD}\sim 1\times10^{-7}~M_\sun$~yr$^{-1}$ 
just below the stability line of hydrogen-shell burning \citep{kat17sh},
the WD mass increases from $1.37~M_\sun$ to 
$M_{\rm Ia}=1.38~M_\sun$ and explodes as a SN~Ia. 
It takes approximately $t_{\rm Ia}\sim 0.01~M_\sun/
1\times10^{-7}~M_\sun$~yr$^{-1} = 1\times10^5$~yr, which is much shorter
than the evolution timescale of the donor (subgiant star).  
Therefore, the WD will explode as a SN~Ia if the core 
consists of carbon and oxygen.  
We suggest that LMC~N~2012a is a promising progenitor of SNe~Ia.


\begin{figure}
\includegraphics[height=8.5cm]{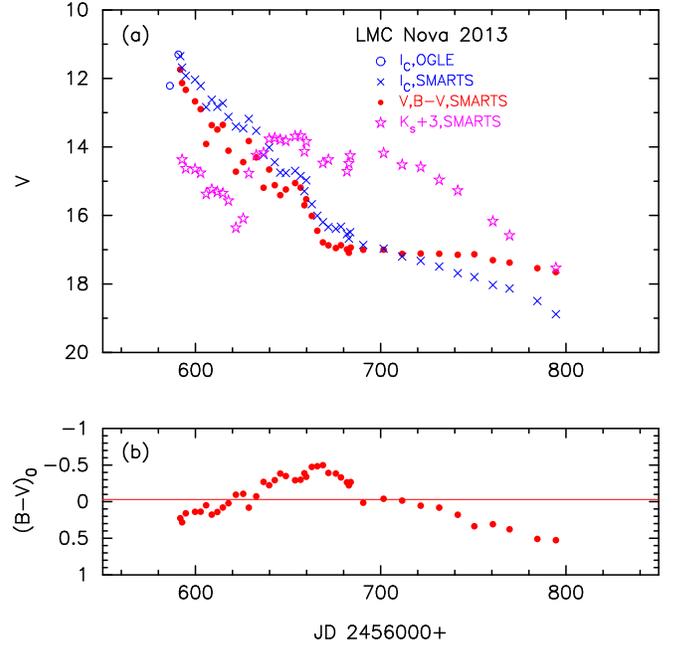}
\caption{
Same as Figure \ref{v407_cyg_v_bv_ub_color_curve}, but for LMC~N~2013.
(a) The $V$ (filled red circles), $I_{\rm C}$ (blue cross), and 
$K_{\rm s}$ (open magenta stars) data are taken from SMARTS.  
The $I_{\rm C}$ (open blue circles) data are from \citet{mro15}.
(b) The $(B-V)_0$ data are taken from SMARTS and 
dereddened with $E(B-V)=0.12$.  
\label{lmc_n_2013_v_bv_ub_color_curve}}
\end{figure}


\begin{figure}
\includegraphics[height=12.5cm]{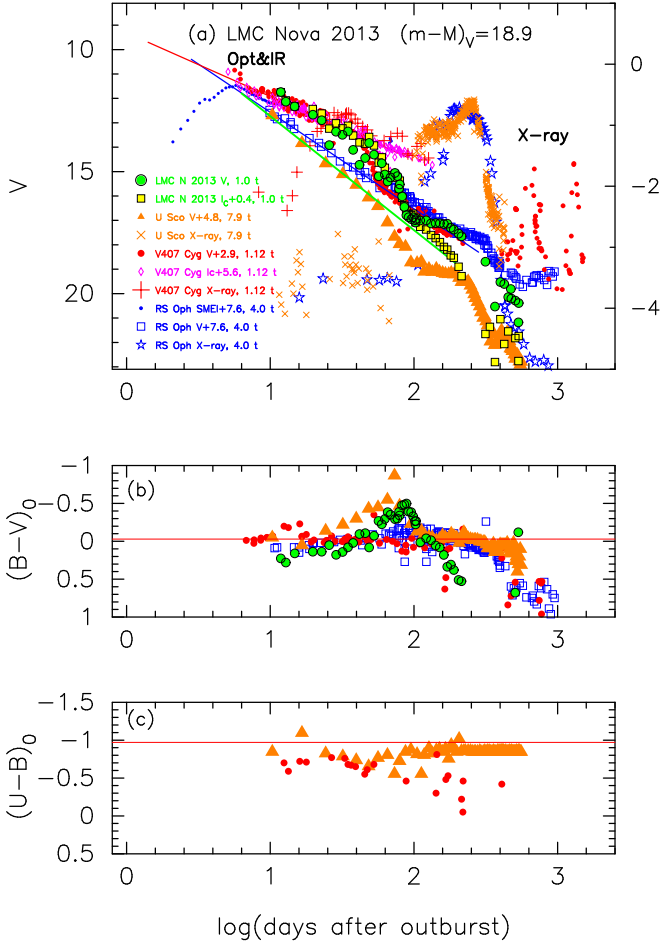}
\caption{
Same as Figure \ref{lmcn2009a_u_sco_v407_cyg_rs_oph_v_bv_ub_logscale_no2},
but for the light/color curves of LMC~N~2013 as well as U~Sco, V407~Cyg,
and RS~Oph.
(a) The filled green circles with black outlines denote
the $V$ magnitudes of LMC~N~2013.
The $I_C$ magnitudes (filled yellow squares with black outlines)
of LMC~N~2013 are added.
The (b) $(B-V)_0$ and (c) $(U-B)_0$ color curves. The $(B-V)_0$ colors 
of LMC~N~2013 are dereddened with $E(B-V)=0.12$.
\label{lmcn2013_u_sco_v407_cyg_rs_oph_v_bv_ub_logscale_no2}}
\end{figure}

\subsection{Nova~LMC~2013}
\label{lmcn_2013}
Figure \ref{lmc_n_2013_v_bv_ub_color_curve} shows the (a) $V$, $I_{\rm C}$,
and $K_{\rm s}$ light and (b) $(B-V)_0$ color curves of LMC~N~2013
on a linear timescale.
The $I_{\rm C}$ light curve decays almost linearly, whereas the $V$ light
curve shows wavy structure with an amplitude of half a magnitude.
The $K_{\rm s}$ magnitude (open magenta stars) sharply increases 
approximately 40 days after the outburst.  
A dust shell formed at this epoch, which was not so optically thick.
The epoch of the $I_{\rm C}$ maximum is determined from the
OGLE \citep[open blue circles,][]{mro15} and SMARTS 
\citep[blue crosses,][]{wal12} data.  
The $I_{\rm C}$ magnitude reaches its maximum 
at $m_{I_{\rm C}, {\rm max}}=11.3$ on JD~2456591.0.
We regard that the $V$ brightness reaches $m_{V, \rm max}=11.5$
on the same day and estimate the decline rates as $t_2=21$ days and 
$t_3=47$ days from this figure.

We adopt $(m-M)_V=18.9$. Then, we determined the vertical shift 
$\Delta V$ and horizontal shift $\log f_{\rm s}$
for the set of LMC~N~2013 and V407~Cyg by the same procedure
as described in Section \ref{yy_dor}. The other timescaling factors
$f_{\rm s}$ and vertical shifts $\Delta V$ are calculated from those
in Figure \ref{u_sco_v407_cyg_rs_oph_v_bv_ub_logscale}. 
Thus, we plot the light/color curves of U~Sco, V407~Cyg, and
RS~Oph against LMC~N~2013 in Figure 
\ref{lmcn2013_u_sco_v407_cyg_rs_oph_v_bv_ub_logscale_no2}.
We apply Equation (\ref{distance_modulus_formula}) to Figure
\ref{lmcn2013_u_sco_v407_cyg_rs_oph_v_bv_ub_logscale_no2}
and obtain the relation:
\begin{eqnarray}
(m&-&M)_{V, \rm LMCN~2013} = 18.9 \cr
&=& (m - M + \Delta V)_{V, \rm U~Sco} - 2.5 \log 7.9 \cr
&=& 16.3 + 4.8 - 2.25 = 18.85 \cr
&=& (m - M + \Delta V)_{V, \rm V407~Cyg} - 2.5 \log 1.12 \cr
&=& 16.1 + 2.9 - 0.13 = 18.87 \cr
&=& (m - M + \Delta V)_{V, \rm RS~Oph} - 2.5 \log 4.0 \cr
&=& 12.8 + 7.6 - 1.5 = 18.9,
\label{distance_modulus_lmcn_2013_rs_oph}
\end{eqnarray}
where we adopt $(m-M)_{V, \rm U~Sco}=16.3$ in Section \ref{u_sco},
$(m-M)_{V, \rm V407~Cyg}=16.1$ in Section \ref{v407_cyg},
and $(m-M)_{V, \rm RS~Oph}=12.8$ in Section \ref{rs_oph}.
Thus, LMC~N~2013 and V407~Cyg satisfy Equations 
(\ref{overlap_brigheness}) and (\ref{distance_modulus_formula}).

The $V$ light curve of LMC~N~2013 shows a similar decline trend to
that of V407~Cyg in the very early phase and then decays to the
trend of RS~Oph in the middle and later phase. Therefore, we regard
LMC~N~2013 as a CSM-shock (RS~Oph) type nova. We plot the MMRD
point of LMC~N~2013 in Figure \ref{max_t3_scale_rs_oph_type}.
The MMRD point of LMC~N~2013 is located in the middle of the MMRD relations
(both the MMRD1 and MMRD2 trends).

Figure \ref{hr_diagram_lmcn2012a_lmcn2013_smcn2016_m31n2008_12a_outburst}(b)
shows the color-magnitude diagram of LMC~N~2013.  
The track of LMC~N~2013 broadly follows the LV~Vul track
(thick solid orange lines) without blue-shifting.
Comparing the LMC~N~2013 track with that of LMC~N~2012a (Figure
\ref{hr_diagram_lmcn2012a_lmcn2013_smcn2016_m31n2008_12a_outburst}(a),
which overlaps with the 0.3 mag blue-shifted template), 
we suppose that LMC~N~2013 has metallicity similar 
to those of galactic novae.

The WD mass of LMC~N~2013 is estimated to be $M_{\rm WD}=1.23~M_\sun$ 
from the linear relation between $M_{\rm WD}$ and $\log f_{\rm s}$
in Figure \ref{timescale_wd_mass} (see Table \ref{wd_mass_recurrent_novae}
for the other WD masses).
Neither the orbital period nor the recurrence period is known.
Even if we assume that the mass transfer rate and WD mass
increasing rate are similar to CI~Aql, 
it takes $t_{\rm Ia} > (1.38 - 1.23)~M_\sun/1.8
\times 10^{-8}~M_\sun$~yr$^{-1}  > 0.83\times10^7$~yr, during which
the companion will lose $\gtrsim 0.83~M_\sun$ and the mass ratio could be
reversed before the WD mass reaches $M_{\rm Ia}=1.38~M_\sun$.
We suppose that the thermal timescale mass transfer
cannot be maintained until a SN~Ia explosion.
We may conclude that LMC~N~2013 is not a progenitor of SNe~Ia.


\begin{figure}
\includegraphics[height=8cm]{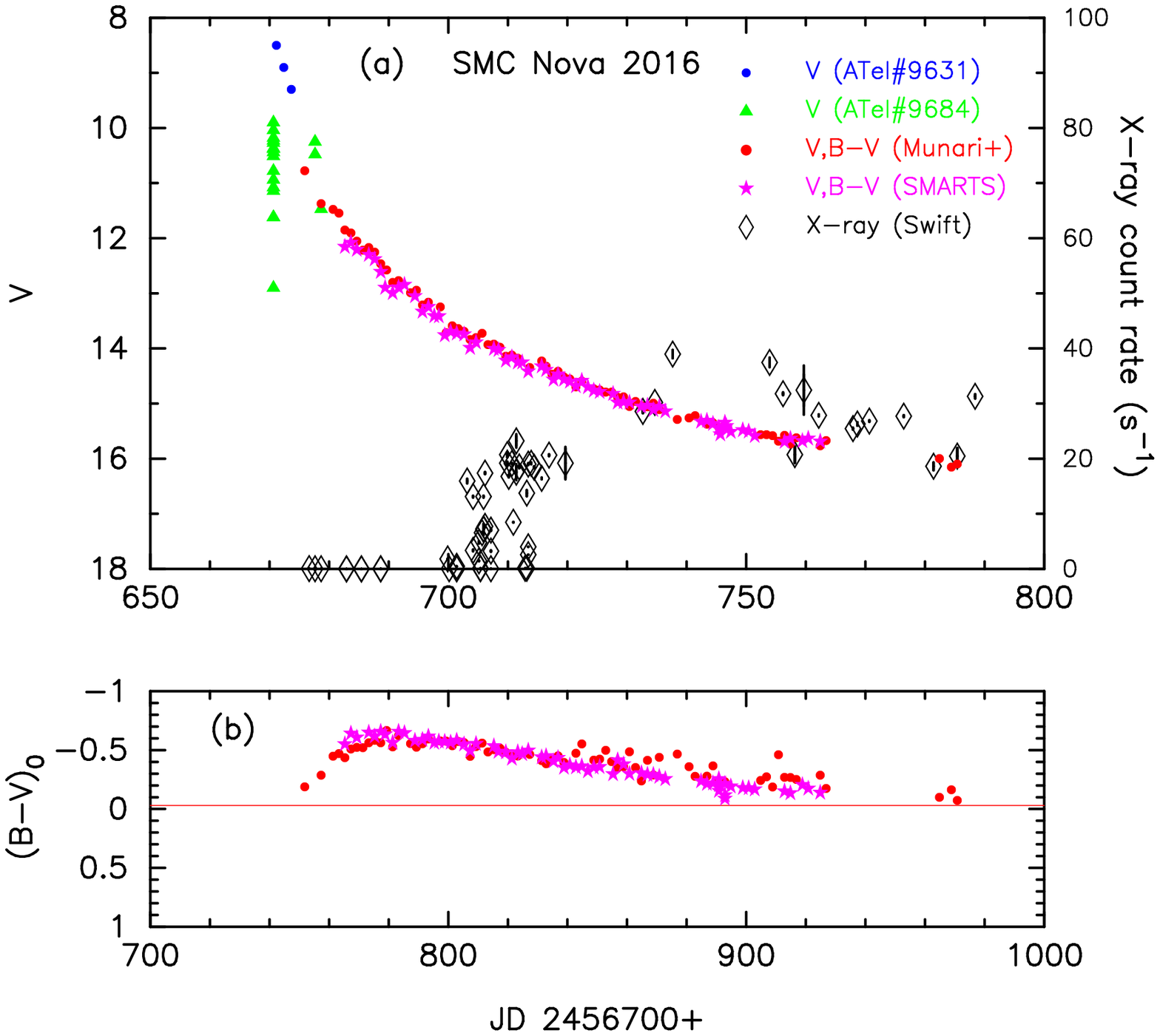}
\caption{
Same as Figure \ref{lmc_n_2013_v_bv_ub_color_curve},
but for SMC~N~2016.
(a) The $V$ (filled red circles) data are taken from \citet{mun17}.
The $V$ (filled magenta stars) data are from SMARTS.
The filled blue circles (ATel No.9631) and filled green triangles
(ATel No.9684) are the pre-discovery brightnesses of SMC~N~2016 and
taken from \citet{lip16} and \citet{job16}, respectively.
We also include the X-ray ($0.3-10$ keV) flux (open black diamonds) 
data taken from the {\it Swift} web site \citep{eva09}.
(b) The $(B-V)_0$ are dereddened with $E(B-V)=0.08$.  
\label{smcn2016_v_bv_ub_color_xray_curve}}
\end{figure}


\begin{figure}
\includegraphics[height=13cm]{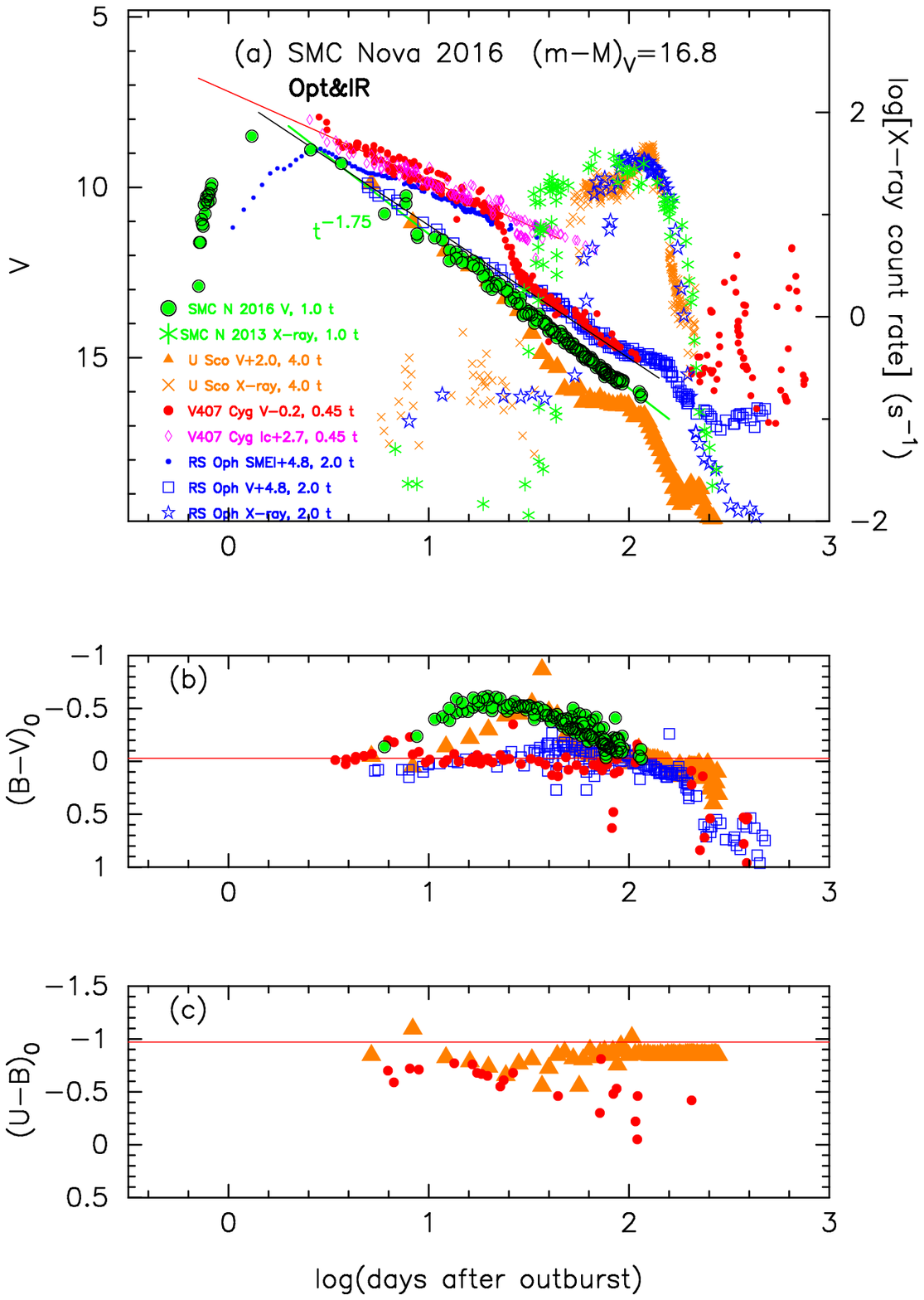}
\caption{
Same as Figure \ref{lmcn2013_u_sco_v407_cyg_rs_oph_v_bv_ub_logscale_no2},
but for the light/color curves of SMC~N~2016.
(a) The filled green circles with black outlines 
denote the $V$ magnitudes of SMC~N~2016.
The (b) $(B-V)_0$ and (c) $(U-B)_0$ color curves. 
\label{smcn2016_yy_dor_v407_cyg_rs_oph_v_bv_ub_logscale_no2}}
\end{figure}


\begin{figure}
\includegraphics[height=12cm]{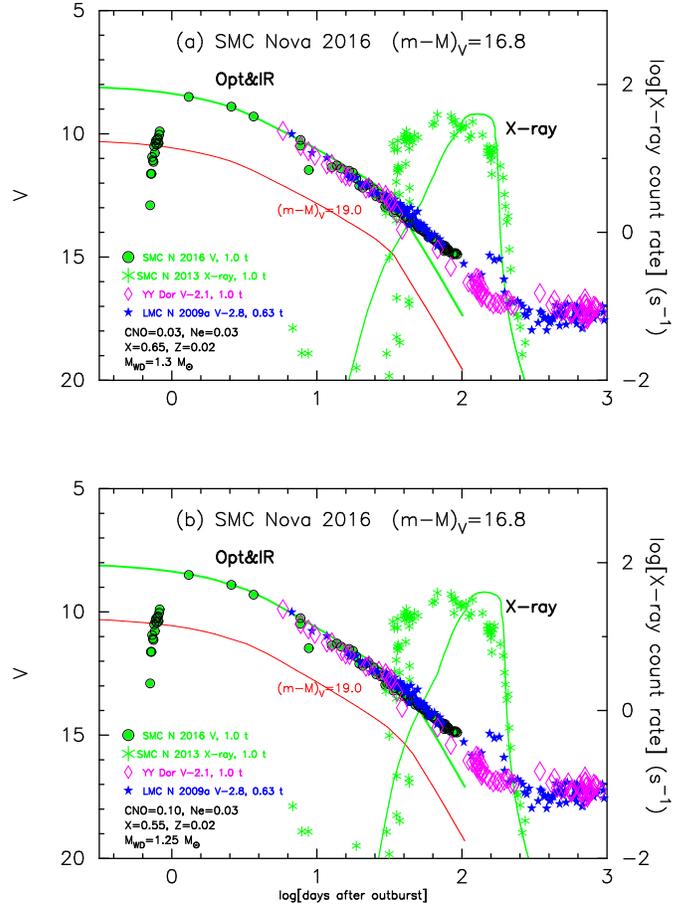}
\caption{
Same as Figure 
\ref{smcn2016_yy_dor_v407_cyg_rs_oph_v_bv_ub_logscale_no2}(a),
but comparison with YY~Dor and LMC~N~2009a.
The solid lines represent the model light curves of 
(a) a $1.3~M_\sun$ WD with the chemical composition of Ne nova 3,
and (b) a $1.25~M_\sun$ WD of Ne nova 2.
The green/red lines correspond to the model $V$ light curves for
the distance modulus of $(m-M)_V=16.8$/19.0.
\label{smcn2016_yy_dor_lmcn_2009a_v_x-ray_logscale}}
\end{figure}


\begin{figure}
\includegraphics[height=8.5cm]{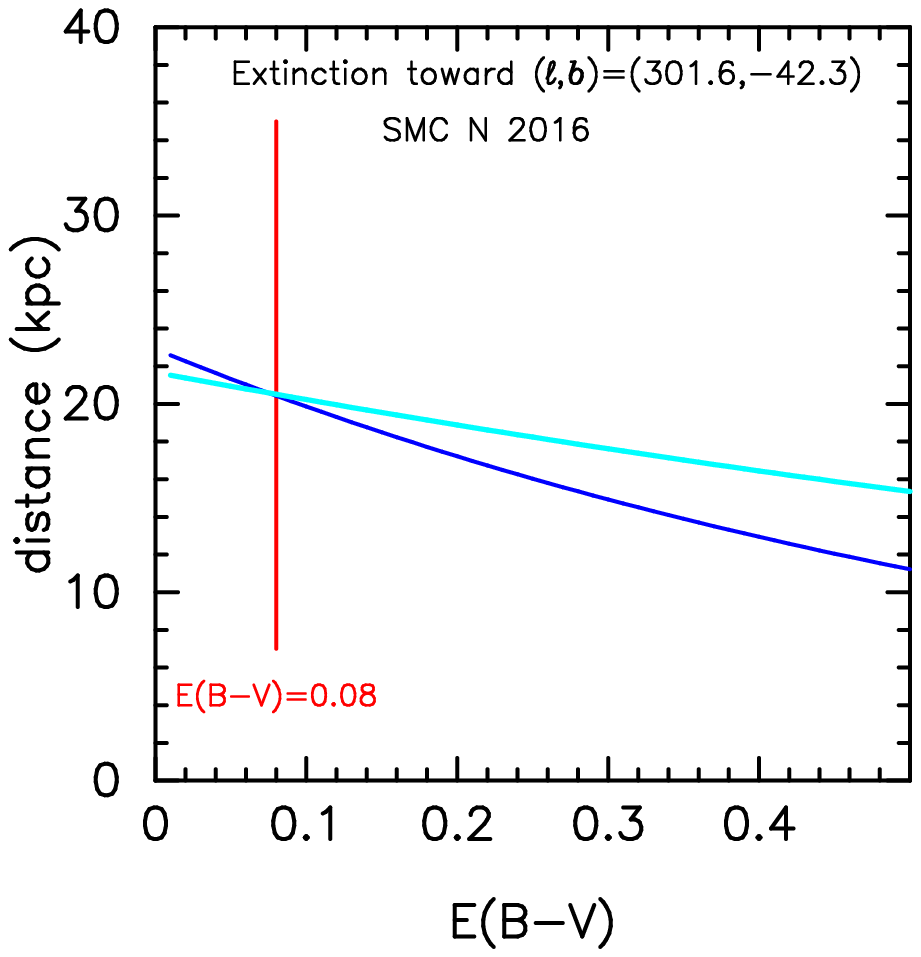}
\caption{
Various distance-reddening relation toward SMC~N~2016 are plotted.
The vertical solid red line denotes the reddening of $E(B-V)=0.08$.
The blue and cyan lines indicate Equations (\ref{v_distance_modulus})
and (\ref{distance_modulus_relation_i}) for $(m-M)_V=16.8$ and 
$(m-M)_I=16.7$, respectively.
\label{distance_reddening_smcn2016_1fig}}
\end{figure}

\subsection{Nova~SMC~2016}
\label{smcn_2016}
SMC~Nova~2016 (or 2016-10a) is located in the direction toward 
the outskirt of SMC,
whose galactic coordinates are $(l,b)=(301\fdg6362, -42\fdg3037)$.
Although \citet{mun17} and \citet{aydi18} assumed that SMC~N~2016 is a
member of SMC, we suppose that SMC~N~2016 is a member of our galaxy,
as discussed below.

Figure \ref{smcn2016_v_bv_ub_color_xray_curve} shows
the (a) $V$ light and (b) $(B-V)_0$ color curves of SMC~N~2016.
This nova was discovered by \citet{shum16} at mag 10.8 (unfiltered)
on UT 2016 October 14.19.
\citet{lip16} and \citet{job16} reported the pre-discovery 
brightnesses of SMC~N~2016. The brightness reached $m_{\rm max}=8.5$ 
\citep[unfiltered,][]{lip16} and the decline rates
were estimated to be $t_2=8.8$ days and $t_3=17.4$ days by \citet{mun17}
or $t_2=4.0\pm1.0$ days and $t_3=7.8\pm2.0$ days by \citet{aydi18}.
\citet{mun17} also estimated the distance to be $d=44.8$~kpc 
and the reddening to be $E(B-V)=0.08$. Then, the distance modulus
in the $V$ band is calculated to be $(m-M)_V=18.5$.
The peak brightness is estimated from the MMRD relations to be
$M_{V, \rm max}=-8.9$ or $-9.3$, depending on which MMRD relation
is adopted. However, \citet{aydi18} obtained 
$M_{V, \rm max}=-10.5\pm0.5$ and $(m-M)_V=19.0\pm0.5$, assuming 
the distance of $d=61\pm10$~kpc and absorption of $A_V=0.11\pm0.06$.

\citet{aydi18} concluded that, if the nova is located in the SMC, it is the
brightest nova ever discovered in the SMC and one of the brightest on record.
\citet{aydi18} also showed that the total luminosity is as high as
$10-20$ times the Eddington luminosity of a $1.2~M_\sun$ WD during
the SSS phase (see their Figure 10).
However, it is very unlikely that the total luminosity significantly exceeds
the Eddington luminosity during the SSS phase.  
Evolution calculations clearly showed that the bolometric luminosity
is close to the Eddington limit in the SSS phase 
\citep[e.g.,][]{sal05, den13, kat17sh}.
This is because such large super-Eddington luminosity pushes 
the envelope against the gravity and expands the envelope.
As a result, the photospheric temperature decreases below 
the supersoft X-ray emitting region. It seems more plausible that 
the distance to the nova is shorter by at least a factor of three,
i.e., approximately $d\sim 20$~kpc, because the luminosity is just 
the Eddington luminosity during the SSS phase.

\subsubsection{Time-stretching method}
Figure \ref{smcn2016_yy_dor_v407_cyg_rs_oph_v_bv_ub_logscale_no2} shows
the same (a) $V$ light and (b) $(B-V)_0$ color curves of SMC~N~2016
as in Figure \ref{smcn2016_v_bv_ub_color_xray_curve} 
but on a logarithmic timescale. We included the $V$ light and $(B-V)_0$
color curves of U~Sco, V407~Cyg, and RS~Oph. We also included the 
supersoft X-ray fluxes of U~Sco and RS~Oph for comparison.
Here, we determine the horizontal shifts $f_{\rm s}$ of each nova
by overlapping the end of the SSS phase.
The $V$ light and $(B-V)_0$ color curves of SMC~N~2016 show
a reasonable overlap with those of U~Sco. 
Then, we determine the set of $\Delta V= 2.0$ and $\log f_{\rm s}=0.6$
($f_{\rm s}=4.0$) for SMC~N~2016 and U~Sco, as shown in Figure 
\ref{smcn2016_yy_dor_v407_cyg_rs_oph_v_bv_ub_logscale_no2}(a). 
We regard that Equation (\ref{overlap_brigheness}) is satisfied
for SMC~N~2016 and U~Sco, because the $V$ light curves of these two novae
nearly overlap with each other.
Thus, we regard SMC~N~2016 as a normal-decline (U~Sco) type nova.
We apply Equation (\ref{distance_modulus_formula}) to all the novae in
Figure \ref{smcn2016_yy_dor_v407_cyg_rs_oph_v_bv_ub_logscale_no2}
and obtain the relation:
\begin{eqnarray}
(m&-&M)_{V, \rm SMC~N~2016} \cr 
&=& (m - M + \Delta V)_{V, \rm U~Sco} - 2.5 \log 4.0 \cr
&=& 16.3 + 2.0 - 1.5 = 16.8 \cr
&=& (m - M + \Delta V)_{V, \rm V407~Cyg} - 2.5 \log 0.45 \cr
&=& 16.1 - 0.2 + 0.88 = 16.78 \cr
&=& (m - M + \Delta V)_{V, \rm RS~Oph} - 2.5 \log 2.0 \cr
&=& 12.8 + 4.8 - 0.75 = 16.75,
\label{distance_modulus_smcn_2016_rs_oph}
\end{eqnarray}
where we adopt $(m-M)_{V, \rm U~Sco}=16.3$ in Section \ref{u_sco},
$(m-M)_{V, \rm V407~Cyg}=16.1$ in Section \ref{v407_cyg}, and
$(m-M)_{V, \rm RS~Oph}=12.8$ in Section \ref{rs_oph}.
Thus, we obtain $(m-M)_V=16.8$ for SMC~N~2016.

Our value of $(m-M)_V=16.8$ yields the distance of $d=20.4$~kpc
from Equation (\ref{v_distance_modulus}) together with
$E(B-V)=0.08$ \citep{mun17}.  
Our distance is much shorter than the SMC distance of 
$d=61\pm10$~kpc \citep{aydi18}.
We contend that this nova belongs to our galaxy.
It should be emphasized that this short distance is consistent with 
the above argument that the luminosity is just the Eddington
luminosity during the SSS phase.  

We check our distance of $d=20.4$~kpc to SMC~N~2016 by comparing with 
the $V$ light curves of LMC novae, YY~Dor and LMC~N~2009a. 
Figure \ref{smcn2016_yy_dor_lmcn_2009a_v_x-ray_logscale} shows that
their $V$ light curves overlap to each other.
We apply Equation (\ref{distance_modulus_formula}) to all the novae
and obtain
\begin{eqnarray}
(m&-&M)_{V, \rm SMC~N~2016} = 16.8 \cr
&=& (m - M + \Delta V)_{V, \rm YY~Dor} - 2.5 \log 1.0 \cr
&=& 18.9 - 2.1 - 0.0 = 16.8 \cr
&=& (m - M + \Delta V)_{V, \rm LMC~N~2009a} - 2.5 \log 0.63 \cr
&=& 19.1 - 2.8 + 0.5 = 16.8,
\label{distance_modulus_v_smcn2016_yy_dor_lmcn2009a}
\end{eqnarray}
where we adopt $(m-M)_{V, \rm YY~Dor}=18.9$ from Section \ref{yy_dor}, 
$(m-M)_{V, \rm LMC~N~2009a}=19.1$ from Section \ref{lmcn_2009a}.
The distance moduli of YY~Dor and LMC~N~2009a are reliable 
because the two novae are members of LMC whose distance modulus
is well constrained.

\subsubsection{Model light curve fitting}
Figure \ref{smcn2016_yy_dor_lmcn_2009a_v_x-ray_logscale} also shows
comparison with our model light curves.
Figure \ref{smcn2016_yy_dor_lmcn_2009a_v_x-ray_logscale}(a) plots
a $1.3~M_\sun$ WD model with the chemical composition of Ne nova 3
as a best-fit one among 1.2, 1.25, and $1.3~M_\sun$ WDs \citep{hac16k}.
For Ne nova 2, we select the $1.25~M_\sun$ WD, which is shown 
in Figure \ref{smcn2016_yy_dor_lmcn_2009a_v_x-ray_logscale}(b), 
as a best-fit one among the three, 1.2, 1.25, and $1.3~M_\sun$ WDs
\citep{hac10k}.
Note that the absolute magnitudes of these $V$ light curves are
already known \citep{hac10k, hac16k}.
If we assume the distance modulus of $(m-M)_V=16.8$, we have a good
fit (solid green line) with the observation.  On the other hand,
if we assume the SMC distance of $d=61$~kpc, that is, $(m-M)_V=19.0$
\citep{aydi18}, the model light curves (solid red lines) are much below
the observation.

\subsubsection{Distance-reddening relation}
We further check our distance of $d=20.4$~kpc by the $I_C$ light curve. 
Figure \ref{yy_dor_lmcn_2009a_smcn2016_u_sco_i_k_logscale_2fig}(a) shows 
the $I_C$ light curves of SMC~N~2016, YY~Dor, U~Sco, and LMC~N~2009a.  
These four novae have a slope of $F_\nu\propto t^{-1.75}$
(green lines) in the $I_C$ light curves and overlap with the same
scaling factors of $f_{\rm s}$ as those of the $V$ light curves.
The distance modulus of SMC~N~2016 in the $I_C$ band is calculated to be
$(m-M)_{I, \rm SMC~N~2016} = 16.8 - 1.6\times 0.08= 16.7$.
We apply Equation (\ref{distance_modulus_formula}) to all the novae
and obtain
\begin{eqnarray}
(m&-&M)_{I, \rm SMC~N~2016} = 16.7 \cr
&=& (m - M + \Delta I_C)_{I, \rm YY~Dor} - 2.5 \log 1.0 \cr
&=& 18.7 - 2.0 - 0.0 = 16.7 \cr
&=& (m - M + \Delta I_C)_{I, \rm U~Sco} - 2.5 \log 4.0 \cr
&=& 15.9 + 2.3 - 1.5 = 16.7 \cr
&=& (m - M + \Delta I_C)_{I, \rm LMC~N~2009a} - 2.5 \log 0.63 \cr
&=& 18.8 - 2.6 + 0.5 = 16.7,
\label{distance_modulus_i_smcn2016_yy_dor_u_sco_lmcn2009a}
\end{eqnarray}
where we adopt $(m-M)_{I, \rm YY~Dor}=18.7$ from Section \ref{yy_dor}, 
$(m-M)_{I, \rm U~Sco}=15.9$ from Section \ref{u_sco}, 
$(m-M)_{I, \rm LMC~N~2009a}=18.8$ from Section \ref{lmcn_2009a}.
Thus, we confirm that the set of SMC~N~2016, YY~Dor, U~Sco, and LMC~N~2009a
consistently overlap with each other in the early phase, i.e., satisfy
the timescaling law of Equation (\ref{overlap_brigheness}) and,
at the same time, satisfy the time-stretching method of 
Equation (\ref{distance_modulus_formula}).

Figure \ref{distance_reddening_smcn2016_1fig} shows various 
distance-reddening relations toward SMC~N~2016.  The vertical solid
red line denotes the reddening of $E(B-V)=0.08$.  The blue and cyan
lines indicate the relations of Equations (\ref{v_distance_modulus})
and (\ref{distance_modulus_relation_i}) for $(m-M)_V=16.8$ and 
$(m-M)_I=16.7$, respectively.
These three lines cross consistently at the point of
$d=20.4$~kpc and $E(B-V)=0.08$, supporting our estimate.

\subsubsection{MMRD relation}
We plot the MMRD point of SMC~N~2016 in Figure \ref{max_t3_scale_u_sco_type}. 
It is located closely to U~Sco, LMC~N~2009a, and YY~Dor,
and in the lower side of the broad MMRD relation.  
Thus, this nova is slightly fainter than the MMRD2 relation, 
not the brightest on record as argued by \citet{aydi18}.
We summarize the results in Tables
\ref{extinction_distance_various_novae}, 
\ref{physical_properties_recurrent_novae}, and
\ref{wd_mass_recurrent_novae}. 

\subsubsection{Color-magnitude diagram}
Figure \ref{hr_diagram_lmcn2012a_lmcn2013_smcn2016_m31n2008_12a_outburst}(c)
shows the color-magnitude diagram of SMC~N~2016.  
Here, we add two tracks of U~Sco (open cyan squares connected by thin 
cyan lines) and T~Pyx (filled green stars). The SMC~N~2016 track
broadly overlaps with those of U~Sco and T~Pyx.  
The start of the nebular phase (large open red square) is located
on the two-headed arrow.  
This coincidence clearly shows that our adopted values of 
$E(B-V)=0.08$ and $(m-M)_V=16.8$ are reasonable and 
demonstrates that SMC~N~2016 is a member of our galaxy.
The distance of $d=20.4$~kpc is consistent with our overall
understanding of the characteristic properties of novae.

\subsubsection{Summary of SMC~N~2016}
The WD mass of SMC~N~2016 is estimated to be $M_{\rm WD}=1.29~M_\sun$ 
from the linear relation between $M_{\rm WD}$ and $\log f_{\rm s}$
in Figure \ref{timescale_wd_mass} (see Table \ref{wd_mass_recurrent_novae}
for the other WD masses).
\citet{aydi18} discussed several constraints and summarized their results as
$M_{\rm WD} > 1.25~M_\sun$ from $t_2$ and $M_{\rm WD}\sim 1.25 - 1.3~M_\sun$
from the SSS duration, being consistent with our estimate both from
the model light curve fitting in Figure
\ref{smcn2016_yy_dor_lmcn_2009a_v_x-ray_logscale} and linear relation of
Figure \ref{timescale_wd_mass}. 
Neither the recurrence period nor the orbital period of the binary
is known. If the evolutionary state is similar to CI~Aql and YY~Dor,
we may conclude that SMC~N~2016 is not a progenitor of SNe~Ia.


\begin{figure}
\includegraphics[height=12.5cm]{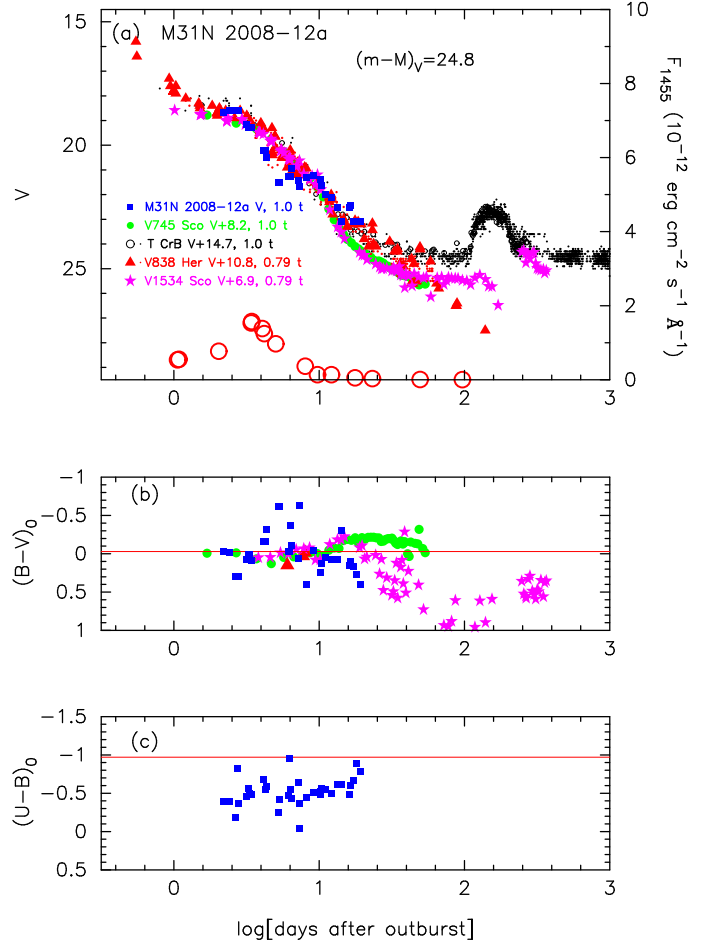}
\caption{
Same as Figure 
\ref{lmcn2012a_v1534_sco_v838_her_t_crb_v745_sco_v_bv_ub_logscale},
but for the light/color curves of M31N~2008-12a.
(a) The filled blue squares denote the $V$ magnitudes of M31N~2008-12a.
The (b) $(B-V)_0$ and (c) $(U-B)_0$ color curves of M31N~2008-12a are
dereddened with $E(B-V)=0.10$. 
\label{m31_12a_v1534_sco_v838_her_t_crb_v745_sco_v_bv_ub_logscale_no2}}
\end{figure}

\subsection{M31N~2008-12a (2015)}
\label{m31n2008_12a}
The 1-yr recurrence period nova M31N~2008-12a is an excellent
example of recurrent novae because the distance and extinction
are well determined \citep{dar15, dar16, hen15, kat15sh, kat16shn, kat17sh}.  
The nova reached $m_{V, \rm max}=18.55$ and its decline rates are
estimated to be $t_2=1.77$ days and $t_3=3.84$ days in the $V$ band
\citep{dar16}, being a very fast nova.
\citet{kat15sh, kat16shn, kat17sh} concluded, 
on the basis of their multiwavelength
light curve model, that the WD mass is close to $1.38~M_\sun$
and a promising candidate of SN~Ia progenitors.
The distance to M31 is $d\approx780$~kpc and the extinction 
toward the nova is $E(B-V)\approx0.1$ \citep[e.g.,][]{dar16}.
Then, the distance modulus is calculated to be $(m-M)_V=24.8$
from Equation (\ref{v_distance_modulus}).

With $(m-M)_V=24.8$ for M31N~2008-12a, the set of vertical shift
$\Delta V$ and horizontal shift $\log f_{\rm s}$ are
uniquely determined with the procedure described in Section \ref{yy_dor}.
We overlap the $V$ light curve of M31N~2008-12a with those of V745~Sco,
T~CrB, V838~Her, and V1534~Sco, as shown in Figure 
\ref{m31_12a_v1534_sco_v838_her_t_crb_v745_sco_v_bv_ub_logscale_no2}.
The $V$ light curve of M31N 2008-12a shows a quick decline
a few days after the optical peak compared with the other rapid-decline
(V745~Sco) type novae.  Although the reason of this behavior is not
identified yet, the decay trend recovers and again follows
the other rapid-decline novae about 10 days after the outburst.
Thus, we regard M31N 2008-12a as a rapid-decline type nova.


From Figure
\ref{m31_12a_v1534_sco_v838_her_t_crb_v745_sco_v_bv_ub_logscale_no2},
we obtain $\Delta V= 8.2\pm0.3$ and $\log f_{\rm s}= 0.0$ ($f_{\rm s}=1.0$)
with respect to V745~Sco by overlapping  the light curve in the
early 1-3 days and late 10-20 days except the mid 4-9 days,
because M31N~2008-12a shows a sharp decline during mid 4-9 days.
We exclude T~CrB and V838~Her from our fitting because these data
show large scatter and are not appropriate.  
The $V$ light curve of M31N~2008-12a does not exactly but broadly
overlap with V745~Sco and V1534~Sco.  Thus,
the vertical overlapping error could be as large as $\pm0.3$ mag.
We regard that these novae satisfy Equation (\ref{overlap_brigheness}).   
Then, we apply Equation (\ref{distance_modulus_formula}) to 
M31N~2008-12a, V745~Sco, and V1534~Sco
in Figure 
\ref{m31_12a_v1534_sco_v838_her_t_crb_v745_sco_v_bv_ub_logscale_no2}
and obtain the relation:
\begin{eqnarray}
(m&-&M)_{V, \rm M31N~2008-12a} = 24.8 \cr
&=& (m - M + \Delta V)_{V, \rm V745~Sco} - 2.5 \log 1.0 \cr
&=& 16.6 + 8.2 + 0.0 = 24.8 \cr
&=& (m - M + \Delta V)_{V, \rm V1534~Sco} - 2.5 \log 0.79 \cr
&=& 17.6 + 6.9 + 0.25 = 24.75,
\label{distance_modulus_m31n_12a_v745_sco}
\end{eqnarray}
where we adopt $(m-M)_{V, \rm V745~Sco}=16.6$ in Section \ref{v745_sco}
and $(m-M)_{V, \rm V1534~Sco}=17.6$ in Section \ref{v1534_sco}. 
Thus, M31N~2008-12a, V745~Sco, 
and V1534~Sco satisfy 
Equations (\ref{overlap_brigheness}) and (\ref{distance_modulus_formula})
within a vertical fitting error of $\pm0.3$ mag. 
We regard M31N~2008-12a as a rapid-decline (V745~Sco) type
nova. We plot the MMRD point of M31N~2008-12a in Figure 
\ref{max_t3_scale_v745_sco_type}. 
Its position is far below both the MMRD1 and MMRD2 relations.
We may conclude that this nova belongs to the faint class, as 
claimed by \citet{kas11}.

Figure \ref{hr_diagram_lmcn2012a_lmcn2013_smcn2016_m31n2008_12a_outburst}(d)
shows the color-magnitude diagram of M31N~2008-12a
(filled green squares with black outlines).
The data for the 2014 outburst are taken from \citet{dar15} and
those for the 2015 outburst from \citet{dar16}.
The peak brightness is approximately $M_V=18.55-24.8=-6.25$,
being fainter than typical classical novae.  
The nova almost follows the track of RS~Oph,
which is a recurrent nova with a RG companion,
although the $(B-V)_0$ data of M31N~2008-12a are rather scattered around
$(B-V)_0= -0.03$.
This resemblance may support our set of $E(B-V)=0.1$ and $(m-M)_V=24.8$
and a RG companion rather than a subgiant companion
\citep[see also discussion of][]{dar16}.

Recently, \citet{dar17} observed M31N~2008-12a in quiescence
with {\it HST} and concluded that a quiescent disk mass accretion rate
is the order of $10^{-6}~M_\sun$~yr$^{-1}$, based on their accretion disk
model.  This large accretion rate, however, clearly contradicts to
the current theoretical understanding of nova outbursts
because this accretion rate is far above the critical accretion
rate for unstable nuclear burning (nova outbursts),
several times $10^{-7}~M_\sun$~yr$^{-1}$ \citep[see, e.g.,][]{nom07skh,
kat14shn}.  \citet{dar17} argued the
possibility that a large part of accreted matter is blown in the
accretion disk wind.  We point out another possible explanation.
Their disk model does not include the irradiation effect by the hot WD,
which emits as much as $\sim 300~L_\sun$ \citep{kat17sh}.
If the irradiation effect by the WD is included in their disk model,
the accretion rate could be much smaller than their adopted value
of $\sim 10^{-6}~M_\sun$~yr$^{-1}$ or more.

\section{Discussion}
\label{discussion}
\subsection{Timescales over the three types}
\label{horizontal_shifts}
We have divided the 14 novae into three groups.
In each group, we have determined the timescaling factor of $f_{\rm s}$
with respect to the template nova of the group. In this subsection, 
we determine the common timescales over the three groups.
We adopt the timescale of V745~Sco as the unit of common timescale, which
is the smallest $f_{\rm s}$ among the 14 novae.

The V745~Sco novae do not have the $F_\nu \propto t^{-1.75}$ slope, as
shown in Figure \ref{v745_sco_u_sco_v407_cyg_rs_oph_v_template}.
To determine the timescale difference between the V745~Sco and U~Sco types,
we use the slope of 
$F_\nu \propto t^{-3.5}$. In U~Sco, this slope started
approximately 10 days after the outburst 
(see Figures \ref{v745_sco_u_sco_v407_cyg_rs_oph_v_template}
and \ref{u_sco_v_bv_ub_xray_radio_logscale}).
We regard that the rapid-decline light curve of V745~Sco
corresponds to the $F_\nu \propto t^{-3.5}$ decline of U~Sco.
This is because the ignition mass of the hydrogen-rich envelope of V745~Sco
is too small to reach the slope of $F_\nu \propto t^{-1.75}$, 
and much smaller than that of U~Sco.  

\begin{figure}
\includegraphics[height=12.5cm]{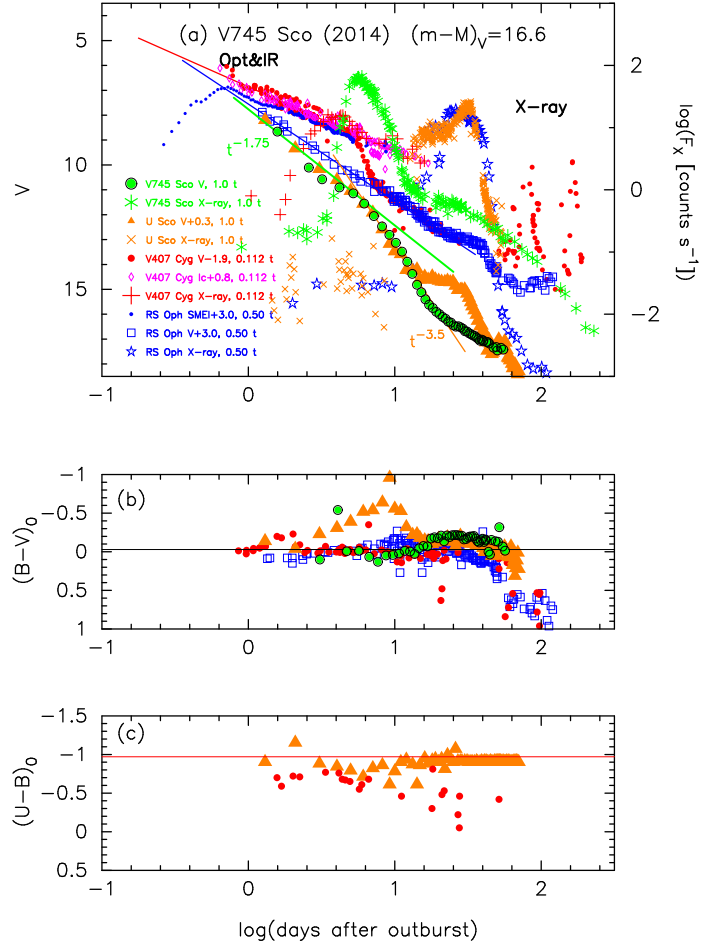}
\caption{
Same as Figure \ref{lmcn2012a_yy_dor_v407_cyg_rs_oph_v_bv_ub_logscale},
but we plot the light/color curves of V745~Sco.
(a) The filled green circles with black outlines 
denote the $V$ magnitudes of V745~Sco and the green asterisks represent
the X-ray ($0.3-2.0$ keV) flux of V745~Sco,
and the other symbols are the same as those in
Figure \ref{lmcn2012a_yy_dor_v407_cyg_rs_oph_v_bv_ub_logscale}.
The (b) $(B-V)_0$ and (c) $(U-B)_0$ color curves. 
The $(B-V)_0$ colors of V745~Sco are dereddened with $E(B-V)=0.70$.
\label{v745_sco_u_sco_v407_cyg_rs_oph_v_bv_ub_logscale}}
\end{figure}

Now, we determine the timescaling factor $f_{\rm s}$ of U~Sco
with respect to V745~Sco with the same procedure as in Sections 
\ref{v407_cyg_timescaling_law} and \ref{yy_dor}.
We overlap the V745~Sco and U~Sco
at the rapid decline trend of $F_\nu \propto t^{-3.5}$ and
obtain the best overlap for the set of $\Delta V= 0.3$ 
and $\log f_{\rm s}= 0.0$ ($f_{\rm s}=1.0$), as shown in
Figures \ref{v745_sco_u_sco_v407_cyg_rs_oph_v_template}
and \ref{v745_sco_u_sco_v407_cyg_rs_oph_v_bv_ub_logscale}. 
The timescaling factor of $f_{\rm s}$ is almost the same between
these two novae, although their WD mass is slightly different.
It is interesting that the start of the optically thin phase of V745~Sco
and the start of the nebular phase (break) of U~Sco roughly overlap.  
From Figures \ref{v745_sco_u_sco_v407_cyg_rs_oph_v_template}
and \ref{v745_sco_u_sco_v407_cyg_rs_oph_v_bv_ub_logscale}, 
we finally obtain
\begin{eqnarray}
(m&-&M)_{V, \rm V745~Sco} = 16.6 \cr
&=& (m - M + \Delta V)_{V, \rm RS~Oph} - 2.5 \log 0.50 \cr
&=& 12.8 + 3.0 + 0.75 = 16.55 \cr
&=& (m - M + \Delta V)_{V, \rm V407~Cyg} - 2.5 \log 0.112 \cr
&=& 16.1 - 1.9 + 2.38 = 16.58 \cr
&=& (m - M + \Delta V)_{V, \rm U~Sco} - 2.5 \log 1.0 \cr
&=& 16.3 + 0.3 + 0.0 = 16.6,
\label{distance_modulus_v745_sco_v407_cyg_rs_oph}
\end{eqnarray}
where we adopt $(m-M)_{V, \rm RS~Oph}=12.8$ in Section \ref{rs_oph}, 
$(m-M)_{V, \rm V407~Cyg}=16.1$ in Section \ref{v407_cyg}, and
$(m-M)_{V, \rm U~Sco}=16.3$ in Section \ref{u_sco}, because
we already know the difference in the timescale
between RS~Oph and U~Sco. 
Figure \ref{v745_sco_u_sco_v407_cyg_rs_oph_v_template} shows the
relations between the template light curves for the four novae 
thus determined. The timescaling factor of each nova against
that of V745~Sco is listed in Tables 
\ref{extinction_distance_various_novae},
\ref{physical_properties_recurrent_novae}, and 
\ref{wd_mass_recurrent_novae}.


\begin{figure}
\includegraphics[height=10.0cm]{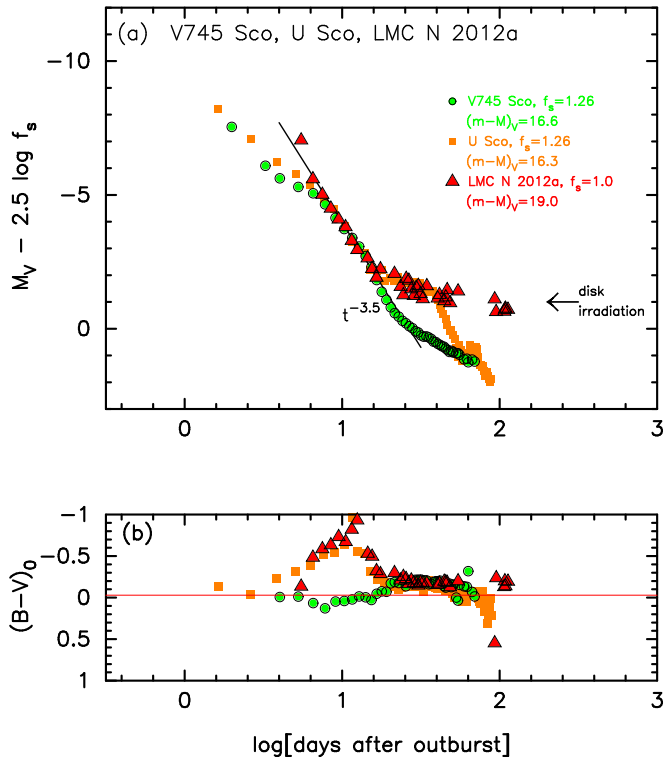}
\caption{
(a) The absolute $V$ light curves of V745~Sco, U~Sco, and
LMC~N~2012a on a logarithmic timescale.  
The ordinate is $M_V - 2.5\log f_{\rm s}$ instead of 
$M_V$.  
(b) The $(B-V)_0$ color curves of V745~Sco, U~Sco, and
LMC~N~2012a.
\label{lmcn2012a_u_sco_v745_sco_abs_v_bv_ub_logscale}}
\end{figure}

We have already tried to bridge the two timescales in the U~Sco and
LMC~N~2012a light curve analysis (Sections \ref{u_sco} and \ref{lmcn_2012a}).  
The above result is supported also by comparing
the timescaling factors between Figures
\ref{lmcn2012a_v1534_sco_v838_her_t_crb_v745_sco_v_bv_ub_logscale} and
\ref{lmcn2012a_yy_dor_v407_cyg_rs_oph_v_bv_ub_logscale}.
The light curve of LMC~N~2012a is similar to that of U~Sco but has only
$F_\nu\propto t^{-3.5}$ slope in the early decline phase.
So we assign this nova to the V745~Sco group.  
Moreover, its absolute magnitude is well determined because of
its LMC membership.  Therefore, LMC~N~2012a is a best example for
bridging the two groups as clearly demonstrated in Section \ref{lmcn_2012a}.
To confirm again the relation among these three novae,
we plot the absolute $V$ light curves and $(B-V)_0$ color curves 
of V745~Sco, U~Sco, and LMC~N~2012a in Figure 
\ref{lmcn2012a_u_sco_v745_sco_abs_v_bv_ub_logscale}.  
The ordinate is $M_V - 2.5\log f_{\rm s}$ instead of $M_V$.
After time-stretch of each nova light curve, the absolute  
$M_V - 2.5\log f_{\rm s}$ magnitude should overlap to each other.
In this particular case, V745~Sco and U~Sco are time-stretched with
$f_{\rm s}=1.26$ against LMC~N~2012a.  
The three $V$ light curves clearly overlap to 
each other on the line of $F_\nu \propto t^{-3.5}$. 
The $V$ light curves of V745~Sco and U~Sco further overlap to each other
even in the very early phase of the outburst.  
The $V$ light curves of U~Sco and LMC~N~2012a show a similar plateau
of $M_V - 2.5\log f_{\rm s}\sim -2$ mag, 
which is caused by a disk irradiation by the central
WD \citep[e.g.,][]{hkkm00}.  Thus, we confirm that our common timescale
between the V745~Sco and U~Sco types supports Equation
(\ref{distance_modulus_formula}) and vice versa.


\subsection{Large deviation from MMRD relations}
\label{deviation_mmrd}
We have already examined the MMRD relation for
many galactic classical novae based on our model light curves
\citep{hac10k, hac15k, hac16k}.
The main conclusions are that (1) the main trend is determined by
the timescaling factor $f_{\rm s}$ (WD mass), 
(2) the scatter from the main trend is due to the difference in the ignition
mass (mass accretion rate on to the WD), and (3) the saturation
of the peak brightness for the longer ($t_2$ or) $t_3$ time regime
(less massive WDs) is caused by the contribution from the photospheric
radiation \citep[see][for more detail]{hac15k}. In this subsection,
we add new reasons for the scatter of MMRD points from the main trend,
especially for recurrent and very fast novae.  

We pointed out that there are three types of nova light curves,
the normal-decline (U~Sco), rapid-decline (V745~Sco), and
CSM-shock (RS~Oph) types. We discuss the three types in this order.  

The light curves of the normal-decline (U~Sco) novae 
closely follow the trend of the universal decline
law, $F_\nu \propto t^{-1.75}$, as shown in Figure 
\ref{v745_sco_u_sco_v407_cyg_rs_oph_v_template}.
Their MMRD points are located in the lower region of the MMRD relations,
as shown in Figure \ref{max_t3_scale_u_sco_type}.
These deviations come from reason (2) mentioned above.
This is because the ignition masses are relatively small in these
novae, i.e., smaller initial envelope mass at hydrogen burning 
(point C rather than point B in Figures \ref{max_t3_scale_u_sco_type},
\ref{all_mass_v1668_cyg_x45z02c15o20_calib_universal_linear_no2}, 
\ref{all_mass_v1668_cyg_x45z02c15o20_calib_universal_no2},
and \ref{max_t3_scale_no4}). 
Therefore, the peak brightness $M_{V, \rm max}$ is relatively
faint, even for the smaller $t_2$ and $t_3$ times.
This trend is the same as that in classical novae.
We explain this effect in more detail in Appendix \ref{mmrd}.

The light curves of the rapid-decline (V745~Sco) type novae 
are characterized by a very short period before the start of SSS phase.
Soon after the optical peak, the $V$ brightness quickly decays 
as $F_\nu \propto t^{-3.5}$, as shown in Figure 
\ref{v745_sco_u_sco_v407_cyg_rs_oph_v_template}.
The start of quick decay from $F_\nu \propto t^{-1.75}$ to 
$F_\nu \propto t^{-3.5}$ roughly corresponds to the break of the
U~Sco $V$ light curve. (See also Figure
\ref{all_mass_v1668_cyg_x45z02c15o20_calib_universal_no2} in Appendix
\ref{model_light_curve} for the break).  This means that
the ignition mass of the hydrogen-rich envelope is too small to reach
the normal-decline segment of $F_\nu \propto t^{-1.75}$.
The $t_2$ and $t_3$ times are much shorter because 
$F_\nu \propto t^{-3.5}$ rather than
$F_\nu \propto t^{-1.75}$. The peak brightness is fainter
because of the smaller ignition mass and smaller wind mass-loss rate.
Thus, this type of novae are located significantly below the MMRD relations,
as shown in Figure \ref{max_t3_scale_v745_sco_type}.
In short, this occurs when the ignition mass is much smaller than
the case of the normal-decline types.
We may conclude that the rapid-decline type novae do not exhibit $F_\nu \propto t^{-1.75}$ (the universal trend of galactic
classical novae) in their early phase $V$ light curves,
but start with $F_\nu \propto t^{-3.5}$ and the two novae,
V1534~Sco and M31N~2008-12a, belong to the faint class 
claimed by \citet{kas11}.

The final case we discuss is the contamination of shock-heating. 
The light curves of the CSM-shock (RS~Oph) type
novae are contaminated by the flux from shock-heating, as shown in Figures
\ref{v745_sco_u_sco_v407_cyg_rs_oph_v_template},
\ref{rs_oph_radio_v_ub_bv_uv_x_logscale_no4},
\ref{v407_cyg_rs_oph_v_bv_ub_logscale}, and
\ref{lmcn2013_u_sco_v407_cyg_rs_oph_v_bv_ub_logscale_no2}. These show
slow decline rates close to $F_\nu \propto t^{-1.0}$
during the strong shock-interaction with the CSM.
Therefore, their $t_2$ and $t_3$ times are longer than those with no shock
contamination, even for the same $M_{V, \rm max}$.
These novae host a massive WD and the ignition mass is large enough 
to start from the decline of $F_\nu \propto t^{-1.75}$.   
Therefore, the naked MMRD points without shock contamination should be
located in the lower side of the broad MMRD relations, like those of 
the normal-decline (U~Sco) types.  With the help of shock-heating, 
novae of this type happen to follow the MMRD relations as shown
in Figure \ref{max_t3_scale_rs_oph_type}.

\subsection{Progenitors of Type Ia supernovae}
\label{sn1a_progenitors}
A typical evolution path of the progenitors
in the SD scenario \citep[e.g.,][]{hkn99, hknu99} is as follows:
When the companion evolves to expand and fills its Roche lobe,
mass transfer begins from the companion to the WD.
The mass transfer rate quickly increases and exceeds 
the critical rate for optically thick wind
\citep[see $\dot M_{\rm cr}$ in Figure 3 of][]{kat14shn}.
The WD evolves from the optically thick wind phase,
through the SSS phase (steady H-burning phase)
and to the recurrent nova phase (H-shell flash phase), and
finally explodes as a SN Ia \citep[e.g.,][]{hac08kn, hac12kn}.
The WD mass substantially increases during the wind and SSS phases
before the WD enters a recurrent nova phase.
In other words, a recurrent nova in the course toward a SN~Ia
has a WD that has already grown up to very close to
the Chandrasekhar mass.  A recurrent nova with a WD mass, say $1.2~M_\sun$,
will not reach the Chandrasekhar mass in the ensuing binary evolution
\citep[e.g.,][]{hac08kn, hac12kn}.

Therefore, a SN~Ia progenitor in a recurrent nova phase
should host a WD close enough to 
$M_{\rm Ia}=1.38~M_\sun$ \citep{nom82} before the SN~Ia explosion,
as mentioned in Section \ref{introduction}. The binary orbital
period is highly likely in the ranges between $\sim$0.3 and $\sim$5 days
for WD+MS binary systems, or between $\sim$100
and $\sim$1000 days for WD+RG binary systems 
\citep[see, e.g., Figure 1 of][]{hac08kn}.
These orbital periods support high mass accretion rates onto the WDs
and sufficient mass retention efficiency of the accreted matter.
The typical mass-increasing rates of WDs 
are $\dot M_{\rm WD}\sim 1\times10^{-7}~M_\sun$~yr$^{-1}$ 
just below the stability line of hydrogen-shell burning 
\citep[see, e.g.,][]{kat17sh}.
If the WD mass increases from $1.35~M_\sun$ to 
$M_{\rm Ia}=1.38~M_\sun$ in the recurrent nova phase and explodes as a SN~Ia, 
it takes approximately $t_{\rm Ia}\sim 0.03~M_\sun/
1\times10^{-7}~M_\sun$~yr$^{-1} = 3\times10^5$~yr, which is shorter
than the evolution timescale of the donor (RG star or 
MS star). It is unlikely that these WDs were born as massive 
as they are ($M_{\rm WD} \ge 1.35~M_\sun$). Therefore, these WDs 
have grown in mass since their births.
This strongly suggests further increase in 
the WD masses in these systems.

Figure \ref{timescale_wd_mass} shows the WD mass against the timescaling
factor of $\log f_{\rm s}$, where the timescaling factor is measured 
based on that of V745~Sco ($f_{\rm s}=1.0$ for V745~Sco).
Various WD masses are estimated from the trend of this $f_{\rm s}$ vs.
$M_{\rm WD}$ relation (solid red lines).
We also plot the WD masses determined by the other methods,
as listed in Table \ref{wd_mass_recurrent_novae}. 
Among these 14 novae, we select eight novae that almost satisfy the conditions
of SN~Ia progenitor mentioned above, i.e.,
T~CrB ($1.38~M_\sun$, RG, $P_{\rm orb}=227.6$~days), 
V838~Her ($1.37~M_\sun$, MS, $P_{\rm orb}=0.2976$~days), 
RS~Oph ($1.35~M_\sun$, RG, $P_{\rm orb}=453.6$~days), 
U~Sco ($1.37~M_\sun$, subgiant, $P_{\rm orb}=1.23$~days), 
V745~Sco ($1.385~M_\sun$, RG, $P_{\rm orb}=$unknown), 
V1534~Sco ($1.37~M_\sun$, RG, $P_{\rm orb}=$unknown), 
LMC~N~2012a ($1.37~M_\sun$, subgiant, $P_{\rm orb}=0.802$~days), and
M31N~2008-12a ($1.38~M_\sun$, RG?, $P_{\rm orb}=$unknown).

\section{Conclusions}
\label{conclusions}
Our results are summarized as follows:
\begin{enumerate}
\item We analyzed 14 fast novae including eight recurrent novae, and divided
them into three types of light curve shapes: the rapid-decline (V745~Sco),
CSM-shock (RS~Oph), and normal-decline (U~Sco) types.  
The rapid-decline type includes V745~Sco, T~CrB, V838~Her, V1534~Sco,
LMC~N~2012a, and M31N~2008-12a;
the CSM-shock type includes RS~Oph, V407~Cyg, and LMC~N~2013;
and the normal-decline type includes U~Sco, CI~Aql, YY~Dor, LMC~N~2009a,
and SMC~N~2016. We obtained
the distances, distance moduli in the $V$ band, and reddenings 
of each nova from various methods.  
The results are summarized in Table \ref{extinction_distance_various_novae}.
\item The normal-decline type novae follow the universal decline
trend of $F_\nu \propto t^{-1.75}$ in a substantial part of their early
$V$ light curves. The CSM-shock type novae have a substantial 
part of $F_\nu \propto t^{-1.0}$ during shock-heating between ejecta and 
CSM. After the shock breakout, the slope changes to $F_\nu \propto t^{-1.55}$. 
The rapid-decline type novae have no or a very short
duration of the universal trend of $F_\nu \propto t^{-1.75}$, but rather
follow a steep trend of $F_\nu \propto t^{-3.5}$. This is because 
their initial envelope masses are too small to reach the normal-decline
segment of $F_\nu \propto t^{-1.75}$, but they only start from a later
phase in which the slope has already changed from $F_\nu \propto t^{-1.75}$
to $F_\nu \propto t^{-3.5}$. 
\item Comparing the five nova light curves of V407~Cyg, RS~Oph, V745~Sco,
T~CrB, and V1534~Sco that have a RG companion, we found that
V407~Cyg is most heavily contaminated by shock-heating
($F_\nu \propto t^{-1.0}$), RS~Oph is
slightly weaker ($F_\nu \propto t^{-1.55}$ 
but $L_{\rm SMEI}\propto t^{-1.0}$)
than V407~Cyg, V745~Sco and T~CrB are much weaker
($F_\nu \propto t^{-3.5}$),
and V1534~Sco is almost uncontaminated ($F_\nu \propto t^{-3.5}$).  
\item In all three types of novae, 
the $V$ light curves follow a timescaling law 
(if we properly stretch or squeeze in the 
time direction and shift up or down each $V$ light curve), 
i.e., they almost overlap each other in the same group.  
We regard that they broadly obey Equation (\ref{overlap_brigheness}). 
Based on the obtained distance moduli,
we confirm that these novae satisfy the time-stretching method, i.e.,   
Equation (\ref{distance_modulus_formula}) when they obey
Equation (\ref{overlap_brigheness}). 
\item We apply our methods, i.e., timescaling law and time-stretching method,
to LMC, SMC, and M31 novae. This is our first attempt to apply
the method to extra-galactic novae. We identified YY~Dor, LMC~N~2009a,
and SMC~N~2016 as the normal-decline type, LMC~N~2013 as the CSM 
shock type, and LMC~N~2012a and M31N~2008-12a as the rapid-decline type.
We confirm that these novae also
satisfy the time-stretching method when they overlap with each other, i.e.,
Equation (\ref{overlap_brigheness}) 
and Equation (\ref{distance_modulus_formula}).
These results support that the time-stretching method 
is applicable, even to M31 and LMC novae. 
\item We find that SMC~N~2016 is not a member of SMC,
but rather a member of our galaxy, because its distance is obtained to be
$d=20\pm2$~kpc.  
\item The rapid-decline type novae do not obey the MMRD relations, i.e.,
their MMRD points are much fainter. This is because their initial
envelope masses, i.e., ignition masses, are too small to reach the
universal decline of $F_\nu \propto t^{-1.75}$. The normal-decline type
novae are located in the lower region of the law of Della Valle \& Livio 
(MMRD2), but far below Kaler-Schmidt's law (MMRD1). 
The CSM-shock type novae happen to be located near the main
trend of the MMRD relations, because the light curves are contaminated
with shock-heating and slowly decay as 
$F_\nu \propto t^{-1.0}$, resulting in longer $t_2$ and $t_3$ times
that compensate the relatively faint $M_{V, \rm max}$.
Thus, in general, the empirical MMRD
relations should not be used to estimate the distance moduli of very
fast and recurrent novae.  
\item The WD mass of V745~Sco is estimated to be $M_{\rm WD}=1.385~M_\sun$
from our model light curve fitting with the supersoft X-ray light curve. 
This WD mass is more massive than $M_{\rm WD}=1.38~M_\sun$ of the 1-yr
recurrence period nova, M31N~2008-12a. This is consistent with the earlier
appearance of the SSS phase of V745~Sco ($t_{\rm SSS-on}\sim 4$~days)
than that of M31N~2008-12a ($t_{\rm SSS-on}\sim 6$~days).   
\item The WD mass of V838~Her is estimated to be $M_{\rm WD}=1.37~M_\sun$
from our model light curve fitting with the UV~1455~\AA\ light curve. 
This is consistent with the timescaling factor $\log f_{\rm s}=0.1$
($f_{\rm s}=1.26$) of V838~Her, which is slightly larger than 
$\log f_{\rm s}=0$ ($f_{\rm s}=1.0$) of V745~Sco
($M_{\rm WD}=1.385~M_\sun$) in the same V745~Sco group.  
\item Among the analyzed 14 novae, 
we select eight novae that almost satisfy the conditions
of SN~Ia progenitors, i.e.,
T~CrB ($M_{\rm WD}=1.38~M_\sun$), 
V838~Her ($1.37~M_\sun$), 
RS~Oph ($1.35~M_\sun$), 
U~Sco ($1.37~M_\sun$), 
V745~Sco ($1.385~M_\sun$), 
V1534~Sco ($1.37~M_\sun$), 
LMC~N~2012a ($1.37~M_\sun$), and
M31N~2008-12a ($1.38~M_\sun$). 
\end{enumerate}

\acknowledgments
We thank the American Association of Variable Star Observers
(AAVSO) and the Variable Star Observers League of Japan (VSOLJ)
for the archival data of various novae.
We are also grateful to the anonymous referee for useful comments
that improved the manuscript.
This research has been supported in part by the Grants-in-Aid for
Scientific Research (15K05026, 16K05289) 
from the Japan Society for the Promotion of Science.

\appendix

\section{Model Light Curves of Novae}
\label{model_light_curve}


\begin{figure}
\plotone{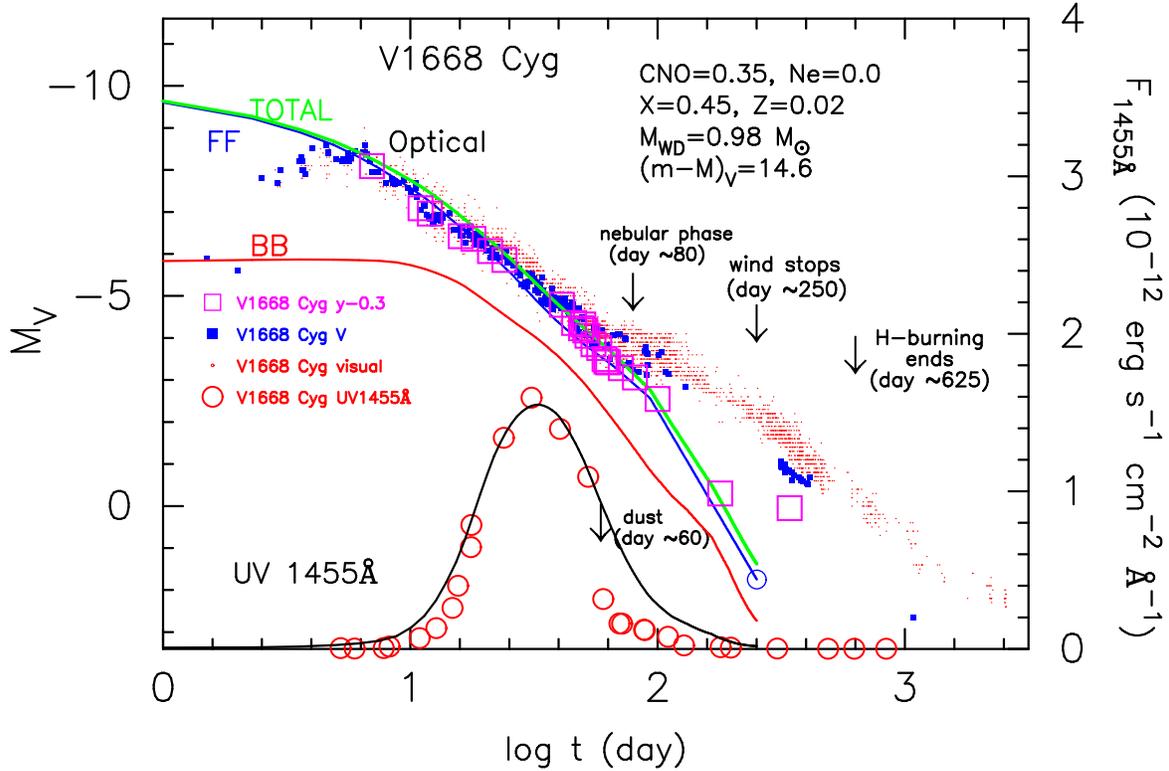}
\caption{
Visual (red dots), $V$ (filled blue circles), and $y$ (open magenta
squares) light curves and UV~1455~\AA\ light curve 
(large open red circles) of V1668~Cyg.
Assuming that $(m-M)_V=14.6$, we plot three model light curves
of a $0.98~M_\sun$ WD. The green, blue, and red
solid lines show the total (labeled ``TOTAL''), free-free
(labeled ``FF''), and blackbody (labeled ``BB'') $V$ fluxes.
The solid black line denotes the UV~1455~\AA\ flux.
An optically thin dust shell formed $\sim60$ days after the outburst
\citep[e.g.,][]{geh80}.
Optically thick winds and hydrogen-shell burning end approximately
250 and 625 days after the outburst, respectively,
for the $0.98~M_\sun$ WD model.
\label{all_mass_v1668_cyg_x45z02c15o20_real_scale_only098_no2}}
\end{figure}


\begin{figure}
\plotone{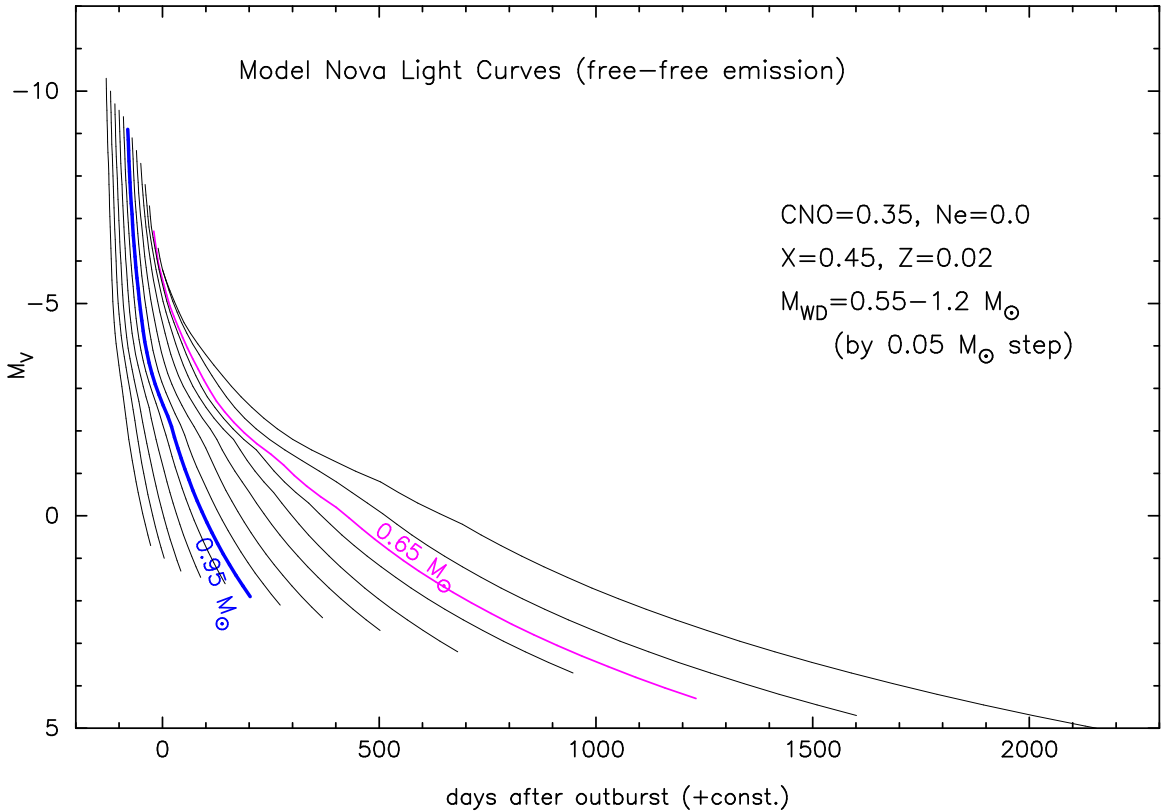}
\caption{
Absolute $V$ magnitudes of free-free emission model light curves
for the chemical composition of CO nova 3 and WD masses of 
$0.55-1.2~M_\sun$ in $0.05~M_\sun$ steps, the numerical data 
of which are taken from Table 3 of \citet{hac16k}.
The mass of hydrogen-rich envelope on the WD gradually decreases
along each light curve because the envelope mass is lost in the wind
and consumed by nuclear burning.   
The right edges of free-free emission model light curves
correspond to the epoch when the optically thick winds stop. 
Two light curves are specified by the thick solid magenta line 
($0.65 ~M_\sun$) and thick solid blue line ($0.95 ~M_\sun$).
A less massive WD declines more slowly, and thus the nova speed class depends
mainly on the WD mass.
\label{light_curve_combine_x45z02c15o20_absmag}}
\end{figure}


\begin{figure}
\plotone{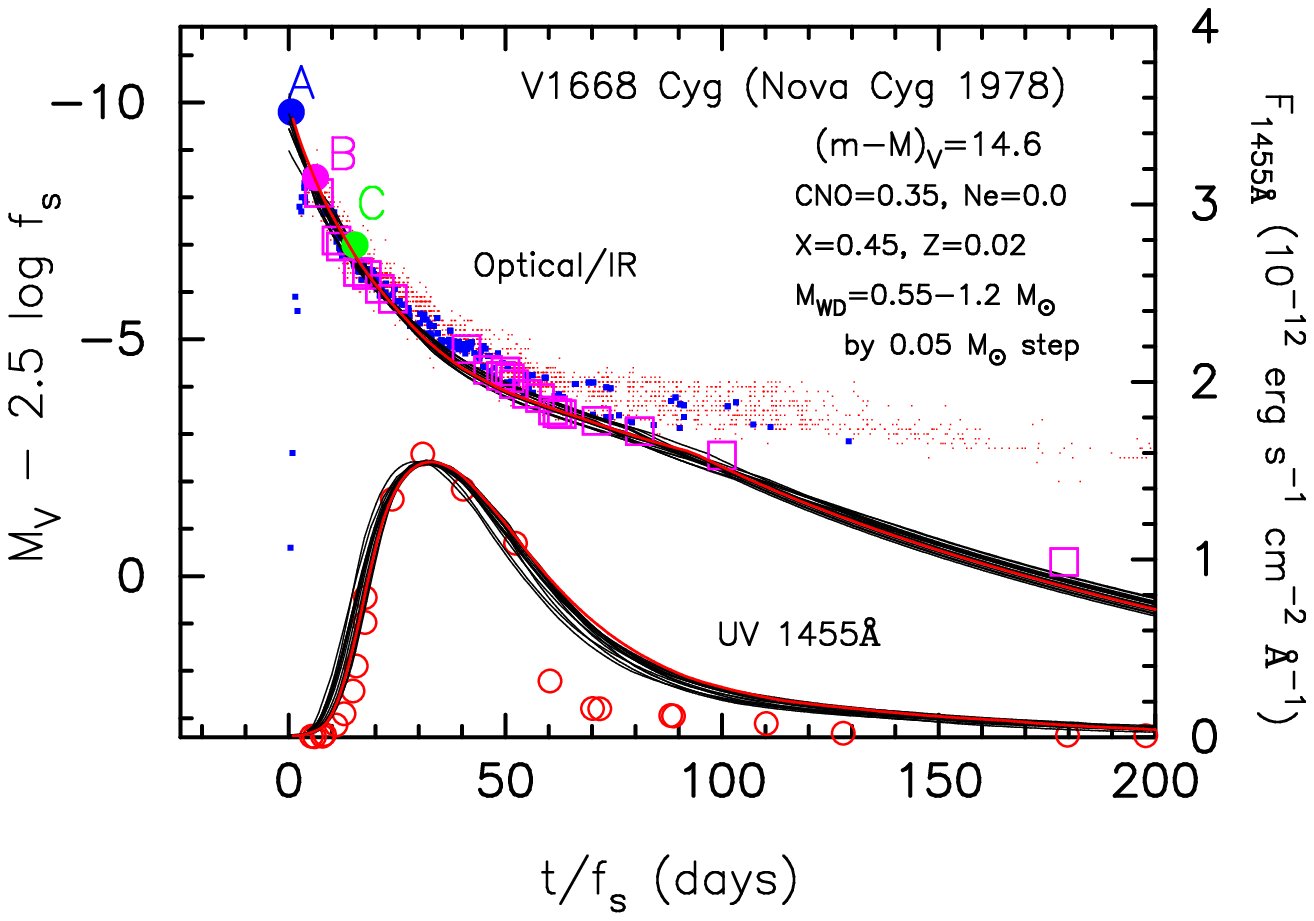}
\caption{
Same model light curves as in Figure 
\ref{light_curve_combine_x45z02c15o20_absmag},
but in the $(t/f_{\rm s})$-$(M_V-2.5\log f_{\rm s})$ coordinates.
We add the UV 1455~\AA\ model light curves.
The optical and UV~1455~\AA\ data of V1668~Cyg are the same as those in
Figure 3 of \citet{hac16k}.
The UV~1455~\AA\ model fluxes are normalized to
the peak value of the UV observation.
We measure the timescaling factor $f_{\rm s}$ of each model
with respect to the V1668~Cyg light curves.
The values of $\log f_{\rm s}$ are tabulated in Table 3 of \citet{hac16k}.
All the free-free emission model light curves
overlap with each other in the $M_V-2.5\log f_{\rm s}$ ordinate
and the $t/f_{\rm s}$ abscissa.
We add a $0.98~M_\sun$~WD model (solid red lines), which simultaneously
reproduce the $V$ (filled blue squares), $y$ (large open magenta squares),
and UV~1455~\AA\ (large open red circles) light curves of V1668~Cyg. 
Three different initial envelope masses of
$M_{\rm env,0} = 2.0 \times 10^{-5}  M_\sun$,
$1.4 \times 10^{-5}  M_\sun$, and
$0.93 \times 10^{-5} M_\sun$
correspond to the three peak brightnesses, i.e., points A, B, and C,
respectively, on the $0.98 ~M_\sun$ WD model with the chemical
composition of CO nova 3 \citep{hac16k}. The peak brightness
of V1668~Cyg, $m_V=6.2~(M_V=6.2 - 14.6=-8.4)$, corresponds to point B. 
}
\label{all_mass_v1668_cyg_x45z02c15o20_calib_universal_linear_no2}
\end{figure}

\subsection{Free-free emission model light curves of novae}
\label{free-free_model}
After hydrogen ignites on a WD, the hydrogen-rich envelope on the WD
expands to giant size and emits optically thick winds 
\citep[e.g.,][]{kat94h, kat17sh}. The optical magnitude attains 
the peak at the maximum expansion of the photosphere (time $t_0$).  
The WD envelope settles in a steady state.
The photospheric radius $R_{\rm ph}$
begins to decrease, whereas the total photospheric luminosity 
$L_{\rm ph}$ is almost constant during the outburst.  
Thus, the photospheric temperature $T_{\rm ph}$ increases with time
because $L_{\rm ph}=4 \pi R_{\rm ph}^2 \sigma T_{\rm ph}^4$.
The main emitting wavelength region moves from optical to UV and
finally, to supersoft X-ray, corresponding to from
$T_{\rm ph} \sim 10^4$~K through $\sim 10^6$~K.
The optically thick winds blow continuously from the very beginning
of the outburst until the photospheric temperature increases to
$\log T_{\rm ph}$~(K)$\sim 5.4$. Just after the optically thick wind
stops (time $t_{\rm wind}$),
the temperature quickly rises and the supersoft X-ray phase
starts \citep[e.g.,][]{kat94h}.

Nova spectra are dominated by free-free emission after optical 
maximum \citep[e.g.,][]{gal76, enn77}. This free-free emission comes
from optically thin plasma outside the photosphere.
\citet{hac06kb} calculated free-free emission model light curves of
novae and showed that theoretical light curves can reproduce observed
NIR/optical light curves of several classical novae from near the peak
to the nebular phase. These free-free emission model light curves are
calculated from the nova evolution models based on the optically thick
wind theory \citep{kat94h}. Their numerical models 
provide the photospheric temperature ($T_{\rm ph}$), radius ($R_{\rm ph}$),
velocity ($v_{\rm ph}$), and wind mass-loss rate ($\dot M_{\rm wind}$)
of a nova hydrogen-rich envelope (mass of $M_{\rm env}$) 
for a specified WD mass ($M_{\rm WD}$) and chemical composition of 
the hydrogen-rich envelope. The free-free emission model light curves
are calculated by Equations (9) and (10) in
\citet{hac06kb} together with these values.

Figure \ref{all_mass_v1668_cyg_x45z02c15o20_real_scale_only098_no2}
shows such an example of the model light curves for the fast nova 
V1668~Cyg \citep[see, e.g.,][]{hac16k}.
Here, we adopt a $0.98~M_\sun$ WD model and the 
chemical composition of CO nova 3
\citep[$X=0.45$, $Y=0.18$, $Z=0.02$, $X_{\rm CNO}=0.35$, $X_{\rm Ne}=0.0$,
see][for the definition of the chemical composition of the hydrogen-rich
envelope]{hac06kb, hac15k, hac16k}.  The green, blue, and red
solid lines show the total (labeled ``TOTAL''), free-free
(labeled ``FF''), and blackbody (labeled ``BB'') $V$ fluxes, respectively.
The blackbody $V$ flux is calculated from $T_{\rm ph}$ and $R_{\rm ph}$,
assuming blackbody emission at the photosphere.
The total flux is the summation of the free-free and blackbody fluxes.
An optically thin dust shell formed $\sim60$ days after the outburst
\citep[e.g.,][]{geh80}. Our model light curves do not include the
effect of dust shell formation.  
Optically thick winds and hydrogen-shell burning end approximately
250 days and 625 days after the outburst, respectively,
for the $0.98~M_\sun$ WD model.  

We should note that strong emission lines such as [\ion{O}{3}]
significantly contribute to the $V$ and visual magnitudes in the
nebular phase. V1668~Cyg entered the nebular phase at $m_V\approx10.4$,
approximately 80 days after the outburst \citep[e.g.,][]{kla80}.
Because our model light curves (free-free plus blackbody) do not
include emission lines, they 
begin to deviate from the observed $V$ and visual
magnitude as shown in Figure 
\ref{all_mass_v1668_cyg_x45z02c15o20_real_scale_only098_no2}.
The intermediate $y$ band (magenta open squares),
which is designed to avoid such strong emission lines,
reasonably follows our model light curve that represents
the continuum spectrum.

This figure also shows the UV~1455~\AA\ 
band fluxes of V1668~Cyg (large open red circles).
The UV~1455~\AA\ band is an emission-line-free narrow band (20~\AA\ width 
centered at 1455~\AA), invented by \citet{cas02} based on the {\it IUE}
spectra of novae. This band represents well the continuum flux at UV
and is useful for light curve fitting. The model flux of UV~1455~\AA\ 
(solid black line) is calculated from $T_{\rm ph}$ and $R_{\rm ph}$,
assuming blackbody emission at the photosphere.
Both the UV~1455~\AA\ and $V$ fluxes simultaneously 
reproduce the observation. 

\citet{hac10k, hac14k, hac15k, hac16k} calibrated the absolute $V$ magnitudes
of these free-free emission model light curves with various novae
whose distance and reddening are well determined.
They obtained a large set of model light curves with 
absolute magnitudes for various WD masses and chemical compositions.
Figure \ref{light_curve_combine_x45z02c15o20_absmag} shows such free-free
model light curves of novae for the chemical composition of CO nova 3.

The decay timescale of a nova light curve depends strongly on the WD mass
and weakly on the chemical composition of hydrogen-rich envelope
\citep{hac06kb}. As the main parameter is the WD mass, model light
curve fitting can constrain the WD mass of a nova. Many nova
WD masses were estimated with this fitting method
\citep[e.g.,][]{hac06kb, hac07k, hac09ka, hac06b, hac07kl, hac08kc}.

\subsection{Universal decline law of nova light curves}
\label{universal_decline_law}
\begin{figure}
\plotone{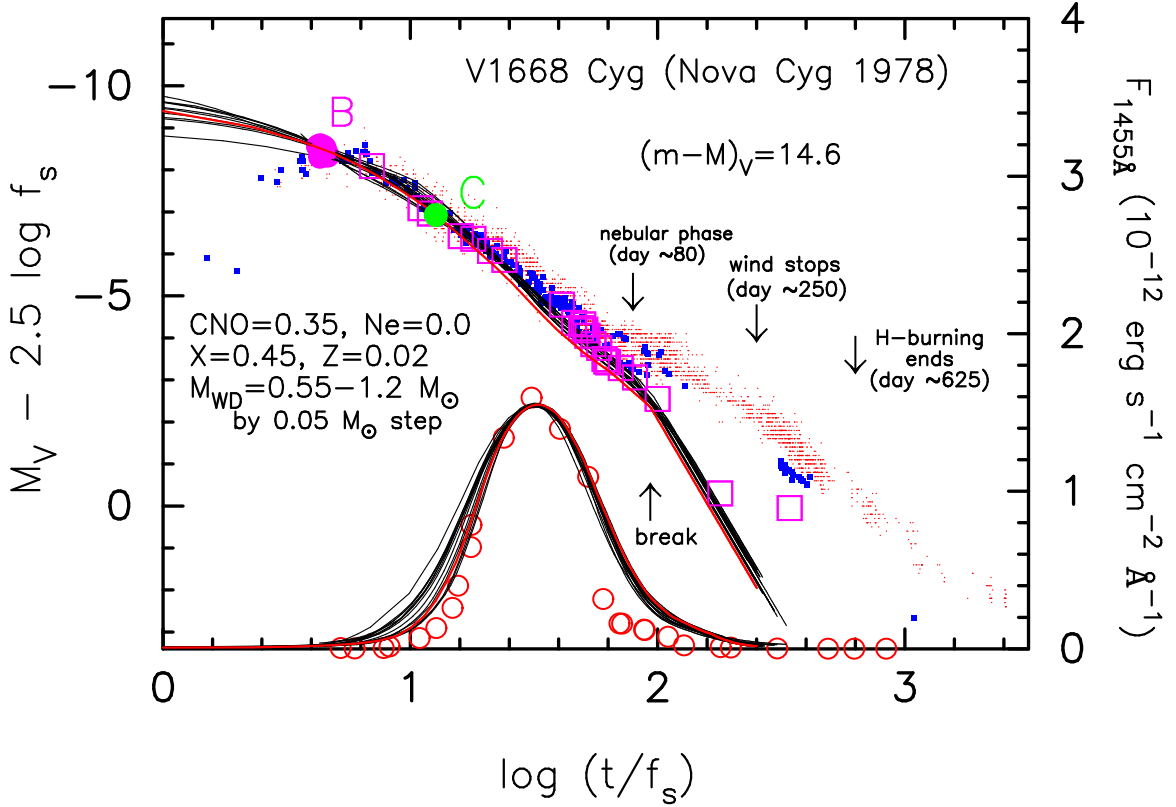}
\caption{
Same as Figure 
\ref{all_mass_v1668_cyg_x45z02c15o20_calib_universal_linear_no2},
but on a logarithmic timescale.
The right edges of free-free emission model light curves
correspond to the epoch when the optically thick winds stop.
Other characteristic epochs are indicated for V1668~Cyg 
and the $0.98~M_\sun$ WD model.
The model light curves show a bend $\sim100$ days after the outburst,
which is labeled ``break.''  This corresponds roughly to the start of
the nebular phase of V1668~Cyg. 
}
\label{all_mass_v1668_cyg_x45z02c15o20_calib_universal_no2}
\end{figure}

\begin{figure}
\plotone{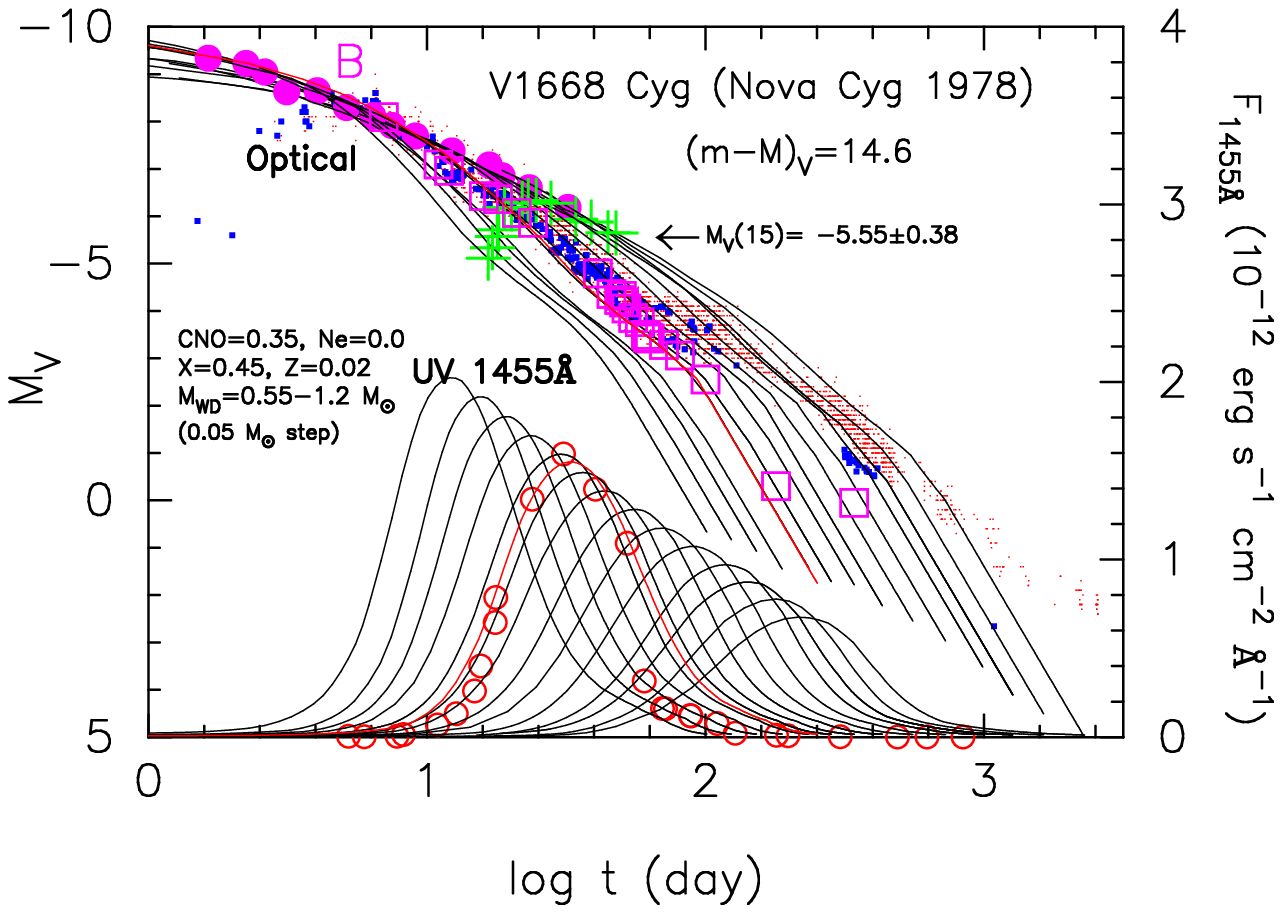}
\caption{
Same as Figure \ref{all_mass_v1668_cyg_x45z02c15o20_calib_universal_no2},
but in the absolute $V$ magnitude vs. logarithmic timescale plane.
The position at point B in Figures
\ref{all_mass_v1668_cyg_x45z02c15o20_calib_universal_linear_no2} and
\ref{all_mass_v1668_cyg_x45z02c15o20_calib_universal_no2}
is indicated by the filled magenta circle on each $V$ light curve.
We also show the absolute magnitude 15 days after the optical maximum,
$M_V(15)$, by the green crosses.
The solid red lines indicate the $0.98~M_\sun$ WD model for V1668~Cyg.
}
\label{all_mass_v1668_cyg_x45z02c15o20_real_scale_universal_no2}
\end{figure}

\citet{hac06kb} found one more important property in nova optical and
NIR light curves; when free-free emission dominates the spectrum, 
there is a universal decline law. 
Figure \ref{all_mass_v1668_cyg_x45z02c15o20_calib_universal_linear_no2}
demonstrates that the fourteen nova light curves (solid black lines) 
in Figure \ref{light_curve_combine_x45z02c15o20_absmag} overlap with each other
if we stretch/squeeze the timescale by a factor of $f_{\rm s}$,
and shift up/down the absolute $V$ magnitude by $-2.5 \log f_{\rm s}$.
\citet{hac06kb} called this property the universal decline law of novae.
They regarded V1668~Cyg as a standard nova and measure the $f_{\rm s}$
of fourteen theoretical light curves against the V1668~Cyg optical/UV 
light curves. They found that a $0.98~M_\sun$ WD model can best 
reproduce the observed light curves of V1668~Cyg, as shown in 
Figure \ref{all_mass_v1668_cyg_x45z02c15o20_real_scale_only098_no2}.  
The free-free fluxes of $0.98~M_\sun$ WD model 
are also shown by the solid red lines in Figures
\ref{all_mass_v1668_cyg_x45z02c15o20_calib_universal_linear_no2},
\ref{all_mass_v1668_cyg_x45z02c15o20_calib_universal_no2},
and \ref{all_mass_v1668_cyg_x45z02c15o20_real_scale_universal_no2},
as a representative of the universal decline law 
among the various WD mass models, i.e., $f_{\rm s}=1.0$ 
both for V1668~Cyg and the $0.98~M_\sun$ WD model.
The $f_{\rm s}$ of each WD mass model is tabulated in Table 3 
of \citet{hac16k} for the chemical composition of CO nova 3.  
Note that the relation between V745~Sco and V1668~Cyg is
$f_{\rm s, V1668~Cyg}/f_{\rm s, V745~Sco}=21$ from
Equation (\ref{distance_modulus_ci_aql_lv_vul}) and Tables
\ref{physical_properties_recurrent_novae} and
\ref{wd_mass_recurrent_novae}.

Figure \ref{all_mass_v1668_cyg_x45z02c15o20_calib_universal_linear_no2} 
also shows the UV~1455~\AA\ light curves 
as well as the observational data of V1668~Cyg \citep{cas02}.
\citet{hac06kb, hac10k, hac15k, hac16k} calculated UV~1455~\AA\  
model light curves for various WD masses 
and chemical compositions of hydrogen-rich envelope,
assuming blackbody emission at the photosphere.
With the same factor of $f_{\rm s}$ as those of optical/NIR light
curves, the model UV~1455~\AA\ light curves are squeezed/stretched
and converged well, as shown in the figure. Here, each
UV~1455~\AA\ peak is normalized to fit with 
the observed peak of V1668~Cyg. 
With simultaneous fit to the $V$ and UV~1455~\AA\ data,
we can specify the WD mass much more precisely. For example,
\citet{hac16k} obtained the WD mass with the accuracy of $\pm0.01~M_\sun$
($M_{\rm WD}=0.98\pm0.01~M_\sun$) for V1668~Cyg.


\subsection{Time-stretching method of nova light curves}
\label{time-stretching_method}
\citet{hac10k} proposed a time-stretching method of nova light curves
based on the universal decline law, and determined the timescaling
factor of $f_{\rm s}$ and distance modulus in the $V$ band,
$\mu_V\equiv (m-M)_V$, for various novae \citep[see also Appendix 
of][]{hac16k}. In this subsection, 
we briefly explain how the time-stretching method works.  

Figure \ref{all_mass_v1668_cyg_x45z02c15o20_calib_universal_no2}
shows the same free-free emission model light curves of Figure
\ref{all_mass_v1668_cyg_x45z02c15o20_calib_universal_linear_no2},
but on a logarithmic timescale, because we manipulate 
the nova light curves much more easily on logarithmic timescales.
Figure \ref{all_mass_v1668_cyg_x45z02c15o20_real_scale_universal_no2}
shows the same model light curves in Figure
\ref{light_curve_combine_x45z02c15o20_absmag} but without time-normalization,
i.e., $M_V$ vs. $\log t$. If we stretch/squeeze these light curves
by $f_{\rm s}$ and shift in the vertical direction by 
$-2.5 \log f_{\rm s}$, we obtain Figure
\ref{all_mass_v1668_cyg_x45z02c15o20_calib_universal_no2}. 

If we pick two light curves in Figure
\ref{all_mass_v1668_cyg_x45z02c15o20_real_scale_universal_no2}
and overlap them by shifting one of the two, horizontally back and forth
and vertically up and down, as in Figure
\ref{all_mass_v1668_cyg_x45z02c15o20_calib_universal_no2}, 
we can measure the timescaling factor of $f_{\rm s}$
from the time-shift of $\log f_{\rm s}= \Delta \log t$ and
the difference in the absolute $V$ magnitude, $\Delta M_V$, from
$\Delta M_V=-2.5 \log f_{\rm s}$ with respect to the other.
The stretching factor $f_{\rm s}$ is obtained for a selected nova 
(target nova) against the other (template nova).
If the absolute magnitude of the template nova is known, the absolute
magnitude of the target nova is calculated from
\begin{eqnarray}
\left( M_V[t] \right)_{\rm template} 
&=& \left(M'_V[t']\right)_{\rm target} \cr 
&=& \left( M_V[t/f_{\rm s}]\right)_{\rm target} - 2.5 \log f_{\rm s}, 
\label{time-stretching_absolute}
\end{eqnarray}
where the brightness $(t,~M_V)$ of the target nova  
is converted to the brightness ($t', ~M'_V$) by time-stretch
$t'=t/f_{\rm s}$ and $M'_V[t']=M_V[t/f_{\rm s}] - 2.5 \log f_{\rm s}$.
Here, we note $M_V[t]$ to demonstrate that $M_V$ is a function of $t$.
This relation is easily understood from
Figures \ref{all_mass_v1668_cyg_x45z02c15o20_calib_universal_no2}
and \ref{all_mass_v1668_cyg_x45z02c15o20_real_scale_universal_no2}.
This equation is equivalent to
\begin{eqnarray}
\left( M_V[t\times f_{\rm s}] \right)_{\rm template} 
&=& \left(M'_V[t]\right)_{\rm target} \cr 
&=& \left( M_V[t]\right)_{\rm target} - 2.5 \log f_{\rm s}. 
\label{time-stretching_absolute_equivalent}
\end{eqnarray}

Using the inverse time-stretch $t'=t\times f_{\rm s}$ and 
$M'_V[t']=M_V[t\times f_{\rm s}] + 2.5 \log f_{\rm s}$ for the template, 
we rewrite Equation (\ref{time-stretching_absolute}) as
\begin{eqnarray}
\left( M_V[t] \right)_{\rm target} 
&=& \left(M'_V[t']\right)_{\rm template} \cr 
&=& \left( M_V[t\times f_{\rm s}]\right)_{\rm template} + 2.5 \log f_{\rm s}. 
\label{time-stretching_absolute_inverse}
\end{eqnarray}

In general, we obtain nova light curves in the apparent magnitude.
The difference between the apparent magnitudes of the two novae is
obtained as
\begin{equation}
(m_V[t])_{\rm target} = (m_V[t\times f_{\rm s}])_{\rm template}
+ (\Delta V)_{\rm template},
\label{differnce-apparent_magnitude}
\end{equation}
where we squeezed the apparent $V$ magnitude light curve of
the template nova horizontally (in the time direction) 
by a factor of $f_{\rm s}$ and shifted vertically  
by $(\Delta V)_{\rm template}$ to overlap with the target nova.
Subtracting Equation (\ref{time-stretching_absolute_inverse}) from
Equation (\ref{differnce-apparent_magnitude}),
we obtain the distance modulus of the target nova as
\begin{equation}
\left( m[t]-M[t] \right)_{V,\rm target} = 
\left( (m[t\times f_{\rm s}]-M[t\times f_{\rm s}] )_V 
+ \Delta V \right)_{\rm template}  - 2.5 \log f_{\rm s}.
\label{distance_modulus_general_temp}
\end{equation}
Because $(m[t]-M[t])_V$ is a constant, we simply write
Equation (\ref{distance_modulus_general_temp}) as
\begin{equation}
(m-M)_{V,\rm target} = \left( (m-M)_V 
+ \Delta V\right)_{\rm template} - 2.5 \log f_{\rm s},
\label{distance_modulus_formula_no2}
\end{equation}
where $(m-M)_{V, \rm template}$ is the distance modulus of the template
nova, which is known.  
Equation (\ref{distance_modulus_formula_no2}) is the same
equation as Equation (\ref{distance_modulus_formula}).
Thus, we can obtain the distance modulus of a target nova
by comparing it with the template nova.
\citet{hac10k} called this method the time-stretching method.

The above mathematical relations are derived from the universal 
decline law, which is based on the free-free emission model
light curves calculated from the optically thick wind model
\citep[see Appendices of][for more detail]{hac10k, hac16k}.
Note that individual nova light curves more or less
deviate from the universal decline law for various reasons
but these relations approximately represent the overall trends 
between nova light curves.  


\begin{figure}
\plotone{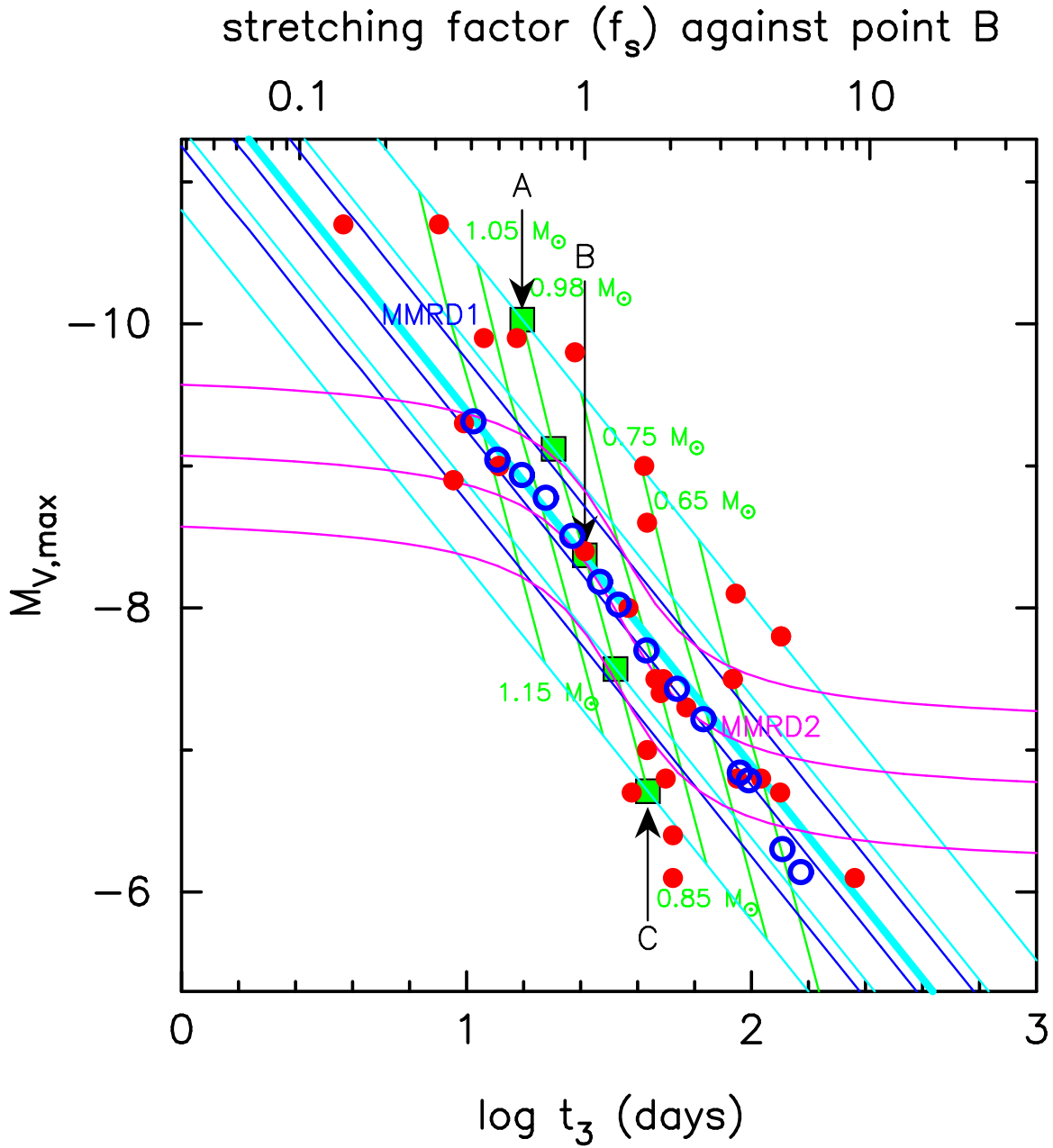}
\caption{
The maximum magnitude vs. rate of decline (MMRD) relation.
The ordinate is the absolute $V$ magnitude and the abscissa is the
$t_3$ time.
The open blue circles are the MMRD points calculated from
our various WD mass models in Figure  
\ref{all_mass_v1668_cyg_x45z02c15o20_real_scale_universal_no2}.
Point B represents the peak brightness of V1668~Cyg
for a nova model on a $0.98 M_\sun$ WD with the chemical
composition of CO nova 3 \citep{hac16k}.
We suppose that B corresponds to a typical initial envelope mass for
a nova of a given WD mass.
The thick cyan line passing through point B
corresponds to a MMRD relation of novae with typical initial
envelope mass but different WD mass.
The cyan line passing through point A is a MMRD relation
for a much larger initial envelope mass than that of point B 
(much lower mass accretion rates).
The cyan line passing through point C is a MMRD relation
for a much smaller initial envelope mass than that of
point B (much higher mass accretion rates).  
The green solid lines connect the same WD mass, i.e., 1.15, 1.05, 0.98,
0.85, 0.75, and 0.65 $M_\sun$.
The filled red circles are MMRD points
for individual galactic novae taken from Table 5 of \citet{dow00}.
The cyan lines envelop these MMRD points studied by \citet{dow00}. 
Two well-known MMRD relations are added:
Kaler-Schmidt's law \citep{sch57} 
indicated by blue solid lines labeled ``MMRD1,''
and Della Valle \& Livio's (1995) law 
indicated by magenta solid lines labeled ``MMRD2.'' 
We add upper/lower bounds of $\pm 0.5$ mag for each of these
two MMRD relations.
}
\label{max_t3_scale_no4}
\end{figure}

\section{Theoretical maximum magnitude vs. rate of decline relations}
\label{mmrd}
The Maximum Magnitude vs. Rate of Decline (MMRD) relation, i.e.,
the relation between $M_{V, {\rm max}}$ and ($t_2$ or) $t_3$,
is often used to estimate the distance to novae 
\citep[e.g.,][]{coh88, del95, dev78, dow00, sch57}.   
The time-stretching method is fundamentally different 
from the MMRD methods, mainly because the time-stretching method does
not require the peak brightness (maximum magnitude). In this subsection, 
we present a theoretical explanation of the MMRD relations not only
on the main trend but also on the large scatter of individual novae
around the main trend, based on the time-stretching method.
The essential part of the theoretical explanation is 
given in Figure 15 of \citet{hac10k}, Figure 36 of \citet{hac15k},
and Figures 47 and 48 of \citet{hac16k}, but here we give a systematic
description on the main trend of MMRD relations and scatter around
the main trend.

The main trend of MMRD relations can be explained as follows.
The model light curves of free-free emission decline along
the decreasing envelope mass, $M_{\rm env}$,
and thus, the peak brightness depends on the initial envelope mass,
$M_{\rm env,0}$, at maximum. In other words, by specifying
the WD mass and initial envelope mass, we can determine the
peak brightness on the light curve. For example, the peak 
brightness of V1668~Cyg corresponds to point B in Figures
\ref{all_mass_v1668_cyg_x45z02c15o20_calib_universal_linear_no2}
and \ref{all_mass_v1668_cyg_x45z02c15o20_calib_universal_no2}.
If we assume that V1668~Cyg has a typical peak brightness, which is
expressed by a point (point B) in
the $\log(t/f_{\rm s})$-$(M_V-2.5\log f_{\rm s})$ plane
(Figure \ref{all_mass_v1668_cyg_x45z02c15o20_calib_universal_no2}),
other speed classes (other WD masses) 
of novae have the typical peak brightness $M_V$ 
at point B (filled magenta circles) 
in the real timescale $\log t$ (Figure
\ref{all_mass_v1668_cyg_x45z02c15o20_real_scale_universal_no2}).
This means that a faster declining nova (smaller $t_2$ or $t_3$ time)
has a brighter maximum magnitude (smaller $M_{V, {\rm max}}$) and
a slower declining nova has a fainter peak.

We read the value of $M_{V, \rm max}$ at point B 
(filled magenta circle) on each model light curve (different WD mass)
and obtain its $t_3$ time in Figure
\ref{all_mass_v1668_cyg_x45z02c15o20_real_scale_universal_no2}.
Then, we plot the positions of $(t_3,~M_{V, \rm max})$
by the open blue circles in Figure \ref{max_t3_scale_no4}, which are
approximately located on the thick straight cyan line.   
This is the main trend of the MMRD relation.  
In short, the main trend of the MMRD relation is explained by the
difference in the WD mass.

Now, we derive a simple theoretical formula for the MMRD relation
based on the universal decline law.  Here, we take the absolute $V$
magnitude light curve of the $0.98~M_\sun$ WD model (solid red lines
in Figures
\ref{all_mass_v1668_cyg_x45z02c15o20_calib_universal_linear_no2},
\ref{all_mass_v1668_cyg_x45z02c15o20_calib_universal_no2}, and
\ref{all_mass_v1668_cyg_x45z02c15o20_real_scale_universal_no2})
as the representative of the universal decline law. 
The apparent brightness, $m_{V, {\rm max}}$, at point B 
of each light curve with different WD masses (different $f_{\rm s}$)
is expressed as $m_{V,{\rm max}} = m'_{V,{\rm max}} + 2.5 \log f_{\rm s}$ 
and $t_3 = t'_3 \times f_{\rm s}$ by time-stretch.
Eliminating $f_{\rm s}$ from these two relations, we have
\begin{equation}
m_{V,{\rm max}} = 2.5 \log t_3 + m'_{V,{\rm max}} - 2.5 \log t'_3.
\label{theoretical_apparent_MMRD_relation}
\end{equation}
Because point B corresponds to the optical peak of V1668 Cyg
($f_{\rm s}=1.0$ in Figure
\ref{all_mass_v1668_cyg_x45z02c15o20_calib_universal_linear_no2}),
we obtain $t'_3 = 26$ days and $m'_{V,{\rm max}} = 6.2$ from the
V1668~Cyg light curve.
Substituting $t'_3 = 26$~days and $m'_{V,{\rm max}}=6.2$
into Equation (\ref{theoretical_apparent_MMRD_relation}), we have
\begin{equation}
m_{V,{\rm max}} = 2.5 \log t_3 + 6.2 - 2.5 \log 26
= 2.5 \log t_3 + 2.7.
\end{equation}
This is our theoretical apparent MMRD relation for a maximum point B
in Figure \ref{all_mass_v1668_cyg_x45z02c15o20_calib_universal_linear_no2}.
Using the distance modulus of $(m-M)_V$, 
we obtain our theoretical MMRD relation as
\begin{eqnarray}
M_{V,{\rm max}} & = & m_{V,{\rm max}}  - (m-M)_V \cr
 & = & 2.5 \log t_3 + m'_{V,{\rm max}} - 2.5 \log t'_3 - (m-M)_V.
\label{theoretical_MMRD_relation}
\end{eqnarray}
Substituting $(m-M)_V = 14.6$ for V1668~Cyg
into Equation (\ref{theoretical_MMRD_relation}), we have
\begin{eqnarray}
M_{V,{\rm max}} &=& 2.5 \log t_3 + 6.2 - 2.5 \log 26 - 14.6 \cr
&=& 2.5 \log t_3 -11.9.
\label{theoretical_MMRD_relation_v1668_cyg}
\end{eqnarray}
Figure \ref{max_t3_scale_no4} shows this theoretical MMRD relation
(passing through point B) for novae by the thick solid cyan line.
We show the value of $f_{\rm s}$ on the upper horizontal axis of 
Figure \ref{max_t3_scale_no4}, which corresponds to the $t_3$ time 
on the lower horizontal axis.

Next, we explain the reason why individual novae deviate from the main
trend of MMRD relation, Equation 
(\ref{theoretical_MMRD_relation_v1668_cyg}).  
If the initial envelope mass, $M_{\rm env,0}$, is larger
or smaller than that at point B, we obtain different MMRD relations.
\citet{hac10k} showed that the peak magnitude is brighter for a
larger initial envelope mass, even for the same WD mass. 
Suppose that a nova follows a similar light curve to V1668~Cyg but
reaches a much brighter peak, e.g., point A in Figure
\ref{all_mass_v1668_cyg_x45z02c15o20_calib_universal_linear_no2}.
Figure \ref{all_mass_v1668_cyg_x45z02c15o20_calib_universal_linear_no2}
shows that $t_3$ time is smaller for a brighter peak (point A) 
compared with $t_3$ for point B.
(See also Figure 2 of \citet{hac10k} for the same trend of $t_3$ time).
This corresponds to a different initial envelope mass for the same
$0.98~M_\sun$ WD and chemical composition.

The solid cyan lines passing through points A, B, and C 
in Figure \ref{max_t3_scale_no4} indicates
the degree of deviation from point B, originating from the difference
in the peak brightness (i.e., the difference in the initial envelope
mass). In other words, novae on the same WD mass show large
scatter from the main trend of the thick solid cyan line 
in both sides of brighter/fainter along the solid green lines of  
the 1.15, 1.05, 0.98, 0.85, 0.75, and $0.65~M_\sun$ WD models.

Now we obtain the region that covers the scatter in  
Figure \ref{max_t3_scale_no4}. Point A $(t'_3 = 16$ days, 
$m'_{V,{\rm max}}=4.6)$ in Figure 
\ref{all_mass_v1668_cyg_x45z02c15o20_calib_universal_linear_no2}
corresponds to a larger envelope mass than point B.
The solid cyan line passing through point A is 
\begin{eqnarray}
M_{V,{\rm max}} &=& 2.5 \log t_3 + 4.6 - 2.5 \log 16 - 14.6 \cr
&=& 2.5 \log t_3 -13.0,
\label{theoretical_MMRD_relation_v1668_cyg_upper}
\end{eqnarray}
from Equation (\ref{theoretical_MMRD_relation}).
The solid cyan line passing through point C
$(t'_3 = 43$ days, $m'_{V,{\rm max}}=7.9)$ has a different
$t_3$--$M_{V,{\rm max}}$ relation
for a much smaller initial envelope mass (at point C in Figure
\ref{all_mass_v1668_cyg_x45z02c15o20_calib_universal_linear_no2}), i.e.,
\begin{eqnarray}
M_{V,{\rm max}} &=& 2.5 \log t_3 + 7.9 - 2.5 \log 43 - 14.6 \cr
&=& 2.5 \log t_3 -10.8,
\label{theoretical_MMRD_relation_v1668_cyg_lower}
\end{eqnarray}
from Equation (\ref{theoretical_MMRD_relation}).
In our theoretical understanding of novae, the smaller the mass accretion rate, the larger the ignition mass. Thus, the envelope mass
at the peak, $M_{\rm env,0}$, is larger 
\citep[see Figure 3 of][for the ignition mass vs. mass accretion 
rate]{kat14shn}.   

Our theoretical MMRD relation of Equation 
(\ref{theoretical_MMRD_relation_v1668_cyg})
is in good agreement with two well-known,
empirically obtained MMRD relations, Kaler-Schmidt's law (MMRD1)
in Equation (\ref{kaler-schmidt-law}) and Della Valle \& Livio's law
(MMRD2) in Equation (\ref{della-valle-livio-law}).
We plot these MMRD relations in Figure \ref{max_t3_scale_no4}.
Kaler-Schmidt's law is denoted by the straight blue line with two attendant
blue lines, corresponding to $\pm 0.5$ mag brighter/fainter cases.
Della Valle \& Livio's law is indicated by the magenta solid lines, 
also with two attendant magenta lines of $\pm 0.5$ mag.  
These two well-known empirical MMRD relations are very close to 
our theoretical MMRD relation.

In Figure \ref{max_t3_scale_no4},
we also add observational MMRD points (filled red circles) for individual
galactic novae, the data of which are taken from Table 5 of \citet{dow00}.
It is clearly shown that the large scatter of individual points 
from the two empirical MMRD relations (blue and magenta lines)
falls into between upper/lower cases of our MMRD relations 
for the largest (point A) and smallest (point C) initial envelope mass
of $M_{\rm env,0}$. This simply means that there is a second parameter to
specify the MMRD points for individual novae.  
As explained above, the main parameter is 
the WD mass, represented by the timescaling factor $f_{\rm s}$. 
The second parameter is the initial envelope mass (or the mass accretion
rate onto the WD). This second parameter can reasonably explain the scatter
of individual novae from all the proposed empirical MMRD relations.
















































\end{document}